\documentclass{aa}

\usepackage{natbib}
\bibpunct{(}{)}{;}{a}{}{,}

\usepackage{graphicx}
\usepackage{caption}
\usepackage{longtable}

\usepackage{txfonts}
\usepackage{lscape}

\newcommand{\kms}{\,km s$^{-1}$}

\newcommand{\mysou}{G328.2551-0.5321}

\newcommand{\unidens}{cm$^{-2}$}
\newcommand{\methanol}{CH$_3$OH}
\newcommand{\dimetheth}{CH$_3$OCH$_3$}

 \def\as     {\ifmmode {\rlap.}$\,$''$\,$\! \else ${\rlap.}$\,$''$\,$\!$\fi}
\def\decsec  {\ifmmode {\rlap.}$\,$^{\rm s}$\,$\! \else ${\rlap.}$\,$^{\rm s}$\,$\!$\fi}\def\decs  {\ifmmode {\rlap.}$\,$^{\rm s}$\,$\! \else ${\rlap.}$\,$^{\rm s}$\,$\!$\fi}

\usepackage{amstext}

\begin{document}

\title{Sulphur-rich cold gas around the hot core precursor \mysou}
\subtitle{An APEX unbiased spectral survey of the 2\,mm, 1.2\,mm, and 0.8\,mm atmospheric windows}
 \author{ {L. Bouscasse}\inst{1}
 \and{T. Csengeri}\inst{2}
 \and{A. Belloche}\inst{1}
 \and{F. Wyrowski}\inst{1}
 \and{S. Bontemps}\inst{2}
 \and{R. Güsten} \inst{1}
 \and{K. M. Menten}\inst{1}
}
 \institute{ {Max-Planck-Institut f\"ur Radiastronomie, Auf dem Hügel 69, 53121 Bonn, Germany} \label{inst1}
\and {Laboratoire d'astrophysique de Bordeaux, Univ. Bordeaux, CNRS, B18N, allée Geoffroy Saint-Hilaire, 33615 Pessac, France} \label{inst2} }
 \date{Received <date> /
       Accepted <date>}
  \abstract
{During the process of star formation, the dense gas undergoes significant chemical evolution leading to the emergence of a rich variety of molecules associated with hot cores and hot corinos. However, the physical conditions and the chemical processes involved in this evolution are poorly constrained; the early phases of emerging hot cores in particular represent an unexplored territory.}
{We provide here a full molecular inventory of a massive protostellar core that is proposed to represent a precursor of a hot core. We investigate the conditions for the molecular richness of hot cores.}
{We performed an unbiased spectral survey towards the hot core precursor associated with clump \mysou\ between 159\,GHz and 374\,GHz, covering the entire atmospheric windows at 2\,mm, 1.2\,mm, and 0.8\,mm.
To identify the spectral lines, we used rotational diagrams and radiative transfer modelling assuming local thermodynamical equilibrium.}
{We detected 39 species plus 26 isotopologues, and were able to distinguish a compact ($\sim$2$^{''}$), warm inner region with a temperature, $T$, of $\sim$100~K, a colder, more extended envelope with $T\sim$20~K, and the kinematic signatures of the accretion shocks that have previously been observed with ALMA. 
We associate most of the emission of the small molecules with the cold component of the envelope, while the molecular emission of the warm gas is enriched by complex organic molecules (COMs). 
We find a high abundance of S-bearing molecules in the cold gas phase, including the molecular ions HCS$^+$ and SO$^+$. The abundance of sulphur-bearing species suggests a low sulphur depletion, with a factor of $\geq$1\,\%, in contrast to low-mass protostars, where the sulphur depletion is found to be stronger. Similarly to other hot cores, the deuterium fractionation of small molecules is low, showing a significant difference compared to low-mass protostars. We find a low isotopic ratio in particular for $^{12}$C/$^{13}$C of $\sim30$, and $^{32}$S/$^{34}$S of $\sim12,$ which are about two times lower than the values expected at the galactocentric distance of \mysou.
We identify nine COMs (CH$_3$OH, CH$_3$OCH$_3$, CH$_3$OCHO, CH$_3$CHO, HC(O)NH$_2$, CH$_3$CN, C$_2$H$_5$CN, C$_2$H$_3$CN, and CH$_3$SH) in the warm component of the envelope, four in the cold gas, and four towards the accretion shocks.  
}
{
The presence of numerous molecular ions and high abundance of sulphur-bearing species originating from the undisturbed gas may suggest a contribution from shocked gas at the outflow cavity walls. The molecular composition of the cold component of the envelope is rich in small molecules, while a high abundance in numerous species of COMs suggests an increasing molecular complexity towards the warmer regions. The molecular composition of the warm gas is similar to that of both hot cores and hot corinos, but the molecular abundances are closer to the values found towards hot corinos than to values found towards hot cores. Considering the compactness of the warm region and its moderate temperature, we suggest that thermal desorption has not been completed towards this object yet, representing an early phase of the emergence of hot cores. 

}

 
\maketitle

\section{Introduction}

The molecular richness of the star-forming gas was first recognised towards sites of high-mass star formation. Hot molecular cores have been identified 
as birthplaces of high-mass stars and clusters where high densities and temperatures prevail. Temperatures above $\sim$100\,K lead to the sublimation of icy dust mantles, which together with the high temperatures give rise to an emergence of molecular complexity \citep[e.g.][]{Kurtz2000, Belloche2016, Jorgensen2020}. The molecular complexity manifests in the detection of O-bearing, N-bearing, and S-bearing deuterated molecules, carbon chains, and complex organic molecules\footnote{Molecules with six or more atoms \citep{Herbst2009}.} (COMs) (for a review, see e.g.\, \citealt{Herbst2009}). Even though hot cores show similarities in their molecular content, they have been found to exhibit a large diversity in terms of relative molecular abundances \citep{Walmsley1993, Bisschop2007, Belloche2013, Widicus2017, Minh2018}.
To reveal the origin of this diversity, the next step is to investigate the molecular composition of recently recognised (high-mass) protostellar envelopes in a young evolutionary stage in great detail, where the early warm-up phase chemistry could reveal the first steps towards a more complex chemical evolution.

Unbiased spectral surveys have been proven to be extremely useful to pinpoint the physical and chemical conditions around (high-mass) protostars. Simultaneously observable large bandwidths grant access to several transitions of a given molecule, allowing us to cover a broad range of upper-level energies to estimate the physical conditions giving rise to the molecular emission. This could potentially reveal the origin of the emitting gas even if the region is unresolved in the beam of single-dish telescopes. The large bandwidth of modern receivers meant that spectral surveys became relatively easy to perform, with a growing number of such programs tracing a range of star formation environments from the quiescent to the extreme regions: for example, the TIMASS survey (The IRAS 16293-2422 Millimeter And Submillimeter Spectral Survey, \citealt{Caux2011}) and ASAI (Astrochemical Surveys At IRAM, \citealt{Lefloch2018}), which targeted mainly low-mass protostars, \citet{Belloche2013}, who targeted the extreme star formation region in Sgr B2, and several others listed in \citet{Herbst2009}. There is a growing number of such studies at high angular resolution as well, such as the PILS survey (The ALMA Protostellar Interferometric Line Survey, \citealt{Jorgensen2016}, SOLIS (Seeds Of Life In Space, \citealt{Ceccarelli2017}), and the EMoCA (Exploring Molecular Complexity with ALMA) and ReMoCA (Re-exploring molecular complexity with ALMA) surveys of Sgr B2(N) \citep{Belloche2016, Belloche2019}.

The high-mass protostar \mysou\ was first identified in the SPARKS (Search for High-mass Protostars with ALMA revealed up to kilo-parsec scales) project \citep{Csengeri2017b,Csengeri2018} based on the ATLASGAL (APEX Telescope Large Area Survey of the Galaxy) survey \citep{Schuller2009, Csengeri2014}. It is located at a distance of 2.5$_{-0.5}^{+1.7}$\,kpc with a systemic velocity of $-43.1$\,\kms \citep{Csengeri2018}. \citet{Csengeri2018} found observational signatures of accretion shocks associated with a putative accretion disk witnessed for the first time towards a high-mass protostar. The physical properties such as the temperature and the molecular characteristics of these accretion shocks were different compared to the rest of the envelope \citep{Csengeri2019}. Towards the inner envelope at $T\sim100$\,K, about ten COMs were identified using the 7.5\,GHz limited bandwidth that is available in a single setup with ALMA. The protostar is expected to be in a stage before the onset of strong ionising radiation \citep[see more details in][]{Csengeri2018}, and the radiatively heated inner region is still compact ($\leq$1000\,au). Because the molecular emission is dominated by the accretion shocks and the cold gas, \citet{Csengeri2019} suggested that this object might be a precursor to a classical, radiatively heated hot core. This object has been found to be isolated from 0.3\,pc down to 400\,au scales with no other compact and bright sub-millimetre source in its vicinity. It is therefore an appropriate target for single-dish observations and is an excellent laboratory for studying the onset of radiative heating in the chemistry of the early warm-up phase.

Here we study the full molecular composition of \mysou\ using the Atacama Pathfinder EXperiment telescope (APEX, \citealt{Gusten2006}). We performed an unbiased spectral survey in the 2mm, 1.2\,mm, and 0.8mm atmospheric bands in order to study the physical conditions and the molecular composition of the dense gas in this hot core precursor. 
The paper is organised as follows: in Sect. \ref{sec:obs} we present the observations and the data reduction. In Sect. \ref{sec:results} we show the results and the molecular composition of the protostellar envelope. In Sect. \ref{sec:source_structure} we describe the envelope structure. In Sect. \ref{sec:light} we discuss the composition of light molecules,
and in Sect. \ref{sec:COMs} we discuss the detection and properties of COMs in the envelope. In Sect. \ref{sec:comparison} we compare this emerging hot core with hot cores from the literature. Finally, in Sect. \ref{sec:conclusion} we present our conclusions.

\section{Observations and data reduction} \label{sec:obs}

We performed an unbiased spectral survey from 159\,GHz to 374\,GHz towards the protostellar object associated with the ATLASGAL clump \mysou\  (R.A. $15$h$57$m$59.791$s and Dec. $-53^\circ58'00.56''$ (J2000); \citealp{Csengeri2018})\footnote{Corresponding to the continuum peak flux position from \citet{Csengeri2018}.} with the APEX telescope on the Chajnantor plateau, in the Chilean Andes.

\begin{figure}[!htpb]
\center
\includegraphics[scale=0.3]{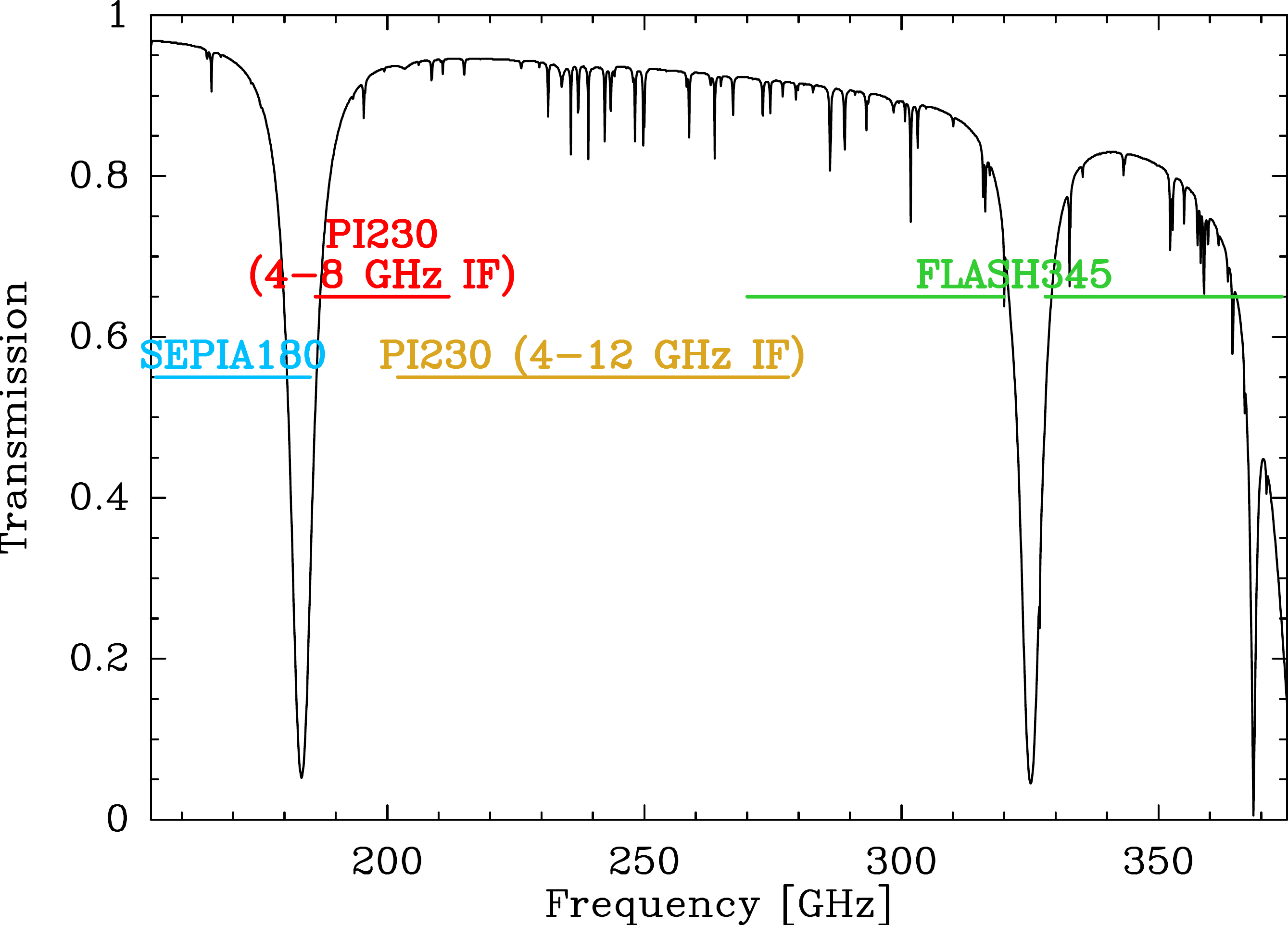}
\caption{Continuous frequency coverage of the spectral survey. The black curve shows the atmospheric transmission as computed by the ATM model written by Juan Pardo \citep{Pardo2001} with a water vapour of 1.2\,mm.}
\label{freq_cov}
\end{figure}

\begin{figure}[!htpb]
\center
\includegraphics[scale=0.3]{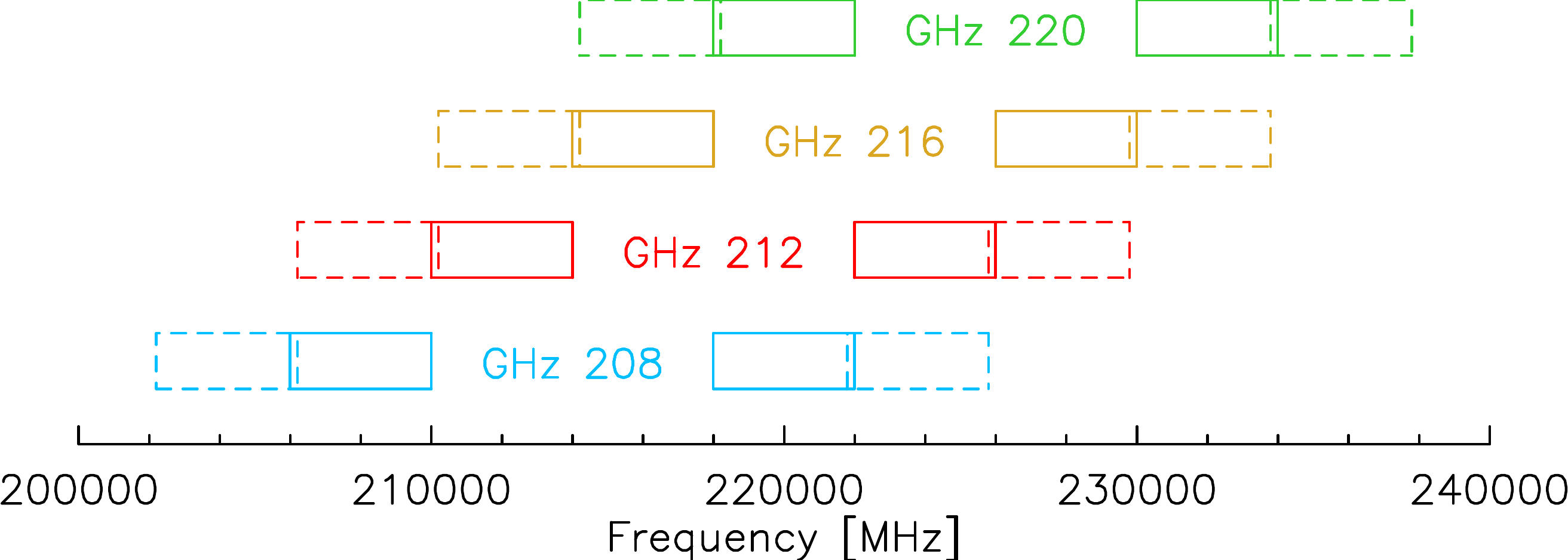}
\caption{Observed spectral windows for the 1\,mm band with the PI230 receiver at APEX. The solid lines represent the inner backend units, and the dashed lines show the outer backend units.}
\label{setup_1mm}
\end{figure}

The observations cover the 0.8\,mm band using the FLASH$^+$ (First Light APEX Submillimeter Heterodyne) receiver \citep{Klein2014}, the 1.2\,mm band with the PI230 receiver, and finally the 2\,mm band with the SEPIA180 (Swedish-ESO PI Instrument for APEX) \citep{Belitsky2018} and PI230 receivers\footnote{Projects M9521A\textunderscore 99, M9512A\textunderscore 102 and M9512B\textunderscore 102.} (Fig.\,\ref{freq_cov}). In addition, we also observed additional setups as a bonus around 408\,GHz and 465\,GHz with the FLASH460 channel in parallel of the FLASH345 observations.
The observations were made in position-switching mode with a reference position offset of 600$''$,$-$600$''$ in right ascension and declination with respect to the position of \mysou. The reference position was chosen to be a nearby position to ensure spectral baseline stability. 
The telescope focus was checked with Mars and Jupiter. The pointing accuracy is $\sim$2", and was checked with regular pointings on IRAS15194-5115, NGC6072 and RAFGL2135. All three receivers are sideband separated with a sideband rejection of 15\,dB for PI230 and FLASH345 (G\"usten priv. comm., \citealt{Klein2014}) and 20\,dB for SEPIA180 \citep{Belitsky2018}. The PI230 receiver was used covering 4--12\,GHz of the intermediate frequency (IF) band between 200\,GHz and 270\,GHz, and 4--8\,GHz between 186\,GHz and 210\,GHz. SEPIA180 and FLASH345 were both used in their original configuration, providing an IF band of 4--8\,GHz.
In order to identify line contamination from the other sideband (ghost lines) and occasionally appearing spikes, we covered the frequency range twice with tunings shifted by 4\,GHz for the PI230 and 2\,GHz for FLASH345, SEPIA180, and PI230 in the 4-8\,GHz configuration, and hence used 12, 24, and 12 setups for the 2\,mm, 1.2\,mm, and 0.8\,mm atmospheric windows, respectively  (Fig. \ref{setup_1mm}). 
The half-power beam width varies from 39\arcsec at 159\,GHz to 16\arcsec at 374\,GHz, corresponding to 0.5\,pc and 0.2\,pc physical scales at the distance of the source, respectively.


The median system temperature is 179\,K over the complete spectral survey, and the noise varies between 11 and 25 mK at 0.7\,\kms\ on the main-beam temperature scale ($T_{\rm mb}$) (Table \ref{parameters_data}), the variation  depending on the receiver temperature and the observing conditions. 
In total we cover a frequency range of 206\,GHz with this spectral survey, which was observed for a total of 7 hours (including the reference position). 

The data were calibrated with the APEX online calibrator. We performed the data reduction with the Aug18 and March19b version of GILDAS\footnote{https://www.iram.fr/IRAMFR/GILDAS/} softwares. For the data reduction, we first averaged each subscan of every backend. Then, we compared the intensities of the lines in each polarisation (for SEPIA180 and PI230), which were then averaged together. For each backend of every setup, we removed a zeroth-order baseline selecting emission free channels. Finally, we compared the two frequency coverages with different tunings and identified about ten ghost lines, which were blanked in the data, while channels with spikes were masked. The FLASH345 data from April 2019 were shifted by 0.25\kms\ to fix a problem related to the receiver. We corrected this shift in the data and then averaged them with the data from September 2019.

We converted the data from antenna temperature ($T_{\rm A}$*) to main-beam temperature ($T_{\rm mb}$) using the telescope efficiencies determined from planet observations that were regularly observed during the project. These efficiencies were then applied to the whole survey using the Ruze formula: $\eta _{\rm mb}=\eta _{\rm 0}\times e^{-(\frac{4 \pi \delta \nu}{c})^2}$, where $\eta _{\rm mb}$ is the main-beam efficiency, $\eta _{\rm 0}$ depends on the receiver, $\delta$ is the surface accuracy, $\nu$ is the frequency, and $c$ is the speed of light. The main-beam efficiencies for some frequencies are given in Table \ref{parameters_data}. A forward efficiency of 95\% was assumed for all receivers. Finally, we smoothed the native velocity resolution to a common resolution of 0.7\kms. The average rms noise level ($\sigma$) over the entire band is 18\,mK. The parameters of the observations are detailed in Table \ref{parameters_data}. 

\begin{table*}[t]
\centering
\caption{Parameters of the observations.}
\label{parameters_data}
\begin{tabular}{c c c c c c c c c c c}     
\hline 
\hline
Band & Receiver & Frequency & Observation & HBPW\tablefootmark{a} &\multicolumn{2}{c}{Spectral resolution}  & <Pwv> & <T$_{\rm sys}$> & <$\sigma$>\tablefootmark{b} & $\eta _{\rm mb}$\tablefootmark{a} \\  
 &  & [GHz] & dates [MM.YY] & [\arcsec] & [kHz] & [\kms] & [mm] & [K] & [mK] & [\%]  \\           
\hline 
2\,mm & SEPIA180 & 159-185 & 09.18 & 36 & 76 &  0.13 & 1.3\tablefootmark{c} & 514\tablefootmark{h} & 11 & 79\\ 
 & PI230 & 186-212 & 09.18 & 31 & 61 & 0.09 &1.2\tablefootmark{d} & 129\tablefootmark{i} & 18 & 71\\
1.2\,mm & PI230 &  202-278 & 03-04.17 & 27 & 61 & 0.08 & 1.9\tablefootmark{e} & 157\tablefootmark{j} & 14 & 71\\ 
0.8\,mm & FLASH345 & 270-374 & 04 and 09.18 & 18 & 38 & 0.03 & 0.7\tablefootmark{f} & 532\tablefootmark{k} & 25 & 69\\ 
0.7\,mm & FLASH460 & 408 \& 464 & 04.18 & 14 & 76 & 0.05 & 0.7\tablefootmark{g} & 480\tablefootmark{l} & 67 & 55\\ 
\hline 
\end{tabular}
\tablefoot{ 
\tablefoottext{a}{Given at the frequency of 180\,GHz, 196\,GHz, 230\,GHz, 345\,GHz, and 408\,GHz for SEPIA180, PI230 (4-8\,GHz IF), PI230 (4-12\,GHz IF), FLASH345, and FLASH460, respectively.}
\tablefoottext{b}{Given for a resolution of 0.7\,\kms on $T_{\rm mb}$ scale. All frequencies were covered twice, except for the one at the edges of the bands, corresponding to the frequencies 159--161\,GHz, 186--188\,GHz, 318--320\,GHz, 328--330\,GHz, and 372--374\,GHz, where the rms is higher by a factor $\sqrt2$.}
\tablefoottext{c}{0.7--1.9\,mm.}
\tablefoottext{d}{0.6--1.6\,mm.}
\tablefoottext{e}{0.3--1.6\,mm.}
\tablefoottext{f}{0.1--1.6\,mm.}
\tablefoottext{g}{0.1--0.5\,mm.}
\tablefoottext{h}{82--293\,K excluding sidebands covering the water line at 183\,GHz.}
\tablefoottext{i}{93--272\,K.}
\tablefoottext{j}{103--377\,K.}
\tablefoottext{k}{146--594\,K excluding sidebands covering water lines.}
\tablefoottext{l}{478 at 408\,GHz \& 829\,K at 464\,GHz.}
}
\end{table*}

\section{Results}
\label{sec:results}

We show the spectrum towards \mysou\ from 252\,GHz to 258\,GHz in Fig.\,\ref{fig:full_spectrum}. The complete spectrum is shown in Appendix \ref{app:SpecSurvey}. In total, 3055 lines were detected over the 206\,GHz covered by the survey above the 3$\sigma$ level. Assuming Gaussian statistics, this gives us $\text{about nine}$\,lines as false-positive detections when we consider a 3$\sigma$ threshold for the peak line temperature. The overall line density is found to be $\sim$~15\,lines/GHz above 3$\sigma$.  This line density is significantly lower than the 3mm band spectrum of the bright hot core, Sgr~B2(N), which shows a line density of 102~lines/GHz\footnote{Measured for a full spectral survey of the 3\,mm band.} above 4$\sigma$ \citep{Belloche2013} with a noise of 22\,mK ($T_{\rm A}^*$) at a spectral resolution of 1.2\,\kms , corresponding to a noise of 28\,mK at a spectral resolution of 0.7\,\kms.

\begin{sidewaysfigure*}
\centering
\includegraphics[width=0.9\linewidth]{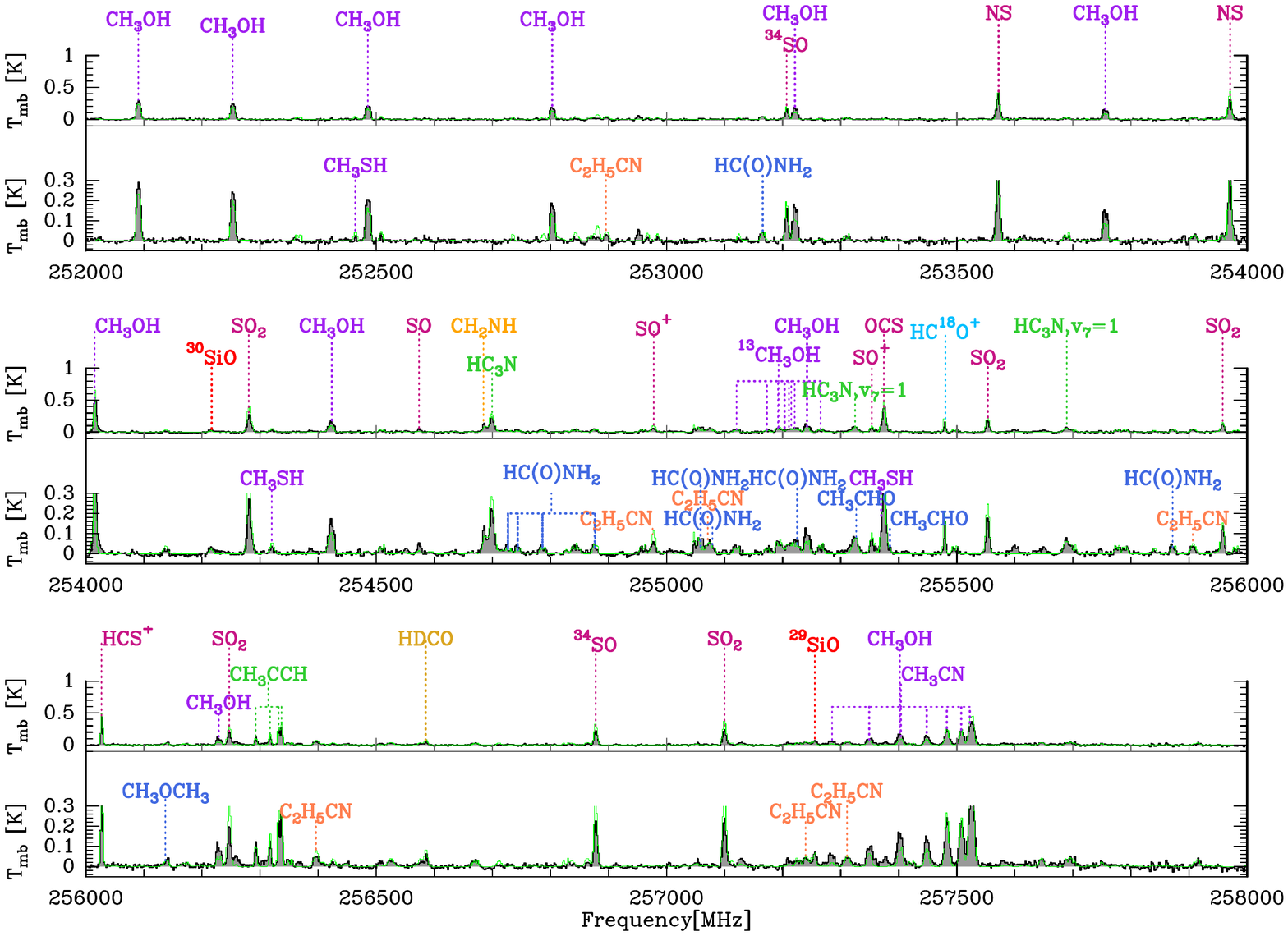}
\caption{Spectrum towards \mysou\ between 252\,GHz and 258\,GHz. The  filled grey histogram shows the spectra, and the coloured labels show the detected transitions of all the light molecules and some transitions of the COMs. The green line represents the LTE model obtained with Weeds. The upper panel for each frequency range shows the labels for the small molecules, and the lower panel corresponds to a zoom in temperature for  better visibility of the COMs which are labeled.}
\label{fig:full_spectrum}
\end{sidewaysfigure*}

\subsection{Line identification} 
We identified the molecules at the origin of the observed spectral lines using the CDMS\footnote{https://cdms.astro.uni-koeln.de/cdms/portal/home} \citep{Muller2005} and JPL\footnote{https://spec.jpl.nasa.gov} \citep{Pickett1998} catalogues with the help of the Weeds package for CLASS \citep{Maret2011}. These two databases provide the rest frequencies, the upper-level energies, $E_{\rm up}$, and the Einstein coefficients, $A_{\rm ij}$, for many molecules. To identify the emitting molecules and their transitions, we first visually identified the brightest lines and looked up  the most probable molecular line with the highest $A_{\rm ij}$ value at the corresponding frequency in the spectroscopic databases. Then we searched for all the transitions from this molecule that we expect to detect considering their upper-level energies and Einstein coefficients, or synthetic local thermal equilibrium (LTE) models (Sect. \ref{LTEmodelsection}) in the band above a minimum signal-to-noise ratio of 3 (Table \ref{tab:detections}). If all the expected transitions were detected, we considered this molecule as identified. We repeated this process with the next brightest line. A final iteration for the robustness of our detections was made against the  LTE synthetic spectrum (Sect. \ref{LTEmodelsection}), where we verified that all expected transitions were detected and no line was present in the synthetic spectrum that was not consistent with the data.

\begin{figure}[!htpb]
    \centering
    \includegraphics[width=0.45\linewidth]{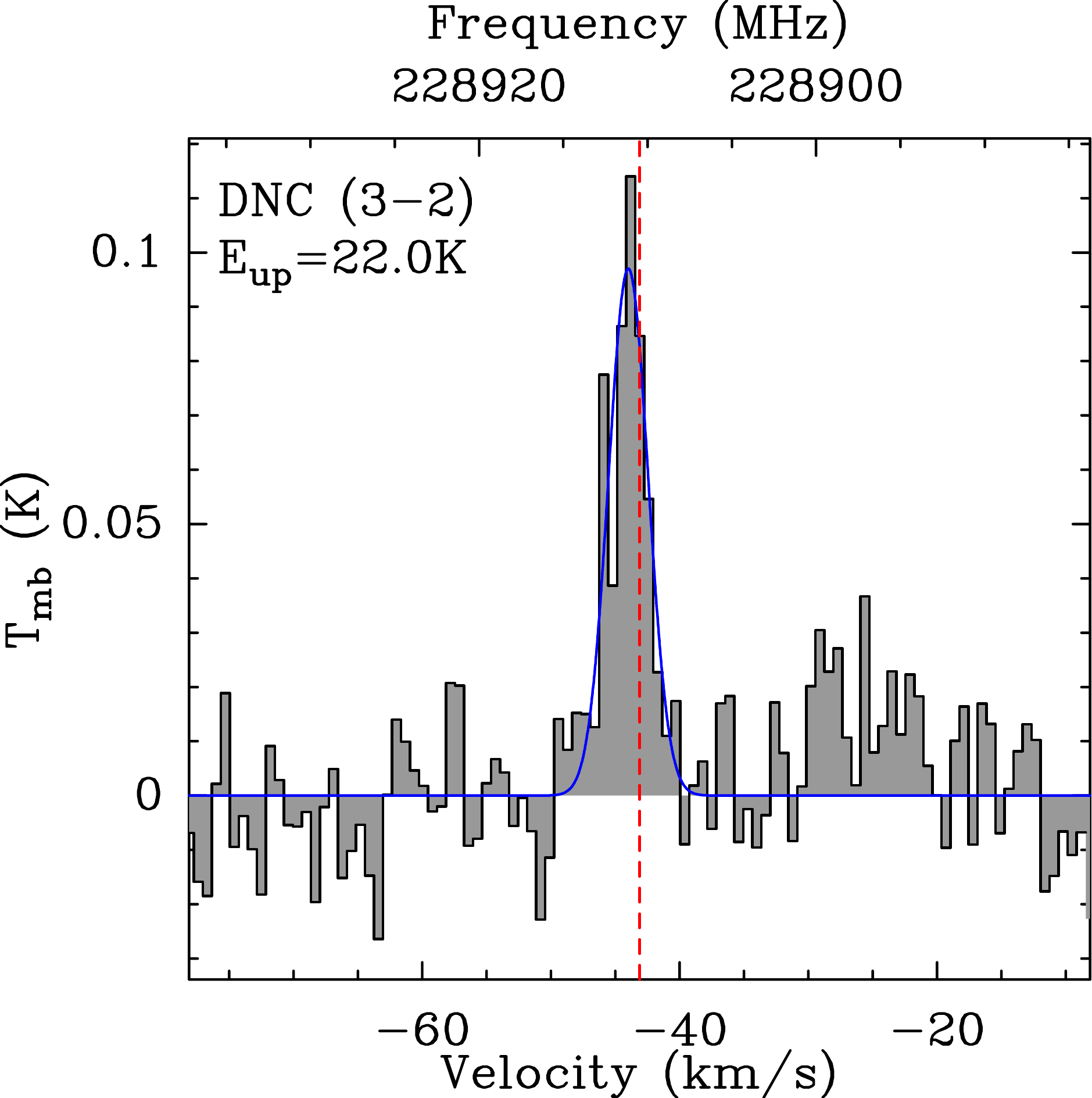}
    \includegraphics[width=0.45\linewidth]{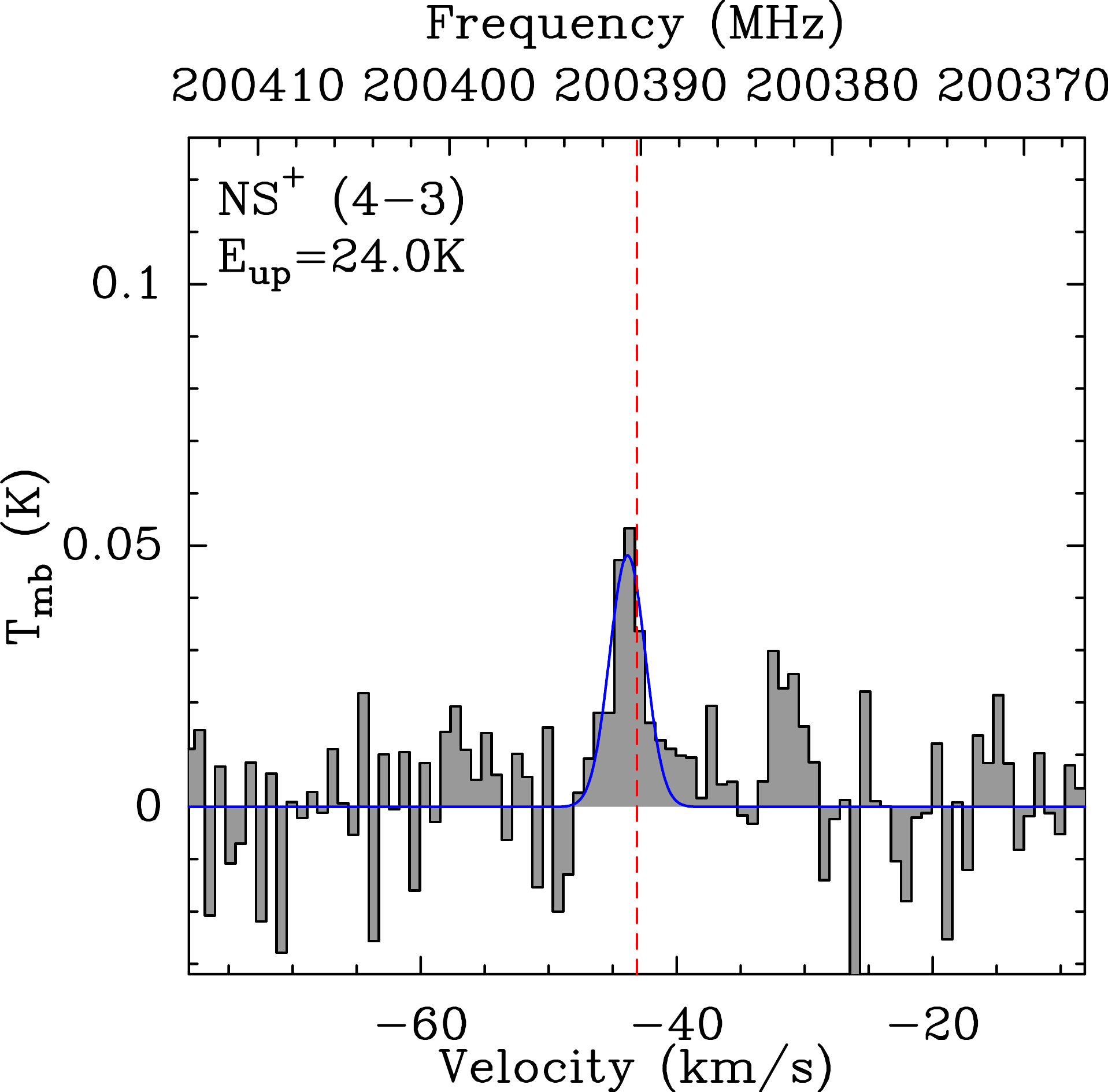}
    \includegraphics[width=0.45\linewidth]{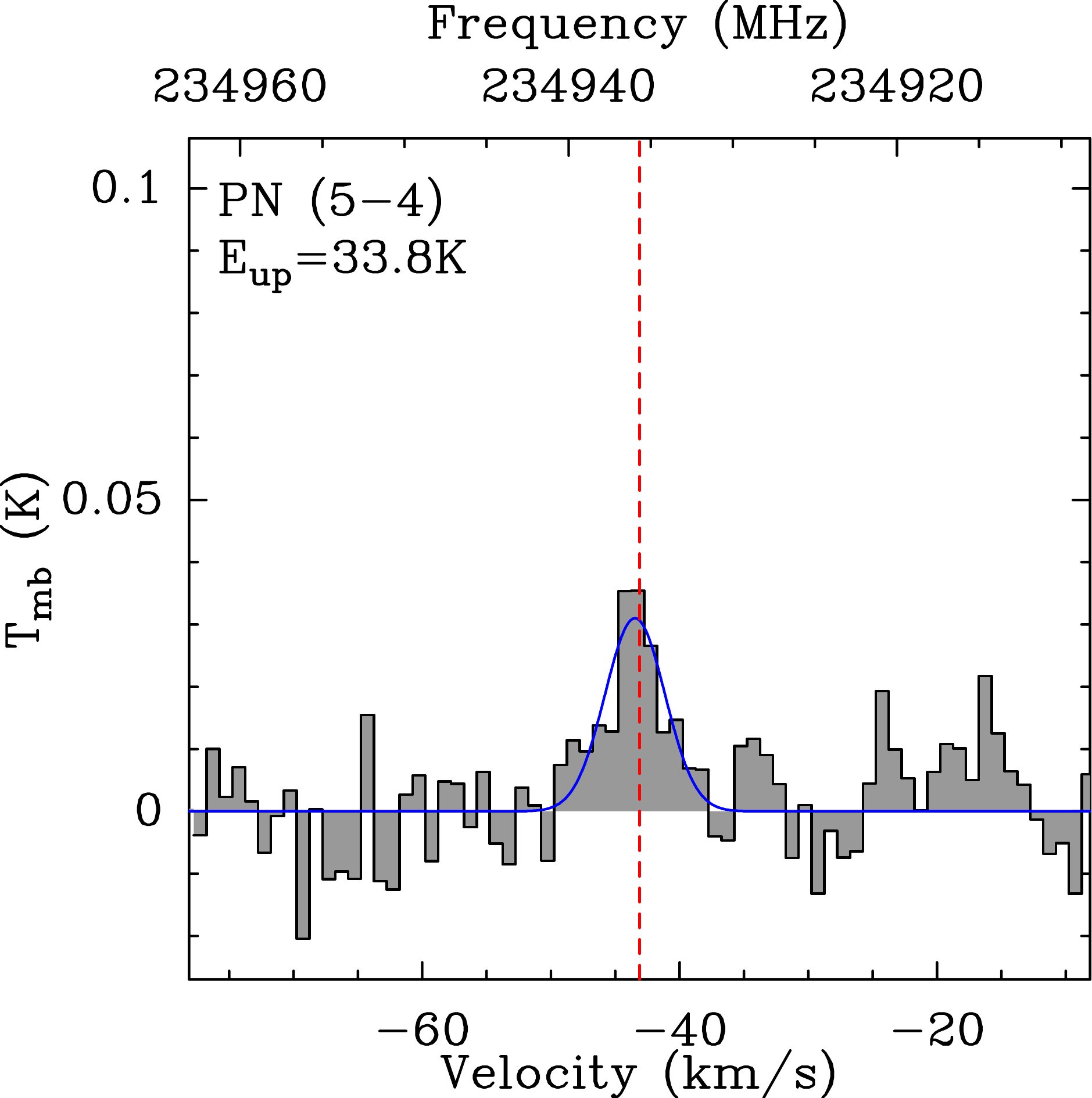}
    \includegraphics[width=0.45\linewidth]{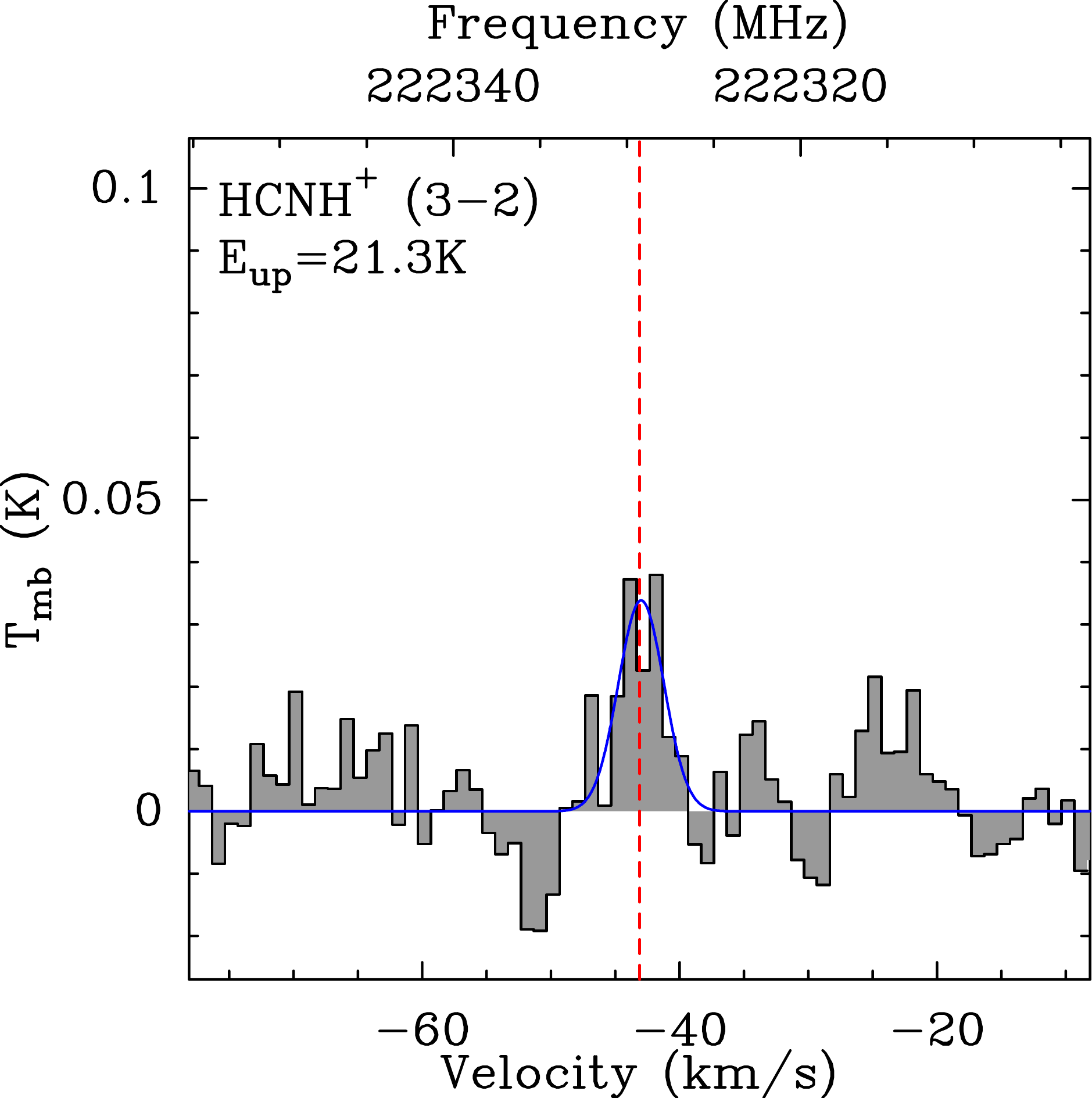}
    \caption{Spectra, in black, of DNC\,(3-2), NS$^+$\,(4-3), PN\,(5-4), HCNH$^+$\,(3-2) shown as filled grey histograms. The dashed red line shows the systemic velocity of the source at $-$43.13\,\kms\ \citep{Csengeri2018}. The blue line represents the best Gaussian fit for each line. $E_{\rm up}$ represents the upper-level energy of the transition in temperature units.}
    \label{fig:tentative_det}
\end{figure}

This procedure was efficient for molecules with a few atoms having up to 60 transitions in the band. 
We found a few molecules with only one transition above our detection threshold of a minimum signal-to-noise ratio of 3. While the transitions of HCNH$^+$\,(3-2), DNC\,(3-2), NS$^+$\,(4-3), and PN\,(5-4) are clearly detected above the 3$\sigma$ threshold  (Fig.\,\ref{fig:tentative_det}), we claim these molecules as tentatively detected because they only have one transition.

The heavier, i.e. higher-mass, COMs have a considerably larger number of transitions in the band due to their large partition function. Therefore, we used the LTE modelling (see Sect.\, \ref{LTEmodelsection}) of their spectra and visually checked whether all the expected lines were detected with intensities consistent with the synthetic spectrum. 
Altogether, we were able to firmly identify 2935 lines from all species. About 120 weak lines above 3$\sigma$ remained unidentified.

In total, we securely detect 39 species plus 22 isotopologues (Table \ref{molecules1} and \ref{molecules}), and we tentatively detect four additional molecules. In Table \ref{molecules1}  we group the molecules according to their atomic content, such as carbon chains, S-bearing, O-bearing, and N-bearing deuterated species, and COMs. The groups with the largest number of molecules are the COMs and S-bearing molecules. About 1600 of the 2935 identified lines are from O-bearing species (including O-bearing COMs), $\sim$300\,lines from S-bearing molecules, $\sim$70\,lines from carbon chains, and $\sim$900\,lines from N-bearing molecules (including N-bearing COMs). Rotational transitions of HC$_3$N from within its first vibrationally excited state ($\varv_7$=1) and CH$_3$OH from within its torsionally excited state ($\varv_t$=1) are detected.

\begin{table*}
\centering
\small
\caption{Detected molecules in \mysou. The molecules listed in {\it \textup{italics}} are tentative detections.}
{\vspace{0.1cm}}
\label{molecules1}
\begin{tabular}{c c c c c c c c c }  
\hline 
\hline
Carbon chains & O-bearing & N-bearing & Deuterated & \multicolumn{2}{c}{S-bearing} & \multicolumn{2}{c}{COMs}& Others \\        
\hline 
CCH          & CO       & CN            & DCN      &  CS         & H$_2$CS & \methanol     & CH$_3$CN      & SiO \\ 
HC$_3$N      & HCO$^+$  & HNC           & DCO$^+$  &  H$_2$S     & HCS$^+$ & CH$_3$CHO     & C$_2$H$_5$CN  & \it{PN}\\ 
CH$_3$CCH    & H$_2$CO  & HCN           & \it{DNC} &  OCS        & & CH$_3$OCH$_3$ & C$_2$H$_3$CN  & \\ 
c-C$_3$H$_2$ & H$_2$CCO & NO            & HDO      &  SO$_2$     &         & CH$_3$OCHO    &               & \\ 
             &          & CH$_2$NH      & HDCO     &  SO         &         & HC(O)NH$_2$   &               &\\ 
             &          & HNCO          & HDCS     &  SO$^+$     &         & CH$_3$SH      &               &\\ 
             &          & \it{HCNH$^+$} &          &  NS         &         &               &               &\\ 
             &          & N$_2$H$^+$    &          & {\it NS$^+$} &         &               &               &\\ 
\hline 
\end{tabular}
\end{table*}

\subsection{Line profiles and fitting}\label{sec:fitting}
\begin{figure}
\begin{center}
\includegraphics[width=0.49\linewidth]{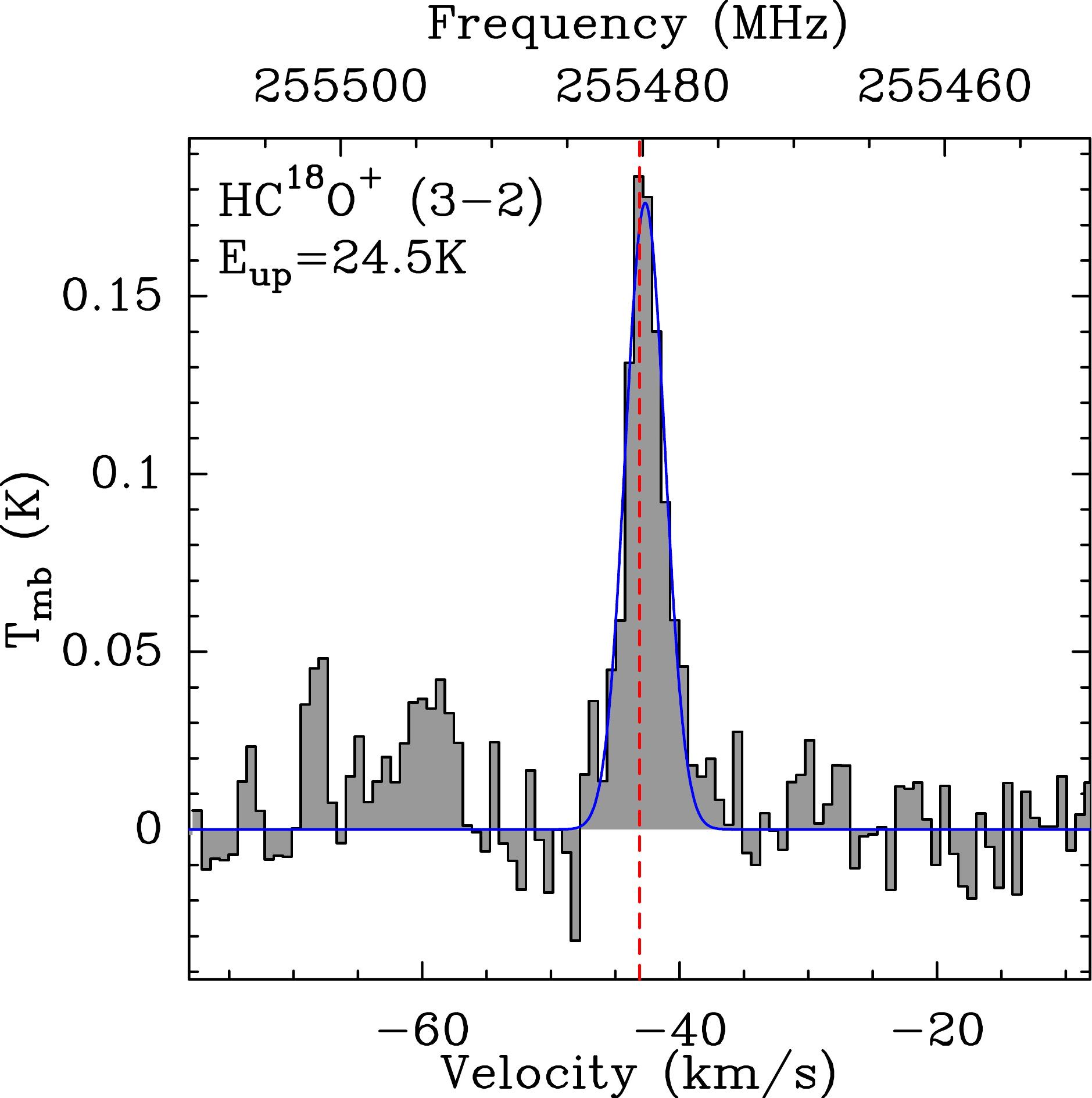}
\includegraphics[width=0.49\linewidth]{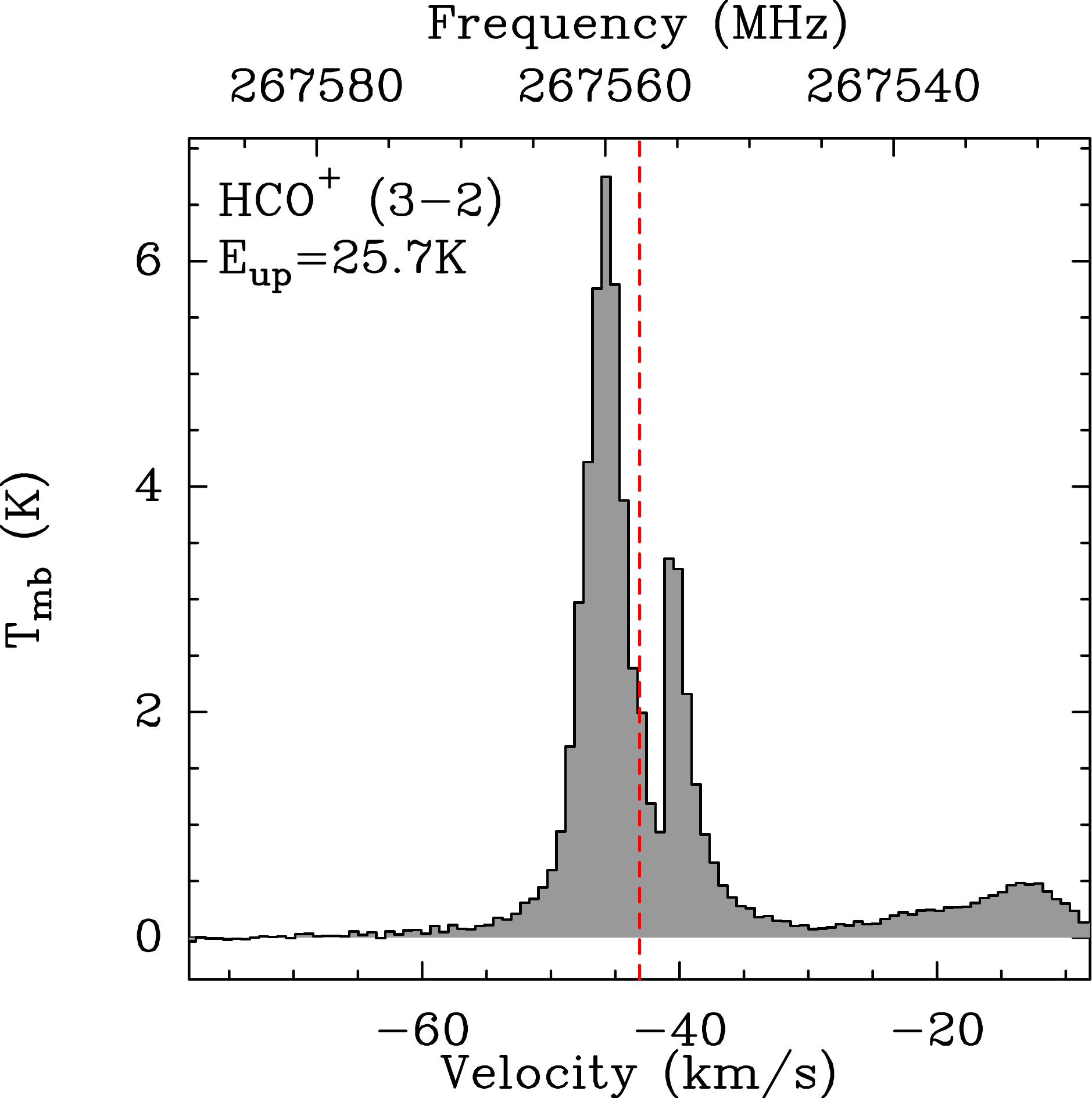}
\includegraphics[width=0.49\linewidth]{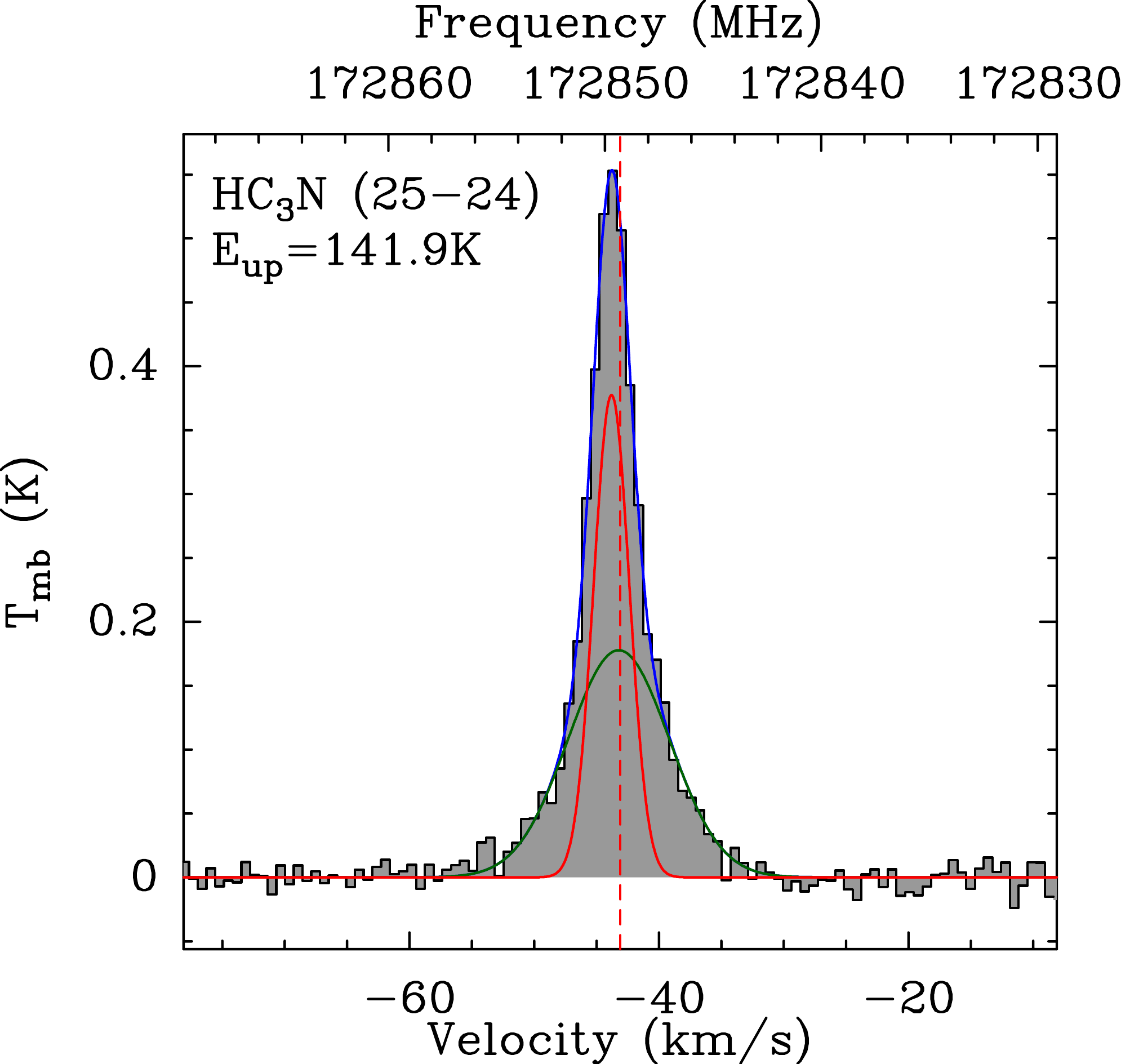}
\includegraphics[width=0.49\linewidth]{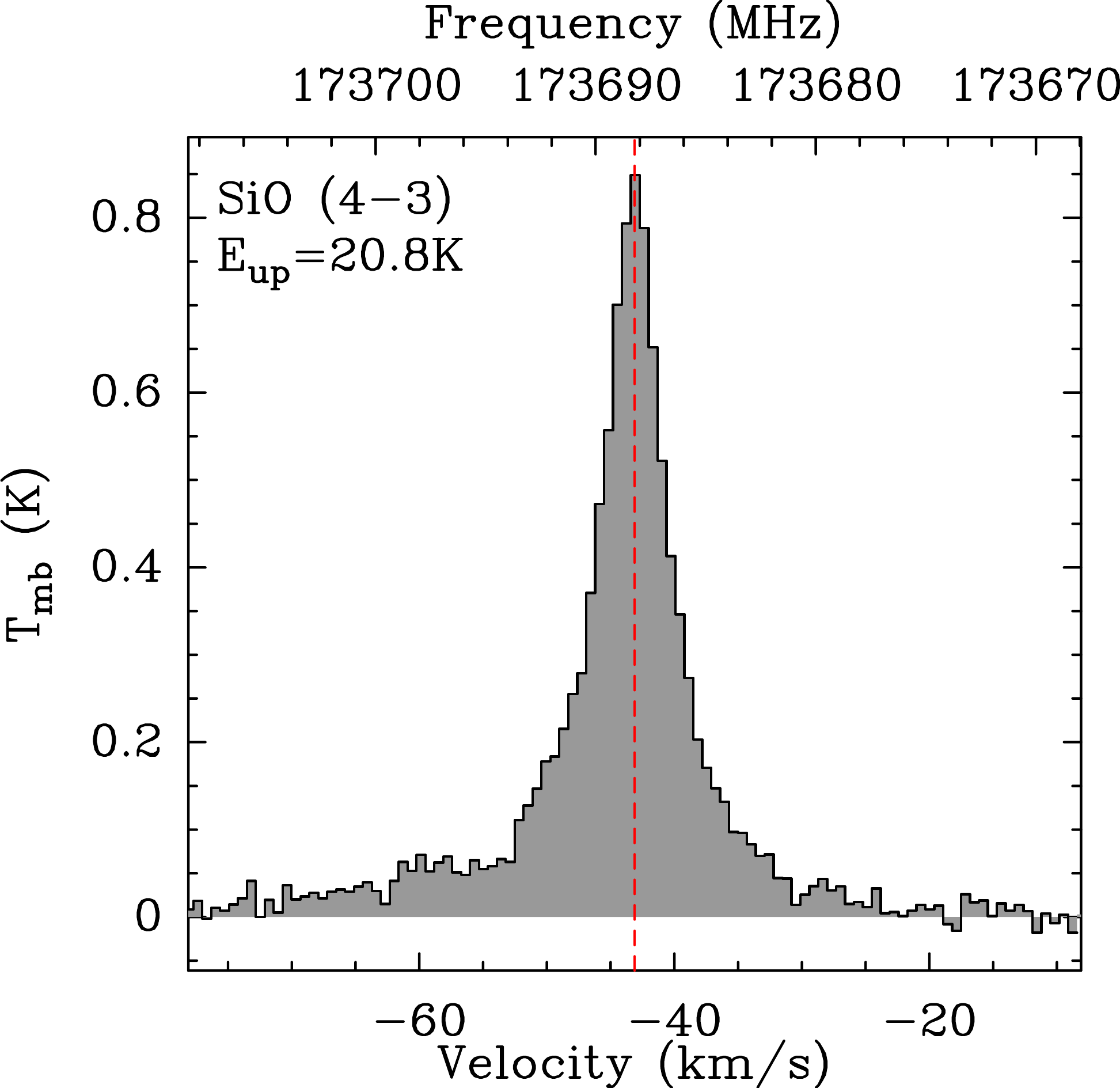}
\includegraphics[width=0.49\linewidth]{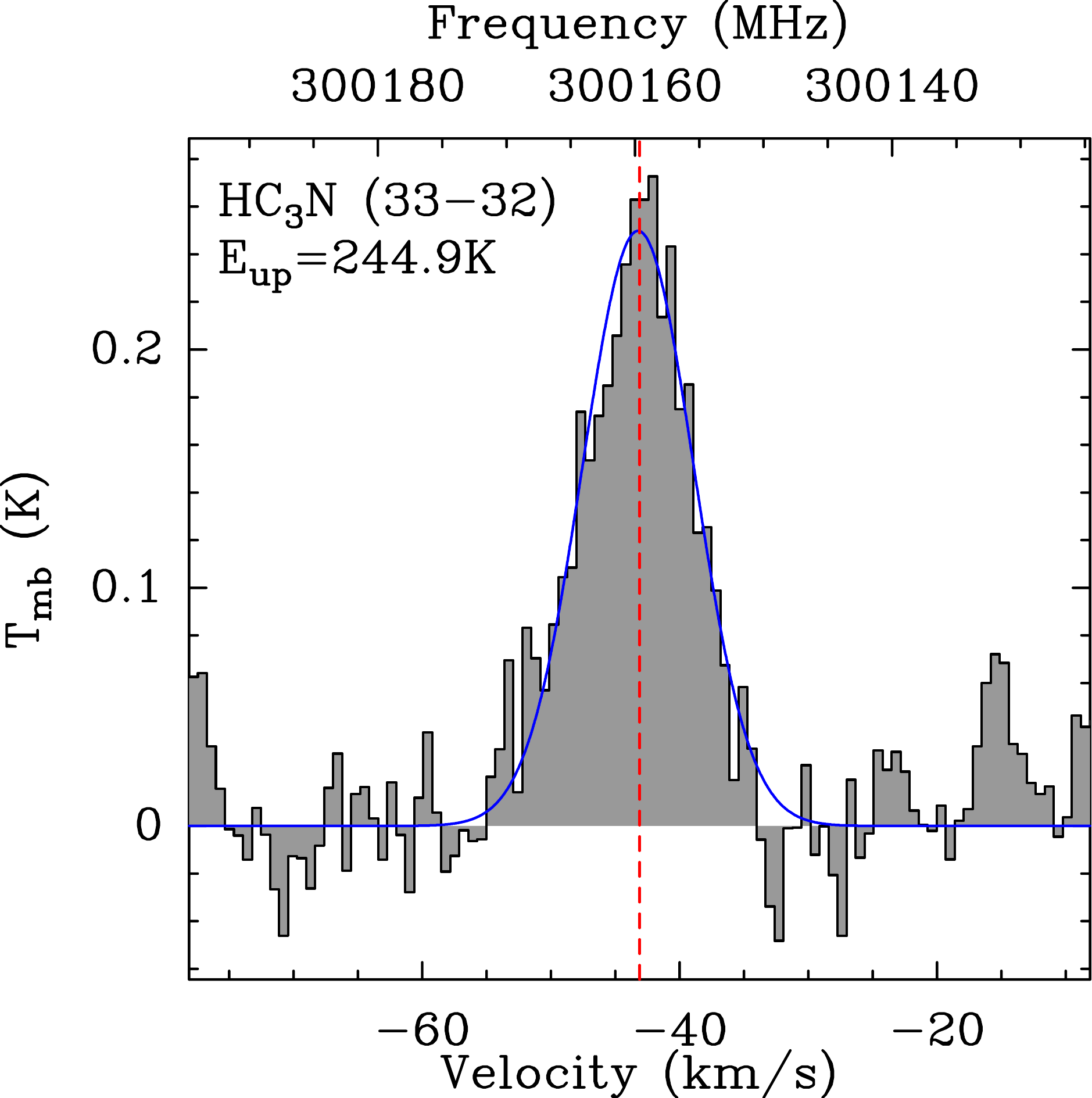}
\includegraphics[width=0.49\linewidth]{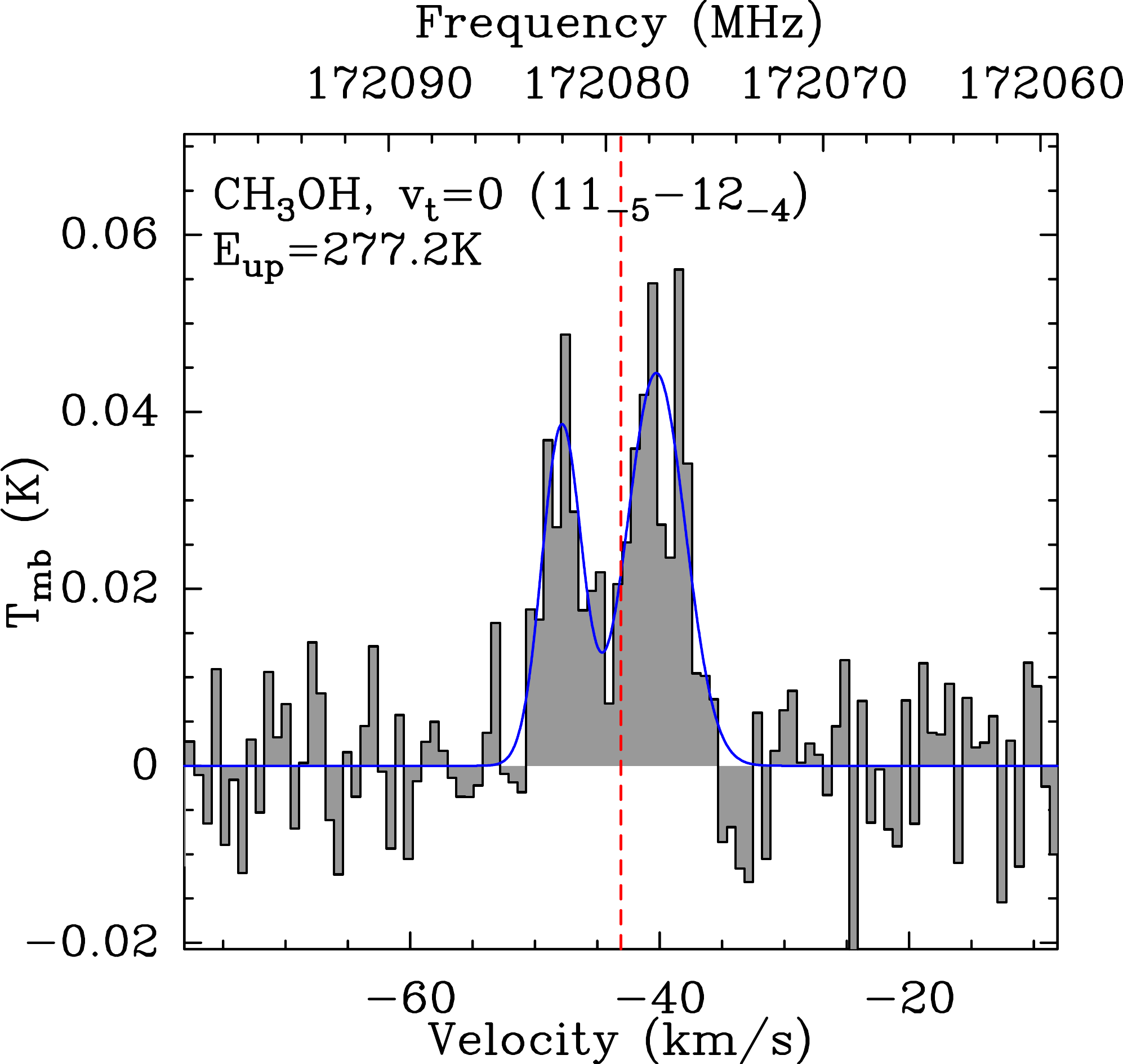}
\caption{Examples of the different line profiles observed in the spectral survey. From the top to the bottom, we show in the left column a narrow line profile for HC$^{18}$O$^+$ (3-2) and HC$_3$N (25-24) with a two-component Gaussian fit and HC$_3$N (33-32) with a single broad component. In the right column, HCO$^+$ (3-2) has a line profile with redshifted self-absorption, SiO (4-3) has a broad component coming from the outflow, and the profile of CH$_3$OH (11$_{-5}$-12$_{-4}$) shows two velocity components tracing the accretion shocks. All the lines are plotted in a window of 70\kms centred on the rest velocity. The dotted red line shows the systemic velocity of the source at $-$43.13\,\kms.}
\label{line_profiles}
\end{center}
\end{figure}

Our spectral resolution of 0.7\kms\ is sufficient to resolve the lines well, and we note a variety of spectral shapes associated with different molecules; some examples are shown in  Fig.\,\ref{line_profiles}. Most of the molecules with optically thin emission, for example HC$^{18}$O$^+$, SO$_2$ , and $^{29}$SiO, show a single Gaussian component. However, a fraction of the lines, such as CH$_3$OH, HNCO, HC$_3$N, and OCS,  requires
a two-component Gaussian fit using a narrow and a broad component. Methanol lines with a high upper-level energy (E$_{\rm up}\geq$150\,K) exhibit a profile consisting of two velocity components that are offset from the source $\varv_{\rm lsr}$ velocity by about $\pm$4\kms, similarly as reported in \citet{Csengeri2018} towards the accretions shocks on the basis of interferometric data. Finally, a few lines show non-Gaussian profiles that are heavily affected by self-absorption, such as the lines of HCO$^+$, HCN, HNC, and H$_2$S. Broad line-wings associated with high-velocity gas are observed in the CO transitions, for example. The outflow emission was associated by eye because the outflow appears as non-Gaussian residual in the fitting process. Table \,\ref{tab:detections} indicates the molecules in which some transitions exhibit outflow wings. We distinguish the cold and warm components in a quantitative way (see Sect.\,\ref{sec:source_structure}).

\begin{figure*}
\begin{center}
\includegraphics[scale=0.8]{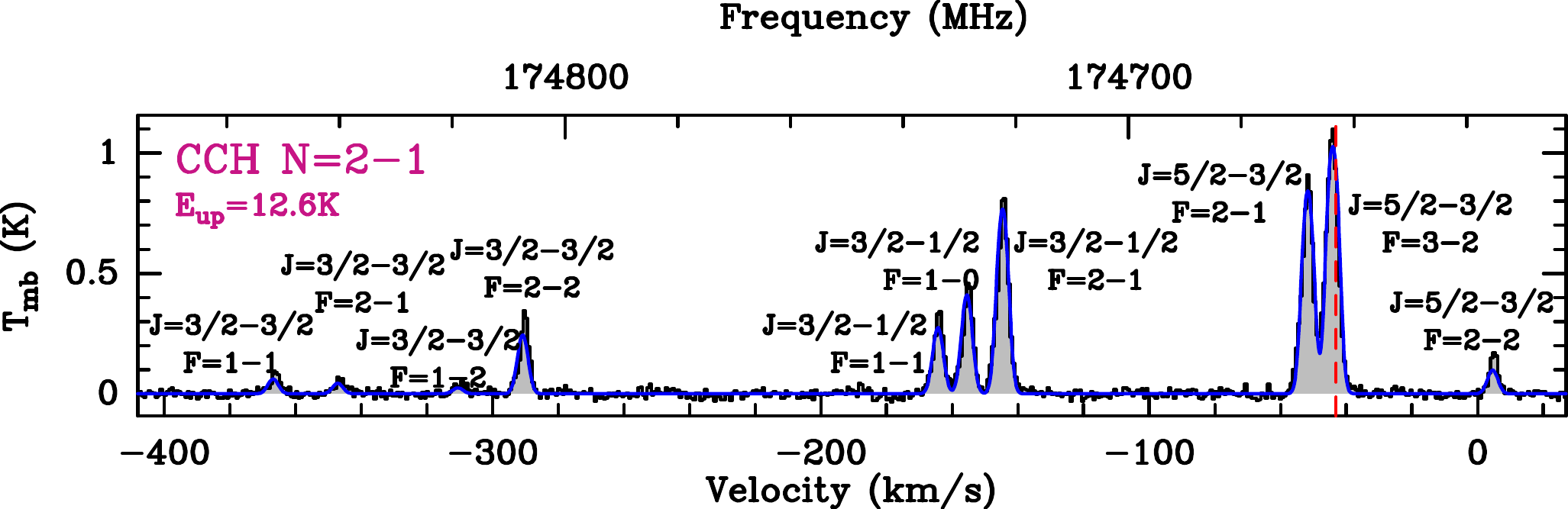}
\caption{Hyperfine structure fitting of the transition CCH (2-1) in the 2\,mm band. The result of the fitting procedure is indicated in blue, and the systemic velocity corresponds to the frequency of the most intense component hyperfine transition. The quantum numbers follow the notation from \citet{Gottlieb1983}, with $N$ corresponding to the rotational levels, $J$ to the spin doubling,  and $F$ to the hyperfine levels.}
\label{hyperfine_CCH}
\end{center}
\end{figure*}

For the lines showing Gaussian  profiles, we used the fitting procedure of CLASS and derived the line peak temperature, $T_{\rm mb}$, velocity at line peak, $\varv$, full width at half maximum, $\Delta \varv$, and integrated intensity, $W$, of each unblended line. Molecules with a resolved hyperfine structure (HFS), such as NS, NO, and CCH (see Fig.\,\ref{hyperfine_CCH} for an example and Table\,\ref{tab:LTE_values} for the list of the fitted molecules with an HFS procedure) were fitted based on the relative intensities of the HFS components from the JPL or CDMS catalogues using the HFS procedure of CLASS. An exception is N$_2$H$^+$, where we marginally resolve the HFS lines and see potentially different velocity components that we do not aim to decompose here and hence used a simplified fitting approach using a single Gaussian. The physical properties of the N$_2$H$^+$ emitting gas and the average line width have considerable uncertainties, therefore we do not use the results obtained for this species in Sect.\,\ref{sec:source_structure}.

\begin{table}
\caption{Summary of the molecules detected in the survey.}\label{tab:detections}
{\vspace{0.1cm}}
\label{molecules}
\scriptsize
\centering
\begin{tabular}{l c c c c}  
\hline 
\hline
Molecule & $E_{\rm up}$  & Detected $E_{\rm up}/k$ & Database & Structure \\
& min-max [K] &  min-max [K] &  & \\
\hline 
$^\dag$CCH               &  13 - 42      & 13-42 & CDMS & c\\
CH$_3$CCH         &  45 - 3367    & 45-237 & CDMS & c\\
c-C$_3$H$_2$      &  18 -2424     & 18-80 & CDMS & c\\
HC$_3$N,v=0       &  75 - 376     & 75-358 & CDMS & c,w\\
HC$_3$N,v$_{\rm 7}$=1      & 396 - 698     & 396-538 & CDMS & w\\
CO                &  17 -  33     & 17 -33 & CDMS & c,o \\
$^{13}$CO         &  16 -  32     & 16 - 32 & CDMS & c,o \\
C$^{18}$O         &  16 -  32     & 16 - 32 & CDMS & c\\
C$^{17}$O         &  16 -  32     & 16 - 32 & CDMS & c\\
$^{13}$C$^{18}$O  &  15 -  30     & 15 - 30 & CDMS & c\\
$^{13}$C$^{17}$O  &  15 -  31     & 15 - 31 & CDMS & c\\
HCO$^+$           &  13 -  43     & 13 - 43 & CDMS & c,w,o \\
H$^{13}$CO$^+$    &  12 -  42     & 12 - 42 & CDMS & c,w,o \\
HC$^{18}$O$^+$    &  12 -  41     & 12 - 41 & CDMS & c,w \\
HNCO              &  38 - 1553    & 38 - 154 & CDMS & c,w \\
H$_2$CO           &  21 - 6212    & 21 - 143 & CDMS & c\\
H$_2\,^{13}$CO    &  20 - 4491    & 20-65 & CDMS & c\\
H$_2$CCO          &  35 - 2682    & 35 - 163 & CDMS & w\\
$^\dag$NO                &  19 - 209     & 19 - 36 & CDMS & c\\
HCN               &  13 - 43      & 13 - 43 & CDMS & c,w,o \\
H$^{13}$CN        &  12 - 41      & 12 - 41 & CDMS & c,w \\
HC$^{15}$N        &  12 - 41      & 12 - 41 & CDMS & c,w \\
HNC               &  13 - 44      & 13 - 44 & CDMS & c, w, o\\
HN$^{13}$C        &  13 - 42      & 13 - 42 & CDMS & c, w\\
N$_2$H$^+$        &  13 - 45      & 13 - 27 & JPL & c \\
CH$_2$NH          &  11 - 2452    & 11 - 56 & CDMS & c\\
HCNH$^+$*         & 21 - 53       & 21 & CDMS &\\
DCN               &  21 - 52      & 21 - 52 & CDMS & c \\
DNC*              &  22 - 37      & 22 & CDMS & \\
DCO$^+$           &  21 - 52      & 21 - 52 & CDMS & c\\
HDO               &  22 -3131     & 22-168 &  JPL & w,s \\
HDCO              &  11 -3514     & 18-56 & CDMS & c\\
H$_2$S            &  8 -2722      & 8-150 & CDMS & c, w\\
H$_2\,^{34}$S     &  28 - 659     & 8-149 & CDMS & c\\
H$_2\,^{33}$S     &  28 - 658     & 8-149 & CDMS & c\\
HCS$^+$           &  20 - 74      & 20 - 74 & CDMS & c\\
H$_2$CS           &  25 - 2942    & 25 - 163 & CDMS & c \\
H$_2$C$^{34}$S    &  24 - 1667    & 24 - 71 & CDMS & c \\
SO                &  15 - 737     & 20 - 120 & CDMS & c, o\\
$^{34}$SO         &  15 - 572     & 33 - 86 & CDMS & c\\
SO$_2$            &  13 - 5355    & 13 - 351 & CDMS & c, w, o\\
$^\dag$NS                &  17 - 389     & 17 - 70 & CDMS & c\\
NS$^+$*           &  24 - 67      & 24 - 50 & CDMS & \\
OCS               &  61 - 271     & 61 - 271 & CDMS & c,w \\
CS                &  24 - 7274    & 24 - 66 & CDMS & c,o\\
$^{13}$CS         &  22 - 1858    & 33 - 80 & CDMS &  c, o\\
C$^{34}$S         &  23 - 1880    & 23 - 65 & CDMS & c\\
C$^{33}$S         &  23 - 1888    & 23 - 65 & CDMS & c\\
$^{13}$C$^{34}$S  &  22 - 79      & 33-46 & CDMS & c\\
SO$^+$            &  17 - 594     & 17-54 &  JPL & c\\
CH$_3$OH, $\varv_{\rm t}=0$    &  17 - 295     & 17-260    &  JPL & c,w,s\\
CH$_3$OH, $\varv_{\rm t}=1$    &  339 - 1971   & 339 - 611 &  JPL & s\\
$^{13}$CH$_3$OH   &  17 - 844     & 17 - 50 &  JPL & w\\
CH$_3$OCHO        &  79 - 757     & 79 - 318    & JPL & w,s \\
CH$_3$OCH$_3$     &  32 - 3106    & 15 - 475    & CDMS & c,w,s \\
HC(O)NH$_2$       &  36 - 940     & 36 - 380  & CDMS & w,s \\
CH$_3$CHO         &  26 - 1063    & 40 - 106  & CDMS & c\\
CH$_3$CN          &  40 - 2747    & 40 - 745  & CDMS & c,w,s\\
C$_2$H$_5$CN      &  34 - 3255    & 43 - 425  & CDMS & w\\
C$_2$H$_3$CN      &  40 - 2695    & 169 - 350 & CDMS & w\\
CH$_3$SH          &  55 - 1317    & 55 - 99 & CDMS & w\\
SiO, $\varv=0$ &  21 - 71      & 21 - 71 & CDMS & c,o \\
$^{30}$SiO        &  20 - 71   & 20 - 43 & CDMS & c \\
$^{29}$SiO        &  21 - 71   & 21 - 43 & CDMS & c \\
PN*               &  23 -  63     & 34 & CDMS & \\
\hline 
\end{tabular}
\tablefoot{The label c represents the cold component of the envelope, w the warm component, s the shocks, and o the outflow. The star indicates a tentative detection. The range of upper-level energies for the COMs is given for $A_{\rm ij}$ above 10$^{-4}$\,s$^{-1}$. $^\dag$ The HFS procedure was used for the line fitting.} 
\end{table}

\subsection{Rotational diagrams}

The large frequency coverage of the survey gives access to a large number of rotational transitions for the different species. Assuming LTE conditions, this allows us to estimate the total column density, $N$, of each species and the rotational temperature, $T_{\rm rot}$. Therefore, in order to obtain a first estimate of the physical and chemical conditions of the emitting gas, we used rotational diagrams to estimate $N$ and $T_{\rm rot}$ that then served as an input for a more detailed LTE modelling using Weeds following a similar approach as described in \citet{Belloche2016} (Sect.\,\ref{LTEmodelsection}).

To do this, we use the integrated intensity, $W$, of each unblended line at a frequency, $\nu$, from the Gaussian fitting.  
The population of each rotational level of a molecule is given by 
\begin{equation}
    N_u^{thin}=\frac{8\pi k\nu^2W}{hc^3A_{\rm ul}B_{\rm dil}}
\end{equation} 
for optically thin lines, where $k$ is the Boltzmann constant, $B_{\rm dil}$ is the beam dilution factor depending on the size of the emitting region, $\theta_{\rm S}$, and the beam size, $\theta_{\rm beam}$, $A_{\rm ul}$ is the Einstein coefficient, $h$ is the Planck constant, and $c$ is the speed of light (see e.g. \citealt{Goldsmith1999}). Because we assumed that most of the emission is unresolved within the beam, the beam dilution factor, defined as $B_{\rm dil}=\frac{\theta_{\rm S}^2}{\theta_{\rm S}^2+\theta_{\rm beam}^2}$, needs to be taken into account and shows a large variation due to the large frequency coverage of our dataset. The size of the emitting region is determined based on the rotational diagrams and the LTE modelling using Weeds in an iterative process (see Sect. \ref{LTEmodelsection}).

In the case of optically thick lines, the line opacity needs to be taken into account, and the population of each level becomes $N_u=N_u^{thin}  C_{\rm \tau}$, where $C_{\rm \tau}=\frac{\rm \tau}{1-e^{-\tau}}$. The parameter $\tau$ corresponds to the total line opacity, from which  $C_{\tau}$, the optical depth correction factor is computed. 
 Following \citet{Goldsmith1999}, to estimate $N_{\rm tot}$ , we used the expression

\begin{equation}
    ln\Big(\frac{N_{\rm u}}{g_{\rm u}}\Big)=ln\Big(\frac{N_{\rm tot}}{Z}\Big)-\frac{E_{\rm up}}{kT_{\rm rot}},
\end{equation}
where $Z$ is the rotational partition function, $g_{\rm u}$ is the statistical weight, and $E_{\rm up}$ is the upper-level energy of each transition. The rotational diagrams are then displayed as $ln\Big(\frac{N_{\rm u}}{g_{\rm u}}\Big)$ versus $E_{\rm up}$, and by fitting a linear function, we derive the total column density $N_{\rm tot}$ and the rotational temperature $T_{\rm rot}$.
The partition function was interpolated from the tables given in the CDMS and JPL databases. We perform a linear least-square fit to the rotational diagrams, and from these parameters, we computed the rotational temperature and the total column density of each temperature component of the molecule. We estimated uncertainties on the rotational temperature and the column density using a Monte Carlo approach. We varied each measurement in the rotational diagram within its error bars assuming a uniform distribution and repeated the least-square fit. The final results correspond to the average of the estimated column densities and rotational temperatures, while their uncertainties correspond to the standard deviation of the values determined with our Monte Carlo method. 

Several of the molecules we cover exist in different symmetries. We treat them together as single species, such as the forms $E$ and $A$ of propyne (CH$_3$CCH), methyl cyanide (CH$_3$CN), and methanol (CH$_3$OH). 

We detect several transitions for C$_2$H$_3$CN, but all except four lines are found to be blended. Therefore, we only used these lines for the rotational diagram covering the range between 160\,K and 230\,K, which represents half of the complete range of detected transitions for C$_2$H$_3$CN (160-350\,K).

\begin{figure}
\begin{center}
\includegraphics[width=0.9\linewidth]{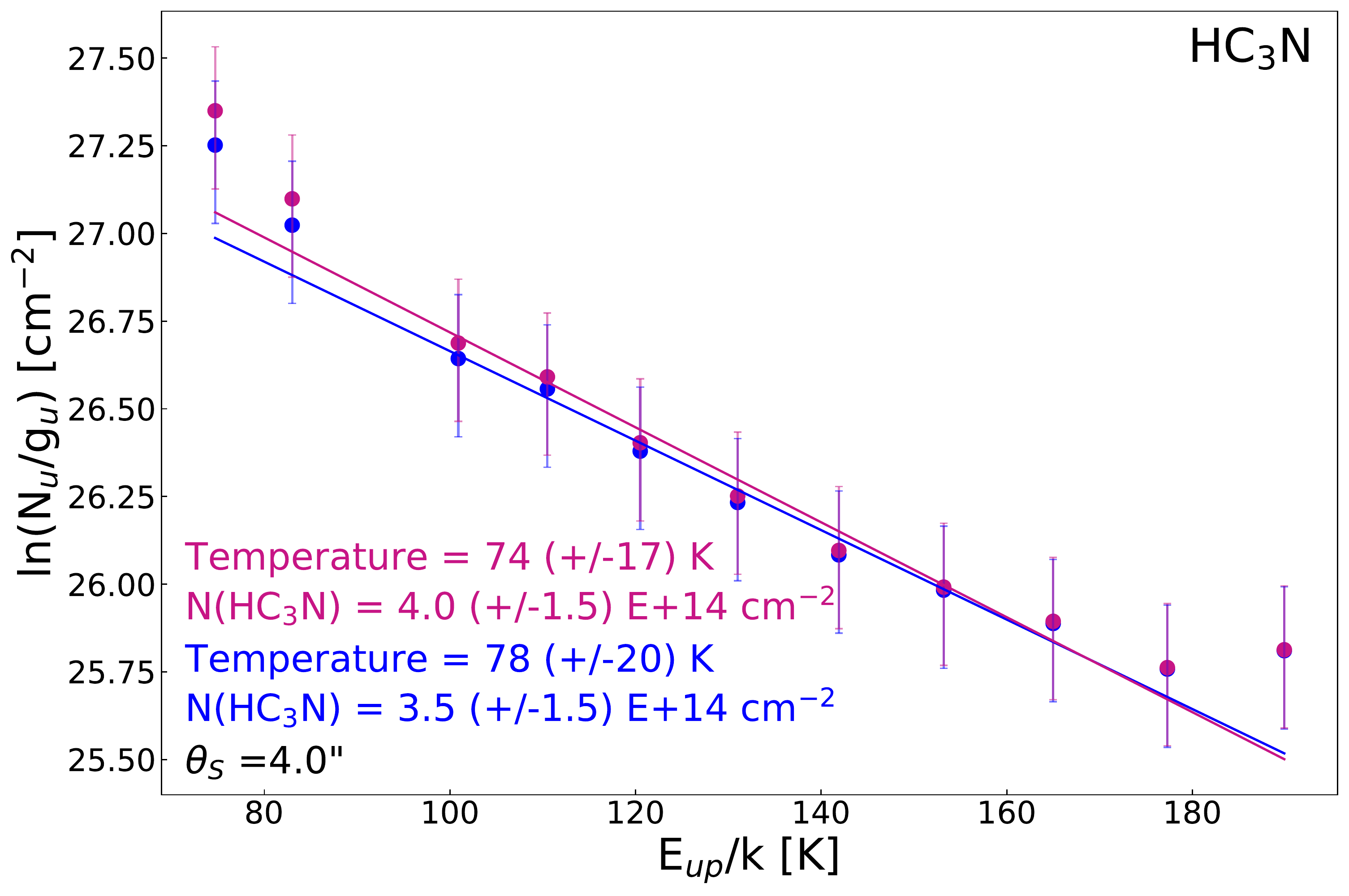}
\caption{Rotational diagram of HC$_3$N for the broad velocity component towards \mysou. The rotational diagram when we assume that HC$_3$N lines are optically thin is shown in blue. We represent the rotational diagram corrected for the opacity of each line in red. Rotational temperatures and column densities are indicated with the same colour code. The beam dilution was calculated for each transition for a source size of 4$''$. The error bars represent a 1$\sigma$ statistical noise plus a 20\% calibration error. To calculate the uncertainties on the rotational temperature and the column density, we used the Monte Carlo method and assumed a uniform distribution of the error for each data point.}
\label{fig:ex_rotdiag}
\end{center}
\end{figure}

As an example, we show the rotational diagram of the HC$_3$N molecule in Fig.\,\ref{fig:ex_rotdiag}, for which the rotational temperature increases from 74\,K to 78\,K, corrected for optical depth effects.  
We show all rotational diagrams used for this study in Appendix\,\ref{app:rot_diag}.

\subsection{LTE modelling}
\label{LTEmodelsection}

\subsubsection{Method}

In order to fit the observed spectrum, we performed an LTE modelling with Weeds \citep{Maret2011}.  The input parameters for each molecule were the column density ($N_{\rm tot}$ [\unidens]), the size of the emission ($\theta_S$ [$''$]), the excitation temperature ($T_{\rm ex}$ [K])\footnote{Formally, this 
is the rotational temperature and we implicitly also assume that all the lines have the same $T_{\rm ex}$. Under LTE conditions this is equal to $T_{\rm kin}$.}, the averaged full-width at half maximum (FWHM, $\Delta \varv$ [\kms]), and the velocity offset ($\varv_{\rm off}$ [\kms]) with respect to the systemic velocity of the source. 

For the light molecules, we first used the excitation temperature and the column density from their rotational diagram  as an input parameter for the modelling. For a first guess of the sizes, we used the ALMA images shown in \citet{Csengeri2019}, which were available for a limited number of molecules. For the remaining molecules, we estimated the sizes where possible based on the iterative approach described below. We used the averaged  FWHM and velocity offset from the Gaussian line-fitting of the unblended transitions.
Using these first-guess parameters, we then visually inspected the obtained model and extracted the residual. When the residual was lower than the 3$\sigma$ threshold, we considered the fit to be satisfactory, otherwise, we modified the input parameters and repeated the fit. To improve the model, we varied the column density and the temperature within the error bars given by the fit to the rotational diagrams. When we failed to obtain a satisfactory fit within this parameter range, we varied the size of the emitting region in the rotational diagrams and used the new column density and temperature as input. With this new set of parameters (column density, temperature, and size), we repeated the process until we found a satisfactory fit within the 3$\sigma$ threshold for all the transitions. 
Our uncertainties depend on the spectroscopic data of each molecule and on the uncertainties due to the source size. For molecules that  transition at similar upper-level energies over the spectral band at different frequencies, for example, CH$_3$OCH$_3$ or SO$_2$, the source size is relatively well constrained. For linear molecules with increasing upper-level energies as a function of frequency, our size estimates are limited to those with optically thin transitions. Molecules with highly optically thick emission in all their transitions, such as HCO$^+$, HCN, HNC, H$_2$S (cold phase), and CS, could not be modelled by this method, hence we used their optically thin isotopologues assuming that they originate from the same gas.

In all cases, the rare isotopologues of each main species were modelled separately in order to determine their column density independently from the main isotopologue, allowing us to derive isotopic ratios (see Sect.\,\ref{sec:iso_ratios}).
Because all SiO transitions show a complex velocity profile, we used the canonical $^{28}$Si/$^{29}$Si ratio (19.6; \citealp{Penzias1981a}) to estimate the SiO column density.

In order to identify all the transitions for each molecule, we first initiated a model with a first guess of the parameters for the complex molecules, identified the unblended transitions, and derived a rotational diagram. We then obtained the LTE model and corrected the list of the detected transitions if necessary in an iterative fashion. We applied this method to all the COMs.

The different envelope components identified by their kinematics (i.e.\,velocity profile) were treated independently at first and then modelled together to inspect the final model of the molecule. 
This approach assumes a uniform spatial distribution of the molecules and ignores their spatial segregation, as suggested by ALMA observations \citep{Csengeri2019}, on scales smaller than our angular resolution.

After we obtained a satisfactory fit within our quantitative criterion (unless stated otherwise, see Sect.\,\ref{sec:exceptions}) for each molecule separately, we modelled all the species together to determine whether some remaining lines were left and if the blended lines were properly modelled. Where necessary,  we adjusted the model of each molecule. 
For each species, we were able to determine the column density, the excitation temperature, and the size of the emission for each part of the envelope from which the emission arises. The results of the modelling are summarised in Table \ref{tab:LTE_values}.

\subsubsection{Exceptions} \label{sec:exceptions}

For molecules with K-ladder transitions (H$_2$CO, H$_2$CS, and CH$_3$CN), we used a different approach. For (ortho and para) H$_2$CO and H$_2$CS, we estimated the temperature following the method of \citet{Mangum1993} using the ratio between the $J$= 3$_{2,1}$--2$_{2,0}$ ($E_{\rm up}/k=68$\,K) and $J$=3$_{0,3}$--2$_{0,2}$ ($E_{\rm up}/k=21$\,K) transitions, and find a temperature of 89\,K, corresponding to the warm component of the envelope. The estimated column density has a larger uncertainty, because the emission sizes are poorly constrained. Therefore, we estimated a column density using sizes of 4-7\arcsec\ and 2-5\arcsec, which is the ortho/para-H$_2$CO and ortho/para H$_2$CS, respectively, which correspond to the typical sizes for the warm component.

We find, however, that this component does not fit all the $K$ transitions well, suggesting the presence of another, cold component for both H$_2$CO and H$_2$CS. We estimated the temperature of this component based on the rotational diagram of the $K$=2 transitions, and similarly then used the LTE models to fit the emission. We took 20-25\arcsec\ and 30-35\arcsec\ for the size of the emitting region for these components, respectively. Similarly as for the warm component, the uncertainty on the size of the emitting region and the column density for these species is considerably. Consequently, we do not use these values for the discussion in Sect.\,\ref{sec:source_structure}.

For CH$_3$CN, we estimated the temperature, column density, and size of the emission using the rotational diagrams for the $K$=0 component. We find a moderately extended size of 15\arcsec\ and a temperature of 40\,K. Our LTE model is not satisfactory, however, suggesting the need of a warm component. Because the APEX data do not give further constraints on this warm component, we simply added a 200\,K temperature component with a size of 1\arcsec, which is constrained by the ALMA data in \citet{Csengeri2019}. This composite model fits the $K$=1-6 transitions rather well and underestimates the higher excitation K=7,8,9 transitions in the $J$=17--16 and $J$=19--18 lines. An additional hot component could explain this excess emission, but we cannot constrain its properties. The $^{13}$CH$_3$CN isotopologue is not detected in our survey.

Although the rotational diagram of C$_2$H$_5$CN shows some scatter (Fig.\,\ref{diag_COMs2}), we fit only one temperature component, which is found to have a temperature around 90\,K. Because the ALMA observations \citep{Csengeri2019} suggest the presence of a compact warm component, we tested whether our LTE fitting using two temperature components (25\,K and 200\,K) instead of a single one (90\,K) would provide a better fit. We found that we cannot distinguish between these models; both of them could be consistent with our data. For simplicity, we therefore used the values of a single-component fit. 

H$^{13}$CN, HC$^{15}$N, and HN$^{13}$C have only three transitions in our band, and the rotational diagram is consistent with a single gas temperature component. However, the LTE model reveals that one temperature component is not satisfactory. Therefore, we need to use a model with two temperature components: one at 10\,K with a size of $\sim$35\arcsec, and a second one at $\sim$80\,K with a size of 3$''$.

The rotational diagrams of HDO and HC(O)NH$_2$ show only one temperature component, but the line profiles reveal the dynamics of accretion shocks similarly to CH$_3$OH (Fig. \ref{line_profiles}). We added a shock component to our model using the excitation temperature and size of the emitting region of CH$_3$OH,$v_{\rm t}$=1.

\longtab{
\setlength{\tabcolsep}{3pt}
\begin{longtable}{l c c c c c c c c c}
 \caption{\label{tab:LTE_values}LTE parameters}\\
\hline\hline
Molecule & $N_{\rm X}$ & $T_{\rm ex}$ & $\theta_s$ & Velocity offset & $\Delta \varv$ & $N_{\rm H_2}$  & $N_{\rm X}$/$N_{\rm H_2}$ & $N_{\rm X}$/$N_{\rm CH_3OH}$ & Comments \\
 &  [cm$^{-2}$] & [K] & [$''$] & [\kms] & [\kms] &  [cm$^{-2}$] &  & &\\
\hline
\endfirsthead

\hline
\multicolumn{6}{l}{Cold component of the envelope}\\
\hline
$^{\dag}$CCH         &     4.0$\times$10$^{14}$ &     13 &      45 &     $-$0.8$\pm$      0.4  &      3.8 $\pm$      0.4 &       8.4$\times$10$^{22}$ &       4.8$\times$10$^{-9}$  &    2.0$\times$10$^{-1}$ & \\
CH$_3$CCH            &     8.7$\times$10$^{15}$ &     35 &       6 &     $-$0.5$\pm$      0.3  &      3.7 $\pm$      0.4 &       5.7$\times$10$^{23}$ &       1.5$\times$10$^{-8}$  &    4.3$\times$10$^{0}$ & \\
c-C$_3$H$_2$         &     9.3$\times$10$^{12}$ &     13 &      20 &     $-$0.5$\pm$      0.4  &      4.5 $\pm$      1.9 &       1.8$\times$10$^{23}$ &       5.0$\times$10$^{-11}$  &    4.6$\times$10$^{-3}$ & \\
HC$_3$N              &     1.6$\times$10$^{14}$ &     24 &      15 &     $-$0.5$\pm$      0.2  &      3.4 $\pm$      0.3 &       2.4$\times$10$^{23}$ &       6.6$\times$10$^{-10}$  &    8.0$\times$10$^{-2}$ & \\
HCO$^+$              &     3.3$\times$10$^{14}$ (6.6$\times$10$^{14}$) &     13 &      26 & -  & - &       1.4$\times$10$^{23}$ &       1.7$\times$10$^{-9}$  &    1.2$\times$10$^{-1}$ & (1) \\
H$^{13}$CO$^+$       &     1.1$\times$10$^{13}$ &     13 &      26 &     $-$0.5$\pm$      0.2  &      3.8 $\pm$      0.2 &       1.4$\times$10$^{23}$ &       7.7$\times$10$^{-11}$  &    5.5$\times$10$^{-3}$ & \\
HC$^{18}$O$^+$       &     1.4$\times$10$^{12}$ &     13 &      26 &      0.7$\pm$      0.9  &      3.4 $\pm$      0.2 &       1.4$\times$10$^{23}$ &       9.8$\times$10$^{-12}$  &    7.0$\times$10$^{-4}$ & \\
o-H$_2$CO            &     1.3$\times$10$^{14}$ &     12 &      32 &     $-$0.4$\pm$      0.5  &      6.5 $\pm$      0.2 &       1.2$\times$10$^{23}$ &       1.1$\times$10$^{-9}$  &    6.5$\times$10$^{-2}$ & \\
p-H$_2$CO            &     3.0$\times$10$^{13}$ &     20 &      35 &     $-$0.3$\pm$      0.2  &      6.0 $\pm$      0.5 &       1.1$\times$10$^{23}$ &       2.8$\times$10$^{-10}$  &    1.5$\times$10$^{-2}$ & \\
H$_2\,^{13}$CO       &     9.7$\times$10$^{12}$ &     25 &      20 &     $-$0.5$\pm$      0.2  &      4.5 $\pm$      1.8 &       1.8$\times$10$^{23}$ &       5.3$\times$10$^{-11}$  &    4.9$\times$10$^{-3}$ & \\
HNC                  &     1.6$\times$10$^{14}$ (3.2$\times$10$^{14}$) &      8 &      35 & -  & - &       1.1$\times$10$^{23}$ &       1.1$\times$10$^{-9}$  &    6.0$\times$10$^{-2}$ & (1) \\
HN$^{13}$C           &     5.4$\times$10$^{12}$ &      8 &      35 &     $-$0.8$\pm$      0.1  &      4.1 $\pm$      0.1 &       1.1$\times$10$^{23}$ &       5.1$\times$10$^{-11}$  &    2.7$\times$10$^{-3}$ & \\
HCN                  &     8.1$\times$10$^{14}$ (1.6$\times$10$^{15}$) &     10 &      29 & -  & - &       1.3$\times$10$^{23}$ &       4.8$\times$10$^{-9}$  &    3.1$\times$10$^{-1}$ & (1) \\
H$^{13}$CN           &     2.7$\times$10$^{13}$ &     10 &      29 &     $-$0.4$\pm$      0.1  &      6.5 $\pm$      0.1 &       1.3$\times$10$^{23}$ &       2.1$\times$10$^{-10}$  &    1.3$\times$10$^{-2}$ & \\
HC$^{15}$N           &     5.9$\times$10$^{12}$ &     10 &      29 &     $-$0.4$\pm$      0.1  &      6.3 $\pm$      0.2 &       1.3$\times$10$^{23}$ &       4.6$\times$10$^{-11}$  &    2.9$\times$10$^{-3}$ & \\
HNCO                 &     1.1$\times$10$^{14}$ &     28 &      12 &     $-$0.2$\pm$      0.1  &      3.6 $\pm$      0.5 &       3.0$\times$10$^{23}$ &       3.7$\times$10$^{-10}$  &    5.5$\times$10$^{-2}$ & \\
$^\dag$NO                   &     6.0$\times$10$^{15}$ &     14 &      14 &     $-$0.8$\pm$      0.4  &      5.8 $\pm$      0.6 &       2.6$\times$10$^{23}$ &       2.3$\times$10$^{-8}$  &    3.0$\times$10$^{0}$ & \\
CH$_2$NH             &     3.0$\times$10$^{14}$ &     43 &       5 &     $-$0.9$\pm$      0.4  &      7.1 $\pm$      1.3 &       6.6$\times$10$^{23}$ &       4.5$\times$10$^{-10}$  &    1.5$\times$10$^{-1}$ & \\
N$_2$H$^+$           &     4.4$\times$10$^{13}$ &     10 &      40 &     $-$0.9$\pm$      0.2  &      5.8 $\pm$      1.1 &       9.4$\times$10$^{22}$ &       4.7$\times$10$^{-10}$  &    2.2$\times$10$^{-2}$ & \\
HDCO                 &     1.4$\times$10$^{13}$ &     28 &      14 &     $-$0.5$\pm$      1.0  &      7.0 $\pm$      2.8 &       2.6$\times$10$^{23}$ &       5.4$\times$10$^{-11}$  &    7.0$\times$10$^{-3}$ & \\
DCO$^+$              &     7.4$\times$10$^{11}$ &     13 &      26 &     $-$0.5$\pm$      0.1  &      3.5 $\pm$      0.3 &       1.4$\times$10$^{23}$ &       5.2$\times$10$^{-12}$  &    3.7$\times$10$^{-4}$ & \\
DCN                  &     7.0$\times$10$^{12}$ &     19 &      29 &     $-$0.4$\pm$      0.2  &      4.6 $\pm$      0.1 &       1.3$\times$10$^{23}$ &       5.4$\times$10$^{-11}$  &    3.5$\times$10$^{-3}$ & \\
DNC                  &  $\leq$ 7.0$\times$10$^{11}$ &     13 &      29 &     $-$0.8  &      3.6  &       1.3$\times$10$^{23}$ &    $\leq$ 5.4$\times$10$^{-12}$  & $\leq$    3.5$\times$10$^{-4}$ & (2) \\
HDCS                 &     2.5$\times$10$^{13}$ &     27 &      14 &     $-$0.8$\pm$      0.1  &      4.0 $\pm$      1.0 &       2.6$\times$10$^{23}$ &       9.6$\times$10$^{-11}$  &    1.3$\times$10$^{-2}$ & \\
HDS                  &  $\leq$ 9.0$\times$10$^{13}$ &     38 &       6 &     -0.0  &      6.0  &       5.7$\times$10$^{23}$ &    $\leq$ 1.6$\times$10$^{-10}$  & $\leq$    4.5$\times$10$^{-2}$ & (3)  \\
N$_2$D$^+$             &  $\leq$ 9.0$\times$10$^{10}$ &     10 &      40 &     -0.9  &      5.0  &       9.4$\times$10$^{22}$ &    $\leq$ 9.6$\times$10$^{-13}$  & $\leq$    4.5$\times$10$^{-5}$ & (3)  \\
CH$_2$DOH            &  $\leq$ 1.0$\times$10$^{13}$ &     20 &      35 &     $-$0.4  &      3.0  &       1.1$\times$10$^{23}$ &    $\leq$ 9.4$\times$10$^{-11}$  & $\leq$    5.0$\times$10$^{-3}$ & \\
CH$_3$OD             &  $\leq$ 4.0$\times$10$^{12}$ &     20 &      35 &     $-$0.4  &      3.0  &       1.1$\times$10$^{23}$ &    $\leq$ 3.7$\times$10$^{-11}$  & $\leq$    2.0$\times$10$^{-3}$ & \\
o-H$_2$S             &     2.2$\times$10$^{16}$ (4.5$\times$10$^{16}$) &     38 &       6 & -  & - &       5.7$\times$10$^{23}$ &       3.9$\times$10$^{-8}$  &    1.1$\times$10$^{1}$ & (1) \\
p-H$_2$S             &     4.0$\times$10$^{15}$ (8.2$\times$10$^{15}$) &     38 &       6 & -  & - &       5.7$\times$10$^{23}$ &       7.1$\times$10$^{-9}$  &    2.0$\times$10$^{0}$ & (1) \\
o-H$_2\,^{34}$S      &     1.8$\times$10$^{15}$ &     38 &       6 &     $-$0.0$\pm$      0.5  &      7.1 $\pm$      2.1 &       5.7$\times$10$^{23}$ &       3.2$\times$10$^{-9}$  &    9.0$\times$10$^{-1}$ & \\
p-H$_2\,^{34}$S      &     3.3$\times$10$^{14}$ &     38 &       6 & -  & - &       5.7$\times$10$^{23}$ &       5.8$\times$10$^{-10}$  &    1.7$\times$10$^{-1}$ & (5)  \\
CS                   &     2.3$\times$10$^{15}$ (4.8$\times$10$^{15}$) &     26 &      15 & -  & - &       2.4$\times$10$^{23}$ &       9.5$\times$10$^{-9}$  &    1.1$\times$10$^{0}$ & (1) \\
C$^{34}$S            &     1.9$\times$10$^{14}$ &     26 &      15 &     $-$0.2$\pm$      0.1  &      4.7 $\pm$      0.2 &       2.4$\times$10$^{23}$ &       7.8$\times$10$^{-10}$  &    9.5$\times$10$^{-2}$ & \\
C$^{33}$S            &     5.2$\times$10$^{13}$ &     25 &      15 &     $-$0.2$\pm$      0.1  &      5.8 $\pm$      1.4 &       2.4$\times$10$^{23}$ &       2.1$\times$10$^{-10}$  &    2.6$\times$10$^{-2}$ & \\
$^{13}$CS            &     1.0$\times$10$^{14}$ &     26 &      15 &     $-$0.3$\pm$      0.1  &      5.0 $\pm$      0.9 &       2.4$\times$10$^{23}$ &       4.1$\times$10$^{-10}$  &    5.0$\times$10$^{-2}$ & \\
$^{13}$C$^{34}$S     &     8.2$\times$10$^{12}$ &     26 &      15 &     $-$0.6$\pm$      0.1  &      5.7 $\pm$      0.3 &       2.4$\times$10$^{23}$ &       3.4$\times$10$^{-11}$  &    4.1$\times$10$^{-3}$ & \\
o-H$_2$CS            &     1.3$\times$10$^{14}$ &     22 &      30 &     $-$0.3$\pm$      0.2  &      4.7 $\pm$      0.6 &       1.2$\times$10$^{23}$ &       1.0$\times$10$^{-9}$  &    6.5$\times$10$^{-2}$ & \\
p-H$_2$CS            &     5.0$\times$10$^{13}$ &     27 &      25 &     $-$0.3$\pm$      0.4  &      5.1 $\pm$      0.9 &       1.5$\times$10$^{23}$ &       3.4$\times$10$^{-10}$  &    2.5$\times$10$^{-2}$ & \\
H$_2$C$^{34}$S       &     1.6$\times$10$^{13}$ &     18 &      18.5 &     $-$0.5$\pm$      0.1  &      3.5 $\pm$      0.8 &       2.0$\times$10$^{23}$ &       8.1$\times$10$^{-11}$  &    8.0$\times$10$^{-3}$ & \\
$^\dag$NS                   &     4.4$\times$10$^{14}$ &     15 &      10 &     $-$0.6$\pm$      0.1  &      5.7 $\pm$      1.3 &       3.6$\times$10$^{23}$ &       1.2$\times$10$^{-9}$  &    2.2$\times$10$^{-1}$ & \\
SO, T=10K            &     5.9$\times$10$^{14}$ &     14 &      30 &     $-$0.2$\pm$      0.3  &      6.1 $\pm$      0.6 &       1.2$\times$10$^{23}$ &       4.7$\times$10$^{-9}$  &    2.9$\times$10$^{-1}$ & \\
SO, T=40K            &     2.9$\times$10$^{15}$ &     48 &       6 &     $-$0.2$\pm$      0.1  &      6.6 $\pm$      0.5 &       5.7$\times$10$^{23}$ &       5.1$\times$10$^{-9}$  &    1.4$\times$10$^{0}$ & \\
$^{34}$SO            &     3.2$\times$10$^{13}$ &     14 &      30 &     $-$0.5$\pm$      0.1  &      6.3 $\pm$      0.5 &       1.2$\times$10$^{23}$ &       2.6$\times$10$^{-10}$  &    1.6$\times$10$^{-2}$ & \\
$^{34}$SO            &     4.5$\times$10$^{14}$ &     48 &       6 &     $-$0.1$\pm$      0.2  &      6.9 $\pm$      0.6 &       5.7$\times$10$^{23}$ &       8.0$\times$10$^{-10}$  &    2.3$\times$10$^{-1}$ & \\
HCS$^+$              &     3.6$\times$10$^{13}$ &     26 &      15 &     $-$0.3$\pm$      0.2  &      3.9 $\pm$      0.3 &       2.4$\times$10$^{23}$ &       1.5$\times$10$^{-10}$  &    1.8$\times$10$^{-2}$ & \\
SO$^+$               &     4.6$\times$10$^{13}$ &     28 &      15 &     $-$0.5$\pm$      0.6  &      6.2 $\pm$      2.6 &       2.4$\times$10$^{23}$ &       1.9$\times$10$^{-10}$  &    2.3$\times$10$^{-2}$ & \\
NS$^+$               &     $\leq$ 2.0$\times$10$^{12}$ &     22 &      15 &     $-$0.1$\pm$      0.7  &      2.7 $\pm$      0.2 &       2.4$\times$10$^{23}$ &     $\leq$   8.2$\times$10$^{-12}$  &   $\leq$  1.0$\times$10$^{-3}$ & (2) \\
OCS                  &     3.3$\times$10$^{15}$ &     38 &       7.5 &     $-$0.4$\pm$      0.1  &      3.6 $\pm$      0.5 &       4.6$\times$10$^{23}$ &       7.1$\times$10$^{-9}$  &    1.6$\times$10$^{0}$ & \\
SO$_2$               &     1.7$\times$10$^{15}$ &     43 &       7.5 &     $-$0.1$\pm$      0.3  &      7.3 $\pm$      1.1 &       4.6$\times$10$^{23}$ &       3.7$\times$10$^{-9}$  &    8.5$\times$10$^{-1}$ & \\
SiO                  &     6.0$\times$10$^{13}$ &     15 &      25 & -  & - &       1.5$\times$10$^{23}$ &       4.0$\times$10$^{-10}$  &    3.0$\times$10$^{-2}$ & (1) \\
$^{29}$SiO           &     1.6$\times$10$^{12}$ &     15 &      25 &     $-$0.4$\pm$      0.2  &      6.1 $\pm$      1.2 &       1.5$\times$10$^{23}$ &       1.1$\times$10$^{-11}$  &    8.0$\times$10$^{-4}$ & \\
$^{30}$SiO           &     8.5$\times$10$^{11}$ &     15 &      25 &      0.3$\pm$      0.1  &      5.3 $\pm$      0.9 &       1.5$\times$10$^{23}$ &       5.7$\times$10$^{-12}$  &    4.2$\times$10$^{-4}$ & \\
CH$_3$OH,$\varv_{\rm t}=0$     &     2.0$\times$10$^{15}$ &     23 &      35 &     $-$0.6$\pm$      0.2  &      3.4 $\pm$      2.0 &       1.1$\times$10$^{23}$ &       1.9$\times$10$^{-8}$  &    1.0$\times$10$^{0}$ &\\
$^{13}$CH$_3$OH      &     7.5$\times$10$^{13}$ &     23 &      35 &     $-$0.4$\pm$      0.8  &      4.0 $\pm$      1.4 &       1.1$\times$10$^{23}$ &       6.8$\times$10$^{-10}$  &    3.8$\times$10$^{-2}$ & (6)  \\
CH$_3$CHO            &     6.9$\times$10$^{13}$ &     25 &      18 &     $-$0.4$\pm$      0.6  &      4.6 $\pm$      1.4 &       2.0$\times$10$^{23}$ &       3.4$\times$10$^{-10}$  &    3.5$\times$10$^{-2}$ & \\
CH$_3$OCHO           &  $\leq$ 9.0$\times$10$^{14}$ &     25 &      18 &     $-$0.0  &      4.0  &       2.0$\times$10$^{23}$ &    $\leq$ 4.4$\times$10$^{-9}$  & $\leq$    4.5$\times$10$^{-1}$ &  (3)  \\
CH$_3$OCH$_3$        &     3.8$\times$10$^{14}$ &     27 &      14 &     $-$0.5$\pm$      0.2  &      5.8 $\pm$      1.0 &       2.6$\times$10$^{23}$ &       1.5$\times$10$^{-9}$  &    1.9$\times$10$^{-1}$ & \\
CH$_3$CN             &     5.0$\times$10$^{13}$ &     40 &      15 &     $-$0.4$\pm$      0.1  &      5.5 $\pm$      0.8 &       2.4$\times$10$^{23}$ &       2.1$\times$10$^{-10}$  &    2.5$\times$10$^{-2}$ & \\
C$_2$H$_5$CN         &  $\leq$ 7.0$\times$10$^{13}$ &      26 &      16 &     $-$0.4  &      7.5  &       2.3$\times$10$^{23}$ &    $\leq$  3.1$\times$10$^{-10}$  & $\leq$    3.5e$\times$10$^{-2}$ & (3)  \\
C$_2$H$_3$CN         &  $\leq$ 2.0$\times$10$^{14}$ &     26 &      16 &      0.0  &      7.5  &       2.3$\times$10$^{23}$ &    $\leq$ 8.8$\times$10$^{-10}$  & $\leq$    1.0$\times$10$^{-1}$ & (3)  \\
HC(O)NH$_2$          &  $\leq$ 9.0$\times$10$^{12}$ &     26 &      16 &      0.0  &      4.0  &       2.3$\times$10$^{23}$ &    $\leq$ 3.9$\times$10$^{-11}$  & $\leq$    4.5e$\times$10$^{-3}$ & (3)  \\
\hline
\multicolumn{6}{l}{Warm component of the envelope} \\
\hline
CH$_3$CCH            &  $\leq$ 1.0$\times$10$^{15}$ &     63 &       3 &      0.0  &      7.0  &       1.0$\times$10$^{24}$ &    $\leq$ 9.9$\times$10$^{-10}$  & $\leq$    2.4$\times$10$^{-2}$ &  (3)  \\
HC$_3$N              &     4.2$\times$10$^{14}$ &     69 &       4 &     $-$0.2$\pm$      0.3  &      9.7 $\pm$      0.7 &       8.0$\times$10$^{23}$ &       5.2$\times$10$^{-10}$  &    1.0$\times$10$^{-2}$ & \\
HC$_3$N              &     6.5$\times$10$^{15}$ &    130 &       0.8 &      0.3$\pm$      0.3  &     10.9 $\pm$      1.2 &       2.1$\times$10$^{24}$ &       3.1$\times$10$^{-9}$  &    1.6$\times$10$^{-1}$ & \\
HC$_3$N,$\varv_{\rm 7}=1$            &     3.0$\times$10$^{16}$ &    128 &       1 &      0.3$\pm$      1.1  &     12.6 $\pm$      2.3 &       2.0$\times$10$^{24}$ &       1.5$\times$10$^{-8}$  &    7.3$\times$10$^{-1}$ & \\
CCS                  &  $\leq$ 2.0$\times$10$^{14}$ &     63 &       3 &      0.0  &      7.0  &       1.0$\times$10$^{24}$ &    $\leq$ 2.0$\times$10$^{-10}$  & $\leq$    4.9$\times$10$^{-3}$ &  (3)  \\
C$_2$O               &  $\leq$ 4.0$\times$10$^{14}$ &     63 &       3 &      0.0  &      7.0  &       1.0$\times$10$^{24}$ &    $\leq$ 4.0$\times$10$^{-10}$  & $\leq$    9.8$\times$10$^{-3}$ &  (3)  \\
C$_3$H               &  $\leq$ 2.0$\times$10$^{14}$ &     63 &       3 &      0.0  &      7.0  &       1.0$\times$10$^{24}$ &    $\leq$ 2.0$\times$10$^{-10}$  & $\leq$    4.9$\times$10$^{-3}$ &  (3)  \\
C$_4$H               &  $\leq$ 2.0$\times$10$^{15}$ &     63 &       3 &      0.0  &      7.0  &       1.0$\times$10$^{24}$ &    $\leq$ 2.0$\times$10$^{-9}$  & $\leq$    4.9$\times$10$^{-2}$ &  (3)  \\
o-H$_2$CO            &     4.0$\times$10$^{15}$ &    130 &       4 &     $-$0.4$\pm$      0.1  &      7.9 $\pm$      0.5 &       8.0$\times$10$^{23}$ &       5.0$\times$10$^{-9}$  &    9.8$\times$10$^{-2}$ & \\
p-H$_2$CO            &     3.0$\times$10$^{14}$ &    100 &       7 &     $-$0.3$\pm$      0.2  &      6.0 $\pm$      0.5 &       4.9$\times$10$^{23}$ &       6.1$\times$10$^{-10}$  &    7.3$\times$10$^{-3}$ & \\
o-H$_2$CCO           &     3.1$\times$10$^{14}$ &     62 &       7 &     $-$0.3$\pm$      0.8  &      8.3 $\pm$      1.9 &       4.9$\times$10$^{23}$ &       6.3$\times$10$^{-10}$  &    7.6$\times$10$^{-3}$ & \\
HNC                  &     6.9$\times$10$^{14}$ &     95 &       3 & -  & - &       1.0$\times$10$^{24}$ &       6.9$\times$10$^{-10}$  &    1.7$\times$10$^{-2}$ & (1) \\
HN$^{13}$C           &     3.0$\times$10$^{13}$ &     95 &       3 &     $-$0.8$\pm$      0.3  &      6.8 $\pm$      1.0 &       1.0$\times$10$^{24}$ &       3.0$\times$10$^{-11}$  &    7.3$\times$10$^{-4}$ & \\
HCN                  &     7.6$\times$10$^{15}$ &     95 &       3 & -  & - &       1.0$\times$10$^{24}$ &       7.5$\times$10$^{-9}$  &    1.9$\times$10$^{-1}$ & (1) \\
H$^{13}$CN           &     3.3$\times$10$^{14}$ &     95 &       3 &     $-$0.4$\pm$      0.1  &      8.9 $\pm$      0.2 &       1.0$\times$10$^{24}$ &       3.3$\times$10$^{-10}$  &    8.0$\times$10$^{-3}$ & \\
HC$^{15}$N           &     1.2$\times$10$^{14}$ &     95 &       3 &     $-$0.3$\pm$      0.1  &      9.3 $\pm$      0.2 &       1.0$\times$10$^{24}$ &       1.2$\times$10$^{-10}$  &    2.9$\times$10$^{-3}$ & \\
HNCO                 &     2.7$\times$10$^{16}$ &     51 &       1.5 &      0.1$\pm$      0.5  &     10.1 $\pm$      1.5 &       1.6$\times$10$^{24}$ &       1.7$\times$10$^{-8}$  &    6.6$\times$10$^{-1}$ & \\
HDO                  &     3.0$\times$10$^{17}$ &     86 &       0.6 &      0.0$\pm$      0.1  &      8.6 $\pm$      2.2 &       2.3$\times$10$^{24}$ &       1.3$\times$10$^{-7}$  &    7.3$\times$10$^{0}$ & \\
CH$_2$DOH            &  $\leq$ 8.0$\times$10$^{14}$ &     78 &       5 &      0.4  &      9.2 &       6.6$\times$10$^{23}$ &    $\leq$ 1.2$\times$10$^{-9}$  & $\leq$    2.0$\times$10$^{-2}$ &  (3)  \\
CH$_3$OD             &  $\leq$ 6.0$\times$10$^{14}$ &     78 &       5 &      0.4  &      9.2  &       6.6$\times$10$^{23}$ &    $\leq$ 9.1$\times$10$^{-10}$  & $\leq$    1.5$\times$10$^{-2}$ &  (3)  \\
o-H$_2$S             &     4.8$\times$10$^{16}$ &    166 &       1 &     $-$0.5$\pm$      0.2  &      9.1 $\pm$      0.3 &       2.0$\times$10$^{24}$ &       2.4$\times$10$^{-8}$  &    1.2$\times$10$^{0}$ & \\
o-H$_2$CS            &     9.2$\times$10$^{15}$ &    100 &       2 &     $-$0.4$\pm$      0.6  &      7.8 $\pm$      0.6 &       1.3$\times$10$^{24}$ &       6.8$\times$10$^{-9}$  &    2.2$\times$10$^{-1}$ & \\
p-H$_2$CS            &     3.0$\times$10$^{14}$ &     90 &       5 &     $-$0.1$\pm$      0.2  &      7.3 $\pm$      1.0 &       6.6$\times$10$^{23}$ &       4.5$\times$10$^{-10}$  &    7.3$\times$10$^{-3}$ & \\
OCS                  &     1.2$\times$10$^{17}$ &     97 &       1.4 &     $-$0.3$\pm$      0.2  &      8.3 $\pm$      0.9 &       1.7$\times$10$^{24}$ &       7.2$\times$10$^{-8}$  &    2.9$\times$10$^{0}$ & \\
SO$_2$               &     1.6$\times$10$^{17}$ &    128 &       1 &      0.1$\pm$      0.5  &      8.3 $\pm$      1.1 &       2.0$\times$10$^{24}$ &       8.1$\times$10$^{-8}$  &    3.9$\times$10$^{0}$ & \\
CH$_3$OH,$\varv_{\rm t}=0$      &     4.1$\times$10$^{16}$ &     74 &       4 &     $-$0.4$\pm$      0.7  &      8.8 $\pm$      1.6 &       7.3$\times$10$^{23}$ &       5.7$\times$10$^{-8}$  &    1.0$\times$10$^{0}$ &\\
${}^{13}$CH$_3$OH      &     8.0$\times$10$^{14}$ &     74 &       4 &     $-$0.4$\pm$      0.8  &     10.0 $\pm$      1.4 &       7.3$\times$10$^{23}$ &       1.1$\times$10$^{-9}$  &    2.0$\times$10$^{-2}$ &  (6)  \\
CH$_3$CHO            &  $\leq$ 3.0$\times$10$^{15}$ &    100 &       1.5 &     $-$0.4  &      4.6  &       1.6$\times$10$^{24}$ &    $\leq$ 1.9$\times$10$^{-9}$  & $\leq$    7.3$\times$10$^{-2}$ &  (3)  \\
CH$_3$OCHO           &     2.0$\times$10$^{16}$ &     86 &       2 &     $-$0.9$\pm$      1.1  &     10.7 $\pm$      1.5 &       1.3$\times$10$^{24}$ &       1.5$\times$10$^{-8}$  &    4.9$\times$10$^{-1}$ & \\
CH$_3$OCH$_3$        &     3.5$\times$10$^{16}$ &     80 &       1.5 &     $-$0.4$\pm$      0.4  &      8.8 $\pm$      1.0 &       1.6$\times$10$^{24}$ &       2.2$\times$10$^{-8}$  &    8.5$\times$10$^{-1}$ & \\
CH$_3$SH             &     1.5$\times$10$^{15}$ &     69 &       5 &     $-$0.6$\pm$      0.2  &      7.0 $\pm$      0.7 &       6.6$\times$10$^{23}$ &       2.3$\times$10$^{-9}$  &    3.7$\times$10$^{-2}$ & \\
g-C$_2$H$_5$SH             &  $\leq$ 1.0$\times$10$^{16}$ &    100 &       1.5 &      0.0  &      8.0  &       1.6$\times$10$^{24}$ &    $\leq$ 6.2$\times$10$^{-9}$  & $\leq$    2.4$\times$10$^{-1}$ & (3) \\
CH$_3$CN             &     1.5$\times$10$^{16}$ &    200 &       1 &     $-$0.2$\pm$      0.3  &      9.5 $\pm$      0.8 &       2.0$\times$10$^{24}$ &       7.6$\times$10$^{-9}$  &    3.7$\times$10$^{-1}$ & \\
C$_2$H$_5$CN         &     5.5$\times$10$^{16}$ &     90 &       0.8 &     $-$1.0  &      9.6 $\pm$      3.0 &       2.1$\times$10$^{24}$ &       2.6$\times$10$^{-8}$  &    1.3$\times$10$^{0}$ & \\
C$_2$H$_3$CN         &     5.3$\times$10$^{15}$ &    100 &       1.7 &      0.1$\pm$      0.6  &      9.7 $\pm$      2.1 &       1.5$\times$10$^{24}$ &       3.5$\times$10$^{-9}$  &    1.3$\times$10$^{-1}$ & \\
HC(O)NH$_2$          &     1.0$\times$10$^{15}$ &     83 &       1.5 &      1.0$\pm$      1.0  &      8.1 $\pm$      1.7 &       1.6$\times$10$^{24}$ &       6.2$\times$10$^{-10}$  &    2.4$\times$10$^{-2}$ & \\
\hline
\multicolumn{6}{l}{Shock 1}\\
\hline
HDO                  &     1.0$\times$10$^{17}$ &    170 &       0.5 &     $-$4.8$\pm$      0.7  &      2.9 $\pm$      0.8 &       2.0$\times$10$^{24}$ &       4.9$\times$10$^{-8}$  &    1.7$\times$10$^{-1}$ & (4) \\
CH$_3$OH,$\varv_{\rm t}=0$      &     6.0$\times$10$^{17}$ &    190 &       1.2 &     $-$4.8$\pm$      0.8  &      2.9 $\pm$      0.8 &       2.0$\times$10$^{24}$ &       2.9$\times$10$^{-7}$  &    1.0$\times$10$^{0}$ &\\
$^{13}$CH$_3$OH      &     5.0$\times$10$^{16}$ &    190 &       1.2 &     $-$4.3$\pm$      0.8  &      2.9 $\pm$      0.8 &       2.0$\times$10$^{24}$ &       2.5$\times$10$^{-8}$  &    8.3$\times$10$^{-2}$ &  (6)  \\
CH$_3$OH,$\varv_{\rm t}=1$           &     5.0$\times$10$^{17}$ &    170 &       1.2 &     $-$4.3$\pm$      0.7  &      3.9 $\pm$      0.7 &       2.0$\times$10$^{24}$ &       2.5$\times$10$^{-7}$  &    8.3$\times$10$^{-1}$ & \\
CH$_3$OCHO           &     8.0$\times$10$^{16}$ &    170 &       0.9 &     $-$4.8$\pm$      0.7  &      2.9 $\pm$      0.8 &       2.0$\times$10$^{24}$ &       3.9$\times$10$^{-8}$  &    1.3$\times$10$^{-1}$ & (4)  \\
CH$_3$OCH$_3$        &     2.0$\times$10$^{17}$ &    176 &       0.7 &     $-$4.8$\pm$      0.7  &      2.9 $\pm$      0.8 &       2.0$\times$10$^{24}$ &       9.8$\times$10$^{-8}$  &    3.3$\times$10$^{-1}$ & \\
HC(O)NH$_2$          &     3.0$\times$10$^{15}$ &    170 &       0.8 &     $-$4.8$\pm$      0.7  &      2.9 $\pm$      0.8 &       2.0$\times$10$^{24}$ &       1.5$\times$10$^{-9}$  &    5.0$\times$10$^{-3}$ & (4) \\
\hline
\multicolumn{6}{l}{Shock 2} \\
\hline
HDO                  &     1.0$\times$10$^{17}$ &    170 &       0.5 &      2.0$\pm$      0.8  &      4.6 $\pm$      1.3 &       1.9$\times$10$^{24}$ &       5.3$\times$10$^{-8}$  &    1.5$\times$10$^{-1}$ & (4) \\
CH$_3$OH,$\varv_{\rm t}=0$      &     6.6$\times$10$^{17}$ &    190 &       1.2 &      2.0$\pm$      0.8  &      5.5 $\pm$      1.3 &       1.9$\times$10$^{24}$ &       3.5$\times$10$^{-7}$  &    1.0$\times$10$^{0}$ &\\
$^{13}$CH$_3$OH      &     6.0$\times$10$^{16}$ &    190 &       1.2 &      4.0$\pm$      0.8  &      5.5 $\pm$      1.3 &       1.9$\times$10$^{24}$ &       3.2$\times$10$^{-8}$  &    9.1$\times$10$^{-2}$ & (6)  \\
CH$_3$OH,$\varv_{\rm t}=1$           &     7.5$\times$10$^{17}$ &    170 &       1.2 &      4.0$\pm$      0.7  &      5.5 $\pm$      0.7 &       1.9$\times$10$^{24}$ &       4.0$\times$10$^{-7}$  &    1.1$\times$10$^{0}$ & \\
CH$_3$OCHO           &     9.0$\times$10$^{16}$ &    170 &       0.9 &      2.0$\pm$      0.8  &      4.6 $\pm$      1.3 &       1.9$\times$10$^{24}$ &       4.8$\times$10$^{-8}$  &    1.4$\times$10$^{-1}$ & (4) \\
CH$_3$OCH$_3$        &     9.0$\times$10$^{16}$ &    176 &       0.7 &      2.0$\pm$      0.8  &      4.6 $\pm$      1.3 &       1.9$\times$10$^{24}$ &       4.8$\times$10$^{-8}$  &    1.4$\times$10$^{-1}$ & \\
HC(O)NH$_2$          &     4.0$\times$10$^{15}$ &    170 &       0.8 &      2.0$\pm$      0.8  &      4.6 $\pm$      1.3 &       1.9$\times$10$^{24}$ &       2.1$\times$10$^{-9}$  &    6.1$\times$10$^{-3}$ & (4)  \\
\hline
\end{longtable}
\tablefoot{(1): Since our results suggest lower isotopic ratios for $^{12}$C/$^{13}$C or $^{32}$S/$^{34}$S, we use the values of 30 and 12, respectively (see Sect. \ref{sec:iso_ratios}), but between parentheses, we indicate the values using the canonical values of 60 and 25. (2) Upper limit is based on a tentative detection.  $^\dag$ the HFS procedure was used for the line fitting. (3): Upper limit is based on a 3$\sigma$ noise level. We fixed the following parameters: size, excitation temperature, line width, and velocity offset.  (4), (5), (6): We used the temperature determined for CH$_3$OH,$\varv_{\rm t}=0$, CH$_3$OH,$\varv_{\rm t}=0$, o-H$_2\,^{34}$S (respectively). }

}
\setlength{\tabcolsep}{6pt}

\subsection{Outflow contribution}
The line profiles of several molecules show high velocity emission. The most pronounced, highest velocity wings are seen in CO, but SiO, H$_2$S, HCO$^+$, HNC, HCN, and CS also indicate non-Gaussian line-wings that are likely associated with the high-velocity outflowing gas. A prominent bipolar molecular outflow has been imaged in the CO $J=3-2$ line by ALMA \citep{Csengeri2018}. 
Owing to their high optical depth, many of these molecules exhibit self-absorbed line profiles. For these molecules, the column densities and the temperature were determined from the rarer isotopologues. 
It is beyond the scope of this work to estimate the properties of the gas associated with the outflowing material. In the following, we therefore omit the outflow components from the LTE modelling and fit a single Gaussian to the line to take only the emission from the envelope into account.

\section{Structure of the \mysou\ protostar} \label{sec:source_structure}
The protostar embedded in the G328.2551-0.5321 clump has been shown to be dominated by a single massive fragment down to a size scale of $\sim$400\,au \citep{Csengeri2018}. Therefore, we assumed that all the emission detected towards this source originates from a single protostellar envelope that is spatially unresolved within the beam sizes of our observations (0.2 to 0.5\,pc at the distance of the source).

\subsection{Observational constraints}
In Sect.\,\ref{sec:fitting} we identified broad and narrow line profiles, outflow wings, and for some species, multiple (typically two) velocity components. This suggests that although it remains spatially unresolved in our observations, the protostellar envelope has an internal structure in which the physical parameters may change.

\begin{figure}
    \centering
    \includegraphics[width=0.9\linewidth]{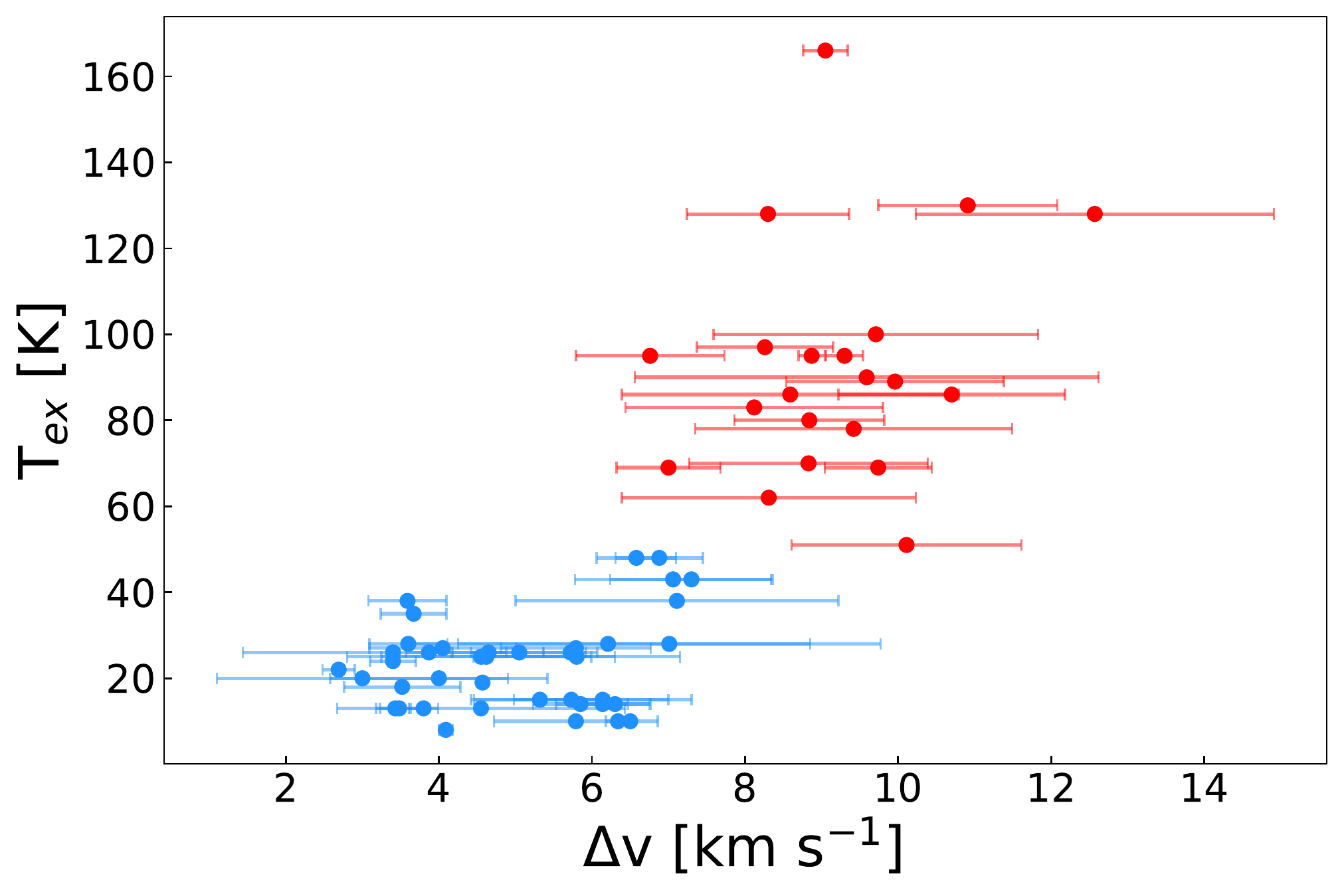}
    \caption{Excitation temperature vs. mean line width. A statistical study of K-Means \citep{Macqueen1967} was performed to determine the limit between the different components of the envelope. The blue data points represent the cold component, and the warm component of the envelope is represented by the red data points. Single transitions fitted with multiple velocity components are considered as individual measurements in this figure.}
    \label{fig:TempVsLineWidth}
\end{figure}

From the LTE fitting of simple species (Sect.\,\ref{LTEmodelsection}), we derived the distribution of measured line widths versus $T_{\rm ex}$ (Fig. \ref{fig:TempVsLineWidth}). We show the results of a $K$-means clustering analysis \citep{Macqueen1967}, which  shows a preference for two groups of line widths corresponding to a low-temperature and more quiescent gas ($T_{\rm ex}<$50\,K, $\Delta v<6$\kms), and a warmer component with typically broader line widths (50\,K$<T_{\rm ex}<$150\,K, 6\kms$<\Delta v<12$\kms). For simplicity, we do not include in Fig. \ref{fig:TempVsLineWidth} the contribution from the accretion shocks, which correspond to high $T_{\rm ex}>$120\,K and exhibit $\Delta v<3-4$\kms. This is discussed in more detail in Sect.\,\ref{sec:acc_shocks}. The broader velocity dispersion close to the protostar can be explained by a combination of larger thermal broadening, a higher level of micro-turbulence due to the mechanical feedback of the protostellar embryo, and to other kinematic effects such as infall and rotation. The contribution of thermal broadening to the line width is estimated by $\sigma_{th}=\sqrt{k_B T/(\mu m_H)}$, where $k_B$ is the Boltzmann constant, $T$ is the gas temperature, $\mu$ is the molecular mass, and $m_H$ is the mass of the hydrogen atom. This gives a thermal velocity dispersion of 0.07\,\kms\ for the second lightest molecule, CCH, in the cold gas and 0.17\,\kms\ for HCN in the warm gas. When $\sigma_{th}$ is converted into the FWHM of a Gaussian line using FWHM = $2\sqrt{2 ln 2} \sigma_{th}$, the thermal broadening is still below the 0.7\,\kms velocity resolution of the final data product. The thermal broadening has only a minor effect in this case. We do not discuss the relative contribution of these effects here.

\begin{figure}
    \centering
    \includegraphics[width=0.9\linewidth]{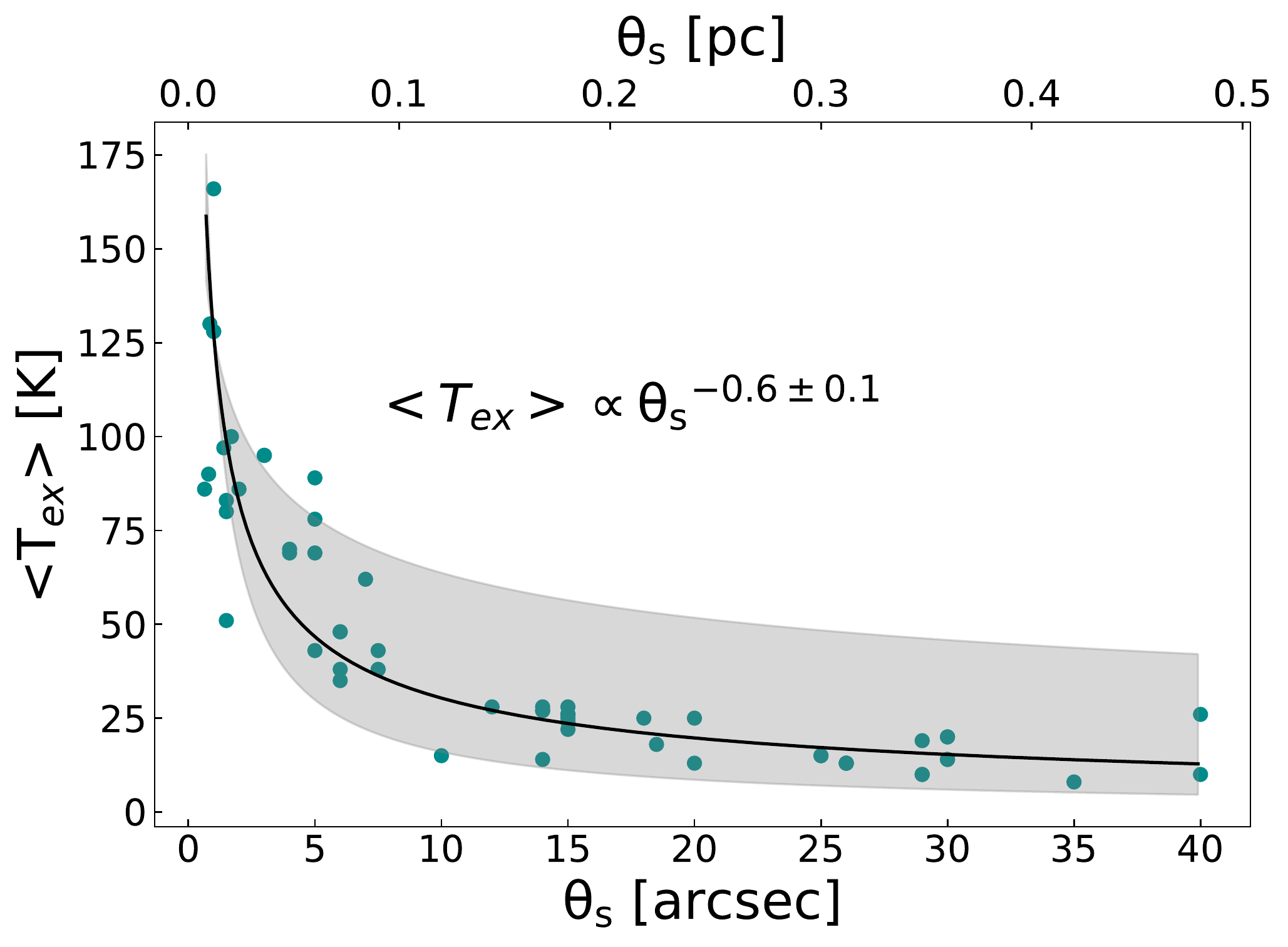}
    \caption{Excitation temperature vs. the size of the emission region for each molecule whose emission arises from the envelope. The black curve represents the power-law fit of the data. The grey area indicates power-law temperature profiles with an exponent between $-$0.3 and $-$0.9. This profile does not include the emission originating from the accretion shocks that correspond to a local increase in temperature. It also excludes species for which no strong constraints on the size of the emitting region could be derived (see Sect.\,\ref{sec:exceptions}).}
    \label{fig:TempVsSize}
\end{figure}

The LTE fitting also provides the distribution of the obtained $T_{\rm ex}$ as a function of the size of the emitting region (Fig.\,\ref{fig:TempVsSize}). Although here <$T_{\rm ex}$> corresponds to a spatially averaged excitation temperature over the size of the emission ($\theta_{\rm s}$), it
is expected to represent the temperature structure of the source because under LTE conditions, $T_{\rm ex}$ is equivalent to the kinetic temperature. We find that even this spatially averaged excitation temperature exhibits a rather gradual decrease in temperature with the radius. We fit a power-law distribution of $<T_{\rm ex}>(\theta_{\rm s})\propto \theta_{\rm s}^{\beta}$, and find a $\beta$ of $-0.6$. This result is intriguing because we did not consider any \textsl {\textup{a priori}} temperature structure in our line fitting and decomposition analysis.

Although we observe a rather gradual decrease in the temperature, we can distinguish two physical components for the bulk emission of the envelope based on
the line-width analysis. Therefore, we discuss in the following the warm component of the envelope (together with the accretion shocks) within a typical size of 0.06\,pc and $T_{\rm ex}>$50\,K and the cold component of the envelope, which extends up to a size of 0.5\,pc and exhibits $T_{\rm ex}<$50\,K. 

\subsection{Toy model for the source structure}\label{sec:toy_model}

\begin{figure}
    \centering
    \includegraphics[width=0.9\linewidth]{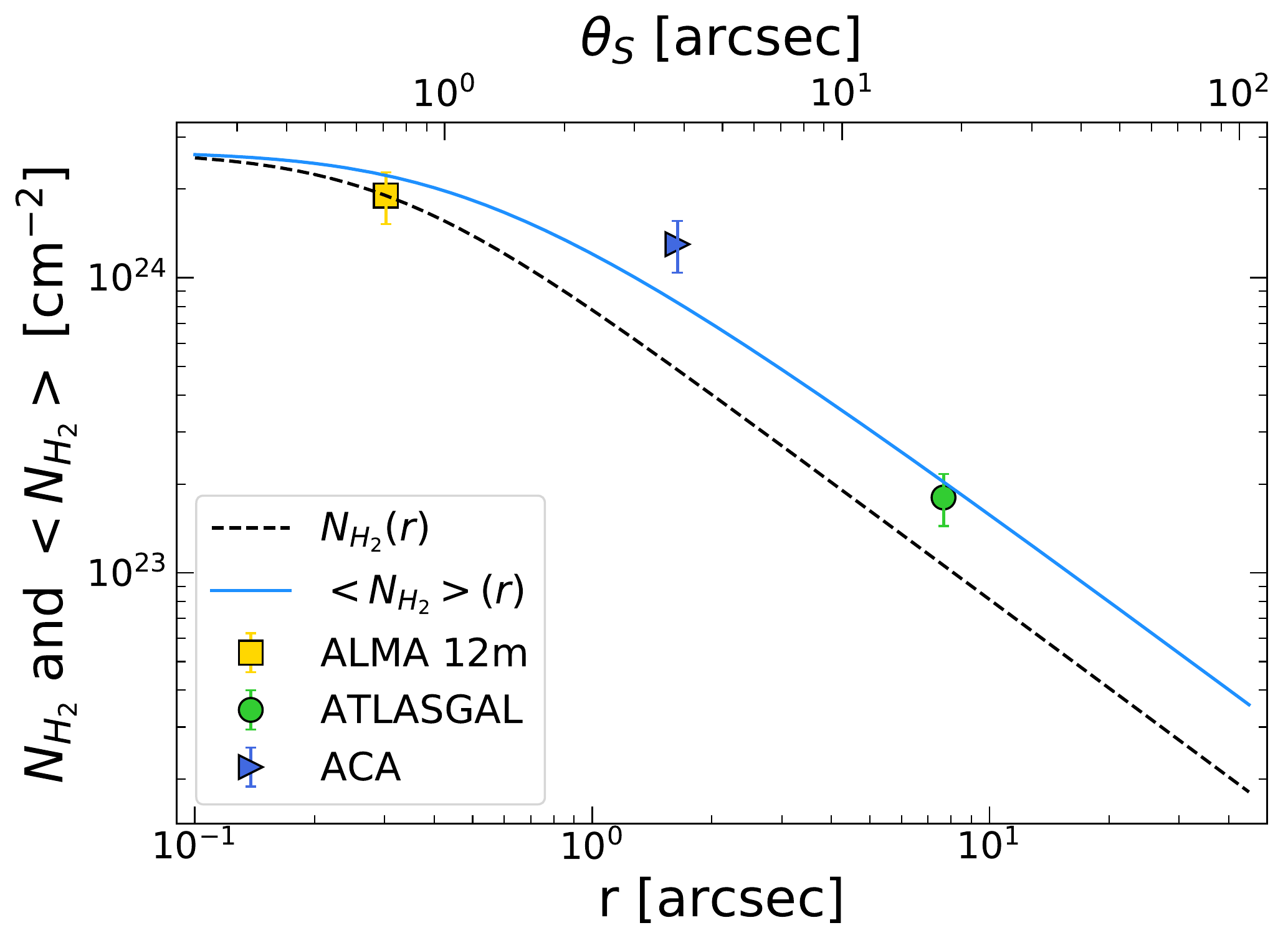}
    \caption{Column density profile (dashed black line) and averaged column density (blue line) vs. radius (r). The blue triangle shows the averaged column density estimate from ACA array observations from \citet{Csengeri2017}. The green circle corresponds to the beam-averaged column density estimated from ATLASGAL \citep{Csengeri2014}. The yellow square shows the column density estimates roughly at the position of the accretion shocks with ALMA \citep{Csengeri2019}. The error bars indicate a 20\% error on these measurements.}
    \label{fig:Nh2}
\end{figure}

To constrain the physical origin and the abundance variations of the molecules detected in this survey, we used a toy model (see Fig.\,\ref{fig:Nh2}) to describe the physical components within the envelope. \citet{Csengeri2018, Csengeri2019} showed that in the close vicinity of the protostellar embryo, a picture emerges that is qualitatively similar to that observed for some low-mass protostars, with a compact accretion disk and shocks at the centrifugal barrier. These components close to the vicinity of the protostar are confined to a compact region of $<$1\arcsec\ that is considerably smaller than our beam, hence we did not model it in detail. For simplicity, we also omit here the discussion of the high-velocity emission originating from the outflowing gas and fast shocks, and instead focus our discussion on the bulk of the envelope and its physical properties. 

\citet{Csengeri2018} reported that the source structure can be fitted with a power-law profile and a compact source, the latter likely corresponding to an accretion disk. Such a profile is consistent with  the results of \citet{Csengeri2019}, who reported similar values of H$_2$ column densities at three positions with a projected distance of $\leq$1000\,au from the protostar, suggesting a flattened inner region. Therefore, we describe the H$_2$ volume density as a function of radius using a Plummer-like function, which corresponds to a flattened density profile towards the innermost regions, and a power-law profile in the outer regions. We used a Plummer-like column density profile of 
\begin{equation}
    N_{\rm H_2}(r)=N_{\rm 0}\times(1+r^2/a^2)^{-1/2}
\end{equation} (Fig.\,\ref{fig:Nh2}), where $a$ and $N_{\rm 0}$ are the Plummer radius and the normalisation factor (respectively). We chose $a$=750\,au, and computed $N_{\rm 0}$ such that the observed $N_{\rm H_2}$ at radius $a$ is consistent with the observed value of $N_{\rm H_2}(a)=$1.9$\times10^{24}$\,cm$^{-2}$ from \citet{Csengeri2019}. This corresponds to a density profile following the form of $n\propto r^{-2}$ at $r\geq a$.
The molecular column density estimates provided by the LTE modelling with Weeds correspond to a  certain size of the emitting region ($\theta_{\rm s}$). To be able to infer molecular abundances relative to H$_2$, we therefore computed the average  H$_2$ column density ($<N_{\rm H_2}>$) for this corresponding size. We computed an average H$_2$ column density,$<N_{\rm H_2}>$, as a function of physical (and angular) radius, r, and $\theta_s$, respectively, as shown in Fig.\,\ref{fig:Nh2}. Weeds assumes a Gaussian brightness distribution for the source, where the size of the emitting region, $\theta_{\rm s}$, corresponds to the $FWHM$ of the Gaussian \citep{Maret2011}. 
We related our source radius, $r$ to $\theta_{\rm s}$ using $r=\theta_s/2.35$.
In Table \ref{tab:LTE_values}, we indicate the source size, $\theta_s$ for which the averaged H$_2$ column density is calculated.

In Fig.\,\ref{fig:Nh2} we compare the average H$_2$ column density of our model to the estimates of the H$_2$ column density from APEX/LABOCA, and ALMA data from the literature. We used a dust emissivity of 0.0185\,cm$^{2}$\,g$^{-1}$ at 345\,GHz (which already includes a gas-to-dust ratio of 100) to compute the H$_2$ column density from the flux density using the standard formulae. Taking the beam-averaged flux density values from the ATLASGAL survey \citep{Csengeri2014} with the temperature of $\sim$22\,K estimated in \citet{Csengeri2018} for the cold gas component, we estimate a beam averaged column density of $N_{\rm H_2}$= 1.8$\times10^{23}$\,cm$^{-2}$ within the APEX beam at 345\,GHz of 19\rlap{.}{\arcsec}2. We performed the same calculations taking the ALMA 7m array measurements from \citet{Csengeri2017} and estimate an averaged column density of $N_{\rm H_2}$= 1.3$\times10^{24}$\,cm$^{-2}$ within the ALMA 7\,m synthesised beam. These estimates are broadly consistent with the $N_{\rm H_2}$ profile we used.

\section{Molecular composition: Simple molecules}\label{sec:light}
\begin{figure*}
    \centering
    \includegraphics[trim=0cm 2.5cm 0cm 3cm,clip,width=0.9\textwidth]{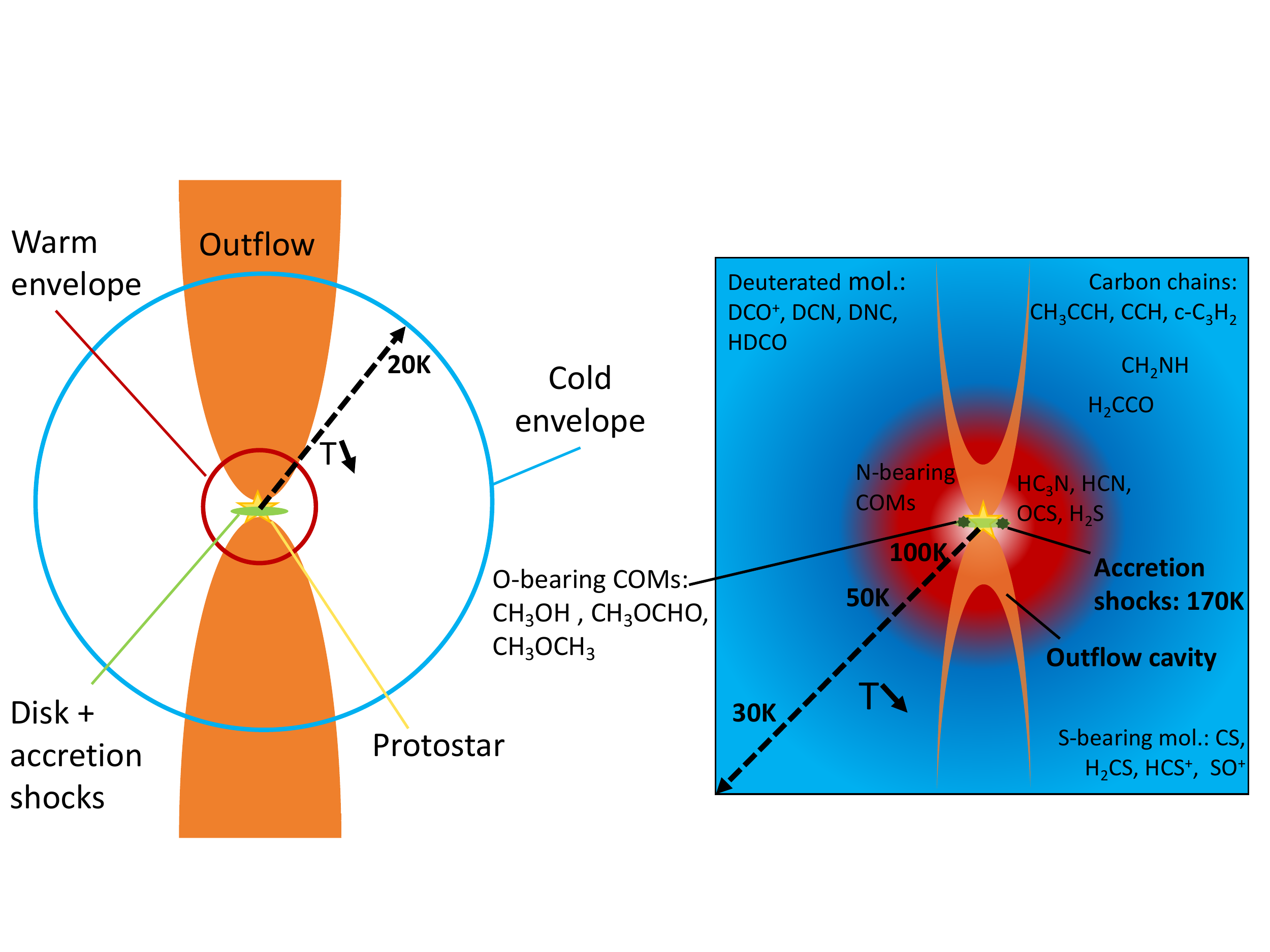}
    \caption{Sketch of \mysou\ ignoring projection effects. The full source structure is represented on the left, and we zoom into the envelope of the protostar on the right.}
    \label{fig:sketch_g328}
\end{figure*}

This unbiased spectral line survey allows us to detect 29 species, 54 including isotopologues, which are simple molecules, that is, have fewer than six atoms. We associate their origin with the physical components of the envelope derived above and investigate the potential chemical origin of these species. In particular, we discuss here the properties of carbon-chain molecules, O-, N-, and S- bearing molecules, and deuterated species in the cold and warm components of the envelope. The sketch of the molecular composition of the various physical components of the envelope is shown in Fig.\,\ref{fig:sketch_g328}.

\begin{figure*}
    \centering
    \includegraphics[width=1\linewidth]{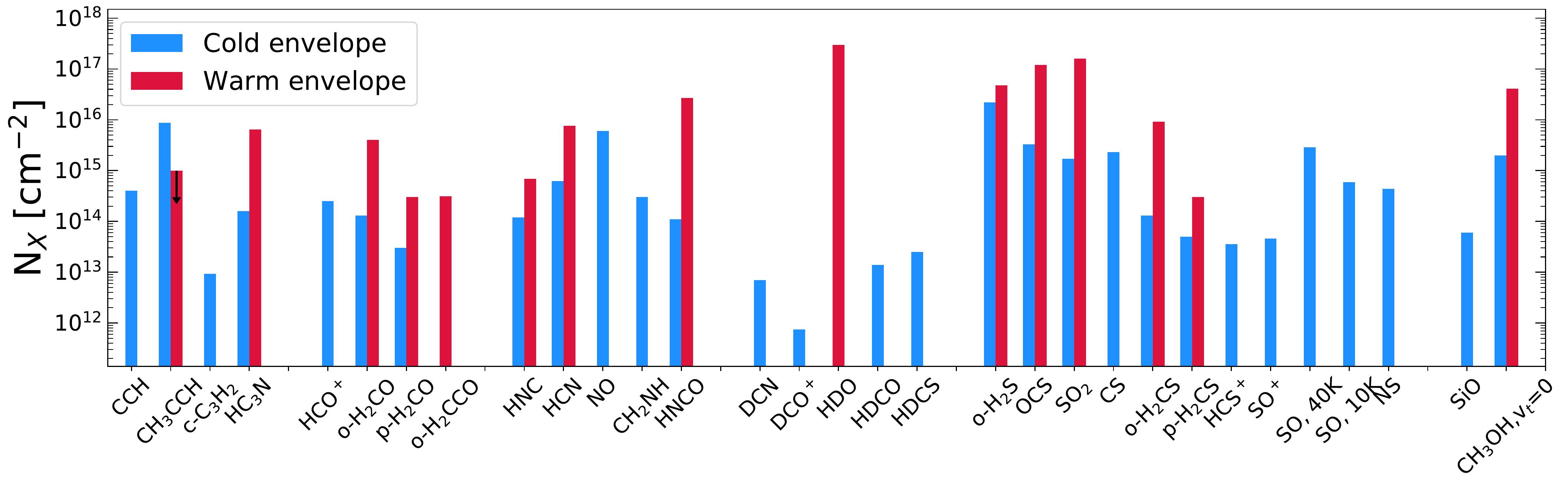}
    \includegraphics[width=1\linewidth]{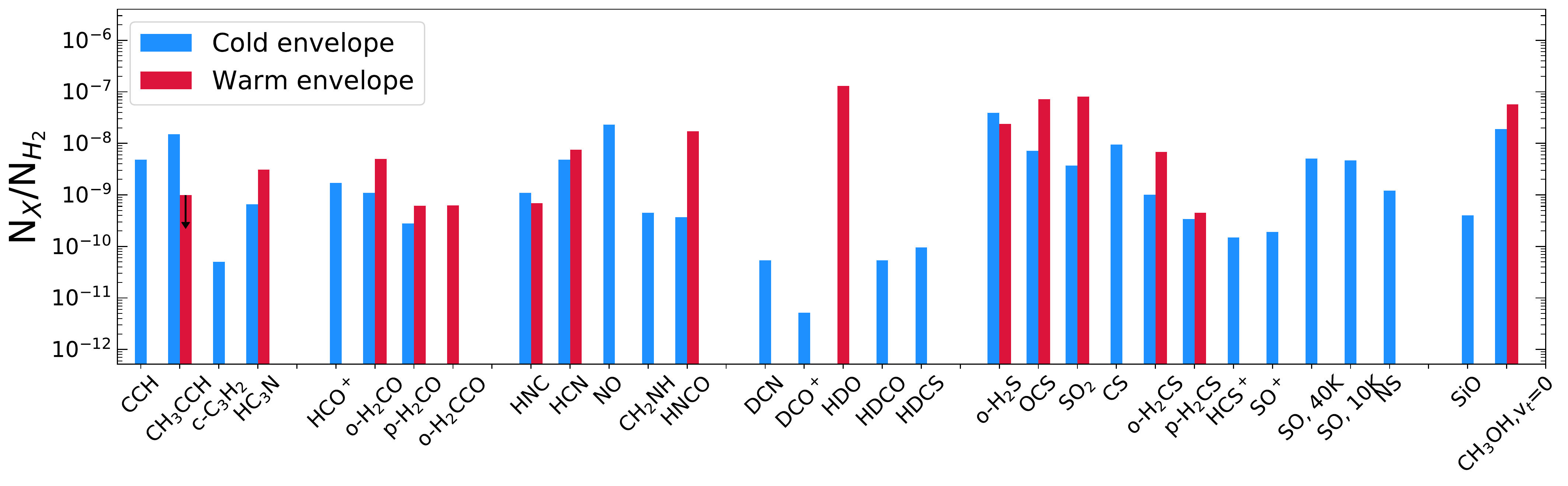}
    \includegraphics[width=1\linewidth]{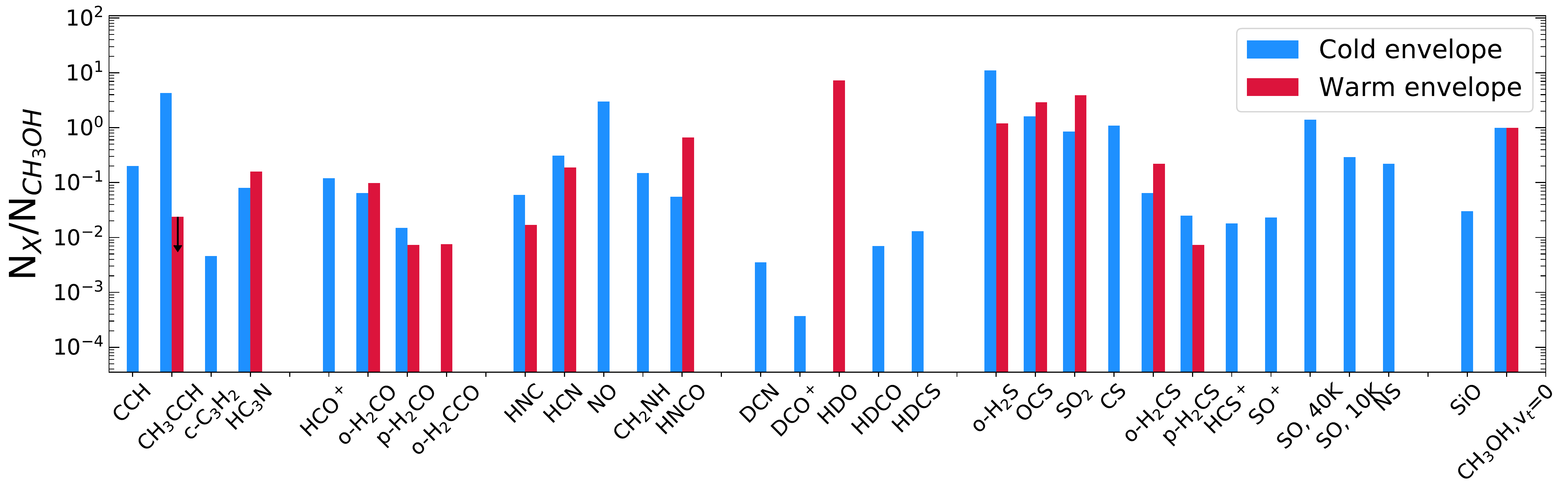}
    \caption{Column densities (upper panel), abundances relative to H$_2$ (middle panel), and relative column densities to methanol (lower panel) of the light molecules and methanol. Only the main isotopologues are represented.}
    \label{fig:N_small}
\end{figure*}

\subsection{Sulphur chemistry}
\begin{figure}
    \centering
    \includegraphics[width=1\linewidth]{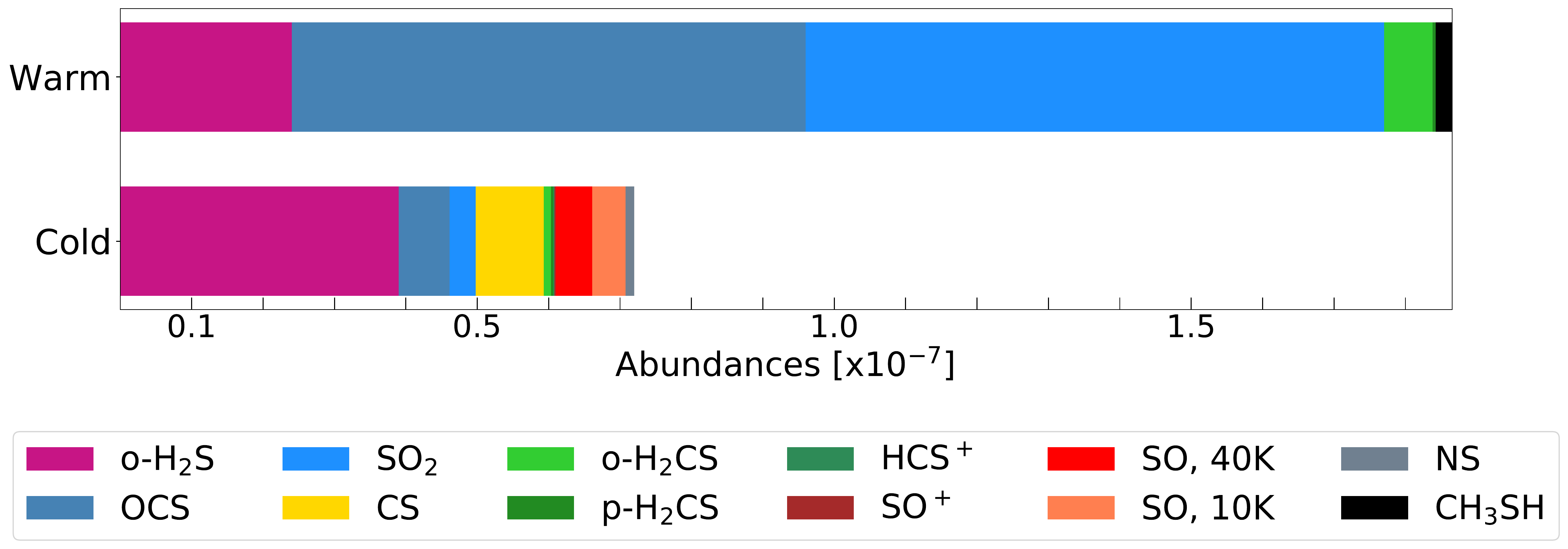}
    \caption{Abundances of the S-bearing molecules relative to H$_2$ in the cold and warm phases. Each molecule is represented by one colour.}
    \label{fig:S_percentage_CvsW}
\end{figure}

\subsubsection{Richness in sulphur-bearing species}\label{sec:S}

One of the most remarkable results concerning the molecular composition of \mysou\ is the detection of a large number of sulphur-bearing molecules (ten species plus eight isotopologues) with high column densities and abundances. In addition to the most abundant S-bearing molecules in star-forming regions, such as SO, SO$_2$, OCS, H$_2$S, CS, and H$_2$CS \citep[e.g][]{Hatchell1998a, Li2015, Weaver2017}, we detect here several transitions from  SO$^+$, HCS$^+$, NS, and CH$_3$SH and show a tentative detection of NS$^+$.
We list the estimated parameters ($T_{\rm{kin}}$, $N$, and $\theta_{\rm s}$) from the LTE modelling (Sect.\,\ref{LTEmodelsection}) in Table\,\ref{tab:LTE_values} and show the column densities and abundances of these species in the cold and warm components of the envelope in Fig.\,\ref{fig:N_small}.

The undepleted sulphur abundance in the diffuse interstellar medium has been found to be around S/H$\sim(3.5\pm1.5)\times10^{-6}$ \citep{Goicoechea2006}. In dark clouds and hot cores, however, the total abundance of S-bearing molecules represents only 0.1\% of the cosmic value \citep{Tieftrunk1994, Charnley1997}. It might be argued that sulphur is present in its atomic form and thus cannot be observed. However, this assumption does not hold in hot core models \citep{Vidal2017}. Towards \mysou,\ the highest molecular column densities are found to be up to $10^{17}$\,cm$^{-2}$ in the warm component of the envelope, and abundances relative to H$_2$ as high as 8$\times$10$^{-8}$ for OCS and SO$_2$ are reached. In the cold gas, the highest column density is found for H$_2$S with a value of 2.6$\times$10$^{16}$\unidens\ and an abundance relative to H$_2$ of 4.6$\times$10$^{-8}$. These high molecular abundances of S-bearing species correspond to $\geq$2\,\% of the cosmic value in the warm component of the envelope and to $\geq$1\,\% of the cosmic value in the cold component of the envelope. Therefore, this object has a less pronounced sulphur depletion than is commonly observed in the dense cold gas. However, high abundances of H$_2$S have been observed towards some high-mass star-forming regions, such as the Orion hot core \citep{Blake1996}, and high values of SO and SO$_2$ were observed in W3 IRS5 \citep{Wang2013}.

Similarly to other species, the sulphur-bearing species show different line profiles, suggesting that these molecules are present in various physical components of the envelope. For example, a line profile with a redshifted self-absorption suggesting infall motions \citep[e.g.][]{Myers1995} is observed in the ground-state transition of ortho-H$_2$S. Broad, non-Gaussian line-wings suggest that CS, SO, and SO$_2$ are also present in the outflowing gas. We associate the Gaussian, relatively narrow velocity components to the bulk of the envelope. 

Based on the kinetic temperatures estimated from the LTE modelling, we distinguish between an origin in the cold or warm components of the envelope. For temperatures below 70\,K, sulphur-bearing molecules might desorb through non-thermal processes, and gas-phase processes dominate the chemical reactions \citep{vanderTak2003, Wakelam2011}. We find that this temperature corresponds rather well to the separation between the cold and warm components of the envelope. The largest variety of sulphuretted molecules (nine, plus eight isotopologues) is detected towards the cold component of the envelope, while in the warm component, we identify only five molecules (Fig.\,\ref{fig:S_percentage_CvsW}). In general, we find higher column densities and molecular abundances in the warm component than in the cold gas, which is consistent with the expected more efficient thermal desorption at higher temperatures. It is very intriguing that 
S-bearing molecules such as OCS, H$_2$S, SO$_2$, CS, and SO exhibit abundances as high as methanol, suggesting a particular enhancement of sulphur chemistry in this object. In particular, methanol is only one order of magnitude more abundant than methanethiol with a ratio CH$_3$OH/CH$_3$SH of 25, while in the well-studied hot core Sgr~B2(N2) \citep{Muller2016}, the methanol abundance relative to H$_2$ is two orders of magnitude higher and the CH$_3$OH/CH$_3$SH ratio is one order of magnitude higher. Towards another classical hot core associated with G327.3$-$0.6, the CH$_3$OH/CH$_3$SH ratio also appears higher, with a value of 60 \citep{Gibb2000}. In the following, we therefore investigate the sulphur-bearing species in the cold and warm components of the envelope in more detail.

\subsubsection{Sulphuretted species in the cold component of the envelope}
The velocity profiles suggest that both the S-bearing ions and the neutral molecules originate from the same gas that is associated with the cold quiescent envelope. Other studies based on mapping of sulphuretted molecules suggest, however, that despite a similar line width compared to quiescent gas tracers (e.g. H$^{13}$CO$^+$, c-C$_3$H$_2$, and HDCO), for example, SO and SO$_2$ do not necessarily originate from the undisturbed gas \citep[e.g.][]{Chernin1994, Bachiller2001,BuckleFuller2002}. Although these species may chemically form purely via gas-phase reactions in the undisturbed gas \citep[e.g.][]{Pratap1997,Dickens2000}, we cannot exclude a contribution from the outflow cavity walls and thus from the disturbed gas \citep{Codella2013} even for the narrow line-width component.
The excitation temperatures in the cold component of the envelope are all around 10--48\,K for sulphur-bearing molecules, and we find a cold-gas component at a temperature of 10\,K traced by SO, $^{34}$SO, and NS. Most of the molecules, such as H$_2$CS, CS, HCS$^+$, and SO$^+$, show a temperature of 25--30\,K, while we also trace a warmer component at about $\sim$40\,K with SO, H$_2$S, SO$_2$, and OCS.

The highest column density in the cold gas is observed for H$_2$S ($2.6\times10^{16}$\,cm$^{-2}$) and corresponds to a molecular abundance relative to H$_2$ of $4.6\times10^{-8}$. This is a significantly higher abundance than the median values calculated for hot cores and hot corinos \citep[e.g.][]{Li2015}, suggesting that a large fraction of sulphur is released into the gas phase here. It is, however, close to the value reported towards Orion KL \citep{Blake1996, Esplugues2014, Crockett2014}. The second most abundant species are CS and OCS, followed by SO and SO$_2$, while the lowest abundances are at the order of a few times $10^{-10}$ , corresponding to molecular ions, such as HCS$^+$ and  SO$^+$. The abundances of S-bearing molecules found in the cold component of \mysou\ are higher than the median values found in hot cores and hot corinos \citep[e.g.][]{Charnley1997,Hatchell1999,Li2015}.

\subsubsection{Evidence for shock chemistry in the cold gas?}\label{sec:shocks}

We also detect the reactive molecular ion SO$^+$ in the cold gas phase with an abundance relative to H$_2$ of 1.9$\times$10$^{-10}$. This molecule has been detected in the interstellar medium in various environments \citep{Turner1992, Turner1994}; for example towards photodissociation regions (PDRs) \citep{Fuente2003}, and star-forming regions as well \citep{Podio2014, Nagy2015}. Towards {\mysou,} we do not detect other species that would be characteristic of PDR chemistry, however, such as HOC$^+$, or CO$^+$ suggesting that the origin of SO$^+$ could be rather related  to ion-neutral chemistry or dissociative shocks \citep{Turner1994}. Further evidence for this scenario is the abundance ratio of $X$(SO$^+$)/$X$(SO$_2$)$\sim$0.05, which  is below the value of 0.4--1 measured for PDRs \citep{Fuente2003, Ginard2012}, and is similar to values determined for other high-mass star-forming regions \citep{Nagy2015}. 
Gas-phase chemical models predict a high $X$(SO$^+$)/$X$(SO) ratio of $\sim$1 for dense and ionised regions \citep{Ginard2012}. Our value of  0.04 is similar to the 0.06--1 found towards MonR2 \citep{Ginard2012} and cold dense cores \citep{Agundez2019} and translucent clouds \citep{Turner1995, Turner1996}. 
\citet{Neufeld1989} have shown that the release and ionisation of sulphur into the gas phase through shocks leads to an enhancement of SO$^+$ with abundances relative to H$_2$ reaching 10$^{-10}$--10$^{-9}$. Shock chemistry could therefore explain the higher abundance of SO$^+$ towards \mysou.

Similarly, HCS$^+$, H$_2$CS, and CS have abundances of about 10$^{-10}$, 10$^{-9}$ , and 10$^{-8}$, respectively, which are higher than what is observed around high-mass protostars \citep{Hatchell1998a,vanderTak2003,Li2015} and are similar to what is observed in shocked regions \citep{Minh1991, Turner1992, Turner1994, Podio2014}. 
Our observations of sulphur-bearing species in the cold component of the envelope therefore suggest a potentially significant impact or contribution from shock chemistry that leads to enhanced abundances of sulphur-bearing molecules.
While the exact origin of shocks is difficult to elucidate because a powerful ejection of material in the form of an outflow has been observed towards this object \citep{Csengeri2018}, the outflow cavity walls, where the velocity profile resembles quiescent gas, might be a possible origin.

\subsubsection{Sulphur in the warm envelope}

In contrast to the cold component of the envelope, the warm gas shows less diversity in sulphur-bearing molecules. We detect SO$_2$, OCS, H$_2$S, H$_2$CS, and CH$_3$SH in decreasing abundance in the warm component of the envelope. The abundances of OCS, SO$_2$, H$_2$CS, and CH$_3$SH relative to H$_2$ are higher in the warm (Fig.\,\ref{fig:S_percentage_CvsW}, upper panel) than in the cold component of the envelope, reaching values as high as $7-8\times10^{-8}$ for SO$_2$ and OCS, which are among the most abundant sulphur-bearing molecules in star-forming regions \citep[e.g.][]{Sakai2010}. Towards typical hot cores, abundances of S-bearing molecules of about 10$^{-8}$ have been reported for SO$_2$, OCS, and CS and abundances of 10$^{-9}$ for H$_2$CS \citep[e.g.][]{Hatchell1998a,vanderTak2003, Herpin2009}. 
The excitation temperatures are similarly high for these molecules, around $\text{}$100\,K, except for H$_2$S, which is at 166\,K. 

Towards Orion KL, the carbon-free sulphur-bearing species show an order-of-magnitude increase in molecular abundances towards the hot regions \citep{Luo2019}. In our case, however, the carbon-sulphur bearing molecule OCS also shows a significant enhancement in molecular abundance towards the warm region. This prevents us from making a similar distinction between the behaviour of carbon-free and carbon-bearing sulphuretted species.

While the main sulphur reservoir on the grains is still 
poorly known, several candidates have been proposed, such as H$_2$S \citep{Charnley1997, Holdship2016, Vidal2017}, SH \citep{Vidal2017}, and/or OCS \citep{Hatchell1998a, vanderTak2003}. On the other hand, SO, SO$_2$, and other simple S-bearing molecules are thought to be formed in the gas phase \citep{Podio2014, Esplugues2014, Holdship2016}. OCS and SO$_2$ have been firmly or tentatively detected in interstellar ices \citep{Palumbo1995, Boogert1997}, but not H$_2$S. 
The high H$_2$S abundances observed in hot cores \citep{Hatchell1998a} and in the present object are inconsistent with the gas-phase formation. Gas-phase models can reproduce observed abundances of $\sim$10$^{-10}$ in the quiescent cold dark cloud TMC-1 \citep{Millar1989}. \citet{Charnley1997} and \citet{Millar1997} suggested that in hot cores, H$_2$S must be produced in significant amount in the grain mantles. When it is released into the cold gas-phase component of the envelope, it leads to the formation of SO and SO$_2$, and ultimately, to CS and H$_2$CS. 
The molecules OCS and SO$_2$ represent the most important reservoir of sulphur among the detected molecules in the warm gas phase, corresponding to 40\,\% each (Fig.\,\ref{fig:S_percentage_CvsW}, upper panel). H$_2$S is the main reservoir in the cold gas phase ($\sim$50\,\%), while it represents only 15\,\% in the warm gas phase. \citet{Chen2015} reported that UV radiation could open a chemical reaction channel that allows the formation of OCS from H$_2$S, which could explain the destruction of H$_2$S in the warm component of the envelope. The weak UV radiation from the protostar could influence the warm component of the envelope without having a significant impact on the cold component of the envelope. Another possible explanation is a reaction on the grain surface of H$_2$S to form ethanethiol (C$_2$H$_5$SH), which was proposed by \citet{Muller2016}. The authors claimed, however, that this reaction has only a minor effect on the H$_2$S abundance. This product is not detected here with an upper limit of 2.7$\times$10$^{-8}$ on its abundance relative to H$_2$. 
\citet{vanderTak2003} argued that OCS is the main sulphur reservoir on the grains, which could naturally provide an explanation for the high abundances of OCS observed in the warm component of the envelope.

\subsubsection{Role of S-bearing species as chemical clocks}

\begin{figure}
    \centering
    \includegraphics[width=0.9\linewidth]{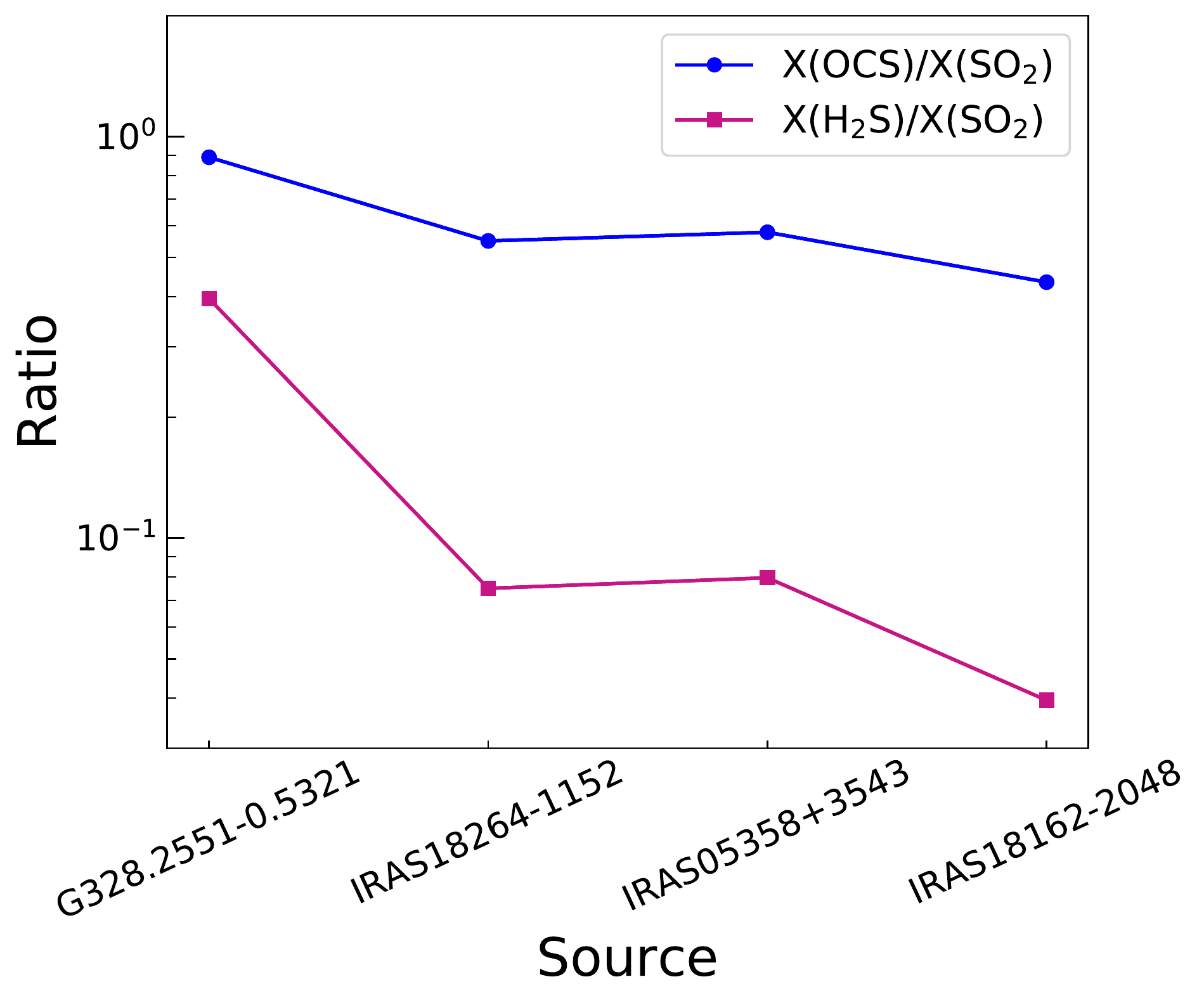}
    \caption{Abundance ratios of OCS and H$_2$S relative to SO$_2$ for each source for the layers at $T$=100\,K for \mysou\ and the sample from \citet{Herpin2009} ordered according to their assumed age sequence, \mysou\ being the youngest and IRAS 18162-2048 the most evolved.}.
    \label{fig:clockS}
\end{figure}

Chemical models predict that the relative abundances of sulphur-bearing species may be used to estimate the ages based on the chemistry of dense clouds \citep{Buckle2003}. 
For this purpose, the abundance ratios of several sulphur-bearing species, for example, OCS/SO$_2$, SO/SO$_2$, and H$_2$S/SO$_2$, have been invoked as tracing the chemical age of the gas \citep[e.g.][]{Hatchell1998a, Herpin2009, Wakelam2011}. Here we investigate these ratios for molecules that we find in the warm component of the envelope, that is, OCS/SO$_2$ and H$_2$S/SO$_2$. The models from \citet{Wakelam2011} predict a decrease in these ratios for sources with more evolved chemistry. In Fig.\,\ref{fig:clockS} we compare our measurements with the values of \citet{Herpin2009}, which were estimated for three sources that are located within a heliocentric distance of 5 kpc. The authors proposed a rough evolutionary sequence from the youngest to the oldest for these sources mainly based on the flux ratio of the hot to cold gas in the spectral energy distribution (SED) of this sample, as of IRAS18264--1152 (infrared quiet dense core with a dynamical age of $\sim$0.5$\times$10$^4$\,yr), IRAS05358+3543 (massive dense core associated with an outflow and a dynamical age of $\sim$3.6$\times$10$^4$\,yr), and IRAS18162--2048 (infrared bright dense core with a powerful outflow and dynamical age of $\sim$10$^6$\,yr). 
This comparison shows moderate variation in the OCS/SO$_2$ ratio, but the ratio of H$_2$S over SO$_2$ is clearly higher towards \mysou, which is\ consistent with a scenario in which it is younger than the other sources.

\subsection{Deuterated molecules}

Deuterium enhancement (deuteration) occurs in cold, dense interstellar clouds, that is, preferentially in prestellar cores \citep{Aikawa2012, Taquet2012, Taquet2014}. It has been invoked as a potential age tracer for pre- and protostellar evolution. 

Towards \mysou, we do not see many deuterated molecules in general, only four deuterated molecules are firmly detected in the cold component of the envelope (DCN, DCO$^+$, HDCO, and HDCS), while in the warm component of the envelope and in the accretion shocks, only HDO is detected. We list their estimated parameters based on LTE modelling (Sect.\,\ref{sec:fitting}) in Table\,\ref{tab:LTE_values} and discuss upper limits for frequently detected deuterated molecules, such as HDS and N$_2$D$^+$, and deuterated methanol in the following.

We derived the deuterium fractionation (D/H ratio) based on the abundance ratios ($X_{\rm XD}/X_{\rm XH}$) of the species for which the non-deuterated form (or one of its rarer isotopologues) is also detected. When the chemical group containing the deuterium atom is composed of several hydrogen atoms, we need to take the different arrangements between H and D within the group into account in order to infer the D/H ratio from the abundance ratios ($X_{\rm XD}/X_{\rm XH}$)  \citep[for more details, see][]{Manigand2019}. In our case, this concerns HDO, HDCO, HDCS, and the upper limits estimated for HDS, where D/H ratio is defined by $(1/2)\times X_{\rm XD}/X_{\rm XH}$. For the upper-limit estimates of CH$_2$DOH, we need to use $\rm{D/H}=(1/3)\times X_{\rm CH_2DOH}/X_{\rm CH_3OH}$. In Table\,\ref{tab:D_ratio} we list the two [XD]/[XH] abundance ratios and the deuterium fractionation (D/H ratio). In the following, we discuss the deuterium fractionation as the D/H ratio.

We find a deuterium fractionation between 0.002 and 0.03 for the detected deuterated molecules in the cold component of the envelope. The strongest deuterium enhancement of 0.03 is seen for HDCS, while DCN is at 0.005--0.01. DCO$^+$ and HDCO exhibit a deuterium fractionation of 0.002--0.003 and 0.02, respectively. The cosmic deuterium abundance is 1.5$\times$10$^{-5}$ \citep{Linsky2003}, hence the deuterium enhancement in general is significant towards this object, although it is typically lower than the values of 0.01--0.3 found for low-mass star-forming regions \citep[e.g.][]{Turner2001,Marcelino2005}.

Deuterium fractionation is efficient in the interstellar medium. The larger proton affinity of D compared to H leads to an increase in the fraction of D in neutrals and ions, but the zero-point energy for D containing molecules is also lower than that of their main isotopologue. One of the main reaction pathways is through the reaction with H$_2$D$^+$ \citep{Millar1989}, which produces for example DCO$^+$ and N$_2$D$^+$ \citep{Aikawa2018}. This reaction is efficient at low temperatures. An alternative route for deuteration to proceed is efficient even at higher temperatures (>60\,K) in the gas phase by reaction with CH$_2$D$^+$ \citep{Turner2001, Roueff2007, Parise2009}. This  chemical pathway indicates that even after the sublimation of CO, deuterium chemistry may remain active. For example both DCN and DCO$^+$ could be formed in an efficient way in the warm component of the envelope \citep{Parise2009}, and they indeed show a high deuteration ratio in the warm gas phase towards hot cores and disks \citep{Ren2012, Zinchenko2012, Favre2015}.
Towards \mysou,\ both DCN and DCO$^+$ are only detected in the cold component of the envelope at low temperatures below 20\,K, favouring a pathway through reaction starting from H$_2$D$^+$.

Similarly, the singly deuterated thioformaldehyde, HDCS, could also form in the gas phase and might even reach a high fractionation of $>$0.02 according to the models of \citet{Minowa1997} and \citet{Marcelino2005}. This is in fact consistent with our observational results. These models also give a similarly high fractionation for HDCO, where our observational results suggest a somewhat higher fractionation of 0.04.

The only deuterated molecule that we identified in the warm component of the envelope is HDO.
While we did not detect H$_2\,^{18}$O, which would allow an estimate of the H$_2$O abundance, our modelling allows us to estimate an upper limit of 1.0$\times$10$^{18}$\unidens\ for H$_2\,^{18}$O in the warm component of the envelope assuming the same size, excitation temperature, and line width as HDO. The canonical $^{16}$O/$^{18}$O ratio of 500 implies a water deuteration of $>$0.0002 towards \mysou. Towards hot cores, this abundance ratio is typically found to be between 5$\times$10$^{-5}$ and 0.0003 \citep{Jacq1990, Comito2003, Liu2013, Coutens2014, Kulczak2016} in the hot envelope and up to 0.001 for Orion-KL \citep{Jacq1990}. Our observations suggest a somewhat stronger enhancement of HDO compared to other high-mass star-forming sites, while it is similar to that of Orion-KL.
HDO can form in the cold gas on the grains (see \citealt{vanDishoeck2013} for a review on its formation) and also by gas-phase reactions at high temperatures or in shocks \citep[e.g.][]{Codella2010}. Water deuteration is therefore strongly linked to physical conditions like the dust temperature during the freeze-out in the quiescent cloud \citep{Cazaux2011}. 

Interestingly, we do not detect the deuterated form of several of the most abundant molecules in the cold envelope, for example H$_2$S and CH$_3$OH. Using the same source sizes, excitation temperatures, and line widths as for their non-deuterated species, we estimate upper limits for the column density of HDS and the deuterated forms of CH$_3$OH (CH$_3$OD and CH$_2$DOH), which gives upper limits of 0.001-0.002 for H$_2$S and 0.002 for CH$_3$OH in the cold gas phase. The upper limits in the warm gas phase for CH$_3$OH are somewhat higher, around 0.007-0.02, and hence are less constraining. The deuterium fractionation for CH$_3$OH in the cold gas is significantly lower than typically observed towards low-mass star-forming regions \citep[e.g.][]{Awad2014}. Interestingly, there are, however, examples of high-mass star-forming regions with considerably higher deuteration. For example \citet{Fontani2015} detected deuterated methanol with a deuteration ratio of $\sim$0.01. 

In the warm gas phase, our upper limit of deuteration is comparable to the threshold for these forms of deuterated methanol found towards the position of source N2 in Sgr(B2) with [XD]/[XH]= 0.0012 \citep{Belloche2016} based on tentative detections of CH$_2$DOH lines. This corresponds to a deuterium fractionation (D/H ratio) of 0.0004 considering the different arrangements between H and D atoms. For the deuterium fractionation of CH$_3$OD towards SgrB2(N), an upper limit of $\leq$7$\times$10$^{-4}$  was estimated, hence the non-detection of a deuterated form of methanol towards \mysou\ is not unexpected. This low deuterium fractionation is a clear difference to the hot corino IRAS16293--2422 \citep{Jorgensen2018}.

\setcounter{table}{4}
\begin{table*}
\centering
\small

\caption{Deuteration ratios in \mysou.}
{\vspace{0.1cm}}
\label{tab:D_ratio}
\begin{tabular}{l l l c c c c}  
\hline 
\hline
Deuterated  & Isotopologue & Envelope & \mysou & \mysou & TMC--1 & IRAS16293--2422 \\        
molecule &  & & X$_{\rm XD}$/X$_{\rm XH}$ & D/H ratio\tablefootmark{a,b} & X$_{\rm XD}$/X$_{\rm XH}$\tablefootmark{c} & X$_{\rm XD}$/X$_{\rm XH}$\tablefootmark{d} \\        
\hline 
DCN        & H$^{13}$CN     & Cold  & ~~~0.01 (0.005) & ~~~~~0.01 (0.005) & 0.008 & 0.013   \\
DCO$^+$    & H$^{13}$CO$^+$ & Cold  & ~~\,0.003 (0.002)      & ~~\,0.003 (0.002)      & 0.013 & 0.009     \\
HDS        & H$_2\,^{34}$S  & Cold  & $\leq$0.004 (0.002)    & $\leq$0.002 (0.001)    & - & 0.1      \\
HDCS       & H$_2$CS        & Cold  & ~~~~~0.07        & ~~~~~0.03        & 0.02 & -    \\
HDO        & H$_2\,^{18}$O  & Warm  & $\geq$1$\times$10$^{-3}$ (4$\times$10$^{-4}$)    & $\geq$5$\times$10$^{-4}$ (2$\times$10$^{-4}$)    & - & 0.25  \\
HDCO       & H$_2$CO        & Cold  & ~~~~~0.04        & ~~~~~0.02        & 0.05 & 0.14    \\
N$_2$D$^+$ & N$_2$H$^+$     & Cold  & $\leq$0.002     & $\leq$0.002     & 0.08 & 0.25\\
CH$_2$DOH  & CH$_3$OH       & Cold  & $\leq$0.005     & $\leq$0.002     & - & - \\
CH$_2$DOH  & CH$_3$OH       & Warm  & ~~~$\leq$0.02    & $\leq$0.007     & 0.012 & 0.3  \\
CH$_3$OD   & CH$_3$OH       & Cold  & $\leq$0.002     & $\leq$0.002     & - & -\\
CH$_3$OD   & CH$_3$OH       & Warm  & ~~~$\leq$0.02    & ~~~$\leq$0.02    & - & 0.02 \\
\hline
\end{tabular}
\tablefoot{Values in parentheses correspond to the uncertainty of the molecular abundances of the main species due to the two different isotopic ratios used for $^{12}$C/$^{13}$C and $^{32}$S/$^{34}$S (see Sect.\,\ref{sec:iso_ratios}).
\tablefoottext{a}{The value of D/H is corrected for the number of equivalent hydrogen atoms in the molecule (see Appendix B in \citet{Manigand2019}).}
\tablefoottext{b}{This work.}
\tablefoottext{c}{\citet{Butner1995,Minowa1997,Turner2001,Markwick2002}}
\tablefoottext{d}{\citet{vanDishoeck1995, Roueff2000, Parise2004, Coutens2013}.}
}
\end{table*}

We also estimate an upper limit for the deuterium fractionation of H$_2$S, which is found to be about 0.001-0.002. The models show a low deuteration of H$_2$S \citep{Hatchell1999, Awad2014}, where the molecule is formed at slightly warmer temperatures on the grain surfaces. This formation route and low deuterium fractionation therefore agrees with our measurements. 

\begin{figure}
    \centering
    \includegraphics[width=0.9\linewidth]{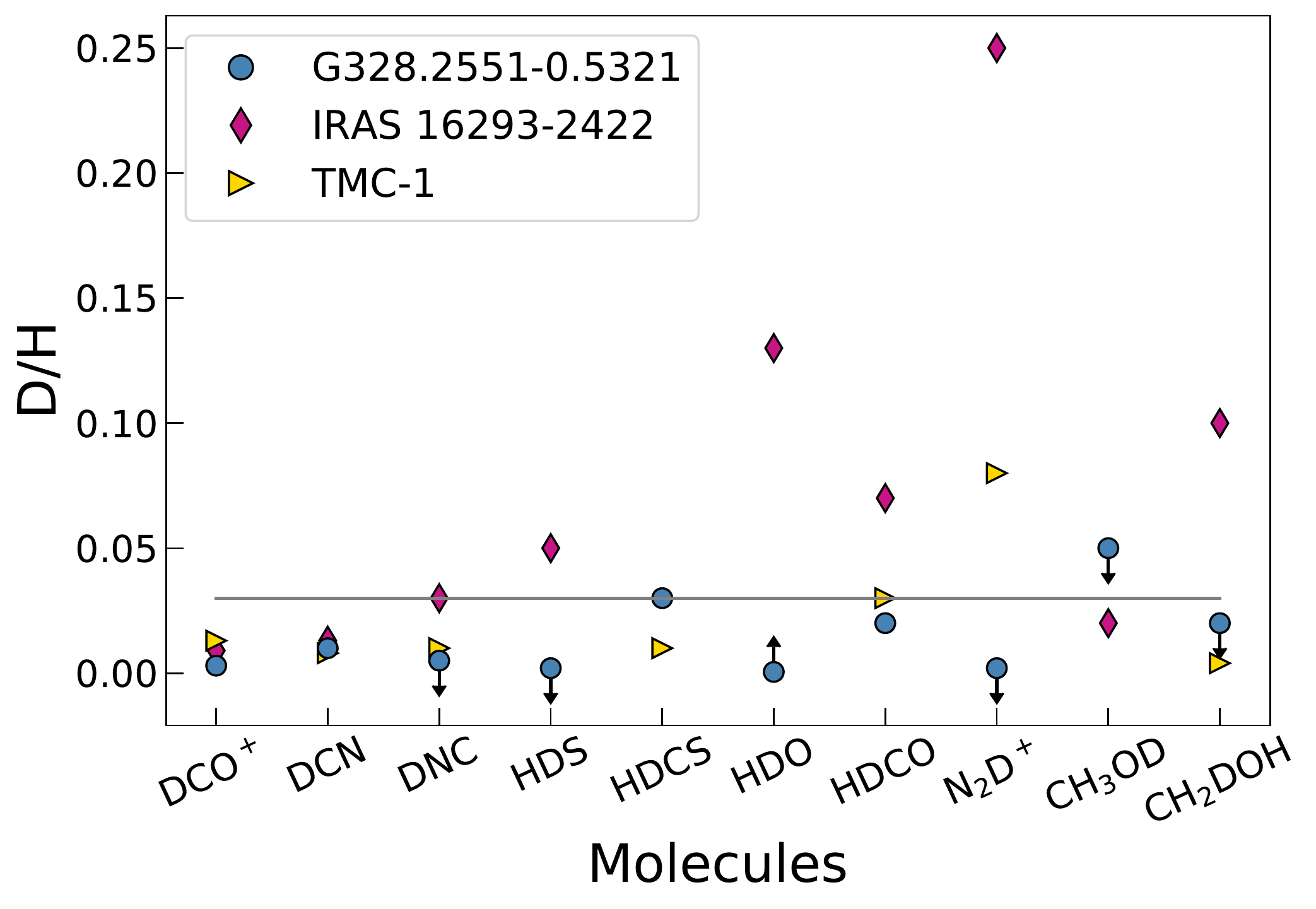}
    \caption{Comparison of the deuteration in \mysou\ (blue circles) and  the dark cloud TMC1 (yellow triangles) and the hot corino IRAS16293--2422 (pink diamonds). See the caption of Table \ref{tab:D_ratio} for the details about the references. The grey line represents the maximum deuteration in \mysou.}
    \label{fig:deut}
\end{figure}

Figure\,\ref{fig:deut} shows a comparison of the D/H ratio of deuterated molecules for \mysou\ and the hot corino, IRAS16293--2422, and the TMC-1 low-mass core. We find that the deuterium fractionation of several molecules (HDS, HDCO, N$_2$D$^+$, and CH$_2$DOH) in the cold envelope of \mysou\ is lower than in IRAS16293--2422, but it is similar to what is observed towards TMC-1 \citep{Butner1995, Minowa1997, Turner2001, Markwick2002}. Other high-mass star-forming regions show similarly low deuterium fractionation \citep{Fontani2011, Rivilla2020}. High deuteration ratios can be achieved by ion-molecule reactions with the deuterated form of H$_3^+$ and CH$_3\,^+$, which efficiently forms in exothermic reactions at low temperatures \citep{Roberts2003, Parise2009}. 

The deuterium chemistry therefore is a good indicator of the conditions at the formation time of deuterated species and reflects the temperature and the time the gas spent at a low temperature.
A longer timescale for the cold collapse phase allows the reaction between HD and H$_2$D$^+$ to lead to a higher deuterium fractionation in the cold gas. The gas temperature during the cold collapse phase also plays a key role because it sets the efficiency of the deuterium exchange reactions. 
Therefore, the generally low deuterium fractionation may suggest that either the temperature during the collapse is higher  than that of low-mass protostars, or that the timescales for the cold collapse phase are shorter.  

\subsection{Isotopic ratios} \label{sec:iso_ratios}
Based on the LTE modelling in Sect.\,\ref{sec:results}, we estimated  the temperature, size of the emitting region, and molecular column density, as well as their abundance for each isotopologue. We did this individually for each molecule as the large bandwidth covered and our sensitivity allows a large enough number of transitions to be detected and fitted, as listed in Table\,\ref{tab:detections}. This allowed us to compute the isotopic ratios for many of these species in the cold gas phase. We discuss here, in particular, the isotopic fractionation for $^{12}$C/$^{13}$C, $^{32}$S/$^{34}$S, and $^{32}$S/$^{33}$S. In Table \ref{tab:isotopic_ratios} we list the molecules we used to estimate the isotopic fractionation, the value of the isotopic ratio, and compare it to the literature values as well as to isotopic ratios towards the Galactic centre. 

\begin{figure*}
    \centering
    \includegraphics[width=0.49\linewidth]{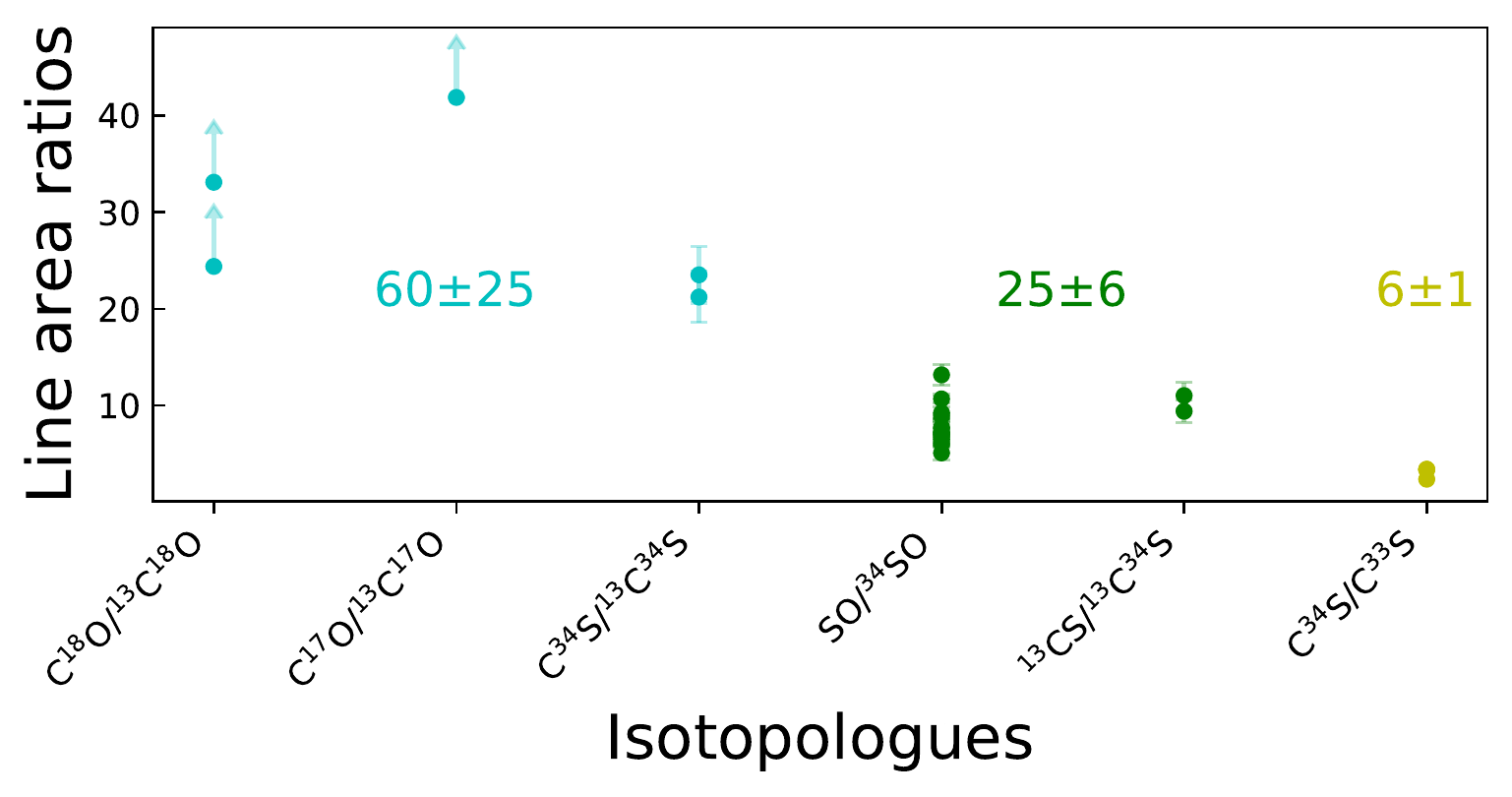}
    \includegraphics[width=0.49\linewidth]{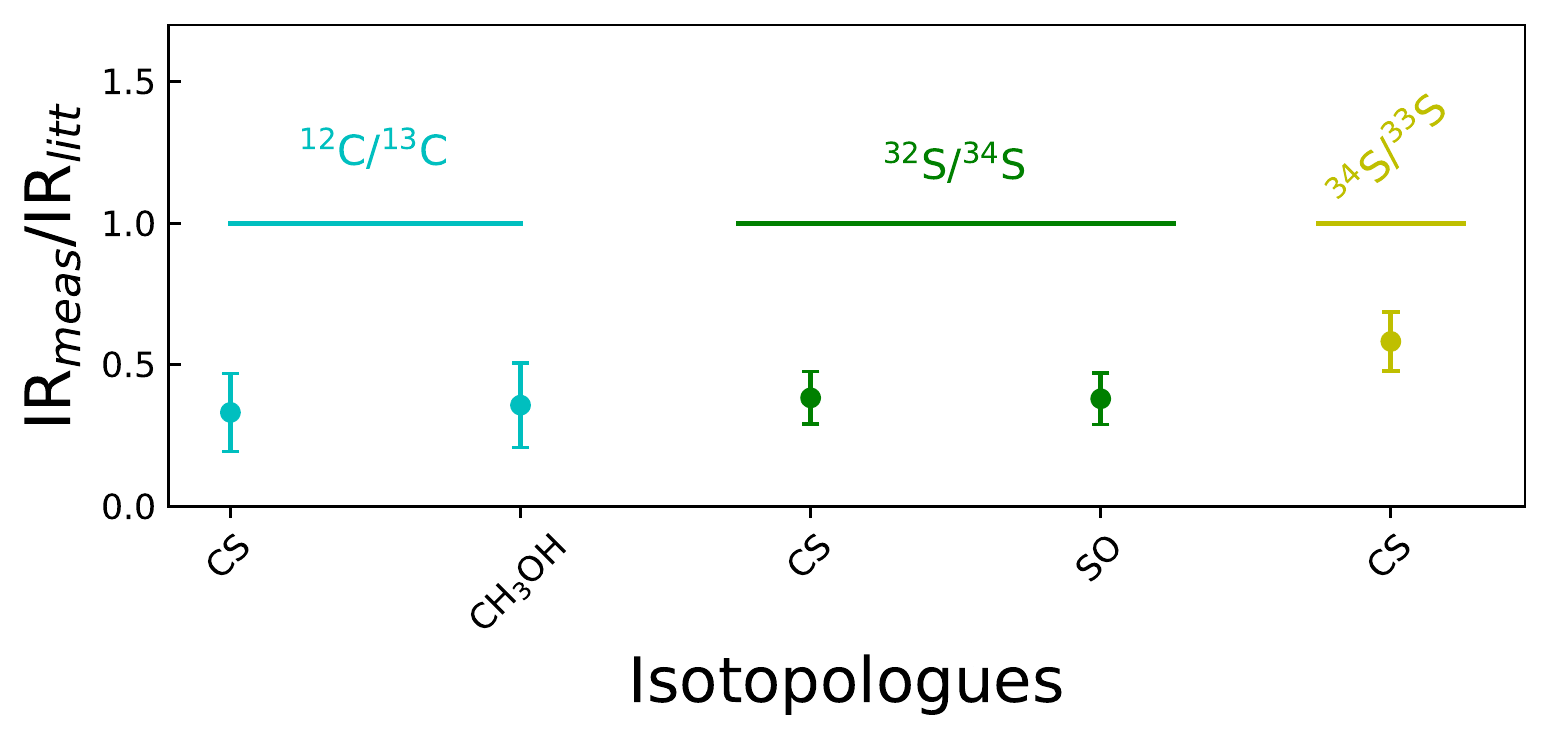}
    \caption{Line area ratio for the pairs of isotopologues (left panel). The ratios expected at the galactocentric distance of \mysou\ from galactocentric variation measurements are indicated (see Table \ref{tab:isotopic_ratios} for references). The error bars represent the errors on the line area given by the Gaussian fitting procedure from CLASS. {\it Right panel:} Ratio of the measured isotopic ratio obtained from the LTE modelling of the column densities and the ratios expected at the galactocentric distance of \mysou\ from galactocentric variation measurements. The error bars represent a 20\% error on the column densities.
    }
    \label{fig:Iso_ratio}
\end{figure*}

In Fig.\,\ref{fig:Iso_ratio} we show two approaches to evaluate the isotopic ratio for the C and S atoms, and compare it to the expected values from the literature (left and right panels, respectively). We first investigated the line area ratios of the two detected transitions ($J$=2--1, $J$=3--2) of C$^{18}$O and $^{13}$C$^{18}$O, while for C$^{17}$O and $^{13}$C$^{17}$O, we only used the $J$=2--1 transition. While lines from isotopologues including $^{13}$C are likely to be optically thin, the transitions from C$^{18}$O could be affected by a higher optical depth. Therefore, these line area ratios, measured over the entire extent of the velocity profile, place lower limits on the $^{12}$C/$^{13}$C ratio, which is about 30 for the $^{18}$O, and 40 for the $^{17}$O containing isotopologues. We expect the latter to be rather close to the real $^{12}$C/$^{13}$C value given the low expected optical depth of the C$^{17}$O $J$=2--1 line.

We also used C$^{34}$S and $^{13}$C$^{34}$S to further investigate the $^{12}$C/$^{13}$C isotopic ratios towards \mysou. Using our Weeds modelling, we selected transitions with $\tau\sim$0.1$-$0.6 and computed line area ratios as shown in the left panel of Fig.\,\ref{fig:Iso_ratio}, where we find a value between 19 and 26 for $^{12}$C/$^{13}$C. Another approach is to use the molecular abundance estimates obtained from the Weeds modelling. We found similar temperatures and source sizes for C$^{34}$S and $^{13}$C$^{34}$S in our modelling, suggesting that they originate from the same gas. Hence, we used their abundance ratios to infer a $^{12}$C/$^{13}$C isotopic ratio of 23, which agrees with our estimates based on line area ratios.

Because these values obtained from CO and CS isotopologues are below the expected value of 60 \citep{Milam2005}, we also checked CH$_3$OH and its $^{13}$C isotopologue. The molecular abundance ratio based on our Weeds modelling for CH$_3$OH and its $^{13}$C isotopologue gives a $^{12}$C/$^{13}$C ratio of around 25, which is very close to the values obtained from the CS isotopologues. While there is scatter in the obtained $^{12}$C/$^{13}$C values between 19 and 26, considering CS and CH$_3$OH, an uncertainty of about 40\% on our Weeds column density estimates is needed to make these estimates consistent with the value of 40 obtained from the  C$^{17}$O/$^{13}$C$^{17}$O measurements. 
However, at least an order of magnitude uncertainty on the column densities obtained from Weeds would be required to make these low isotopic ratios consistent with the expected literature value of 60. Therefore, we conclude that the $^{12}$C/$^{13}$C ratio is lower than the literature value of 60 towards \mysou. To estimate column densities from highly optically thick isotopologues, such as HCN, HNC, HCO$^+$, and H$_2$S, we used a $^{12}$C/$^{13}$C of 30, corresponding to the rough average of our estimates. In parentheses, we list the values corresponding to the canonical $^{12}$C/$^{13}$C of 60 for completeness. 

Similarly to the above, to estimate $^{32}$S/$^{34}$S, we used optically thin transitions of $^{13}$CS and $^{13}$C$^{34}$S as well as SO and $^{34}$SO. These molecules consistently give a value of $^{32}$S/$^{34}$S=12 (left panel of Fig.\,\ref{fig:Iso_ratio}). Similarly, the ratio of $^{34}$S/$^{33}$S could be determined by using C$^{34}$S and C$^{33}$S and is found to be around 3.7. 

In the right panel of Fig.\,\ref{fig:Iso_ratio} we show the ratio of the literature values and our estimates and find that our isotopic ratios for $^{12}$C/$^{13}$C, $^{32}$S/$^{34}$S, and $^{32}$S/$^{33}$S  are all about a factor of 2--3 lower than expected at the galactocentric distance of \mysou.
Isotopic ratios may exhibit variations as a function of galactocentric radius due to the variations of nucleosynthetic processing of the interstellar medium over the disk of the Milky Way \citep[e.g.][]{Penzias1981a,Penzias1981b,Wilson1994,Milam2005,Wouterloot2008,Giannetti2014,Jacob2020}. The galactocentric variations for the $^{12}$C/$^{13}$C and $^{32}$S/$^{34}$S isotopic ratios are given by $^{12}$C/$^{13}$C=$(8\pm 2)D_{\rm GC}+(8\pm 13)$ and $^{32}$S/$^{34}$S=$(3.3\pm 0.5D_{\rm GC}+(3.1\pm 4.1)$ \citep{Milam2005, Chin1996}. Following \citet{Roman-Duval2009}, we computed the galactocentric distance as $D_{\rm DG}=\sqrt{{\Big{(}}R_{\rm 0} cos(l)-d{\Big{)}}^2+R_{\rm 0}^2sin^2 (l)}$, where $R_{\rm 0}$ is the distance to the Galactic centre, $l$ is the longitude of the source, and $d$ is its heliocentric distance. With an estimated distance of 2.5\,kpc, \mysou\ is located at a distance of 6.4\,kpc from the Galactic centre, hence its expected isotopic ratio is about 60$\pm$25 for $^{12}$C/$^{13}$C \citep{Milam2005} and 25$\pm$6 for $^{32}$S/$^{34}$S \citep{Chin1996}, at least two times higher than suggested by several $^{13}$C and $^{34}$S bearing molecules. Instead, the carbon and sulphur isotopic ratios resemble that of the Galactic centre. Assuming a galactocentric variation, we find that the observed values resemble the expected ratios around a galactocentric distance of 2\,kpc, which is inconsistent with the location of our source, suggesting that these low values are intrinsic to our source.

While a scatter in isotopic abundances is often observed \citep[cf.][]{Giannetti2014}, this deviation from the expected values suggests an anomalous isotopic fractionation in \mysou. Recent observations of hot cores reported such anomalies of unusually low $^{12}$C/$^{13}$C ratio towards IRAS20126$+$4104 and G31.41$+$0.31, where $^{12}$C/$^{13}$C as low as 10 and 30 was found based on CH$_3$OH, and CH$_3$CN (and also CH$_3$OCHO) molecules in the warm gas, respectively \citep{Palau2017, Beltran2018}.

Because we estimate these isotopic ratios for the cold component of the envelope, it is unlikely that internal effects due to the emerging protostar play a role. Fractionation effects caused by UV photons have been taken into account in chemical models for example in \citet{vanDishoeck1988}, who showed that UV photons can significantly impact the isotopic ratios. However, the UV photons from the protostar are unlikely to be strong enough to lead to the decrease in isotopic ratios. 
Cosmic rays are also found to influence the isotopic fractionation chemistry. The recent study of \citet{Colzi2020} indicates that a cosmic ionisation rate higher than 1.3$\times$10$^{-17}$s$^{-1}$ tends to decrease the $^{12}$C/$^{13}$C ratio. 
Additionally, low temperature isotopic exchange reactions have been considered in chemical networks \citep{Roueff2015, Colzi2020, Loison2020}. These reactions lead to an enhancement in the $^{13}$C isotopologues in molecules formed from CO and C$_3$ (when it does not react with atomic oxygen) and a decrement in molecules forming from atomic carbon \citep{Loison2020}. The models from \citet{Loison2020} show that isotopic fractionation for CO, CS, methanol, but also CCH, HC$_3$N, CN, for example, is time dependent. Additionally, their models present different isotopic fractionation for each molecules, in particular for CO, CS, and CH$_3$OH. The carbon isotopic fractionation appears to be stronger in CO than in CS and CH$_3$OH. 

Isotopic fractionation of S-bearing molecules through low-temperature reactions is found to be low \citep{Loison2019}, suggesting that $^{34}$S-bearing molecules are good tracers of their $^{32}$S main isotopologue. However, our results towards {\mysou} suggest an elevated sulphur fractionation based on CS, similarly to \citet{Tercero2010}, and especially on SO molecules, which is an intriguing result in light of these models. Similarly as above, for molecules with highly optically thick lines, such as CS and H$_2$S, we used an isotopic ratio for $^{32}$S/$^{34}$S of 12, and give in parentheses the column densities using the canonical value of 25 as well.

\begin{table*}
\centering
\small
\caption{Isotopic ratios in \mysou.}
{\vspace{0.1cm}}
\label{tab:isotopic_ratios}
\begin{tabular}{c c c c c c c}  
\hline
\hline
Atoms & Molecules & \multicolumn{2}{c}{Isotopic ratios} & Adopted & Isotopic ratios & Galactic  \\
 & & Line area & LTE model & isotopic ratios & from the literature\tablefootmark{a} & centre\\
\hline
$^{12}$C/$^{13}$C &  C$^{18}$O/$^{13}$C$^{18}$O & 30-35 & -- & 30 & 60$\pm$25\tablefootmark{b} & 24\tablefootmark{d} \\
& C$^{17}$O/$^{13}$C$^{17}$O & 40 & -- &30 & 60$\pm$25\tablefootmark{b} & -- \\
& C$^{34}$S/$^{13}$C$^{34}$S & 22 &23 & 30 & 60$\pm$25\tablefootmark{b} & 22\tablefootmark{e}\\
& CH$_3$OH/$^{13}$CH$_3$OH &-- & 25 & 30 & 60$\pm$25\tablefootmark{b} & 22\tablefootmark{f}\\
$^{32}$S/$^{34}$S & $^{13}$CS/$^{13}$C$^{34}$S & 12 & 12 & 12 & 25$\pm$6 \tablefootmark{c} &18\tablefootmark{e}\\
& SO/$^{34}$SO & 8 & 12 & 12 &32$\pm$7 \tablefootmark{c} &  -- \\
$^{34}$S/$^{33}$S & C$^{34}$S/C$^{33}$S & 3.8 & 3.7 & 3.7 & 6.3 $\pm$ 1.0\tablefootmark{c} & 4.5\tablefootmark{e}\\
\hline 
\end{tabular}
\tablefoot{
\tablefoottext{a}{These values were calculated for a source located at 6.4\,kpc from the Galactic centre using the Galactic gradient from the papers in this column.}
\tablefoottext{b}{\citet{Milam2005}}
\tablefoottext{c}{\citet{Chin1996}}
\tablefoottext{d}{\citet{Langer1990}}
\tablefoottext{e}{\citet{Humire2020}}
\tablefoottext{f}{\citet{Muller2016}}
}
\end{table*}


\section{Molecular composition: COMs} \label{sec:COMs}
\begin{figure*}
    \centering
    \includegraphics[width=0.49\linewidth]{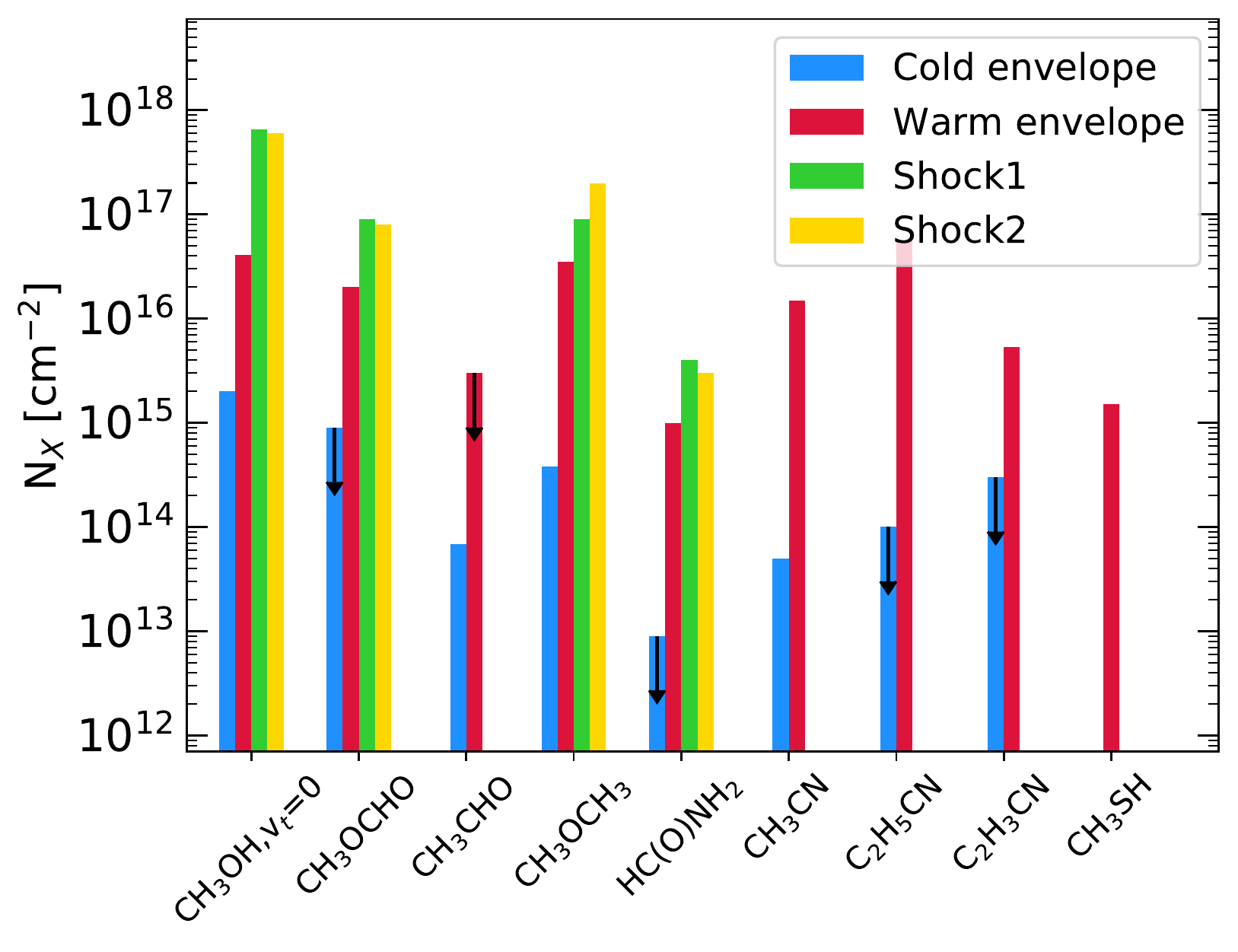}
    \includegraphics[width=0.49\linewidth]{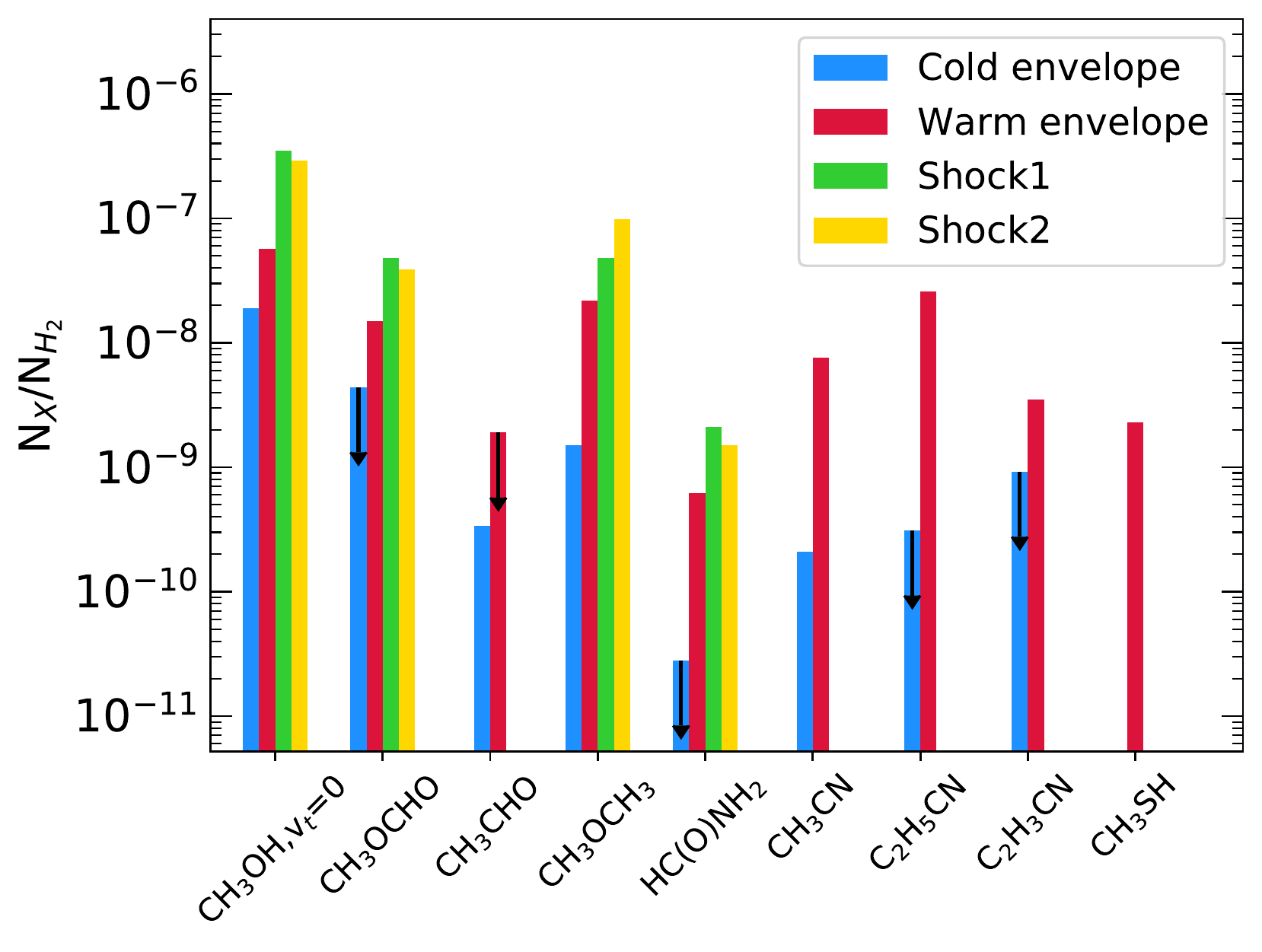}
    \includegraphics[width=0.49\linewidth]{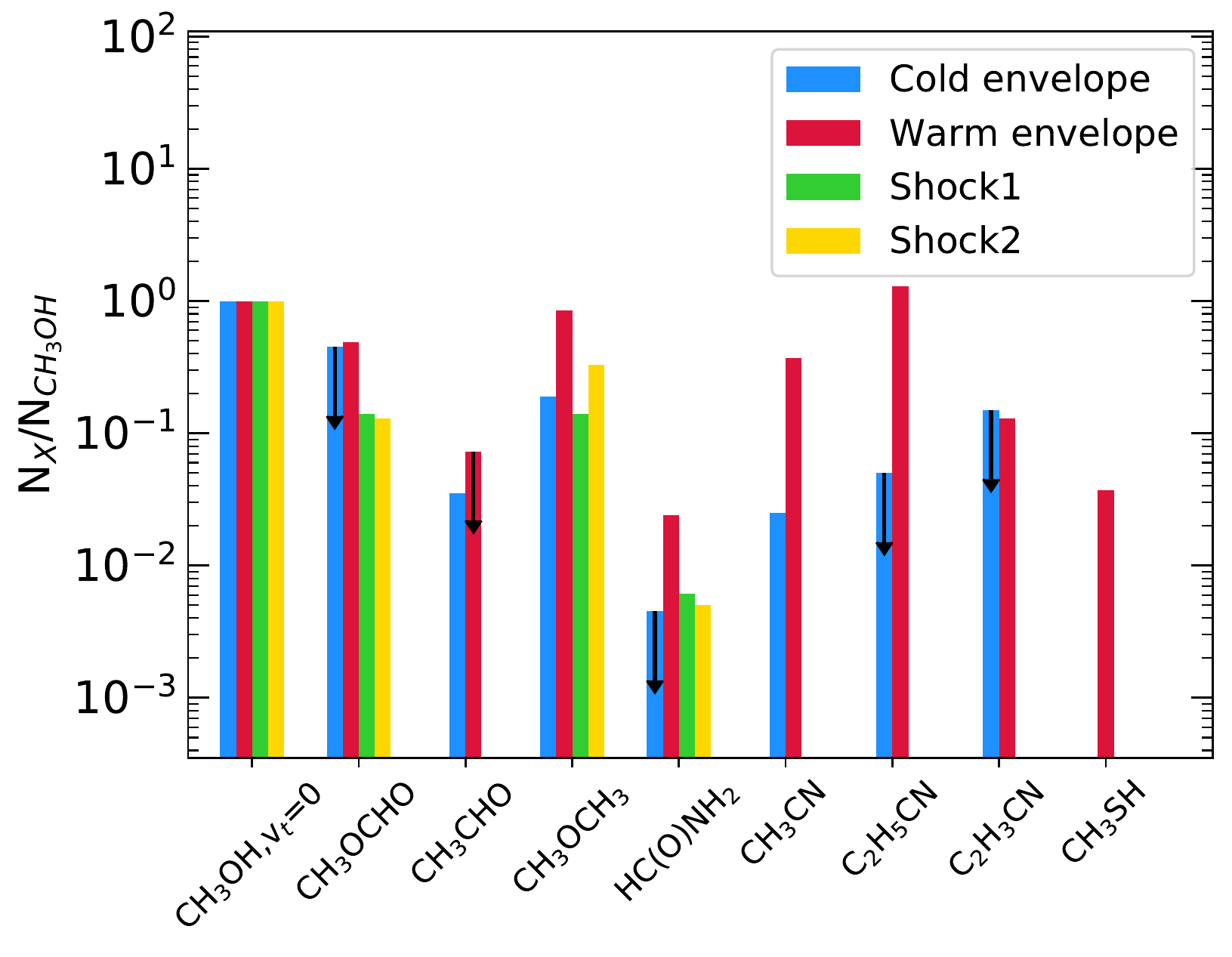}
    \includegraphics[width=0.49\linewidth]{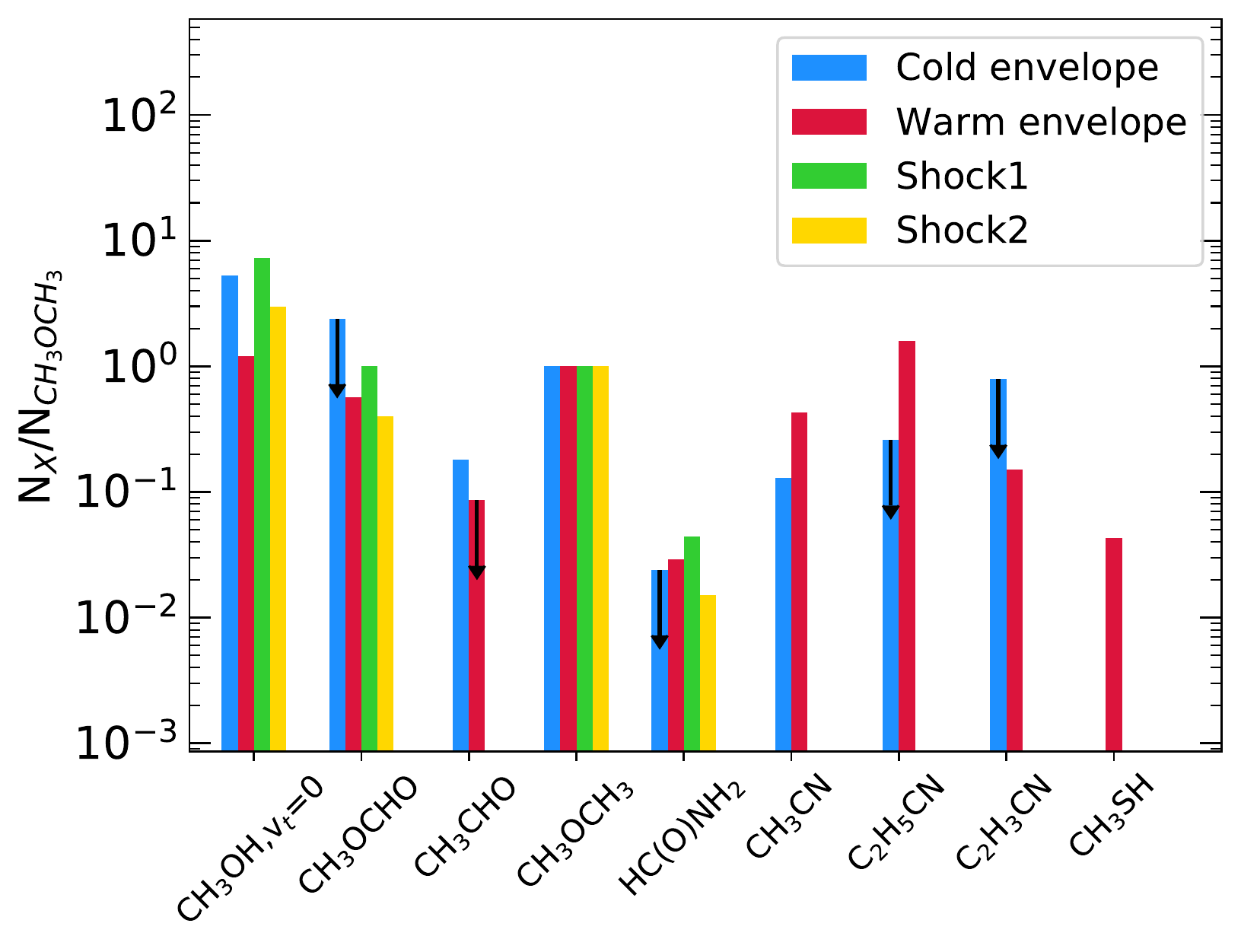}
    \caption{COM column densities (top left panel) and abundances relative to H$_2$ (top right panel), methanol (bottom left panel), and dimethyl ether (bottom right panel). Only the main isotopologues are represented. The cold and warm components of the envelope are described by the blue and red histograms. The two shocks are represented in yellow and green. The upper limits are indicated by the black arrows.}
    \label{fig:N_COMs}
\end{figure*}

The complete 206\,GHz frequency coverage of this study offers a large range of upper-level energies for the different COMs.
We detect nine COMs towards this object: methanol (CH$_3$OH), methyl formate (CH$_3$OCHO), dimethyl ether (CH$_3$OCH$_3$), formamide (HC(O)NH$_2$), acetaldehyde (CH$_3$CHO) (O-bearing COMs), methyl cyanide (CH$_3$CN), ethyl cyanide (C$_2$H$_5$CN), vinyl cyanide (C$_2$H$_3$CN) (N-bearing COMs), and methyl mercaptan (CH$_3$SH) (S-bearing COM). All these molecules except for acetaldehyde are detected in the warm component of the envelope, four (CH$_3$OH, CH$_3$OCHO, CH$_3$OCH$_3$, and HC(O)NH$_2$) are also found to be present in the accretion shocks based on their spectral profile (Fig.\,\ref{line_profiles}), and four are also identified in the cold component of the envelope (CH$_3$OH, CH$_3$CHO, CH$_3$OCH$_3$, and CH$_3$CN).
Compared to the study of \citet{Csengeri2019} using ALMA, we detect here two additional COMs, CH$_3$OCH$_3$ and CH$_3$SH. This is because the limited spectral coverage of these ALMA observations of 7.5\,GHz around 345\,GHz was not suitable for the detection of these two molecules.  
We do not detect ethylene glycol ((CH$_2$OH)$_2$), formic acid (HCOOH), acetone (CH$_3$COCH$_3$), and ethanol (C$_2$H$_5$OH), however, which are the species with the most compact emission found in \citet{Csengeri2019}, and are thus heavily diluted in the beam of the APEX telescope. 
The column densities, abundances relative to H$_2$, methanol, and dimethyl ether are reported in Table \ref{tab:LTE_values} and Fig. \ref{fig:N_COMs}.
We also list upper limits for several COMs in the cold gas component, where we assumed a size of 26\arcsec\ and a temperature of 16\,K, which corresponds to an average of the physical properties of COMs detected in the cold gas component.

\subsection{Accretion shocks in \mysou} \label{sec:acc_shocks}
The accretion shocks reported in \citet{Csengeri2018} using methanol, and then typically O-bearing COMs \citep{Csengeri2019} show a velocity offset of $\sim\pm4.2$\,\kms\ relative to the source $\varv_{\rm lsr}$ (Fig. \ref{line_profiles}). Interestingly, in high-excitation transitions of several COMs, such as CH$_3$OH, CH$_3$OCHO, CH$_3$OCH$_3$, HC(O)NH$_2$, and HDO, we also see these velocity components even with a single-dish telescope. In addition, we detect several rotational transitions of methanol from within its torsionally excited state, $v_{\rm t}$=1, which only trace the accretion shocks \citep{Csengeri2018}. 

Because we have more transitions in our spectral survey than \citet{Csengeri2019}, we attempted to place more stringent constraints on the gas temperature based on our LTE modelling. Our models provide parameters for the two velocity components that have been spatially resolved in \citet{Csengeri2019}. These velocity components are blended with the envelope component in several of these cases, hence we find a larger dispersion for the peak velocity of the accretion shocks. We estimate a gas temperature between 170-190\,K for the accretion shocks (Fig.\,\ref{diag_COMs1}) based on all the molecules that show these velocity components. This is consistent with the range of values around 180\,K found by  \citet{Csengeri2018} based on CH$_3$OH alone. The column densities and hence the abundances are about an order of magnitude lower compared to the spatially resolved measurements based on ALMA, however, which is expected because our observations do not spatially resolve these regions.

\subsection{COMs in the cold component of the envelope} \label{sec:COMsfroid}
We were able to detect four COMs, \methanol, \dimetheth, CH$_3$CHO, and CH$_3$CN, in the cold component of the envelope with a gas excitation temperature between 20 and 34\,K. Towards low-mass cores, COMs were found at low temperatures \citep{Oberg2010, Bacmann2012, Jaber2014, Vastel2014}, but in the vicinity of hot cores, most studies reveal COMs in the warm component of the envelope \citep[cf.][]{Bisschop2007, Weaver2017}. 
The derived column densities in the cold phase ($2\times10^{15}$\,\unidens\ for methanol and $5\times10^{13}$--10$^{15}$\,\unidens\ for the other molecules) are about two orders of magnitude lower than in the warm phase, while the abundances relative to H$_2$ (1.9$\times$10$^{-8}$ for methanol and $3.4\times10^{-10}$--$1.5\times10^{-9}$ for the other molecules) are about one order of magnitude lower in the cold component of the envelope (Fig. \ref{fig:N_COMs}). 

The formation of COMs on the surface of dust grains and ices at low temperatures was shown to be possible by \cite{Garrod2008}. From the grain mantles, they are then liberated into the gas phase through non-thermal processes caused either by shocks, supersonic turbulence \citep{Requena-Torres2006}, energy release from exothermic reactions on the grains \citep{Garrod2007}, UV radiation originating from the protostar propagating through the outflow cavity \citep{Oberg2010, Drozdovskaya2015}, or by cosmic ray radiation or secondary UV photons (produced by the electron recombination after the impact of cosmic rays on H$_2$ molecules) \citep{Bacmann2012}. The high abundances of sulphur-bearing species suggest the presence of shocks (Sect. \ref{sec:shocks}), and the presence of COMs in the cold component of the envelope could be also explained by the sputtering of grain mantles due to such shocks. No strong UV radiation can be expected from the protostar itself because it does not show a strong radio continuum emission, and energetic photons from the protostar could escape through the cavity formed by a collimated outflow. 
Chemical desorption or low-velocity shocks could also give rise to non-thermal desorption, leading to the observed cold-phase COMs. It then remains to understand to which extent this composition of COMs reflects that of the grain surfaces. The cold component of the envelope exhibits a diverse molecular composition of simple molecules including a large number of molecular ions. This suggests rather active chemical processes in the cold gas phase, which is expected to represent the pristine, pre-collapse gas not affected by the heating of the central object. Low-temperature gas-phase chemistry could therefore have an impact on the molecular composition of COMs. 

\subsection{COMs in the warm envelope}
After methanol, the second most abundant COM in the warm component of the envelope is dimethyl ether. Both molecules are detected in the cold and warm components of the envelope as well as in the accretion shocks. Because they are chemically linked \citep{Garrod2008}, they are expected to be found in the same gas components. We also detect methyl formate at an abundance relative to H$_2$ of 1.5$\times$10$^{-8}$ and temperature of 86\,K. We  detect it in the accretion shocks and in the warm component of the envelope, but the frequency range of our survey is not suitable to detect it in the cold component of the envelope. Similarly, CH$_3$CHO is chemically linked to these species, which is present in the warm component seen by ALMA \citep{Csengeri2019}, but our sensitivity is not good enough to detect it because it originates form a very compact region in the vicinity of the protostar.
Methyl formate and dimethyl ether are only an order of magnitude less abundant than methanol.

In \mysou, COMs with a CN functional group (vinyl and ethyl cyanides) are detected in the warm gas (90--100\,K) with high abundances of 2.6$\times$10$^{-8}$ for C$_2$H$_5$CN and 3.5$\times10^{-9}$ for C$_2$H$_3$CN. They exhibit a slightly higher temperature than O-bearing COMs, but this difference would be even more pronounced if we had used a two-component model for ethyl cyanide (see Sect. \ref{sec:exceptions}), suggesting that the gas temperatures in the immediate vicinity of the protostar are still moderate, again pointing to a very compact radiatively heated inner region. Gas temperatures of these cyanides are found to be typically higher than for O-bearing COMs \citep{Widicus2017}, and towards hot cores, they often trace the hot gas component above 150\,K \citep{Belloche2016}. We find that the case of CH$_3$CN is somewhat particular. We detect several high-energy transitions from the K=7-9 lines. Our model with a temperature of 200\,K, a size of 1\arcsec\ , and a high column density of $1.5\times10^{16}$\,cm$^{-2}$ provides a relatively good fit to the K=0--6 lines, but it underestimates the K=7--9 components, suggesting that a hot compact component traced by CH$_3$CN could be present. A hot envelope component with temperatures up to 280\,K has already been observed with single-dish telescopes towards hot cores using the CH$_3$CN line \citep[e.g.][]{Weaver2017,Bisschop2007}. Similar energy levels are available in our frequency coverage for methanol as well, but they do not show this hot component.

We also detect the sulphur-bearing COM methanethiol (CH$_3$SH) towards the warm component of the envelope at a temperature of 69\,K and with an abundance relative to H$_2$ of $3.4\times10^{-9}$. CH$_3$SH represents only a small fraction of the total sulphur reservoir.

\subsection{Origin of COMs in the envelope of \mysou: Direct sublimation of grain-surface products?} \label{sec:COMs_chem}
We have shown that based on their typical temperature,  line profile, and size of the emitting region, three different physical components can be distinguished in the envelope of \mysou, corresponding to the cold and warm gas components and the accretion shocks. 
The overall molecular composition of the cold and warm envelope components is very diverse, likely pointing to a different chemical history of the gas. Based on COMs detected both in the cold and warm gas phases, we investigate here their potential formation based on their abundance ratios in the cold and warm gas listed in Table\,\ref{tab:COMS_ratio}. Because only a few COMs are detected both in the cold and in the warm gas, we are limited to perform only a qualitative comparison to models, for which we chose the gas-grain chemical models by \citet{Garrod2006, Garrod2008} and \citet{Garrod2017}.

In the cold gas, we detect methanol, dimethyl ether, and acetaldehyde at temperatures between 20 and 27\,K with molecular abundances with respect to H$_2$ of $1.9\times10^{-8}$, $1.5\times10^{-9}$, and $3.4\times10^{-10}$, respectively (Sect.\,\ref{sec:COMsfroid}). 
Pure gas-phase chemistry in the cold gas phase is not sufficient to explain such high abundances for COMs, although its role has been emphasized in recent, revised chemical models of low-mass prestellar cores \citep{Vasyunin2017}.
\citet{Garrod2006} demonstrated that methanol, methyl formate, and dimethyl ether may form efficiently on the grain surfaces, which then sublimate to the gas phase when the temperature increases. 
Therefore, the detection of COMs in the cold gas towards \mysou\ suggests grain surface formation and subsequent desorption to the gas phase. In the cold envelope, this invokes non-thermal desorption processes that need to be efficient to lead to the observed  molecular abundances of these COMs. 

The warm gas typically exhibits molecular abundances in methanol, dimethyl ether, and methyl formate that are an order of magnitude higher than the cold gas, demonstrating that the inner, warmer regions become richer in saturated oxygen-bearing COMs. The difference of abundances relative to H$_2$ between the cold and warm components of the envelope could be explained by a considerably higher efficiency of the thermal desorption in the warm gas phase.
Observations suggest a tight chemical connection between methanol, methyl formate, and dimethyl ether \citep[e.g.][]{Garrod2006, Brouillet2013, Jaber2014}, and indeed we detect these molecules in high abundance in the warm gas and in the accretion shocks. Their abundances in the warm gas are qualitatively consistent with the models of  \citet{Garrod2006} and \citet{Garrod2008, Garrod2017}, who predicted similar abundances around 10$^{-8}-10^{-7}$ for methyl formate and dimethyl ether, while the peak abundance of methyl formate decreases in their models as a function of warm-up time. Our observations suggest that the abundances of methanol are an order of magnitude lower than in these models, therefore the abundance ratios of methyl formate and dimethyl ether to methanol are about an order of magnitude lower only. This is consistent with observational results of other hot cores \citep[e.g.][]{Taquet2015}, but it suggests some discrepancies to these chemical models. 

The abundance ratio of CH$_3$OCH$_3$ to CH$_3$OH in the cold and warm components of the envelope are different by a factor of only $\sim$4 (Table \ref{tab:COMS_ratio}). These ratios are representative of the difference of chemistry and complexity between the cold and warm components of the envelope. The higher CH$_3$OCH$_3$/CH$_3$OH ratio in the warm component of the envelope may suggest a higher degree of molecular complexity.

Formamide is a potential precursor to heavier COMs, such as amino acids \citep[cf.][]{Lopez-Sepulcre2015}. While observationally it is detected in various environments, its formation conditions remain debated. On the grains, the successive hydrogenation of HNCO is thought to be efficient \citep{Charnley1997, Mendoza2014, Lopez-Sepulcre2015}, but the link between HNCO and HC(O)NH$_2$ is still debated in experiments \citep[e.g.][]{Noble2015,Haupa2019}. During the cold phase on the grains, further hydrogenation can lead to more complex molecules \citep{Raunier2004}. At $\sim$100\,K, formamide desorbs and is present in the gas phase. Towards \mysou, the ratio of HC(O)NH$_2$ to HNCO is 0.04 in the warm gas phase at a temperature of $\sim$80\,K, slightly below its desorption temperature. This ratio is somewhat lower than what was found towards high-mass protostars \citep{Lopez-Sepulcre2015}, and it is considerably lower than the extreme star-forming region, SgrB2(N2), for which \citet{Belloche2017} reported a ratio of HC(O)NH$_2$/HNCO of 1.8. When comparing the molecular abundance of HC(O)NH$_2$ with respect to H$_2$ in the sample of \citet{Bisschop2007}, we see that the abundance of \mysou\ is a factor of a few lower than that of these hot cores. Considering the lower temperatures in the warm component of the envelope, this could suggest that the thermal desorption has not been complete and only part of the formamide has been liberated from the grains, if it formed on the grain surfaces. 

Another formation route of formamide is in the gas phase from H$_2$CO and NH$_2$ \citep{Skouteris2017}. This reaction has an energy barrier of 26.9\,K \citep{Barone2015}, which makes it efficient even at the low temperatures of the cold gas towards  \mysou,\ where the bulk of the gas is at 20\,K. The models of \citet{Vasyunin2017} show that including this reaction can lead to abundances relative to H$_2$ of 5$\times$10$^{-12}$ in the cold component of the envelope. However, \citet{Belloche2017} showed that when this reaction is included in SgrB2(N2), it leads to an overestimation of the abundance of formamide in the warm component of the envelope, favouring a grain surface production. Such an abundance is in any case below our detection limit and is therefore consistent with the non-detection of formamide in the cold gas phase.

\begin{table}
\centering
\small
\caption{Abundance ratios of COMs in the cold and warm components of the envelope of \mysou.}
{\vspace{0.1cm}}
\label{tab:COMS_ratio}
\begin{tabular}{l c c }  
\hline
\hline
Ratios & Cold & Warm  \\
 & envelope & envelope  \\
\hline
H$_2$CO/CH$_3$OH & 0.07 & 0.1  \\
CH$_3$OCHO/CH$_3$OH & $\leq$0.2 & 0.2 \\
CH$_3$OCH$_3$/CH$_3$OH & 0.08 & 0.3 \\
CH$_3$CHO/CH$_3$OH & 0.02 & $\leq$0.03 \\
CH$_3$OCHO/CH$_3$OCH$_3$ & $\leq$2 & 0.7 \\
CH$_3$CHO/CH$_3$OCHO & $\geq$0.08 & $\leq$0.1 \\
CH$_3$CHO/CH$_3$OCH$_3$ & 0.2 & $\leq$0.08 \\
CH$_3$CHO/HC(O)NH$_2$ & $\geq$13 & $\leq$3 \\
H$_2$CO/HC(O)NH$_2$ & $\geq$16 & 12 \\
HC(O)NH$_2$/HNCO & $\leq$0.1 & 0.04 \\
CH$_3$SH/H$_2$CS & - & 0.2 \\
CH$_3$OH/CH$_3$SH & - & 25 \\
C$_2$H$_5$CN/CH$_3$CN & - & 3 \\
C$_2$H$_5$CN/C$_2$H$_3$CN & - & 6 \\
n-C$_3$H$_7$CN/C$_2$H$_5$CN & - & $\leq$0.1 \\
\hline
\end{tabular}
\end{table}

The chemical network leading to the formation of cyanides has been modelled for example by \citet{Belloche2009} and \citet{Garrod2017}, who suggested that C$_2$H$_5$CN can be formed on the grains by the hydrogenation of C$_2$H$_3$CN formed from HC$_3$N during the collapse and the warm-up phase. C$_2$H$_5$CN then sublimates to the gas phase at a temperature of $\sim$100\,K \citep{Garrod2017}. Subsequent gas-phase reactions lead to the destruction of most of the C$_2$H$_5$CN formed on the grains, hence the ratio of  ethyl to vinyl cyanide has been discussed as a potential chemical clock (\citealp[cf.][]{Allen2018}, but see also \citealp{Caselli1993})  reflecting the minimum age of the protostar. Towards \mysou, we find here a ratio of ethyl to vinyl cyanide of $\sim$ 6, which is broadly consistent with the ratio of 2.4-3.4 found by \citet{Csengeri2019}, who based their abundance estimates on adopting the kinetic temperature of methanol for all COMs. Our results confirm that saturated nitriles are more abundant than unsaturated nitriles in this source, which is expected as they are chemically more stable. On the other hand, comparing the ratio of ethyl to vinyl cyanide to the values of other hot cores from \citet{Fontani2007}, we find here a somewhat higher value, implying a larger amount of C$_2$H$_5$CN, suggesting that the source could be considered as younger than the hot cores in their sample.

We do not detect any of these complex nitriles in the cold envelope, although \citet{Garrod2017} showed that vinyl cyanide can be formed efficiently through gas-phase reactions between CN and C$_2$H$_4$ up to 100\,K, leading to gas-phase abundances of $\sim10^{-11}$, which is below our detection limit. Therefore, our observations within the limitation of the assumptions for the upper limits given in Table\,\ref{tab:LTE_values} are not sensitive enough to distinguish between the models of \citet{Garrod2017}.

We saw in Sect. \ref{sec:S} that the ratio of  methanol and methanethiol is lower than what is observed in SgrB2(N2), OrionKL, or G327.3 or by chemical modelling, suggesting an enhancement of the S-bearing molecules in this object. 
CH$_3$OH and CH$_3$SH are thought to be formed by successive hydrogenation of H$_2$CS and H$_2$CO on the ices and then released to the gas phase \citep{Gibb2000, Majumdar2016}. In the warm component of the envelope, the ratio CH$_3$SH/H$_2$CS(o+p) is equal to 0.2, while CH$_3$OH(A+E)/H$_2$CO(o+p) is close to two orders of magnitude higher, which is in agreement with the chemical modelling from \citet{Muller2016}. This can be explained by subsequent reactions in parallel specific to the sulphur and oxygen chemistry. For instance CH$_3$SH can be formed through reactions between the two radicals CH$_3$ and SH, and H$_2$CS through reactions between CH$_2$ and S \citep{Muller2016}. CH$_3$SH/H$_2$CS is lower by an order of magnitude than what is observed in IRAS16293--2422 ($\sim$3) and in comet 67P ($\sim$3) \citep{Drozdovskaya2018}.

In the cold component of the envelope, the only molecule that is well reproduced by the models from \citet{Garrod2008} is acetaldehyde, for which we estimated an abundance of 3.4$\times$10$^{-10}$. Acetaldehyde forms on the grains like other complex molecules, but can desorb at low temperatures.

To summarise, the cold component of the envelope of \mysou\ is richer in small molecules than the warm gas phase. In contrast, the warm component of the envelope shows a richness in COMs that suggests an increasing chemical complexification at higher temperatures, although several heavier COMs that are expected to form on the grains remain undetected. This suggests chemical differences in the cold and warm components of the envelope, favouring a scenario in which the molecular composition of the gas in the cold component of the envelope is chemically evolved, but it is poor in COMs due to its low temperature. The warm component of the envelope is expected to inherit the molecular composition of the grain mantles, which have reached the sublimation temperature only recently as the protostar is young. At the estimated age of the protostar, no physical and chemical equilibrium could be reached.

\section{Comparison of \mysou\ to hot corinos and hot cores} \label{sec:comparison}

\citet{Csengeri2019} proposed \mysou\ to be a precursor of hot molecular cores because its radiatively heated region is more compact and is at a moderate temperature compared to hot cores. Its richness in COMs has been first attributed to the impact of accretion shocks, which has been shown to lead to an increased abundance of O-bearing COMs. Here we can disentangle the contribution of the accretion shocks from the molecular composition of the warm inner region, which in addition to being among the most compact of hot cores \citep[i.e.][]{Allen2018,Ginsburg2018}, also shows a typically colder temperature of about 100~K compared to the classical hot cores, which reach higher temperatures \citep{Widicus2017}, typically $>$150~K within a size scale above $\sim$0.01\,pc \citep[e.g.][]{SanchezMonge2014}. Here we aim to place the molecular composition of the cold and warm gas phase of \mysou\ into context by comparing it to the archetypical hot corino IRAS16293--2422 and hot cores from \citet{Bisschop2007}.

To do this, we compared our detections with the findings of the ASAI \citep{Lefloch2018} and TIMASS \citep{Caux2011} surveys. In Sect.\,\ref{sec:results} we have classified the 2935 identified lines into the categories: O-, S-, and N-bearing and carbon chain molecules. The larger number of O-bearing lines compared to that of carbon chains confirms that the molecular content of \mysou\ is more similar to hot cores and hot corinos than to WCCC objects \citep{Lefloch2018}. The ratio of O-bearing and N-bearing lines is similar to that of IRAS16293--2422 \citep{Caux2011}. 
The proportion between the different types of molecules also highlights a large number of lines from S-bearing molecules that is more similar to shocked regions, such as L1157-B1 \citep{Lefloch2018}.

\begin{figure}
    \centering
    \includegraphics[width=0.9\linewidth]{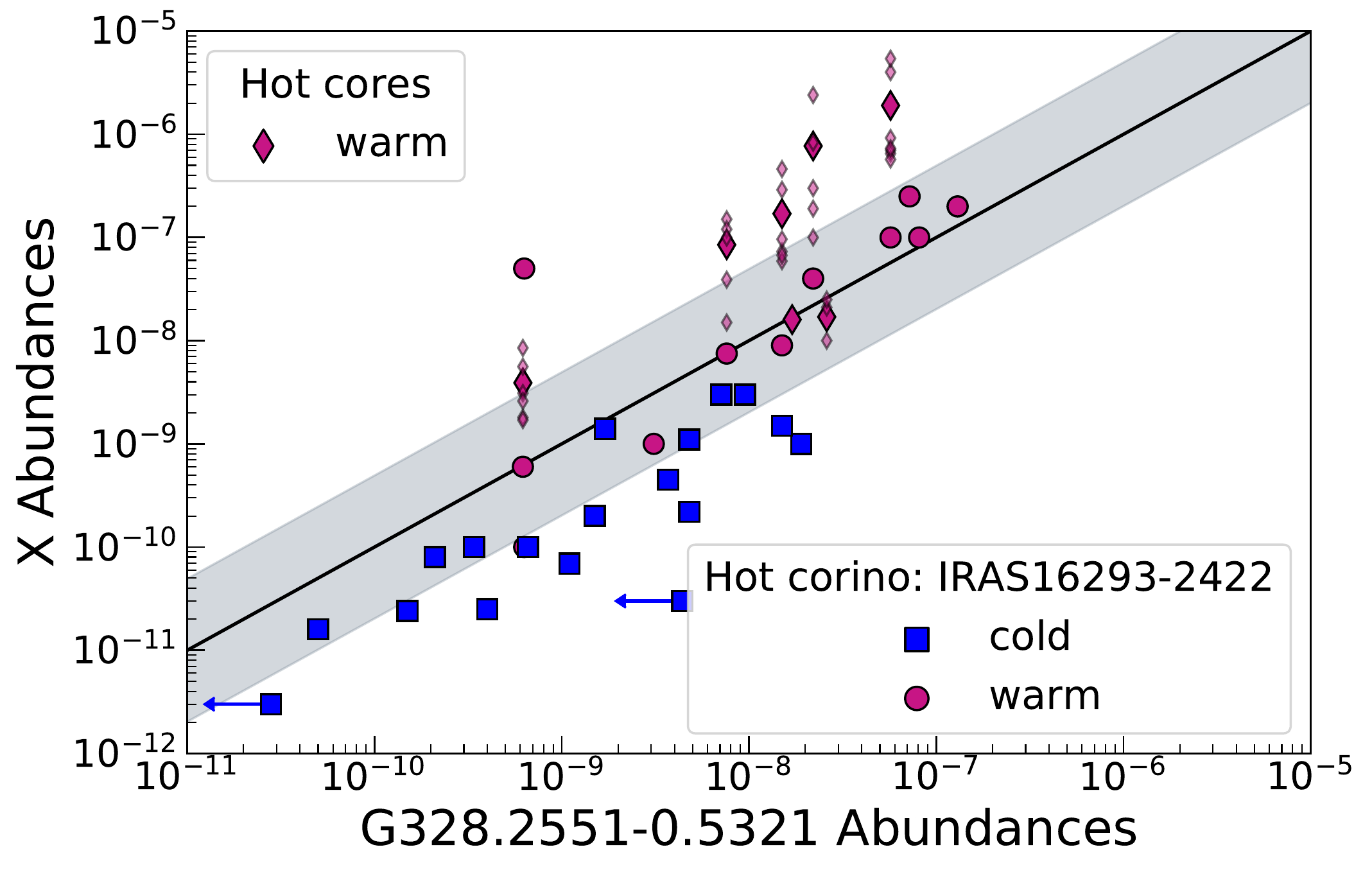}
    \caption{Molecular abundances relative to H$_2$  towards IRAS16293--2422 \citep{Schoier2002,Maret2005,Jaber2014,Taquet2014, Majumdar2016} and hot cores from \citet{Bisschop2007} vs. that of \mysou. The diamond represents the COMs in the hot cores. The small symbols are used to show each hot core. The larger symbol shows the averaged value. The blue data points represent the cold component of the envelope, and red represents the warm component of the envelope. The grey shaded region corresponds to a factor of 5 scatter from the one-to-one relation.}
    \label{fig:comp}
\end{figure}

We show a quantitative comparison of the molecular abundances of \mysou\ and IRAS16293--2422 in  Fig. \ref{fig:comp},  based on single-dish observations of the O-bearing COMs and the light molecules \citep{Schoier2002,Maret2005,Jaber2014,Taquet2014, Majumdar2016}, where the cold and warm components of the envelope have been distinguished. In the warm envelope, we were able to compare the O-bearing molecules as well as HDO, HC$_3$N, and S-bearing molecules. For the comparison of the cold component of the envelope, we could select S-, N-, O-molecules, carbon chains, but also the four COMs detected in the cold gas, which are detected in both objects. As Fig.\,\ref{fig:comp} illustrates, the cold gas shows higher molecular abundances for all species in \mysou, and we find abundances typically higher by an order of magnitude than IRAS16293--2422. This enhancement is not specific to one type of molecules, but to the ensemble of the cold component of the envelope. However, molecules in the warm gas exhibit similar abundances in both objects, with values between 10$^{-9}$ and 10$^{-7}$. 

Finally, we also compare our object to hot cores from \citet{Bisschop2007}, where molecules originating from  the cold and warm components of the envelope could be distinguished. We find that the warm gas towards \mysou\ has an abundance of COMs lower by an order of magnitude relative to H$_2$ compared to their sample of hot cores.

Overall, we detect in \mysou\ similar families of molecules compared to hot corinos and hot cores, although both hot cores and hot corinos themselves show some diversity in their molecular composition. Quantitatively, however, using single-dish molecular abundance measurements with distinction between the warm and the cold components of the envelope for these samples, we find that the molecular abundances of COMs with respect to H$_2$ resemble that of the hot corino associated with IRAS16293--2422 rather than hot cores, which typically exhibit higher molecular abundances of COMs in their warm gas phase than \mysou.

\section{Summary and conclusions} \label{sec:conclusion}
We performed a spectral survey with the APEX telecope between 159\,GHz and 374\,GHz towards the high-mass protostar, \mysou.  We detected 39 species (61 including isotopologues), and report tentative detections for the NS$^+$, PN, DNC, and HCNH$^+$ species. The detected species include carbon chains, O-, N-, S-bearing, deuterated molecules, and COMs. Using an iterative process relying on rotational diagrams and LTE modelling with Weeds, we estimated the physical conditions (e.g. column density, excitation temperature, and size of the emitting region), and based on a simple source model, we inferred the molecular abundances of these species.  Our main results are listed below

\begin{itemize}
 \item We identify 2935 lines from O-, S-, and N-bearing and carbon chain molecules. The proportion between the different types of molecules suggests that the molecular content of \mysou\ is more similar to hot cores and hot corinos than to WCCC objects, but we find high abundances relative to H$_2$ from sulphur-bearing species.
 
    \item While most lines show a Gaussian profile, we also find kinematic signatures of outflow and infall motions. Several transitions require a two-component Gaussian fitting, corresponding to a narrow and broad velocity component characterised by $\Delta$v<6\kms, and $\Delta$v>6\kms, 
    where the position of the line peak coincides with the source $\varv_{\rm lsr}$. The quiescent gas component is found to show $T_{\rm ex}$<50\,K (for $\Delta$v<6\,\kms), while the broader velocity component exhibits $T_{\rm ex}$>50\,K (for $\Delta$v>6\,\kms). These two components correspond to a warm and cold gas phase in the envelope.
    \item Based on their line profiles, we also identify emission from the accretion shocks that show a velocity offset of $\sim$ $\pm$4\,\kms\ with respect to the source $\varv_{\rm lsr}$. Rotational transitions from the first torsionally excited state of CH$_3$OH exclusively pinpoint the accretion shocks, while we find that these velocity components are also present in the rotational transitions from the torsional ground state of CH$_3$OH, and rotational transitions from CH$_3$OCH$_3$, CH$_3$OCHO, HC(O)NH$_2$, and HDO. The accretion shocks exhibit the narrowest line profiles with  $\Delta\varv\sim$3\,\kms, and are associated with gas at $T_{\rm ex}=170-190$\,K.
  
    \item The molecular composition of the cold gas is characterised by small molecules and four COMs, CH$_3$OH, CH$_3$OCH$_3$, CH$_3$CHO, and CH$_3$CN, while among the detected molecules, the COMs (CH$_3$OH, CH$_3$OCH$_3$, CH$_3$OCHO, CH$_3$CHO, HC(O)NH$_2$, CH$_3$CN, C$_2$H$_5$CN, C$_2$H$_3$CN, CH$_3$SH) dominate the chemical composition of the warm envelope.
    
    \item The cold component of the envelope is particularly rich in sulphur-bearing molecules with abundances as high as $3.9\times10^{-8}$ for H$_2$S, which is even higher than the CH$_3$OH abundance of $1.9\times10^{-8}$. Sulphur-bearing ions such as HCS$^+$ and SO$^+$ have abundances of $1.5-1.9\times10^{-10}$, respectively. Overall, we find that the abundances of sulphur-bearing species are higher than typically observed towards other star-forming regions, suggesting that sulphur is either not as strongly depleted as towards other objects with a depletion greater than 1\,\%, or that efficient mechanisms liberate sulphur-bearing species into the gas phase.
   
    \item We detect DCN, DCO$^+$, HDCS, and HDCO towards the cold envelope component, while HDO is found in the warm component of the envelope as well as towards the accretion shocks. Similarly to other hot cores, in these molecules, \mysou\ exhibits a  deuterium fractionation of 0.2\%--3\%, which is lower than for hot corinos. However, the deuteration values in \mysou\ are found to be close to what is observed in dark clouds, suggesting that the prestellar stage might be shorter or warmer than the low-mass star formation scenario. 
    \item We investigate the isotopic ratios within the cold gas and find that \mysou\ exhibits an unusually low isotopic ratio of $^{12}$C/$^{13}$C of $\sim$30 based on column density estimates of C$^{34}$S and $^{13}$C$^{34}$S. Based on $^{13}$CS and $^{13}$C$^{34}$S, we also find a lower than expected value for the $^{32}$S/$^{34}$S isotopic ratio of 12, which is similarly about half of its expected value. The $^{34}$S/$^{33}$S isotopic ratio is about half of its expected value.

       \item
        Consistent with the conclusion of \citet{Csengeri2019}, we identify here emission from the warm component of the envelope, which is found to be compact (typically $<2$\arcsec) and exhibits a moderate temperature of $\sim$63--166\,K that is lower than that of the accretion shocks. Despite the presence of warm gas, 
 the global molecular emission is dominated by the cold component of the envelope, with a temperature around 20\,K.

    \item Four COMs are identified in the cold gas phase of the envelope (CH$_3$OH, CH$_3$OCH$_3$, CH$_3$CHO, CH$_3$CN) showing molecular abundances similar to that of chemical models of grain surface production and non-thermal desorption. 
    \item The richness in COMs within the warm gas suggests an increase in chemical complexity. The differences in molecular abundances between the cold and warm gas favour different chemical processes in action. The observed temperatures in the warm gas are found to be close to the thermal desorption temperatures of several COMs, suggesting that the molecular composition of the warm gas may reflect that of the  grains.
    \item Comparing the molecular richness to the archetypical hot corino, IRAS16293--2422, we find that it exhibits a similar richness in the number of molecules detected in its cold gas. A quantitative comparison between the cold and warm gas phases of the archetypical hot corino IRAS16293--2422 and \mysou\ shows that the cold component of the envelope has molecular abundances that are an order of magnitude higher in \mysou, while the warm gas exhibits similar molecular abundances in both types of objects. While the molecular emission of the warm gas is similar to that of hot cores, the molecular abundances of COMs relative to H$_2$ towards typical hot cores are found to be an order of magnitude higher than that of \mysou\ and IRAS16293--2422. 
    The moderate temperature of the warm gas phase together with abundances of COMs that are an order of magnitude lower than expected may suggest that thermal desorption has not yet liberated all grain surface products to the gas phase.
\end{itemize}

This unbiased spectral survey allowed us to investigate the molecular composition of \mysou\ in great detail. The number of detected molecules suggests an emerging molecular complexity in this envelope, in which  a cold and warm gas phase can be distinguished. Altogether, this points to a picture in which the overall molecular diversity of the source is similar to that of hot cores and hot corinos, but the gas in the vicinity of the high-mass protostellar embryo resembles that of a hot corino in the warm gas phase, rather than a hot core. A statistical study of high-mass protostars at a similarly early evolutionary stage should reveal whether this deeply embedded hot corino stage is a common phenomenon in the early evolutionary phase of high-mass star formation.

\begin{acknowledgements} 
We thank the referee for comments that helped improve the manuscript. This publication is based on data acquired with the Atacama Pathfinder Experiment (APEX). APEX is a collaboration between the Max-Planck-Institut f\"ur Radioastronomie, the European Southern Observatory, and the Onsala Space Observatory. L.B. thanks the following institutions for support: the International Max-Planck-Research School (IMPRS) for Astronomy and Astrophysics at the Universities of Bonn and Cologne, the Bonn-Cologne Graduate School (BCGS) and the Collaborative Research Centre 956 (sub-project B6) funded by the Deutsche Forschungsgemeinschaft (DFG). T. Cs. has received financial support from the French State in the framework of the IdEx Université de Bordeaux Investments for the future Program.

\end{acknowledgements}

\bibliographystyle{aa} 
\bibliography{G328p25}

\begin{appendix}

\section{Simple molecules}

\subsection{Carbon chain molecules}

Carbon chains  are unsaturated molecules, for example C$_n$H, c-C$_3$H$_2$, CH$_3$CCH, and HC$_{2n+1}$N. Because we do not cover the 3mm atmospheric window, we do not have access to the heavier carbon chain molecules in the cold phase, such as HC$_{2n+1}$N, where $n>1$, because their spectral lines in the range of our survey are too weak to be detected at low temperature. While prevalent in quiescent molecular clouds, they have also been studied in detail towards nearby protostellar objects \citep{Turner1991, Beuther2008, Sakai2008, Sakai2010}, which led to the definition of protostellar envelopes with \textsl{\textup{warm carbon chain chemistry}} (WCCC), which are characterised by being particularly rich in unsaturated hydrocarbons (e.g.\,\citealp{Sakai2008}). This molecular composition is different from \textsl{\textup{hot cores}} and \textsl{\textup{hot corinos}} (e.g.\,\citealp{Ceccarelli2000, Ceccarelli2007}), which are instead dominated by saturated hydrocarbons and exhibit a rich emission from COMs. However, there is at least one exception, the low-mass protostar L483, which exhibits the characteristics of both chemical types \citep{Oya2017}. Because carbon chain molecules are predicted to form in larger amounts in  gas heated to lukewarm temperatures of around 30\,K \citep{Sakai2013}, the difference in chemistry for hot corinos and WCCC sources has been explained by a long versus short starless-core phase, implying different physical conditions in the pre-collapse gas.

Towards \mysou, we detect the carbon chain species CCH, c-C$_3$H$_2$ , and CH$_3$CCH in the cold component of the envelope, typically exhibiting low excitation temperatures of 13-35\,K. We also identify emission from HC$_3$N in the cold component of the envelope at a similar temperature. However, this molecule is also detected in the warm component of the envelope with a similar abundance. We estimate an upper limit for c-C$_3$H$_2$ and CH$_3$CCH in the warm envelope that is at least an order of magnitude lower than what is observed in the cold phase. Heavier unsaturated species such as C$_4$H, l-C$_3$H, and C$_2$S, C$_2$O seen towards the archetypical WCCC source, L1527, \citep{Sakai2008} are not detected. The most promising lines of these molecules have their low upper-level energy transitions at lower frequencies than our survey, but their non-detection in our survey suggests that they are not present in the warm component of the envelope, with an upper limit on their abundances relative to H$_2$ of $\sim$10$^{-10}$ (10$^{-9}$ for C$_4$H). These results are consistent with previous observations suggesting that  the lightest carbon chain species are frequently detected towards high-mass star-forming regions.

\citep{Sakai2008, Sakai2010, Sakai2013}.

Recent observations suggest that some carbon chain species, such as CCH and HC$_3$N, could be as abundant as CH$_3$OH \citep{Taniguchi2018,Taniguchi2021}.
The molecular abundances in the cold material resemble what is observed towards other hot cores and hot corinos \citep{Aikawa2008, Sakai2010}. 
The main formation path for unsaturated carbon chain molecules is mostly through reactions of CH$_4$ and C$^+$ \citep{Aikawa2008, Sakai2013}. The desorption of CH$_4$ from the grains is expected at a temperature of $\sim$25\,K, which then reacts with C$^+$ in the gas phase. This formation route is possible only for low temperature. When the temperature increases above $\geq30$\,K, other radicals desorb from the grain surfaces \citep{Garrod2008}, and C$^+$ reacts preferentially with them to form COMs. Therefore, WCCC develops preferentially when the temperature is just enough to desorb CH$_4$ from the grains at a temperature of $\sim$25\,K because this molecule is not strongly bound to the ices. In \mysou\ the carbon chains in the cold component of the envelope might be formed through these reactions, which would explain their observed temperatures. 
HC$_3$N is found in similar abundances in the cold and warm components of the envelope, but our upper limits show that other carbon chains have an abundance at least an order of magnitude lower in the warm component of the envelope, which suggests different chemistry in the cold and warm components of the envelope. Therefore, HC$_3$N is more likely to form at high temperature through the reaction between H$_2$ and C$_3$N \citep{Hassel2011}, which could explain the presence of HC$_3$N in the hot gas phase.

\subsection{N-bearing molecules}

We detect several N-bearing molecules towards \mysou. Most of them are composed of H, N, and C atoms, CN, HNC, HCN, N$_2$H$^+$, and CH$_2$NH, and two with an oxygen atom, HNCO and NO. All these molecules are important precursors of a more complex chemistry.

Methylenimine, CH$_2$NH (together with CH$_3$NH$_2$, CH$_3$CN, and HC(O)NH$_2$) is considered as a prebiotic molecule and potential precursor of glycine \citep[e.g.][]{Holtom2005}. We detect it in the cold component of the envelope at a temperature of 43\,K with a column density of 3.0$\times$10$^{14}${\unidens} and an abundance relative to H$_2$ of 4.5$\times$10$^{-10}$. This temperature is similar to that of other COMs in the cold gas phase (see Sect.\,\ref{sec:COMs}).

HNCO is thought to be a potential precursor of formamide \citep{Mendoza2014, Lopez-Sepulcre2015}, but this link is debated based on different laboratory experiments \citep[e.g.][]{Noble2015, Haupa2019}. We detect HNCO both in the cold and warm gas phase, but with a difference of two orders of magnitude in column density. 
The abundance relative to H$_2$ of 3.7$\times$10$^{-10}$ in the cold component of the envelope is similar to what is found in the envelope of low-mass protostars \citep{Lopez-Sepulcre2015}, and the abundance relative to H$_2$ of 1.7$\times$10$^{-8}$ in the warm component of the envelope is similar to what is observed towards hot cores \citep{Bisschop2007}.

We also detect NO in the cold component of the envelope with a column density of 6.0$\times$10$^{15}$\,cm$^{-2}$ and an abundance relative to H$_2$ of 2.3$\times$10$^{-8}$ at a temperature of 14\,K. This value is similar to what is found in quiescent gas towards low-mass protostars \citep{Akyilmaz2007, Codella2018} and dark clouds like TMC-1 \citep{Gerin1993}.

\section{Rotational diagrams}
\label{app:rot_diag}
\subsection{Carbon chains}
\begin{figure*}[!h]
    \centering
    \includegraphics[scale=0.25]{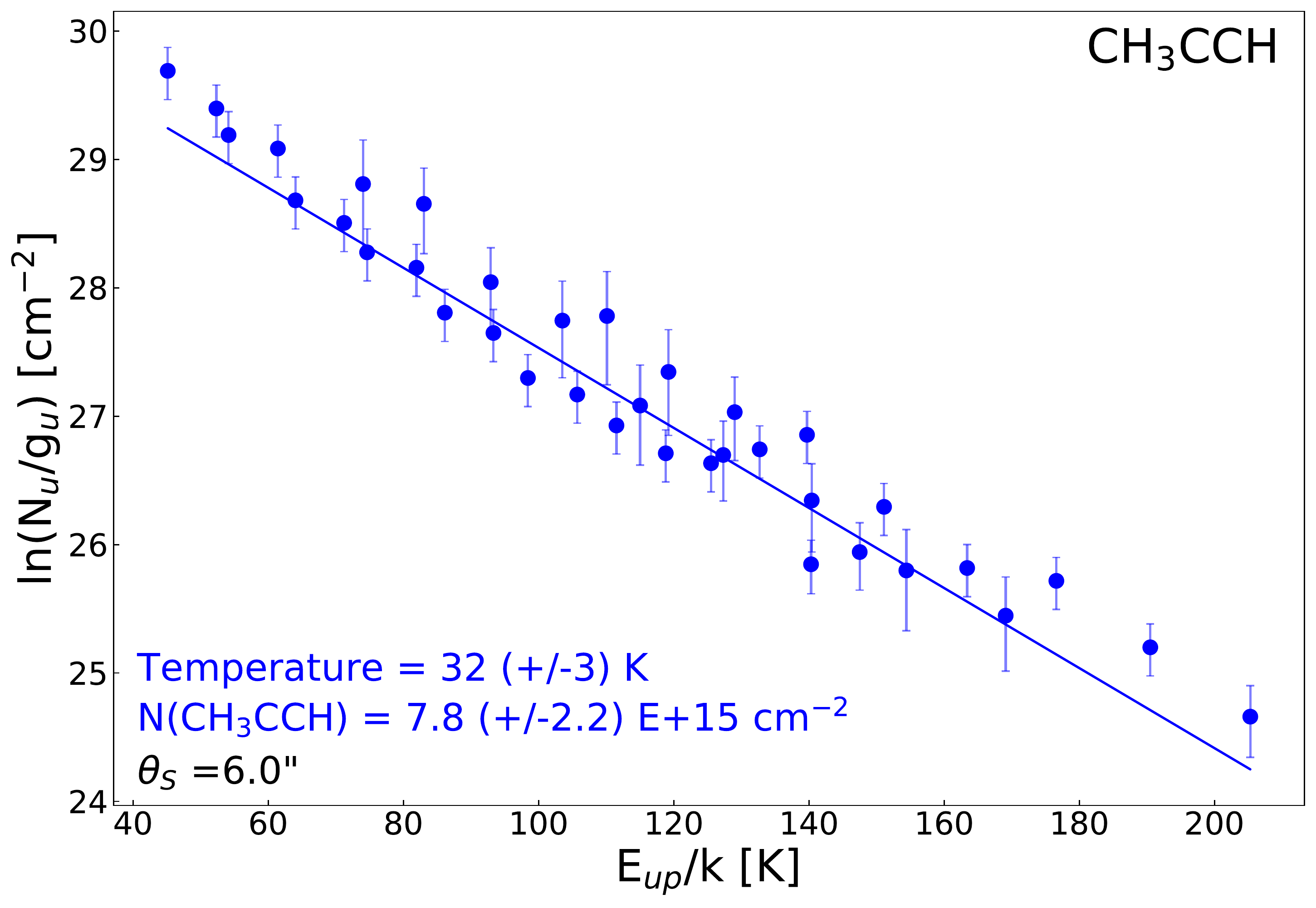}
    \includegraphics[scale=0.25]{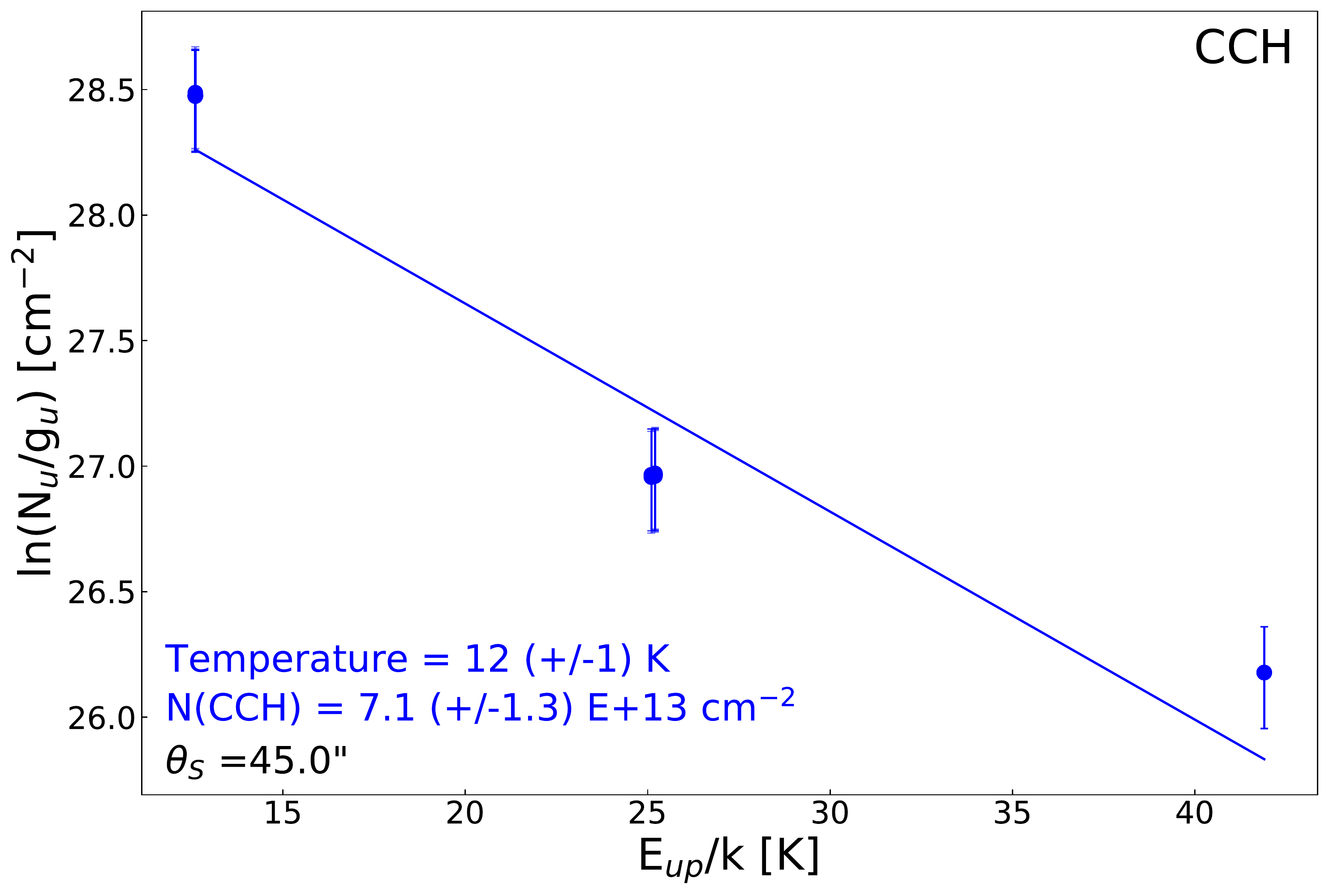}
    \includegraphics[scale=0.25]{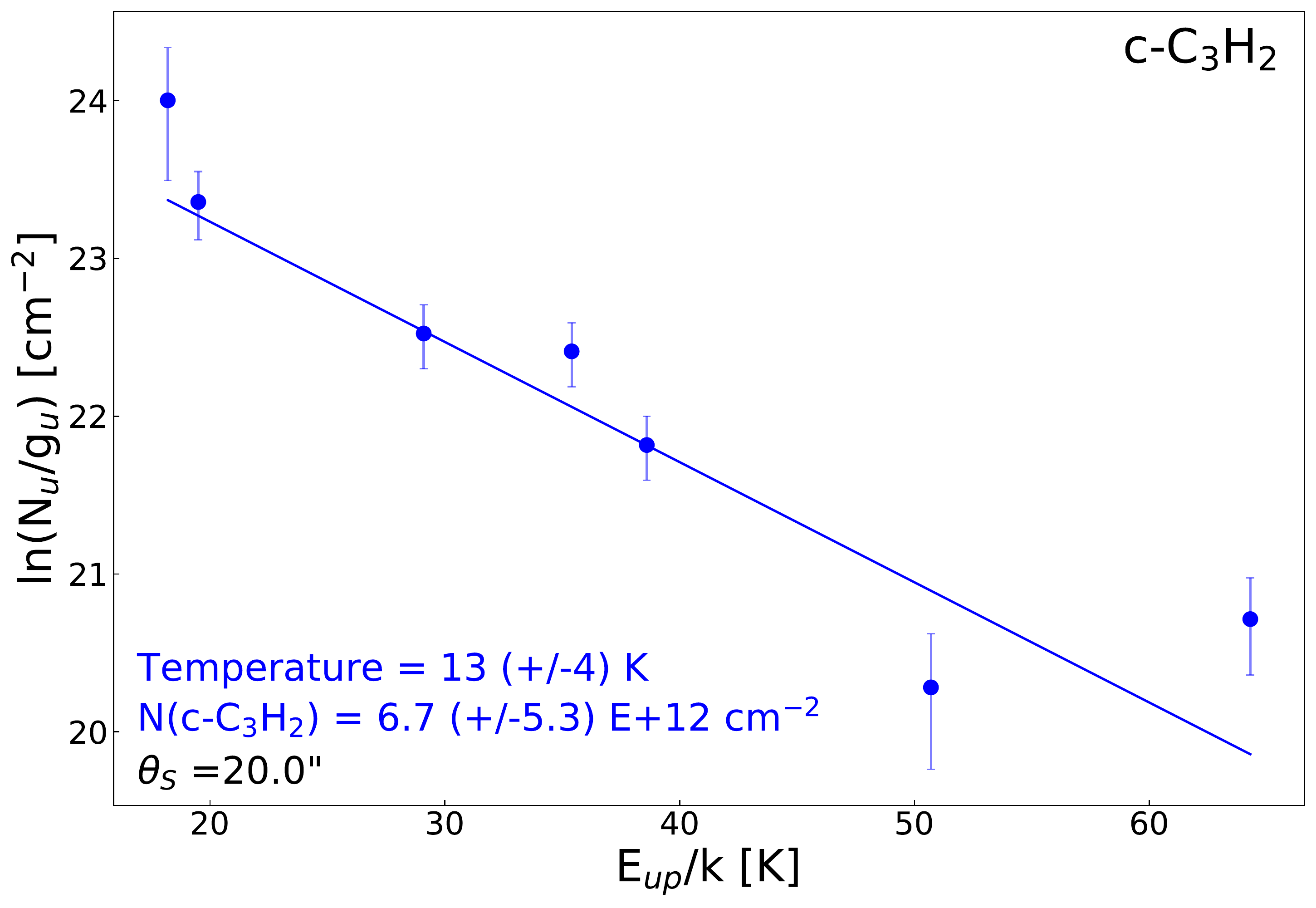}
    \includegraphics[scale=0.25]{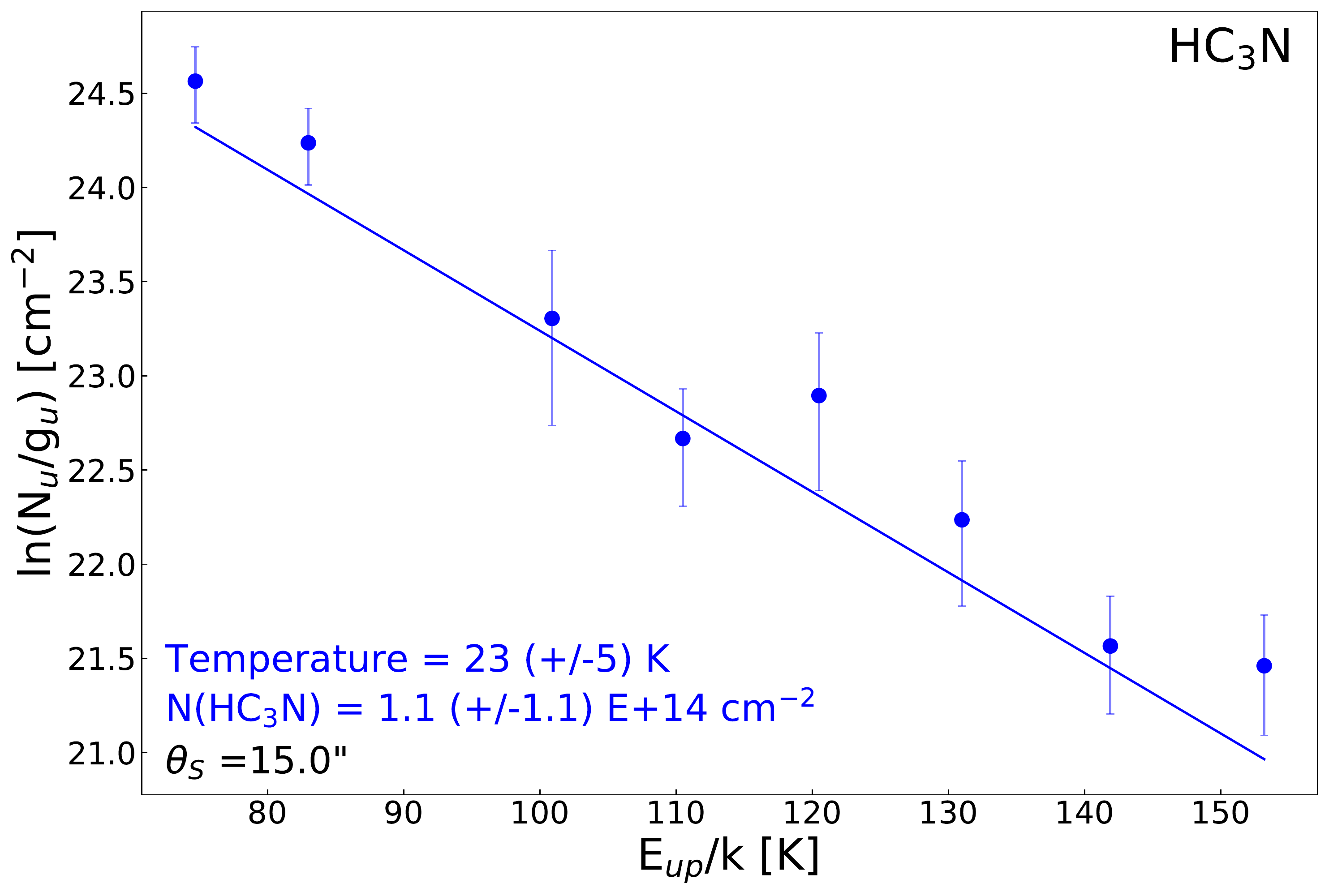}
    \includegraphics[scale=0.25]{HC3NL1_opacity.pdf}
    \includegraphics[scale=0.25]{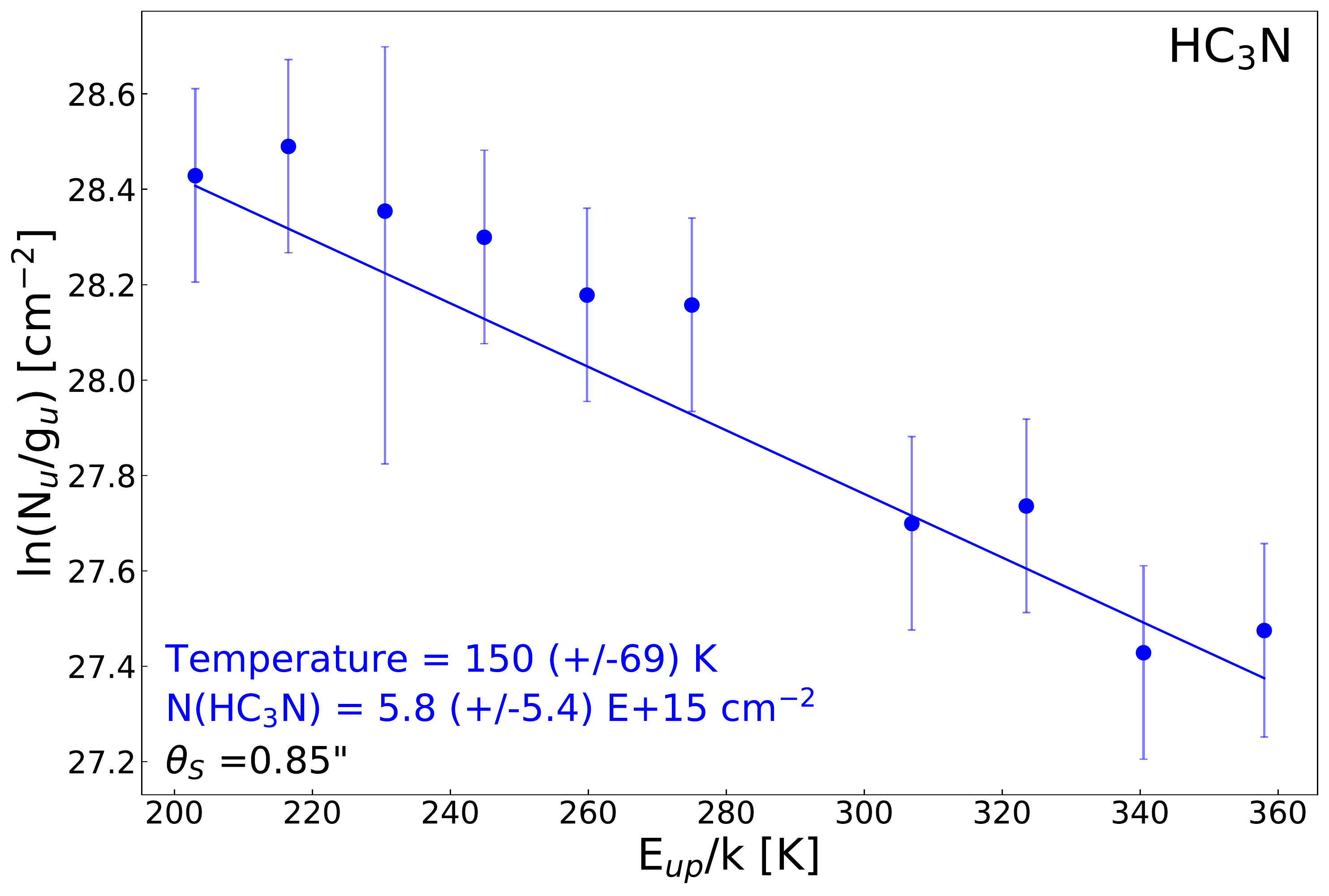}
    \includegraphics[scale=0.25]{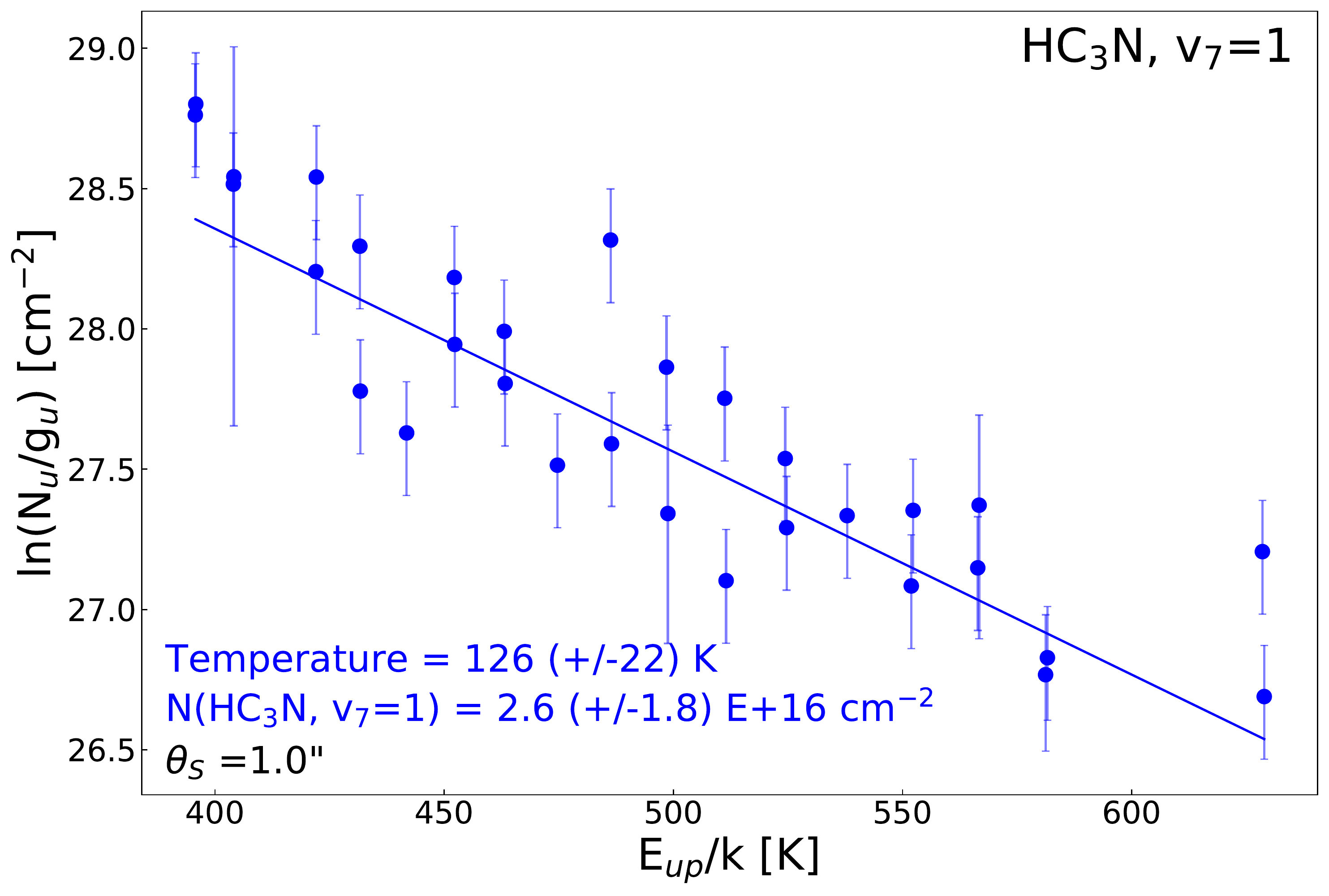}
    \caption{Rotational diagrams of the carbon chains CH$_3$CCH, CCH, c-C$_3$H$_2$, and HC$_3$N. HC$_3$N is described by three rotational diagrams showing the three temperatures traced by HC$_3$N. The first rotational diagram is for the narrow component and low temperature. The two following rotational diagram are the two rotational diagrams for the two temperature components covered with the larger line width. Finally, we show the vibrationally excited state of HC$_3$N, which is a different entry and represents the column density of the molecule present in this vibrationally excited state.}
    \label{diag_CC}
\end{figure*} 

\subsection{O-bearing molecules}

\begin{figure*}[!h]
    \centering
    \includegraphics[scale=0.25]{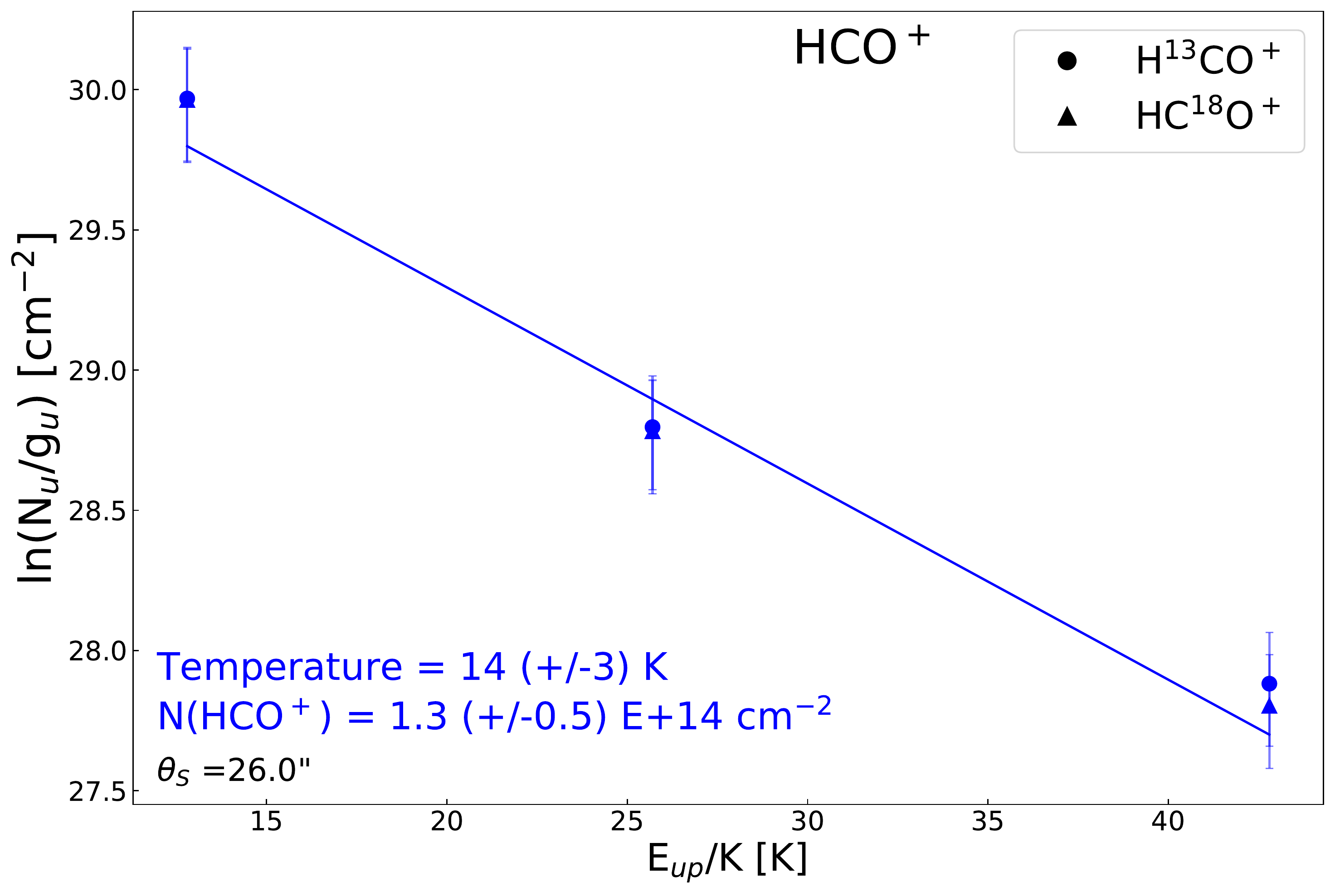}
    \includegraphics[scale=0.25]{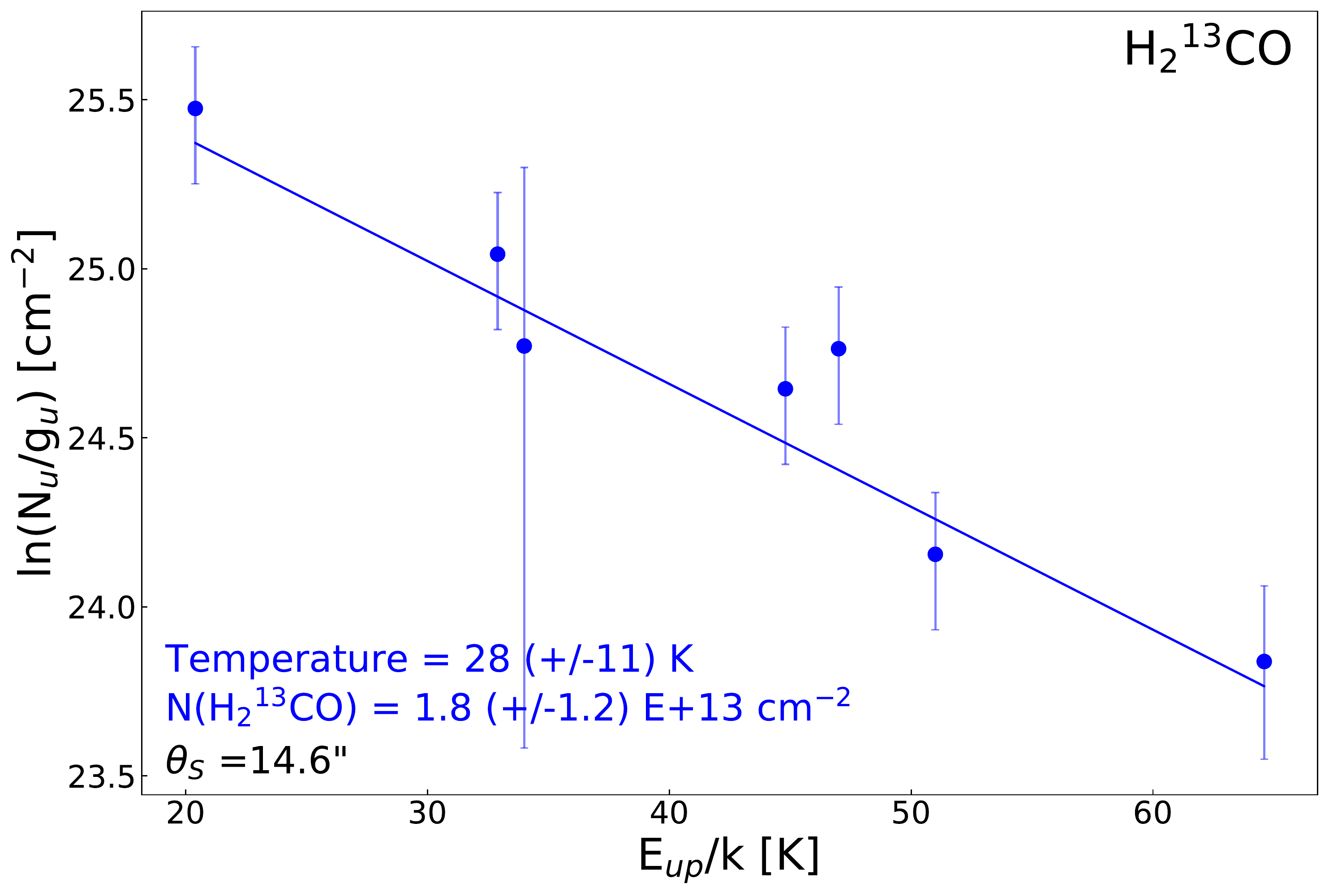}
    \includegraphics[scale=0.25]{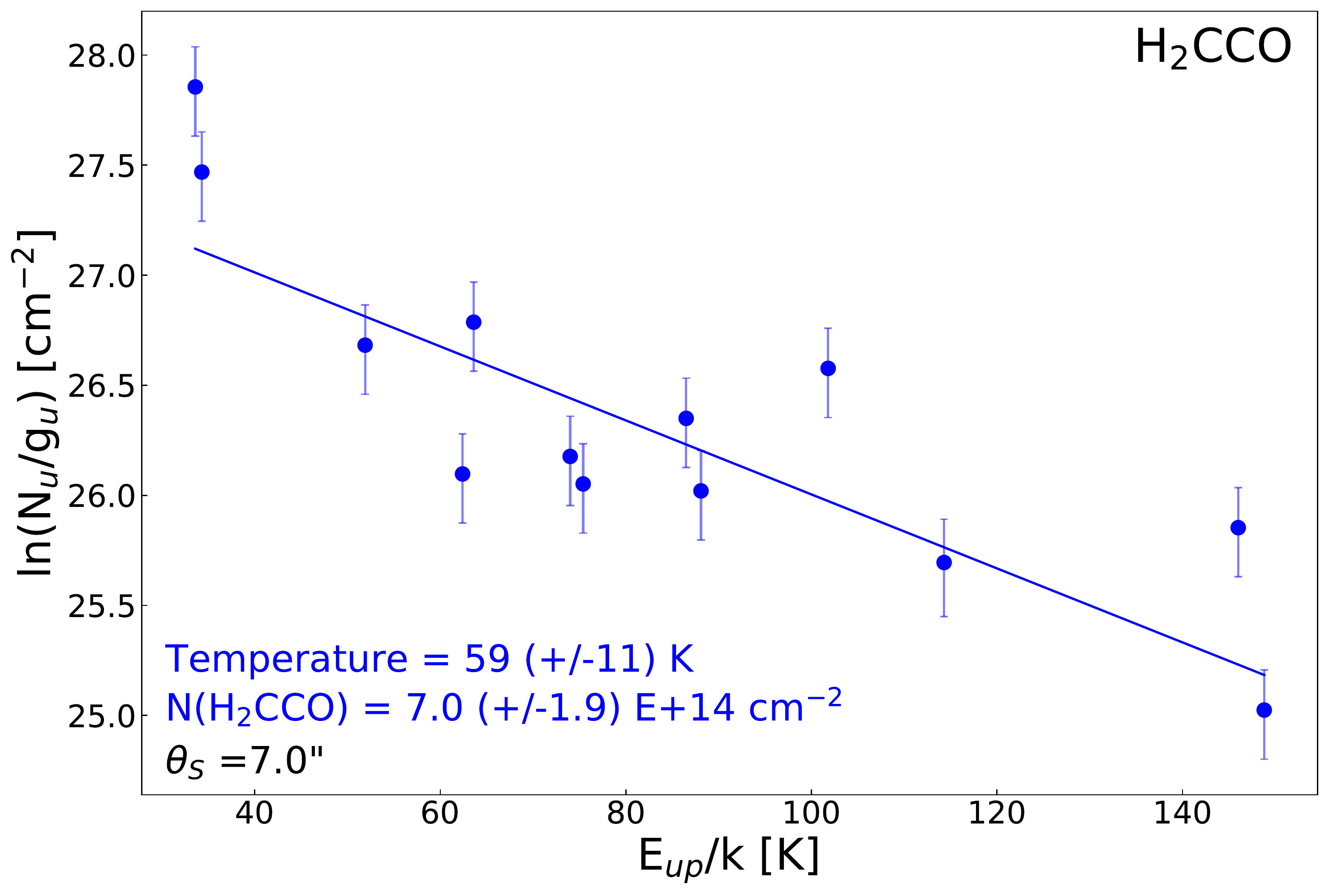}
    \caption{Rotational diagrams of the O-bearing molecules HCO$^+$, H$_2\,^{13}$CO, and H$_2$CCO.}
    \label{diag_O}
\end{figure*}

\subsection{N-bearing molecules}
\begin{figure*}[!h]
    \centering
    \includegraphics[scale=0.25]{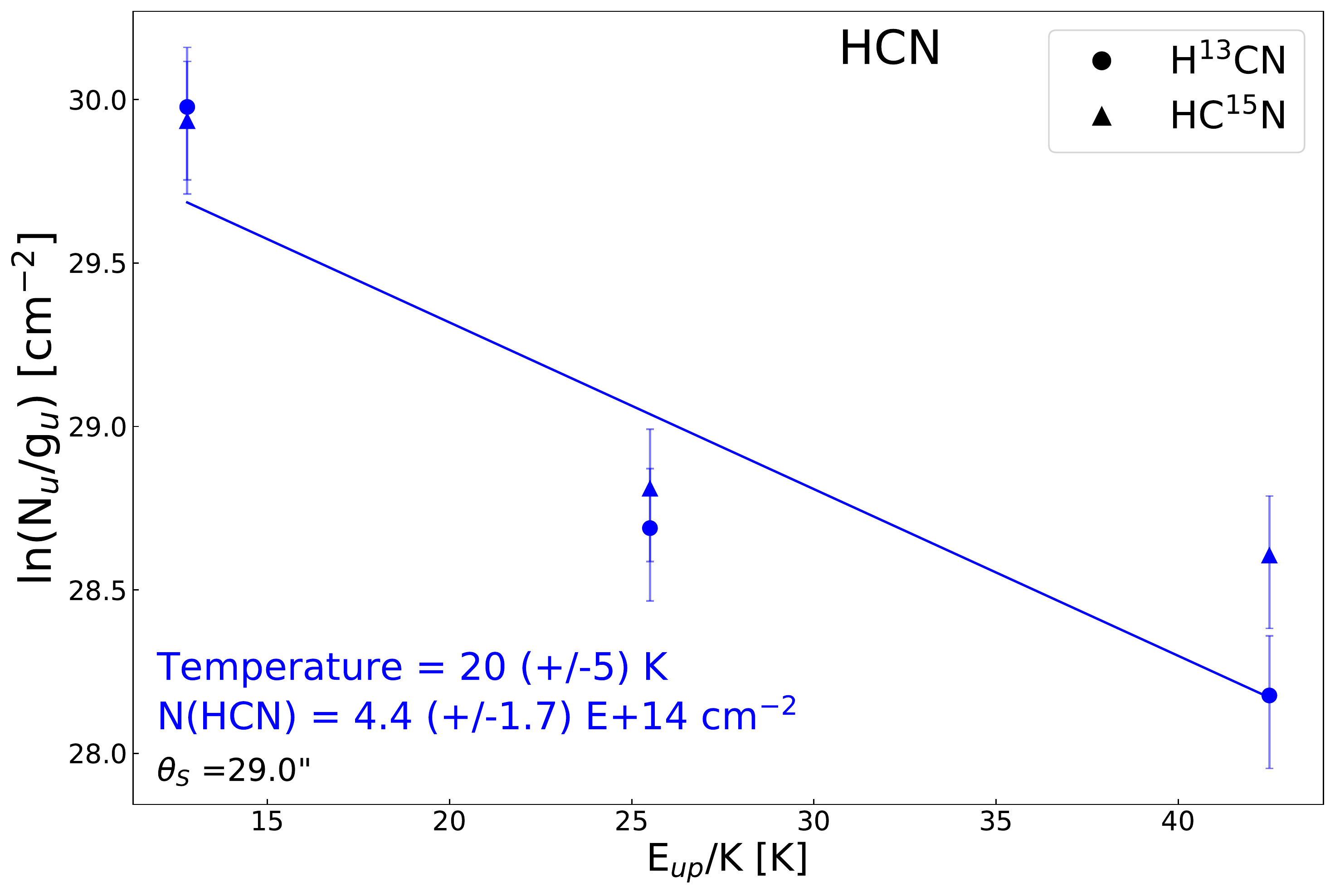}
    \includegraphics[scale=0.25]{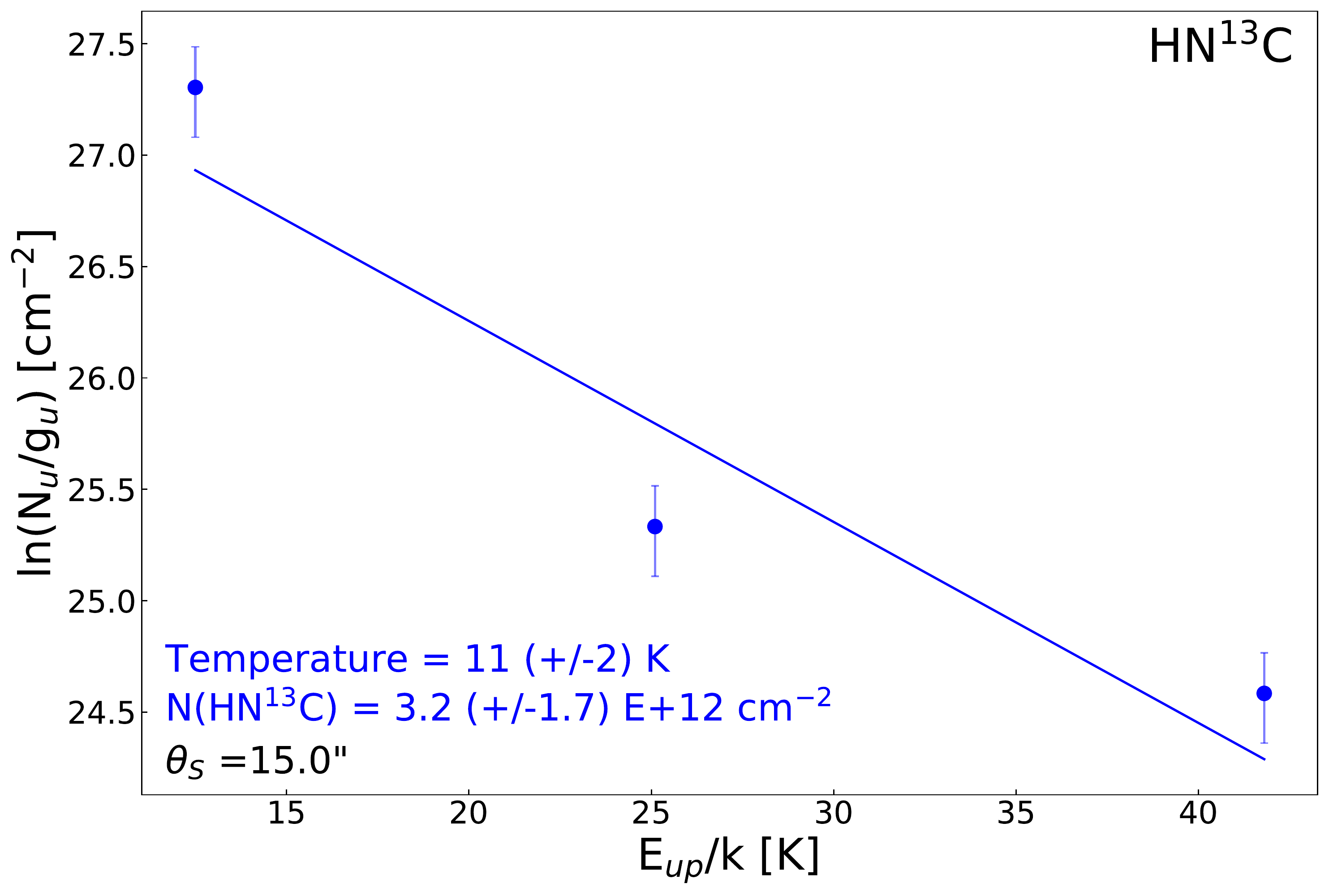}
    \includegraphics[scale=0.25]{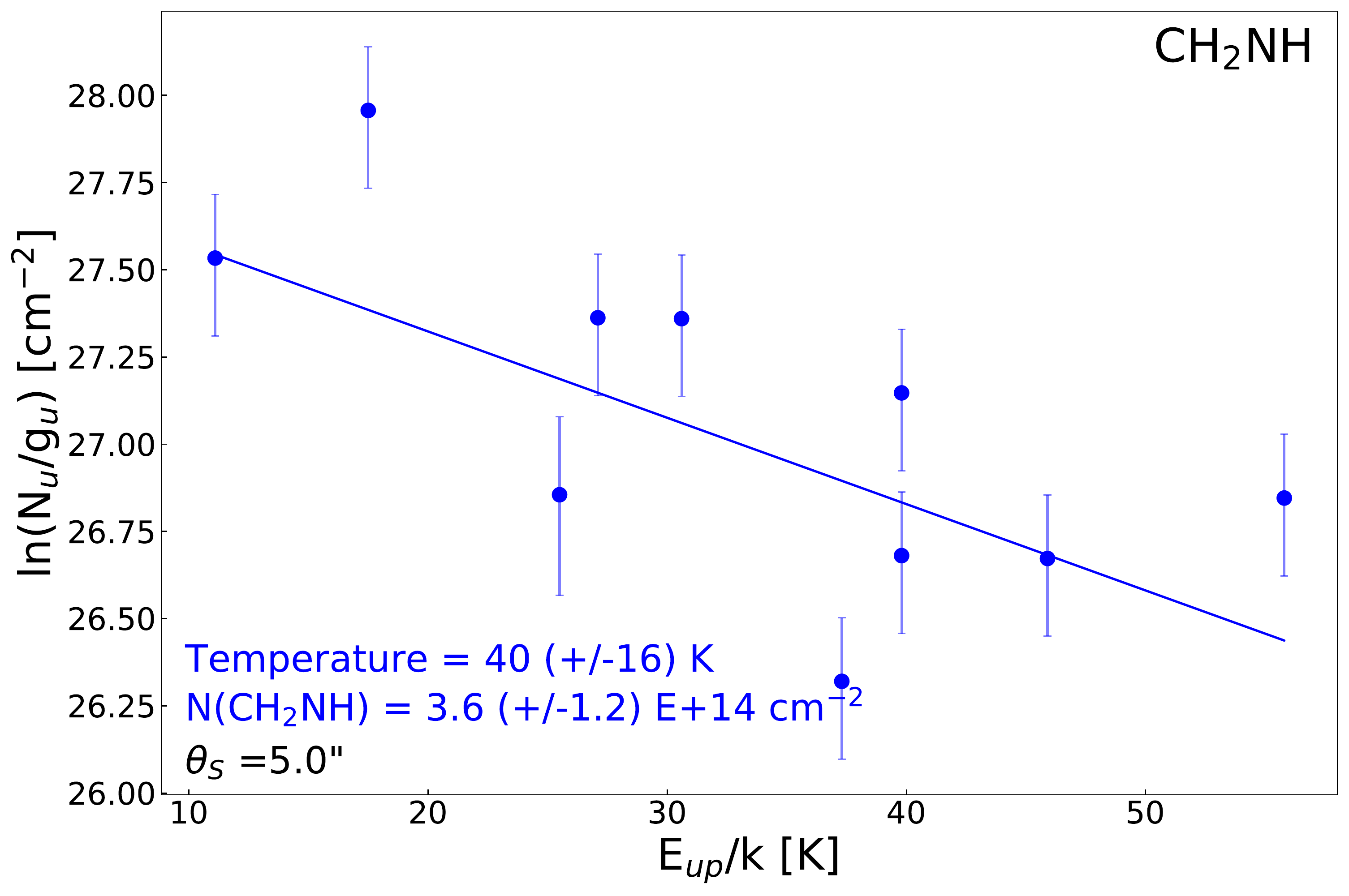}
    \includegraphics[scale=0.25]{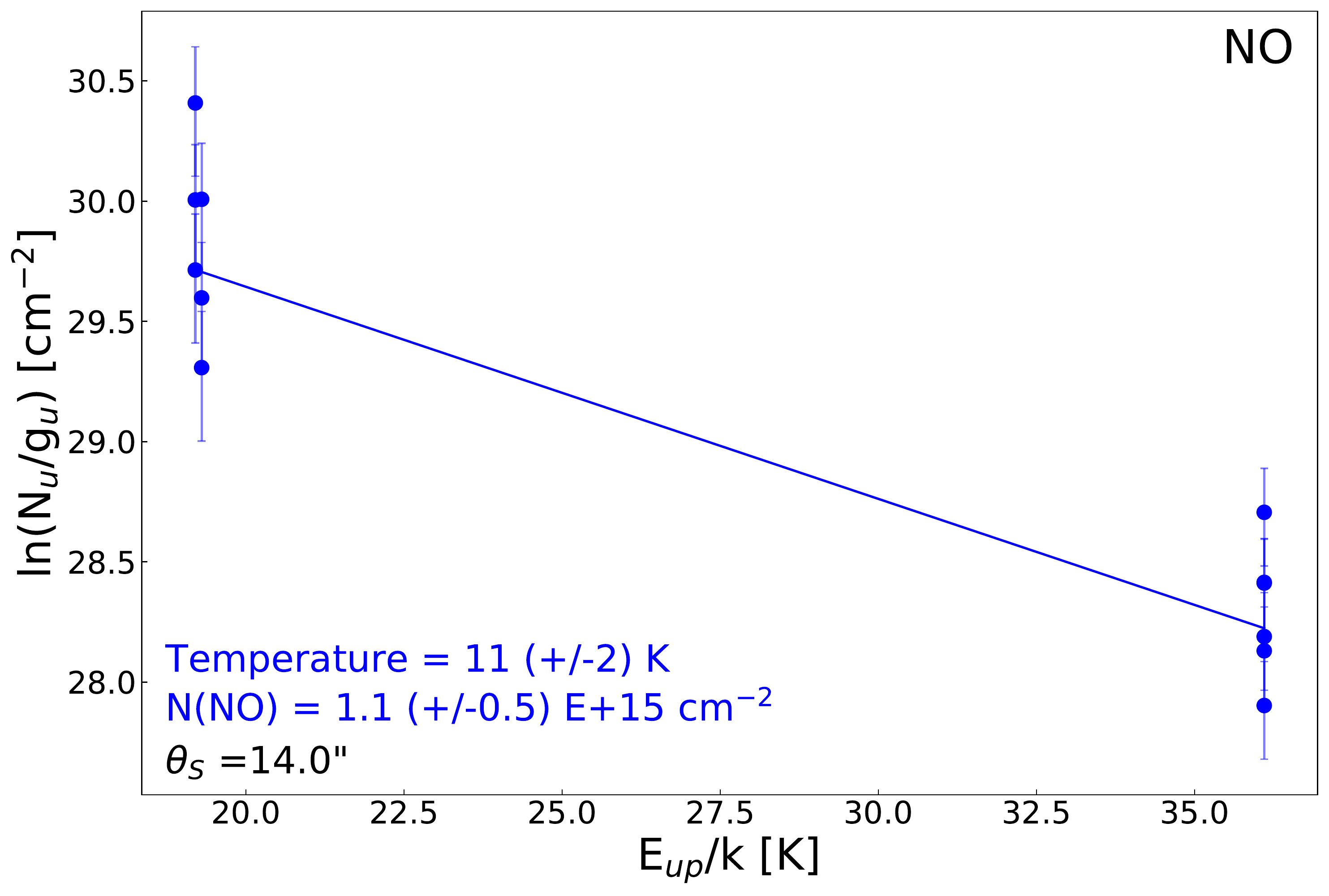}
    \includegraphics[scale=0.25]{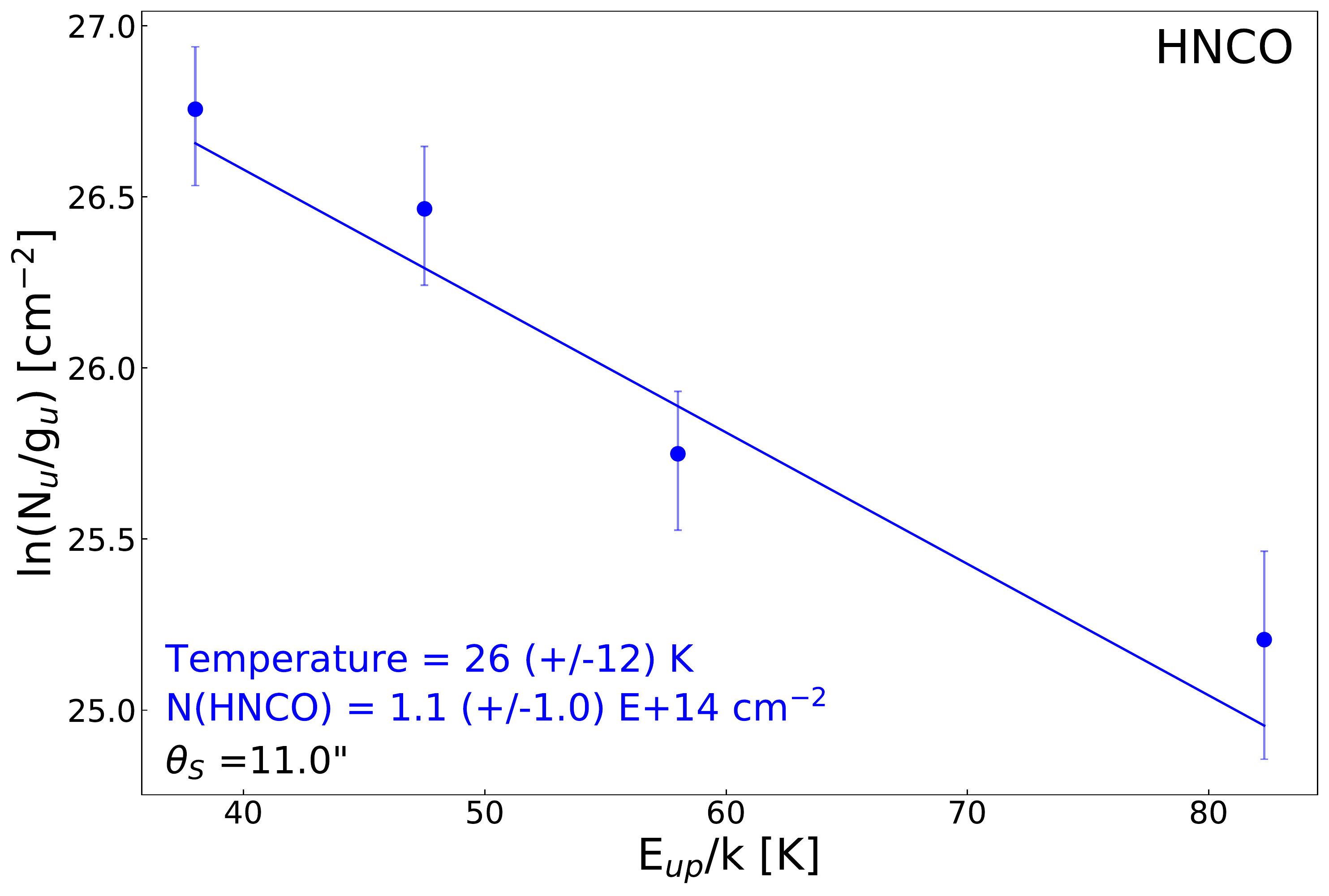}
    \includegraphics[scale=0.25]{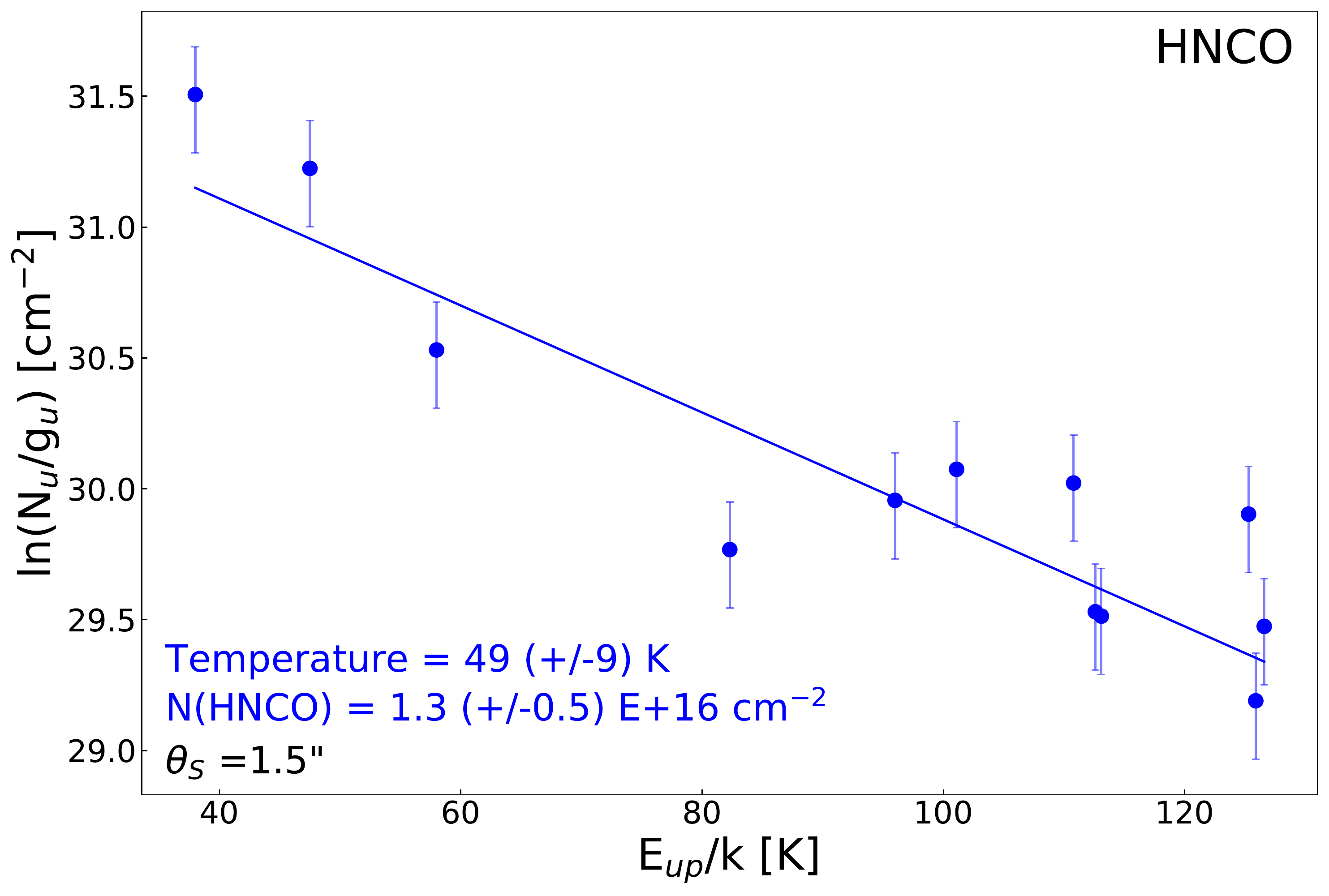}
    \includegraphics[scale=0.25]{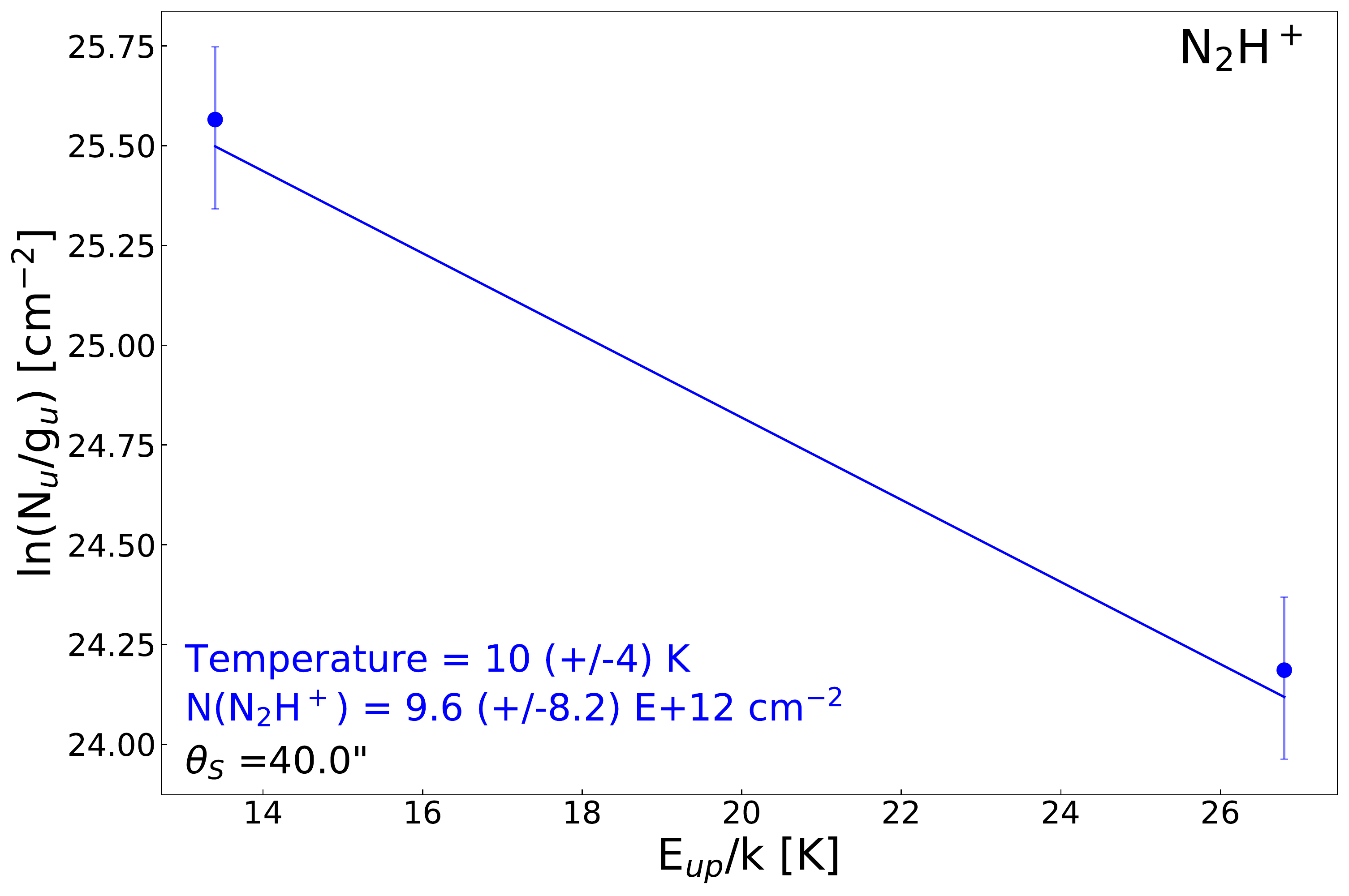}
    \caption{Rotational diagrams of the N-bearing molecules HCN, HN$^{13}$C, CH$_2$NH, NO, HNCO, and N$_2$H$^+$. The two rotational diagrams of HNCO represent the narrow and the broader components of the Gaussian fit.}
    \label{diag_N}
\end{figure*}

\subsection{S-bearing molecules}

\begin{figure*}[h!]
    \centering
    \includegraphics[scale=0.25]{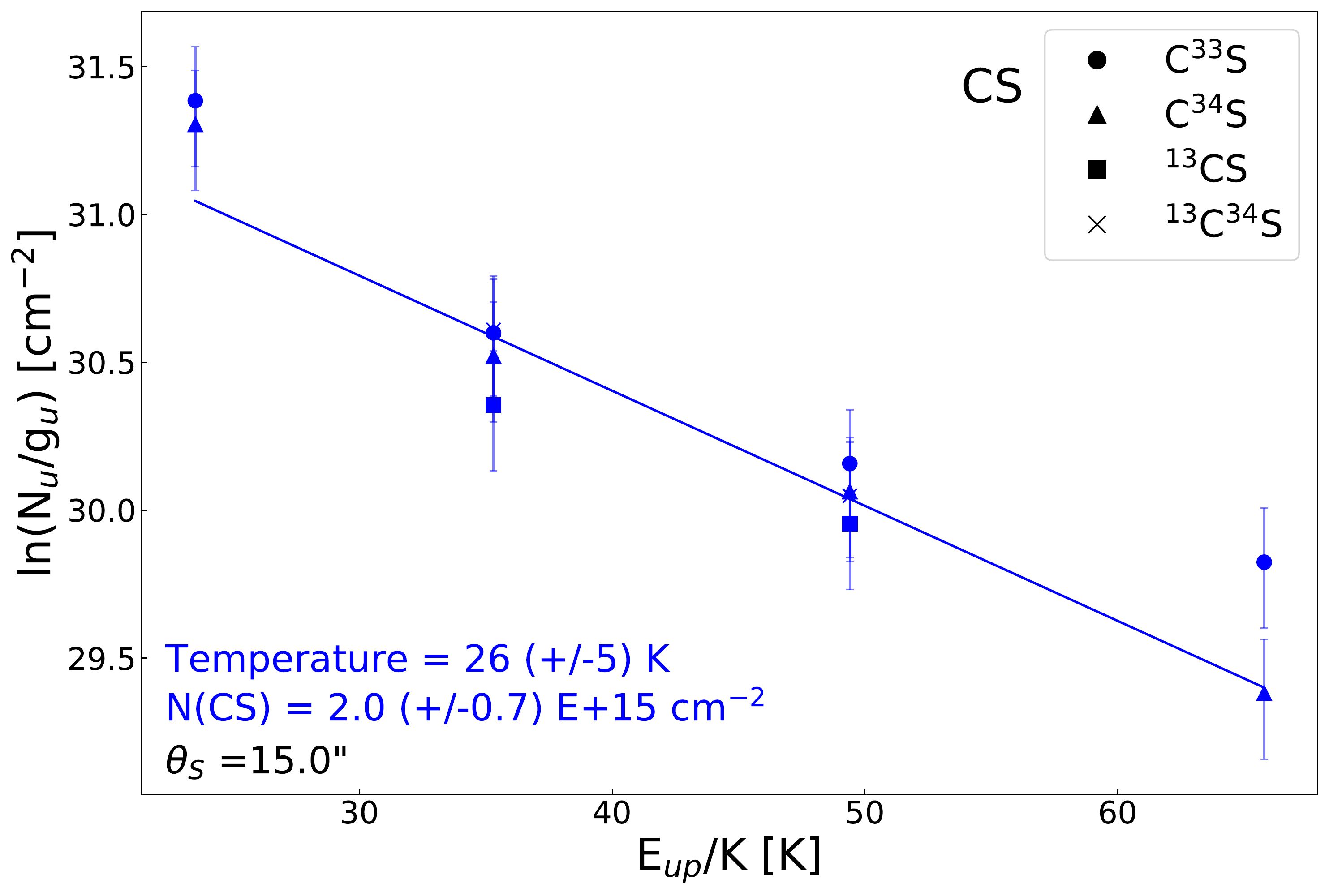}
    \includegraphics[scale=0.25]{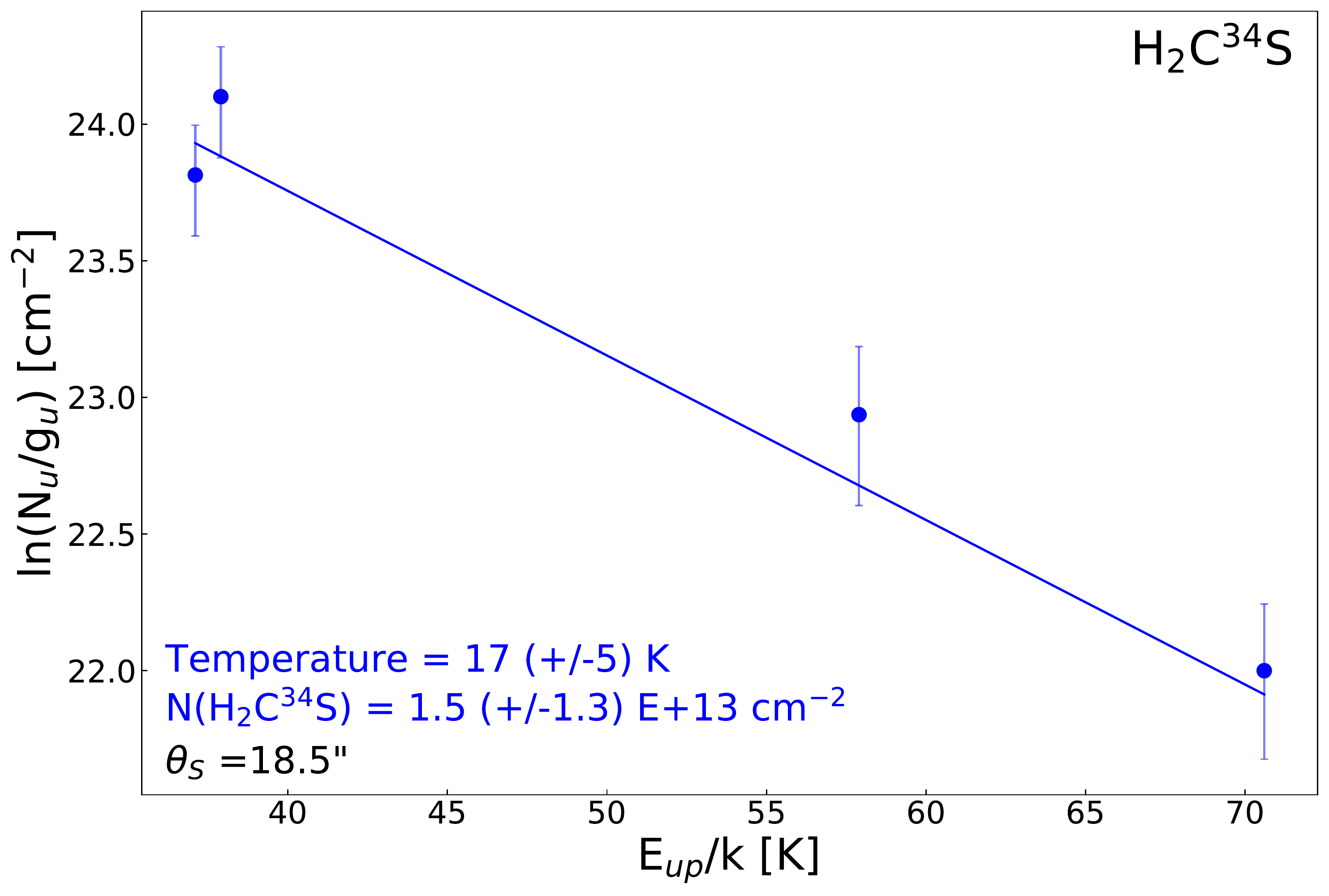}
    \includegraphics[scale=0.25]{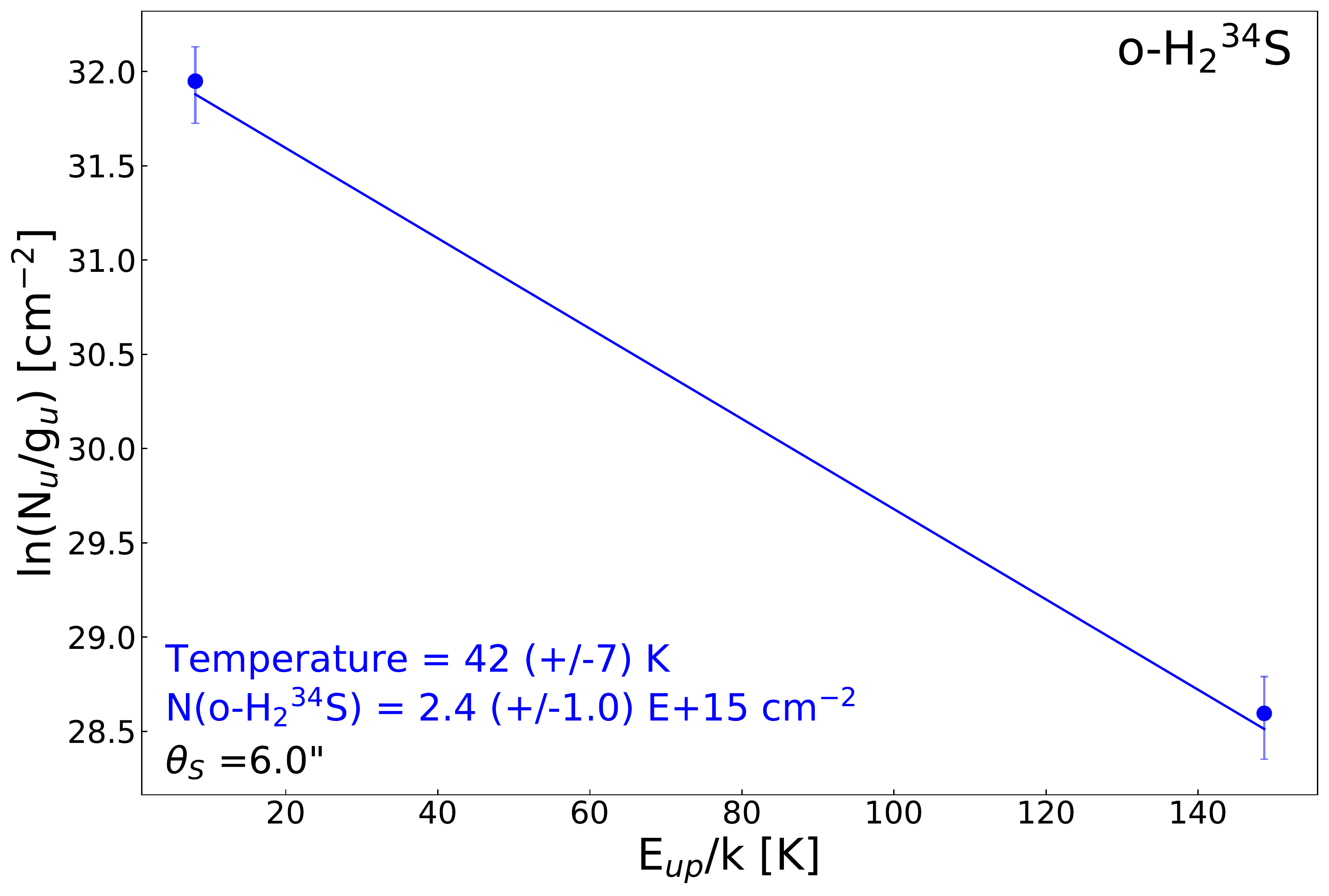}
    \includegraphics[scale=0.25]{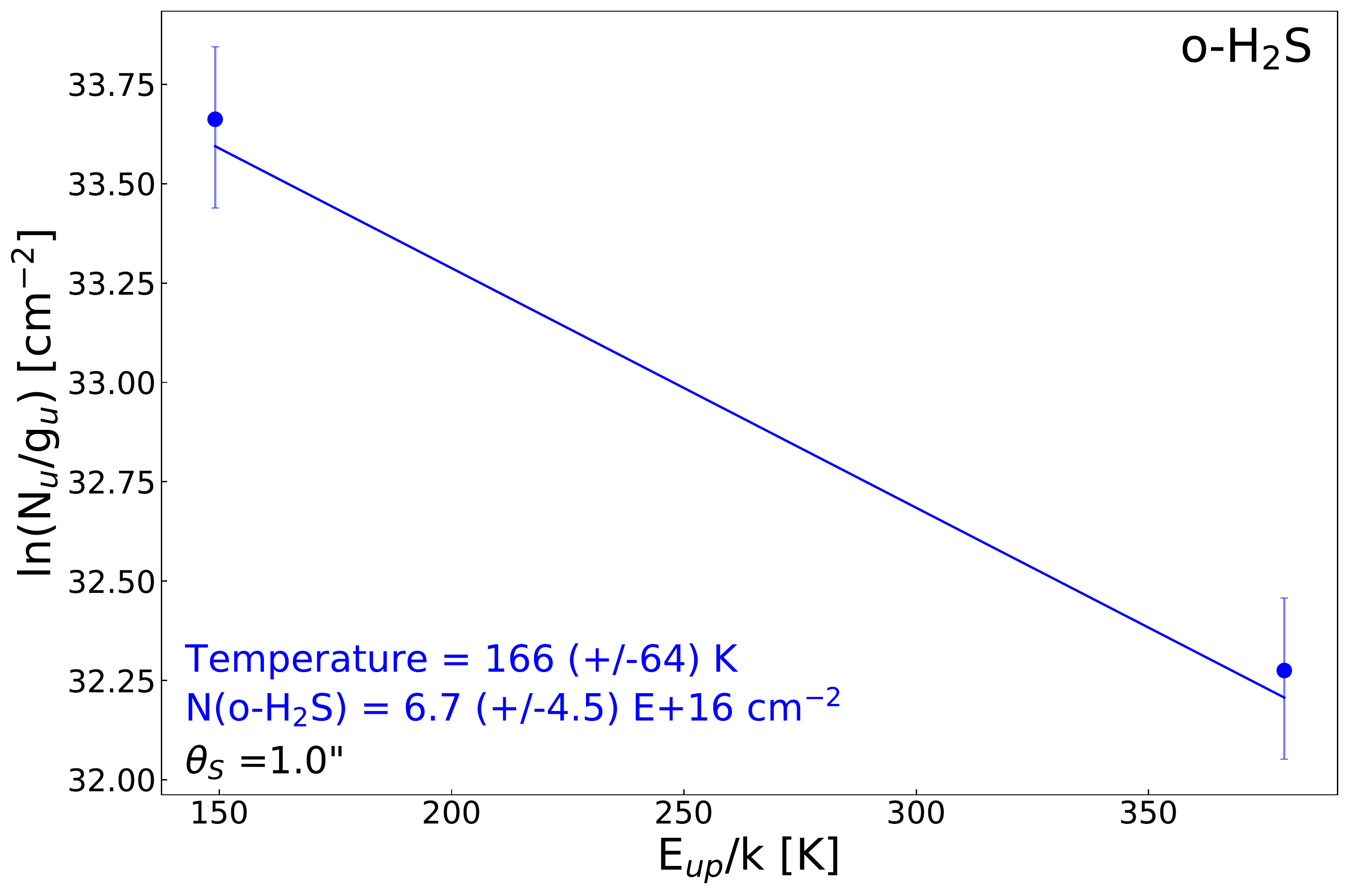}
    \includegraphics[scale=0.25]{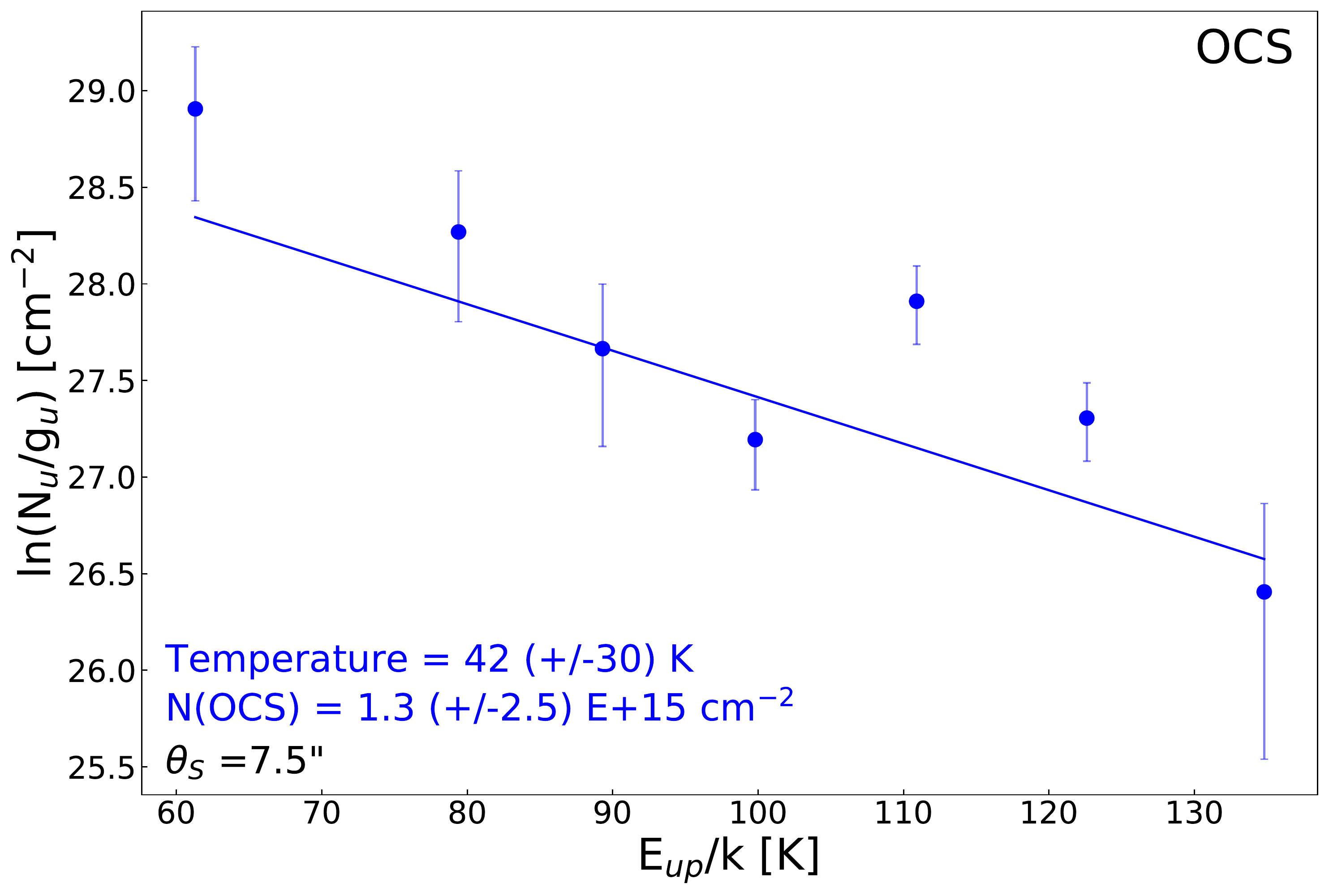}
    \includegraphics[scale=0.25]{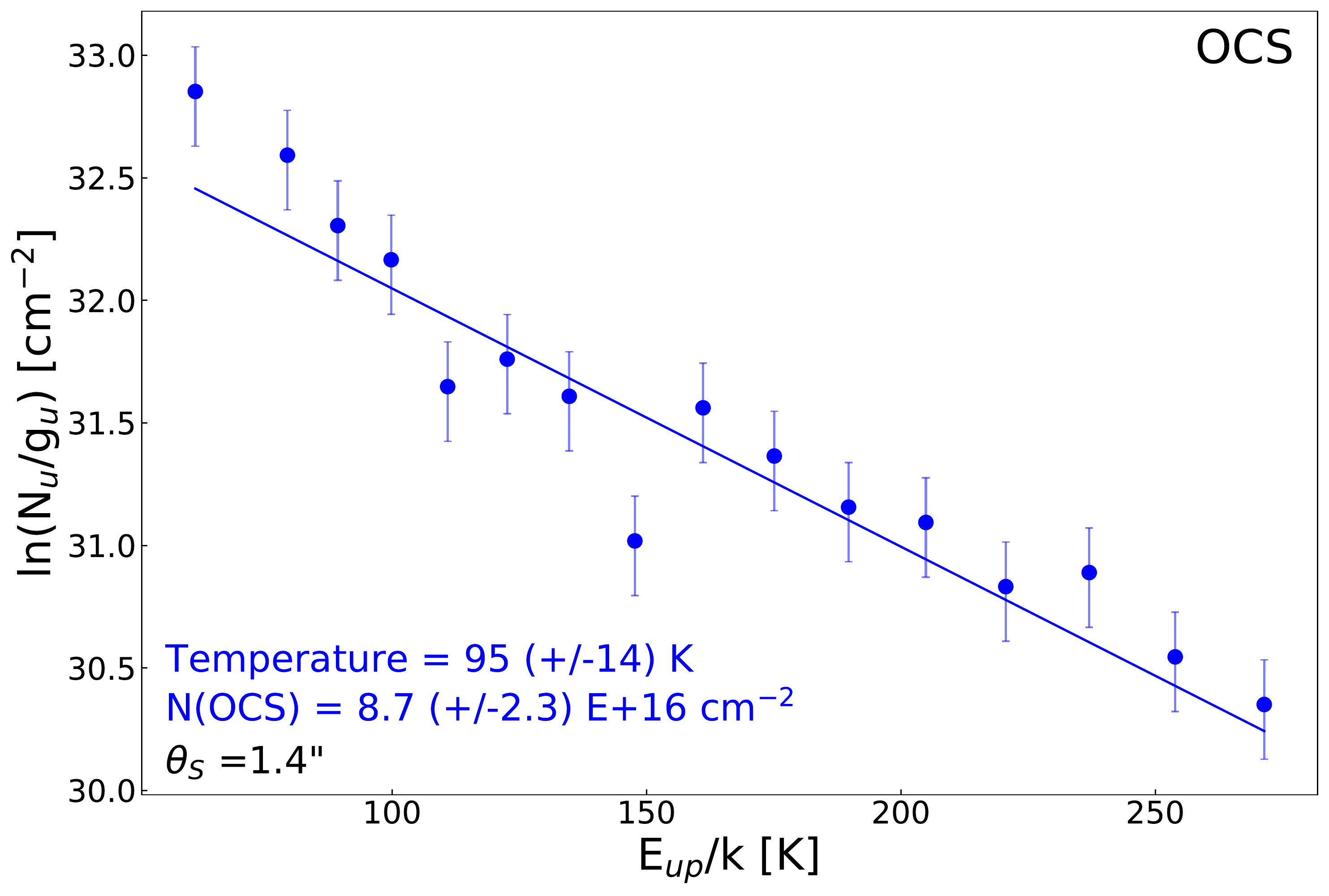}
    \includegraphics[scale=0.25]{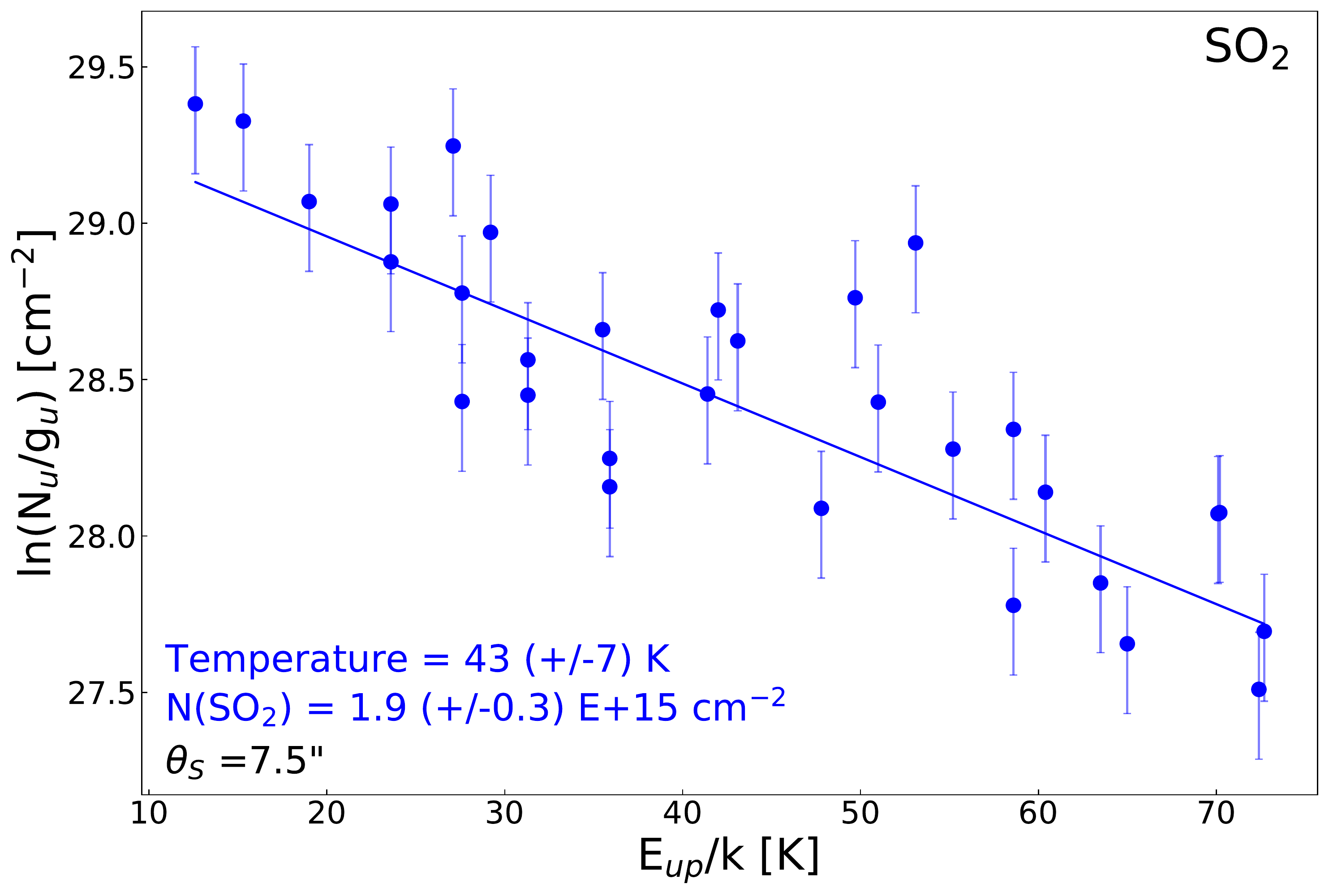}
    \includegraphics[scale=0.25]{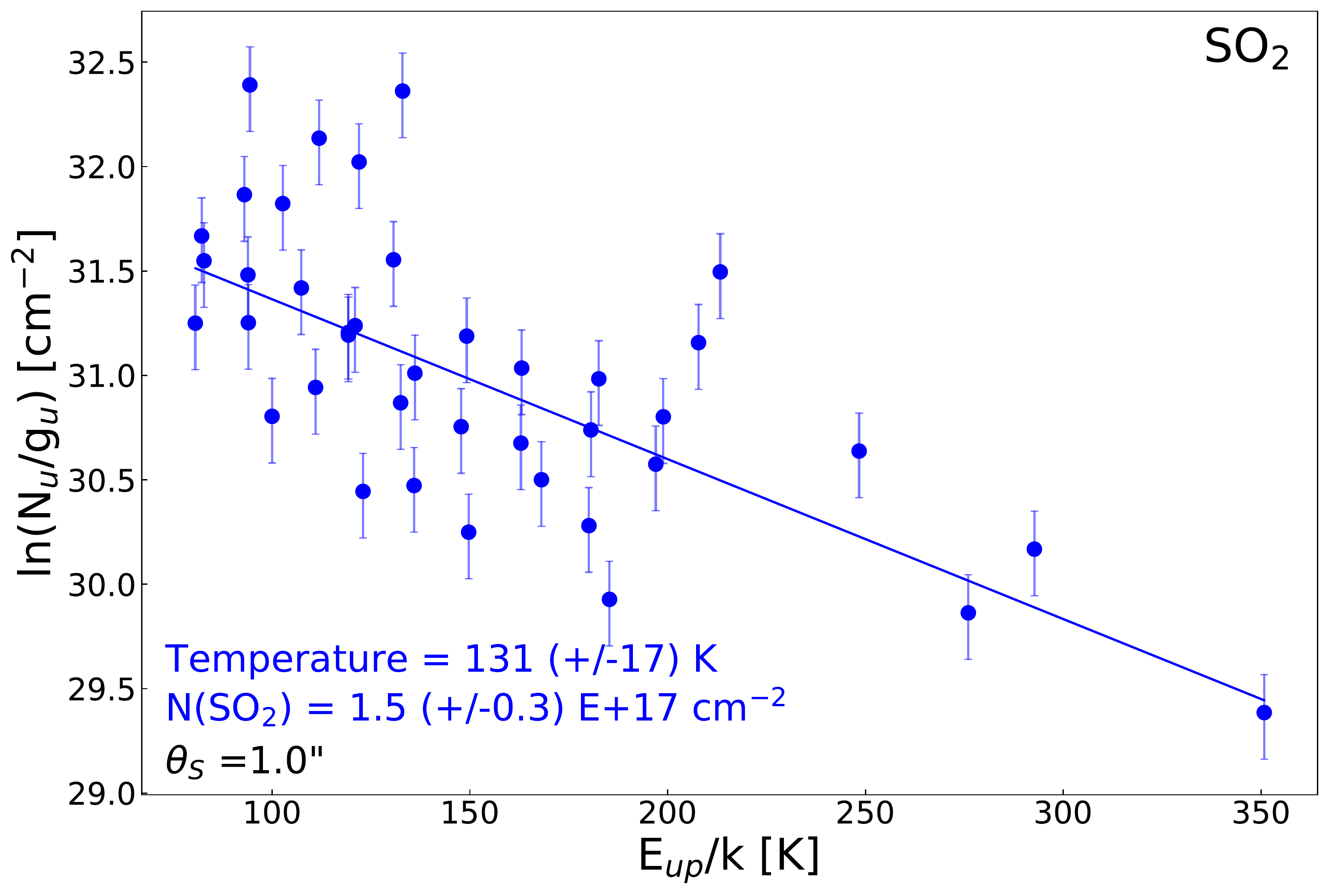}
    \includegraphics[scale=0.25]{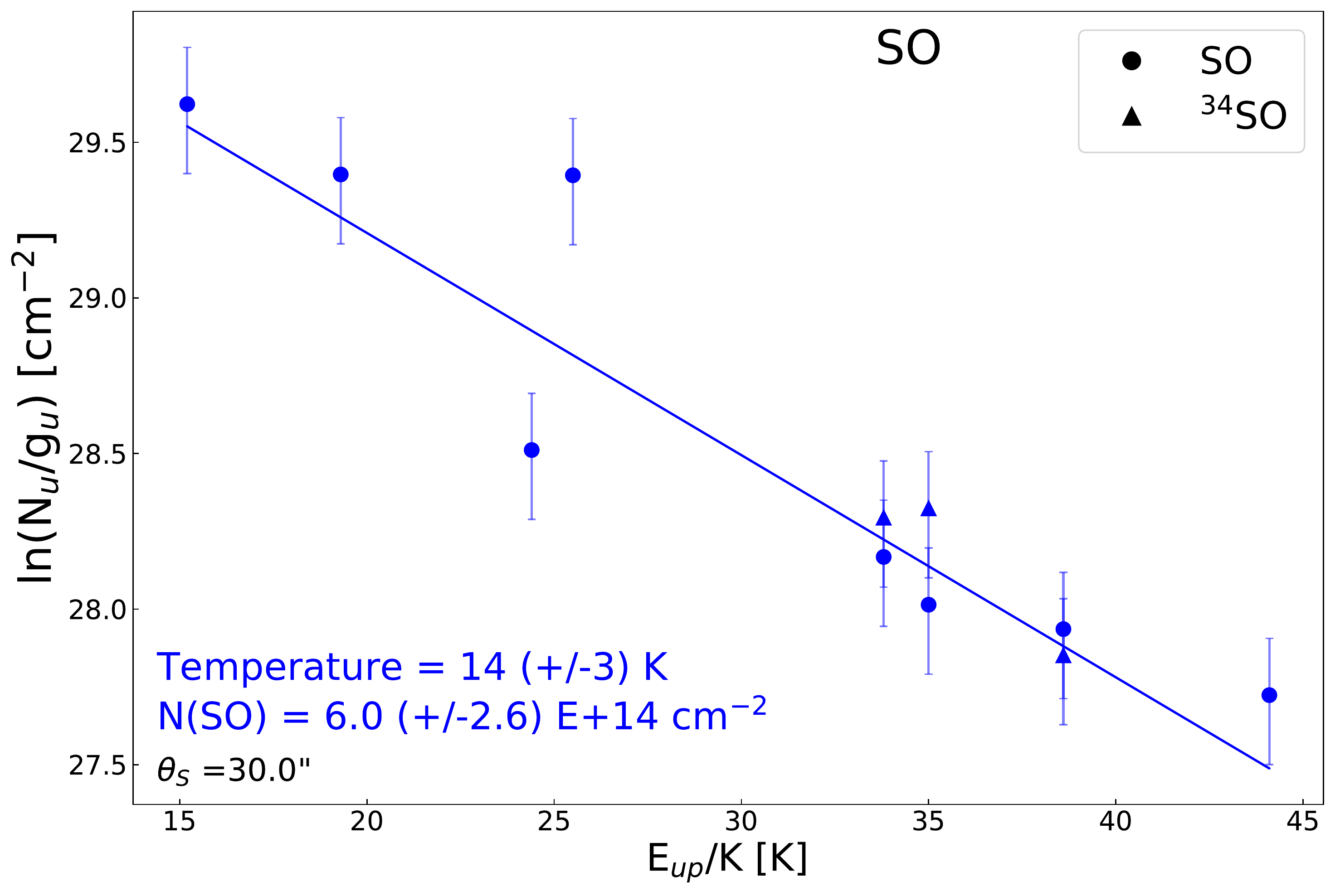}
    \includegraphics[scale=0.25]{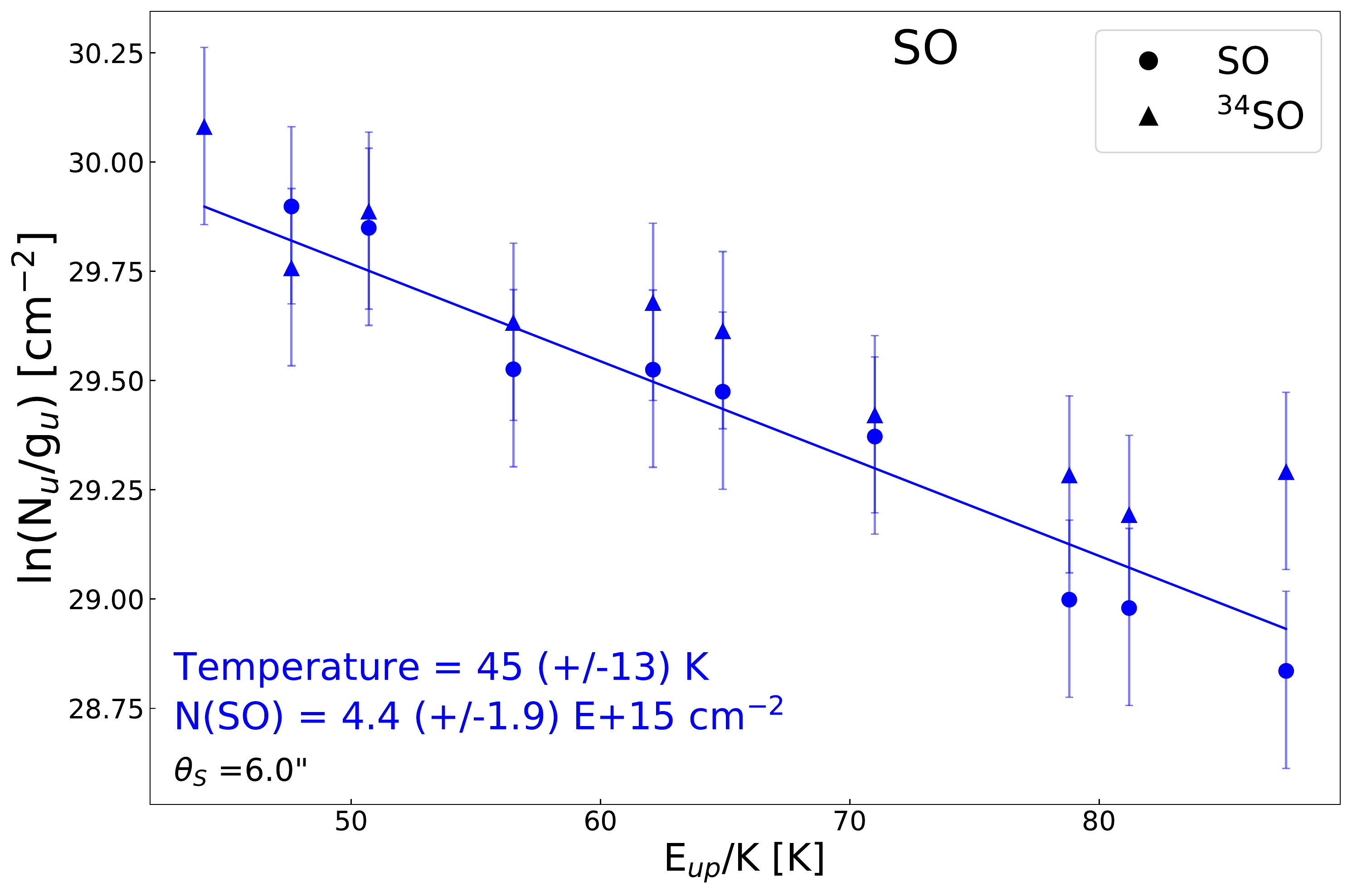}
    \caption{Rotational diagrams of the S-bearing molecules CS, H$_2\,^{34}$S, ortho-H$_2$S, OCS, SO$_2$, and SO. The emission from the cold component of the envelope for H$_2$S is described by the rotational diagram of o-H$_2\,^{34}$S. The two rotational diagrams of OCS, SO$_2$, and SO represent the narrow and the broader components of the Gaussian fit.}
    \label{diag_S2}
\end{figure*}
\begin{figure*}[h!]
    \centering
    \includegraphics[scale=0.25]{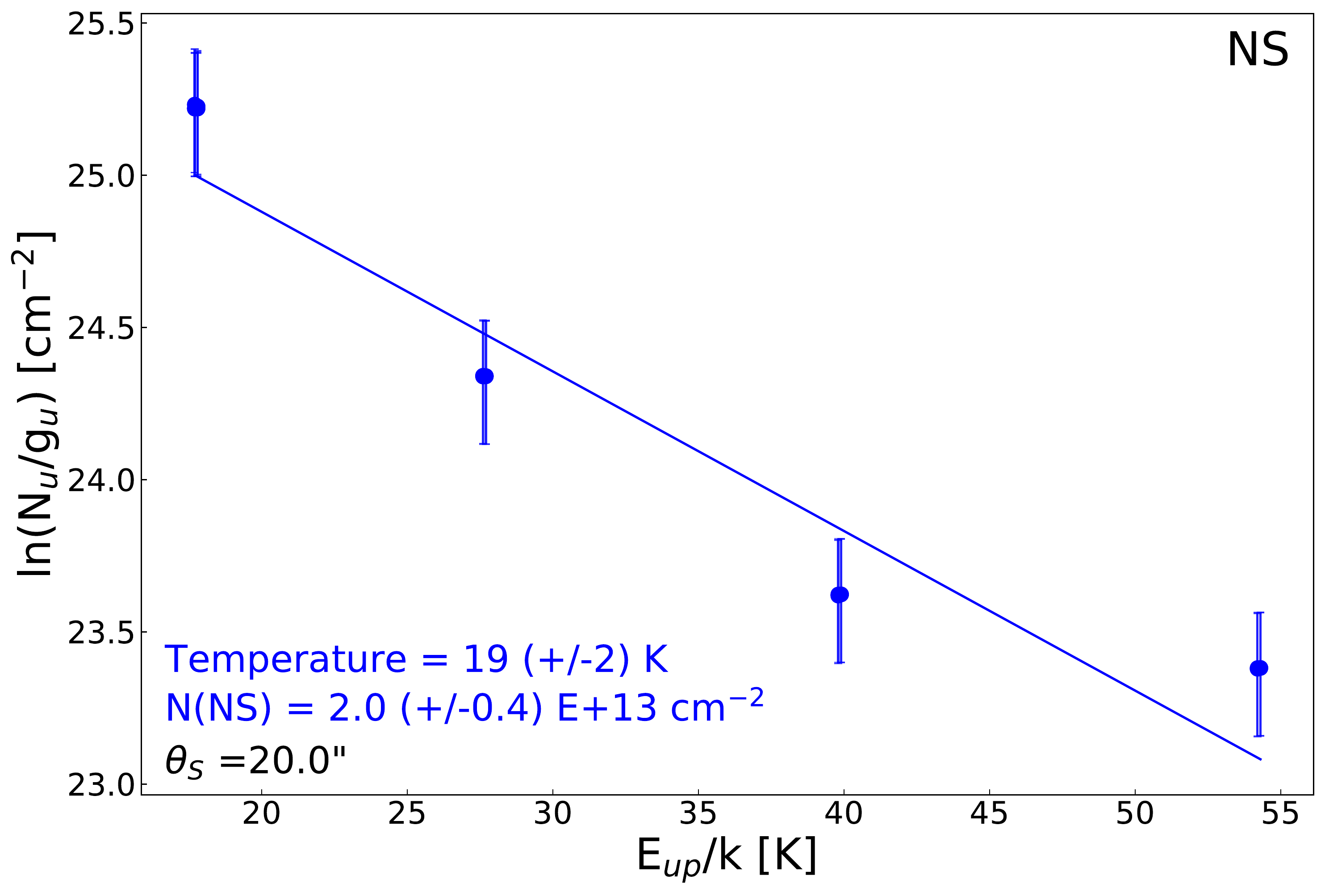}
    \includegraphics[scale=0.25]{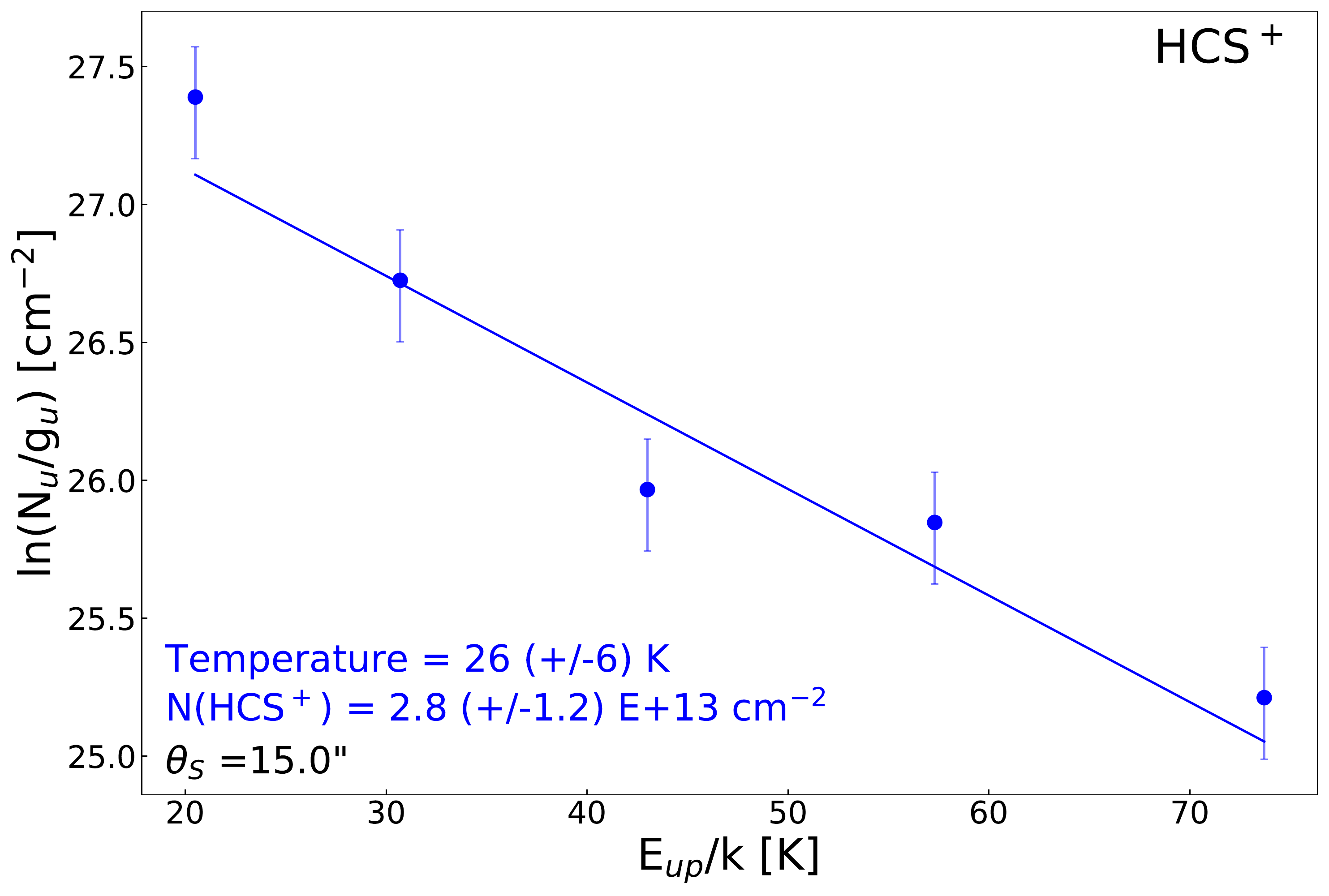}
    \includegraphics[scale=0.25]{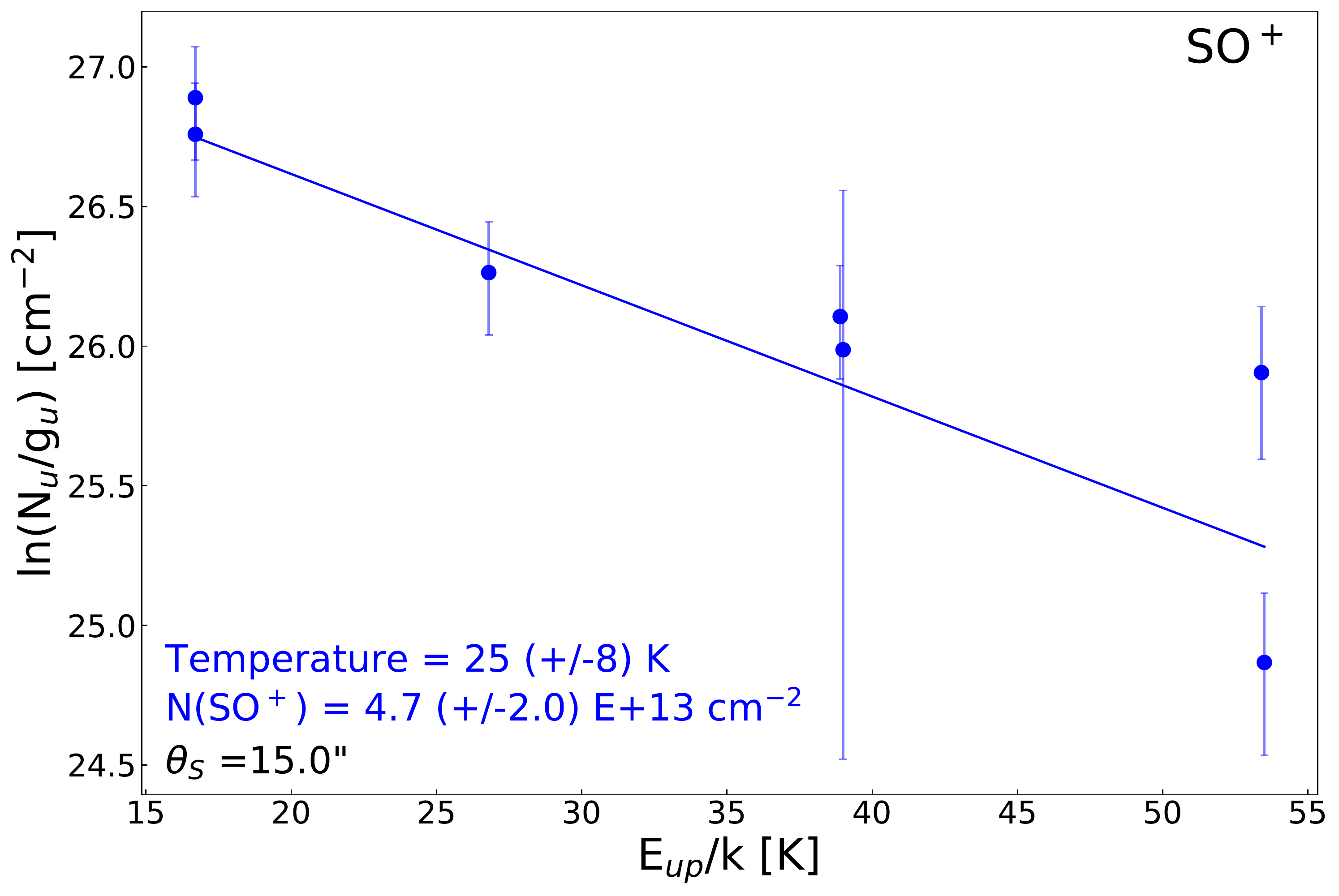}
    \caption{Rotational diagrams of the S-bearing molecules NS, HCS$^+$, and SO$^+$.}
    \label{diag_S1}
\end{figure*}

\subsection{Deuterated molecules}
\begin{figure*}[!h]
    \centering
    \includegraphics[scale=0.25]{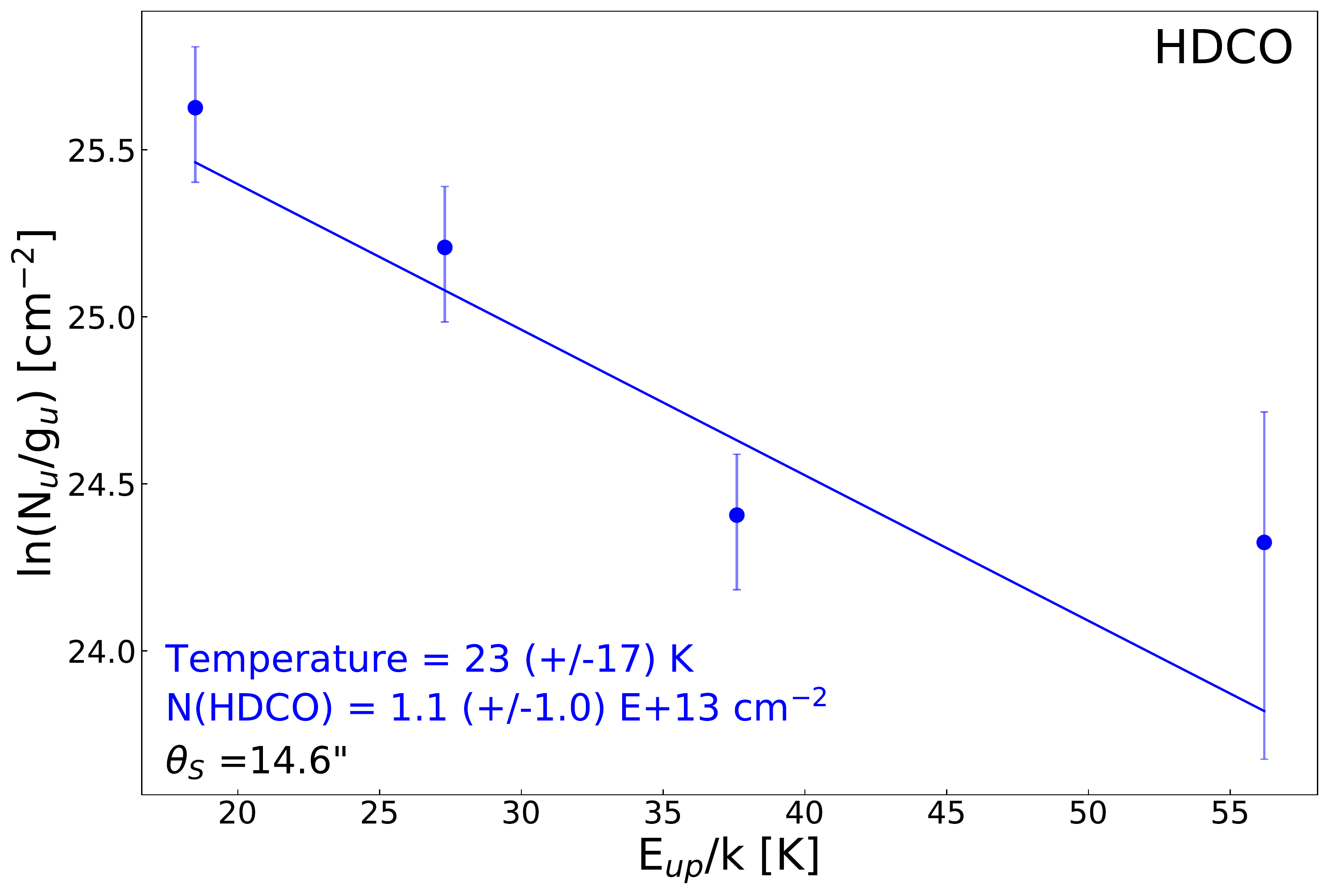}
    \includegraphics[scale=0.25]{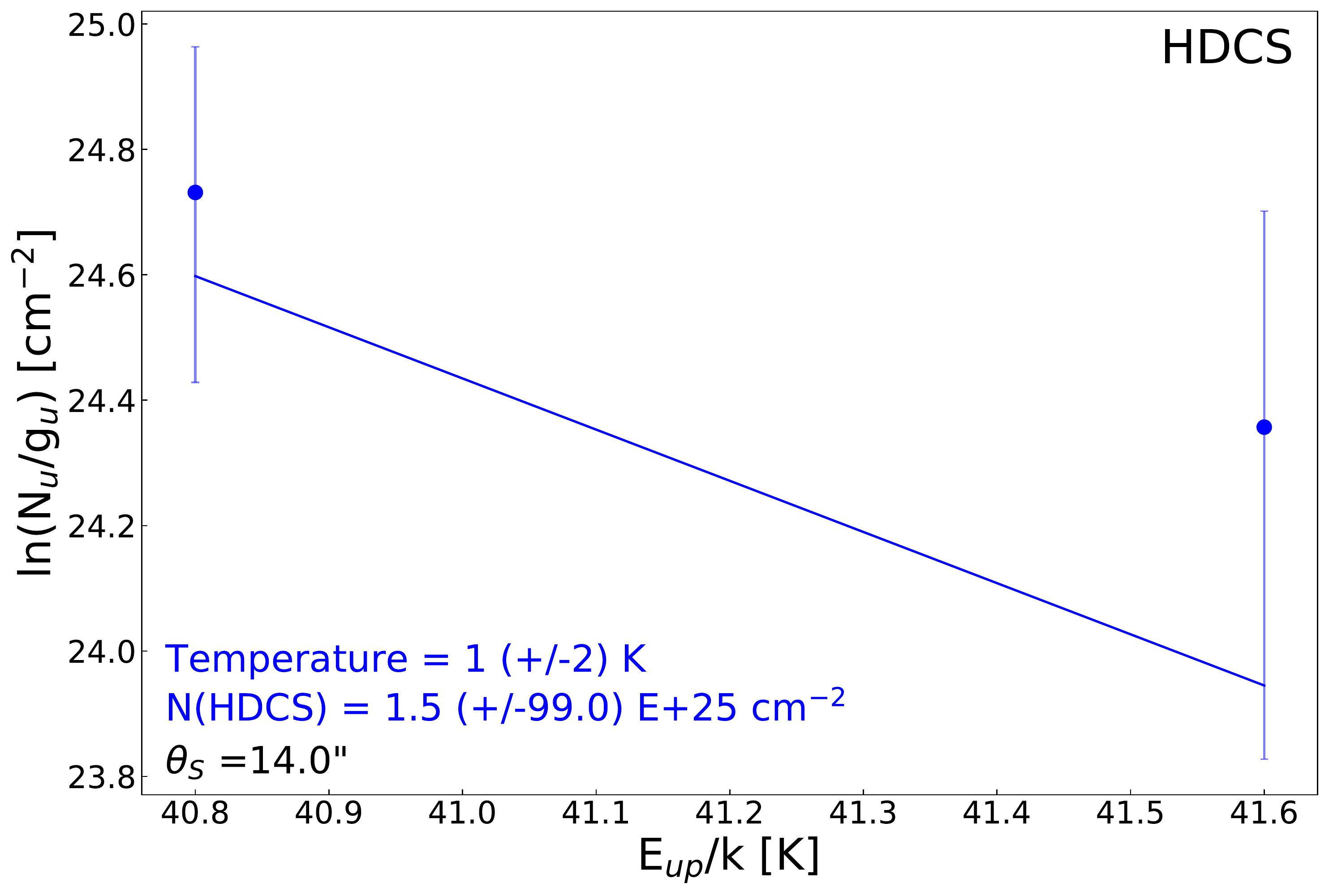}
    \includegraphics[scale=0.25]{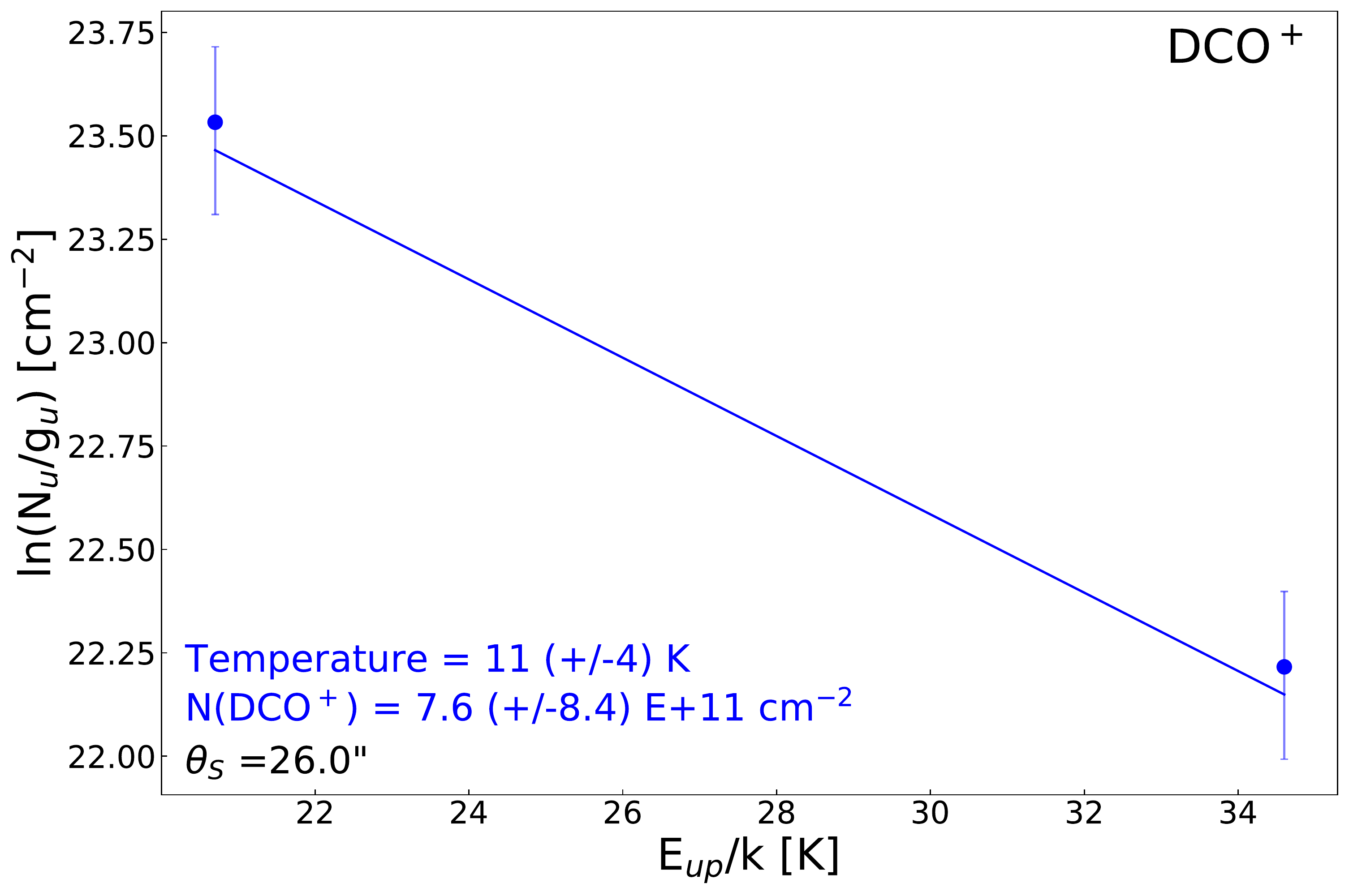}
    \includegraphics[scale=0.25]{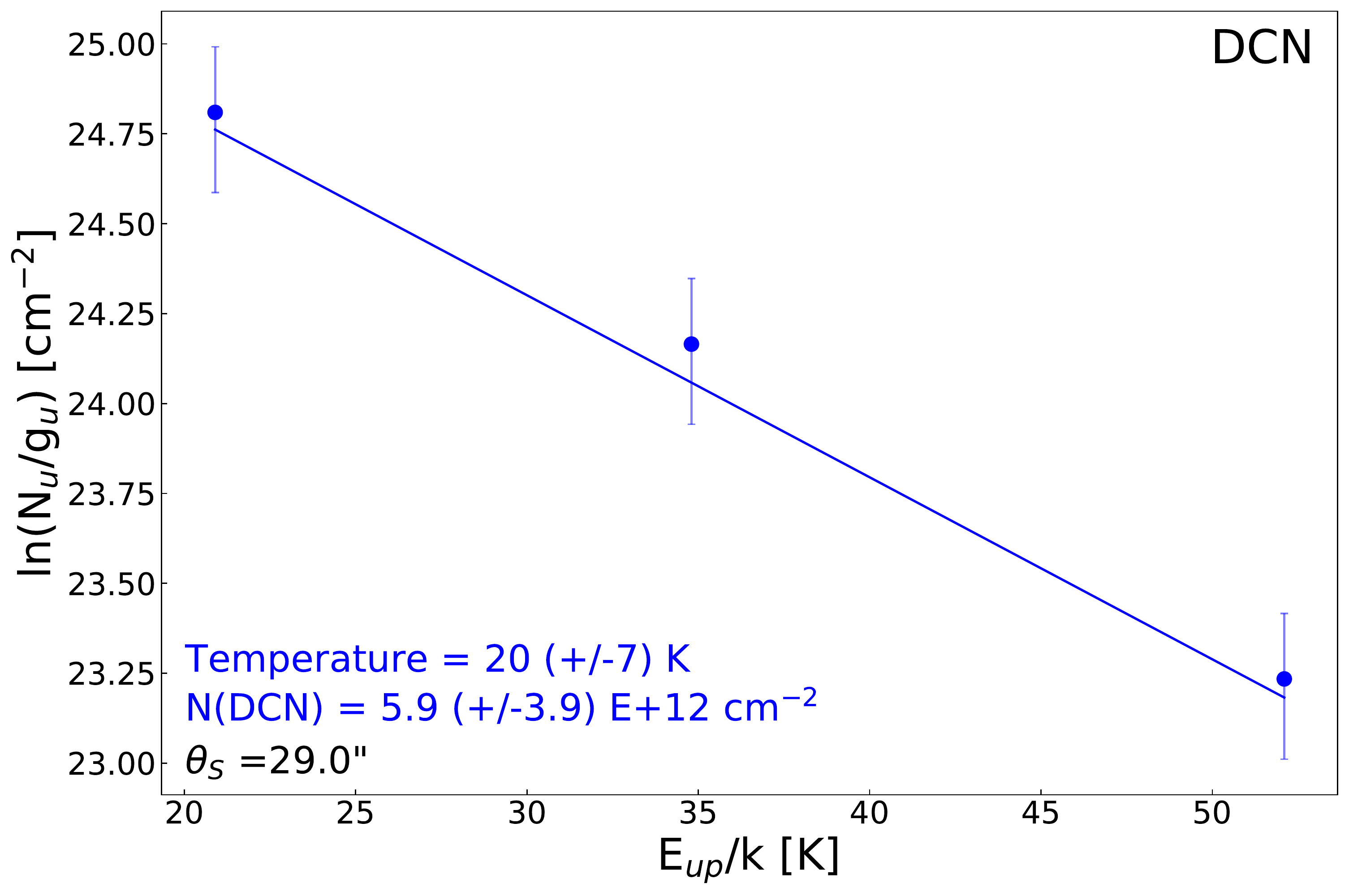}
    \includegraphics[scale=0.25]{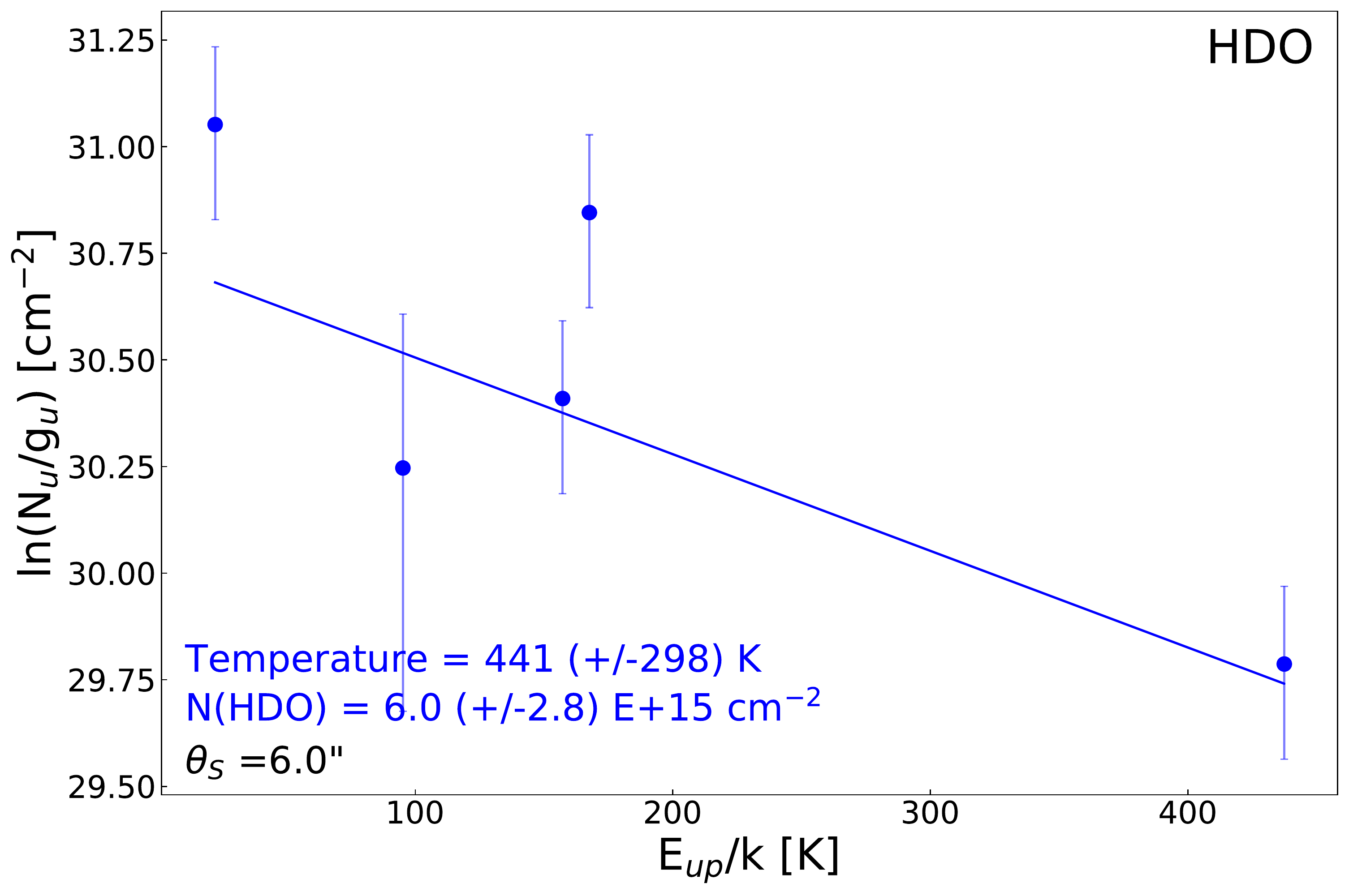}
    \caption{Rotational diagrams of the deuterated molecules HDCO, DCO$^+$, DCN, and HDO.}
    \label{diag_HDCO}
\end{figure*}

\subsection{Others}
\begin{figure*}[h!]
    \centering
    \includegraphics[scale=0.25]{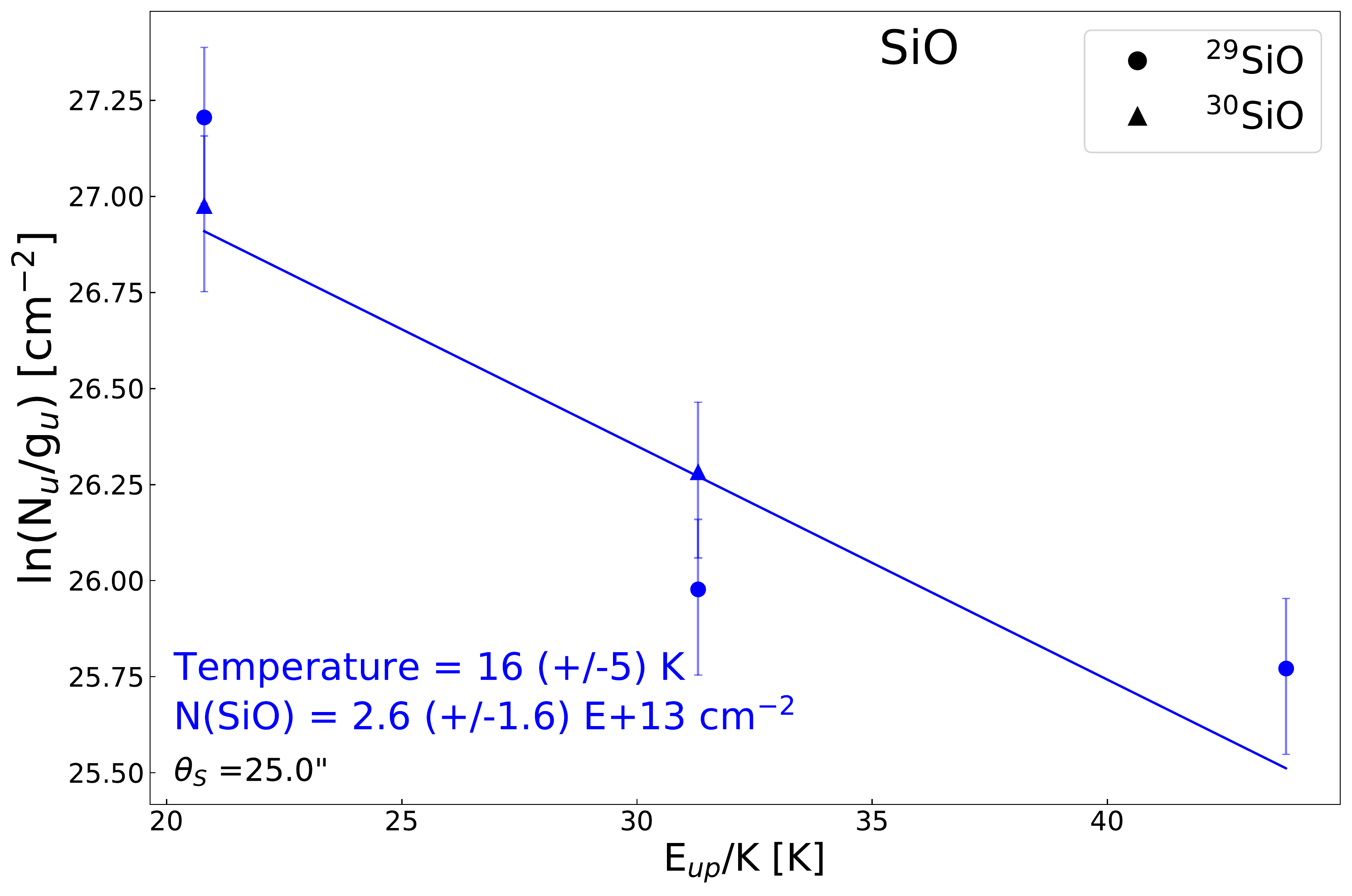}
    \caption{Rotational diagrams of SiO.}
    \label{diag_SiO}
\end{figure*}

\subsection{COMs}

\begin{figure*}[h!]
    \centering
    \includegraphics[scale=0.25]{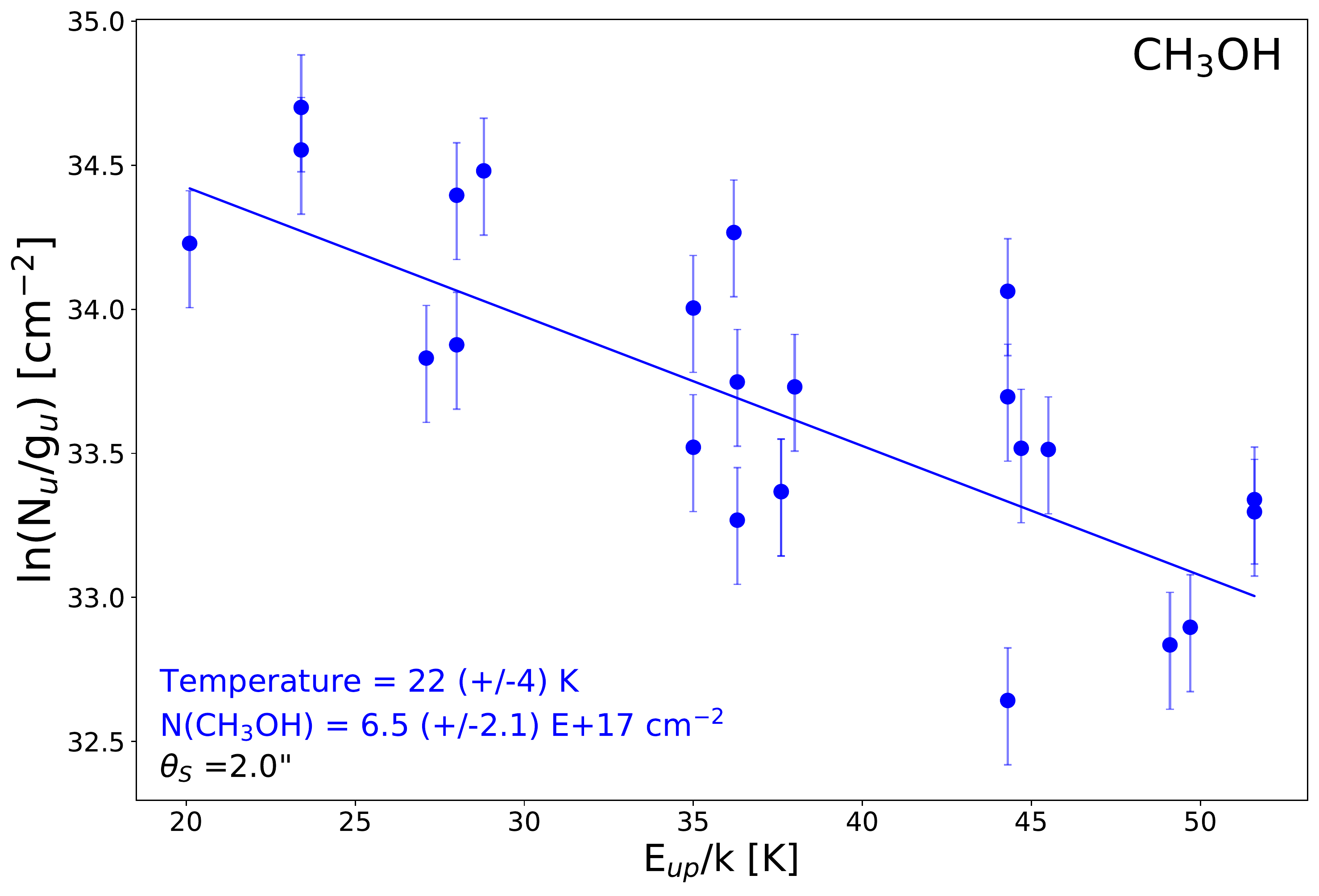}
    \includegraphics[scale=0.25]{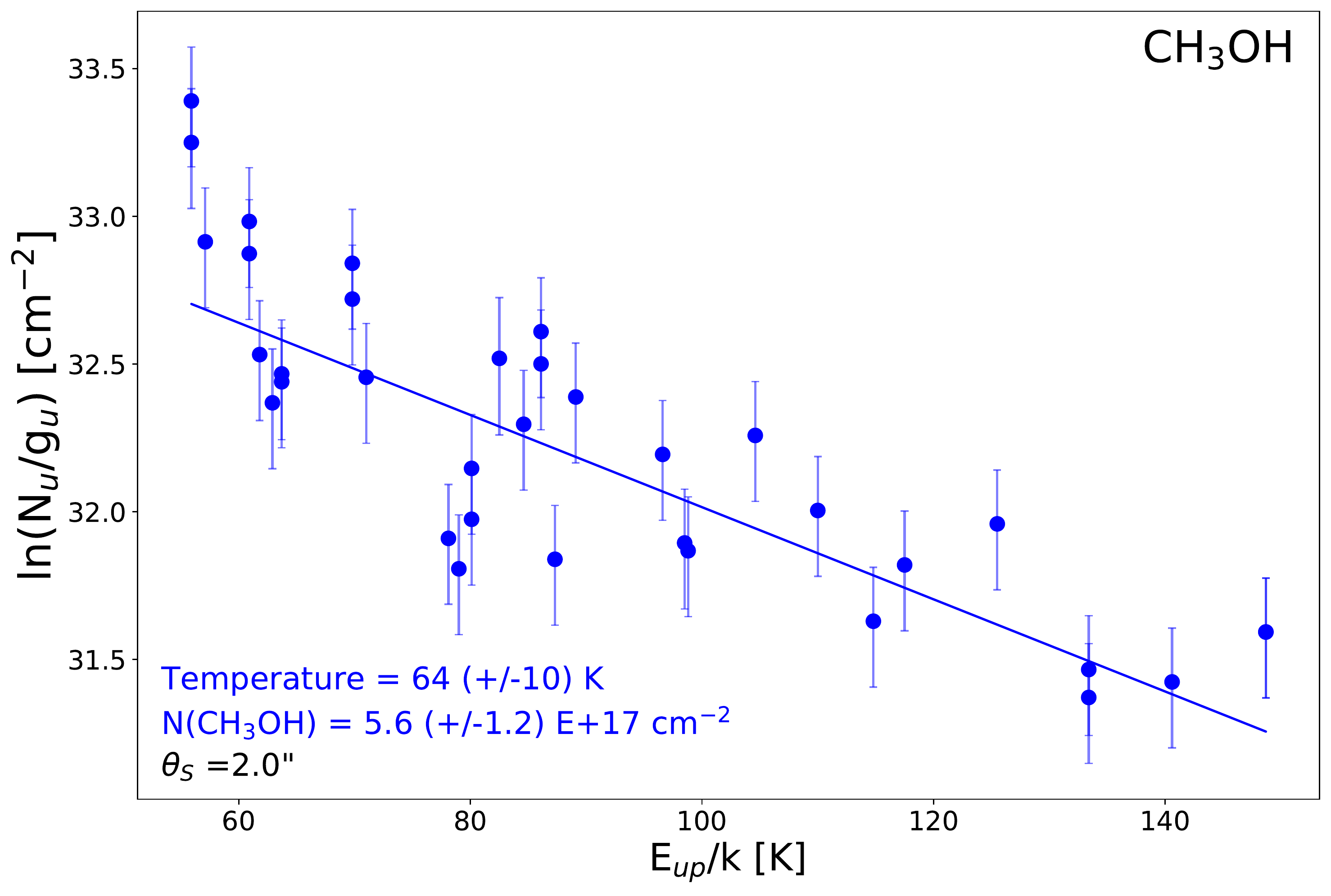}
    \includegraphics[scale=0.25]{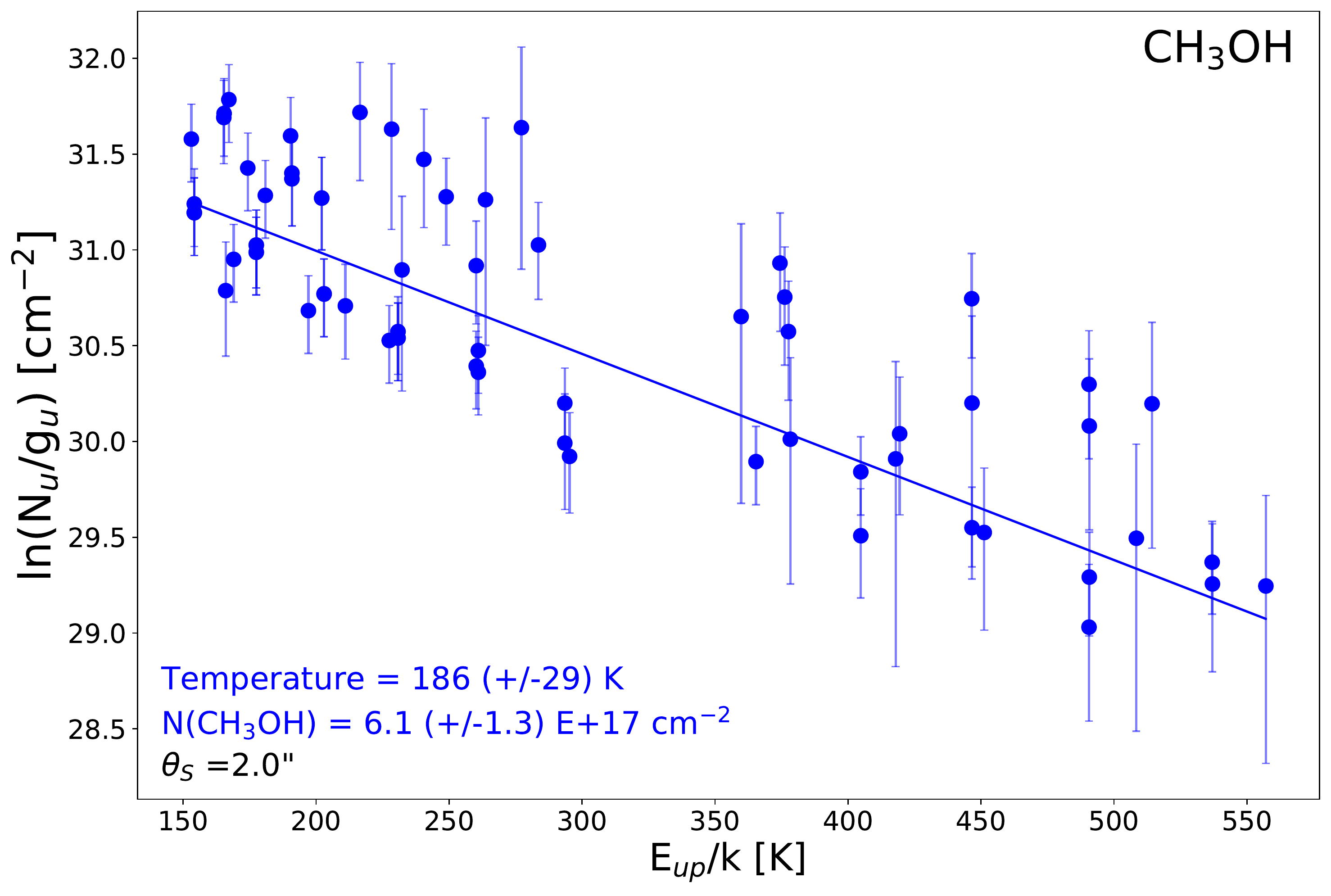}
    \includegraphics[scale=0.25]{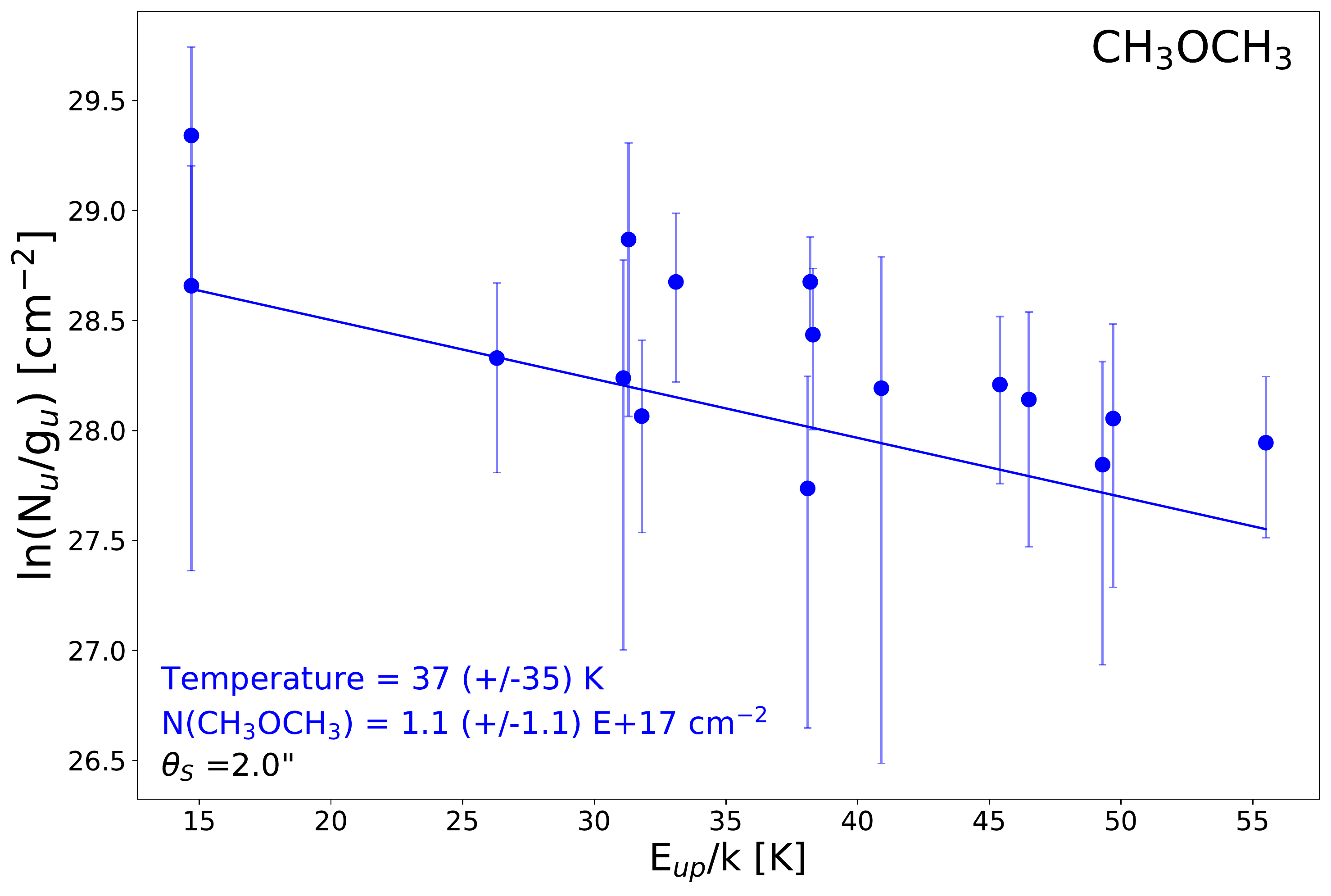}
    \includegraphics[scale=0.25]{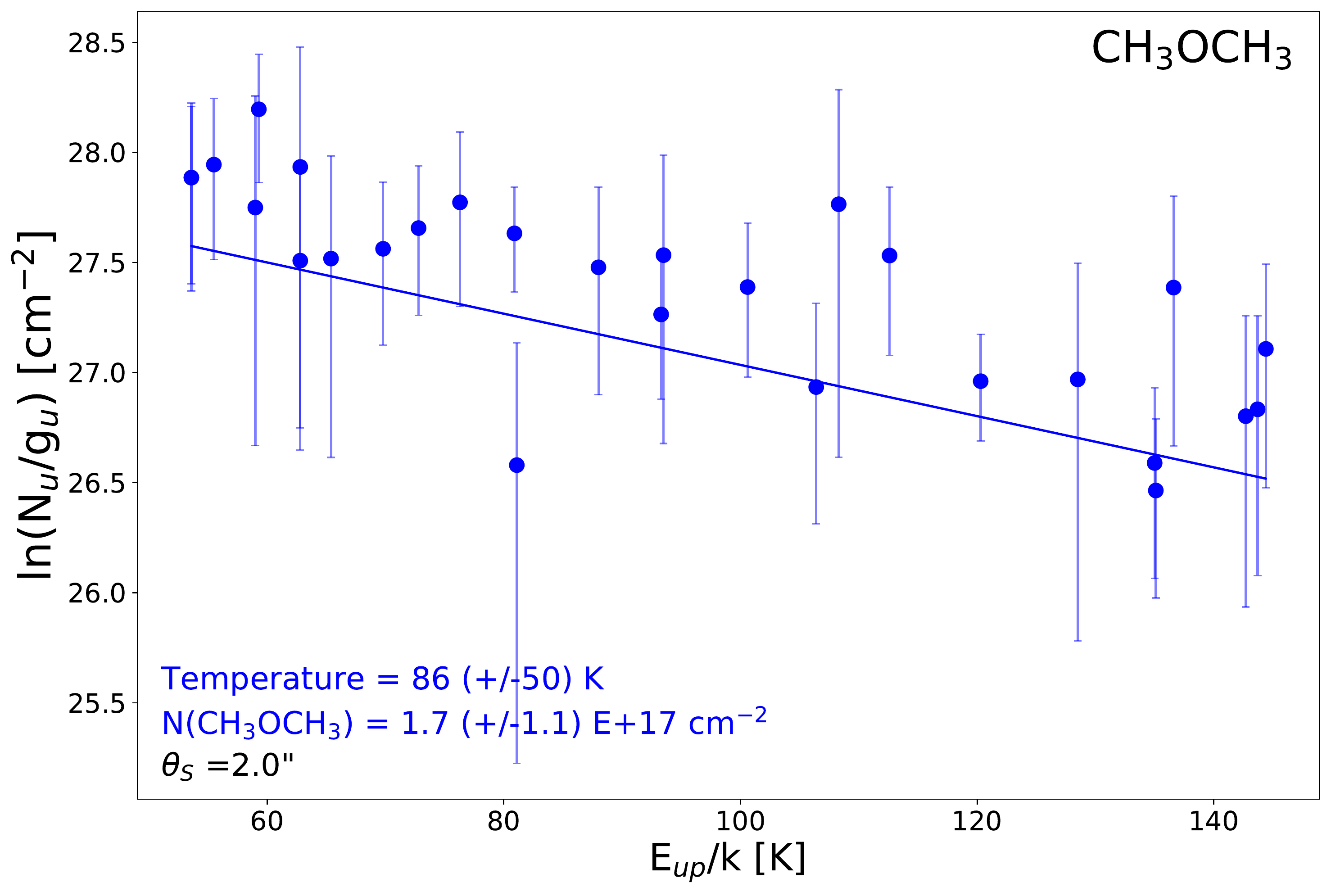}
    \includegraphics[scale=0.25]{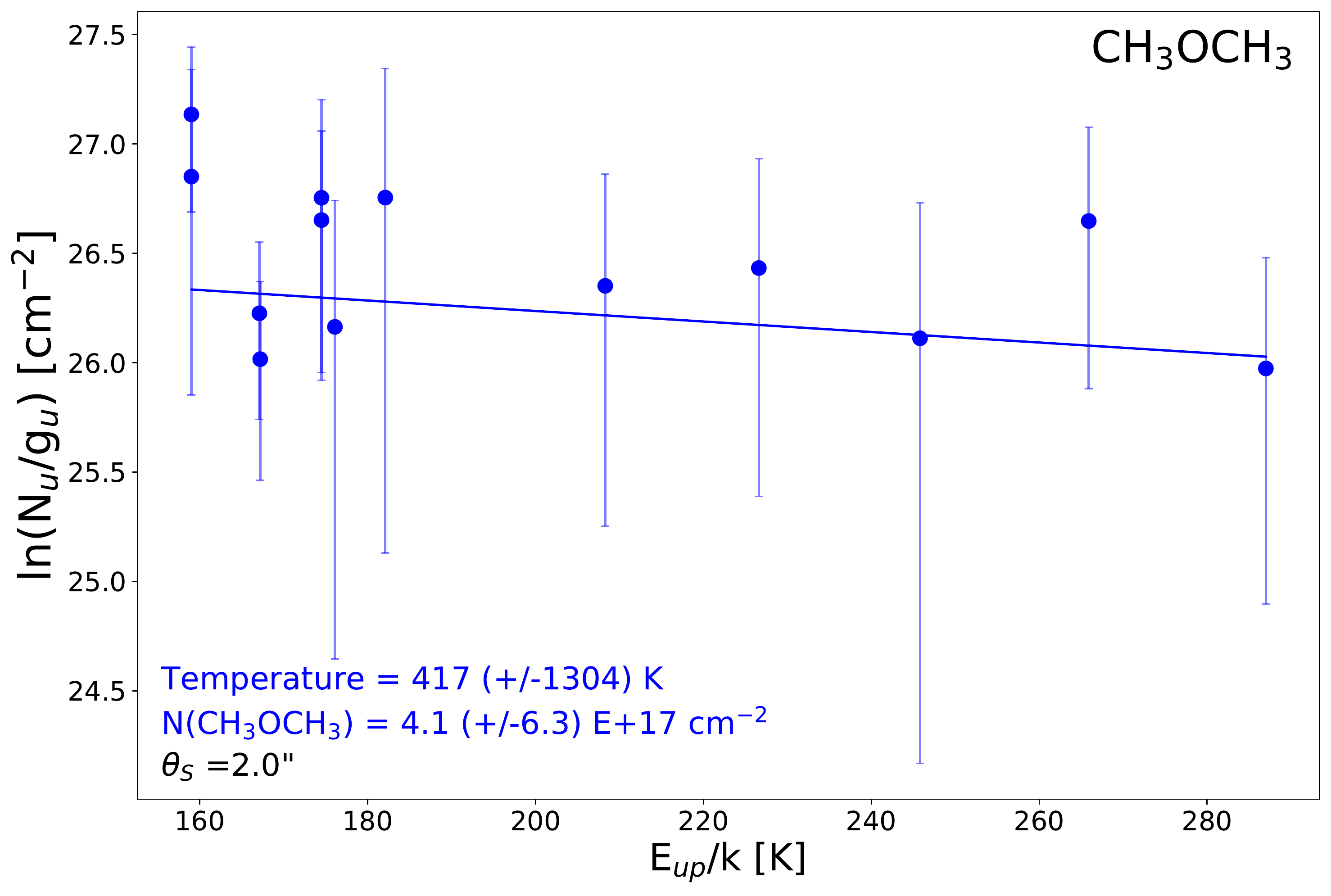}
    \includegraphics[scale=0.25]{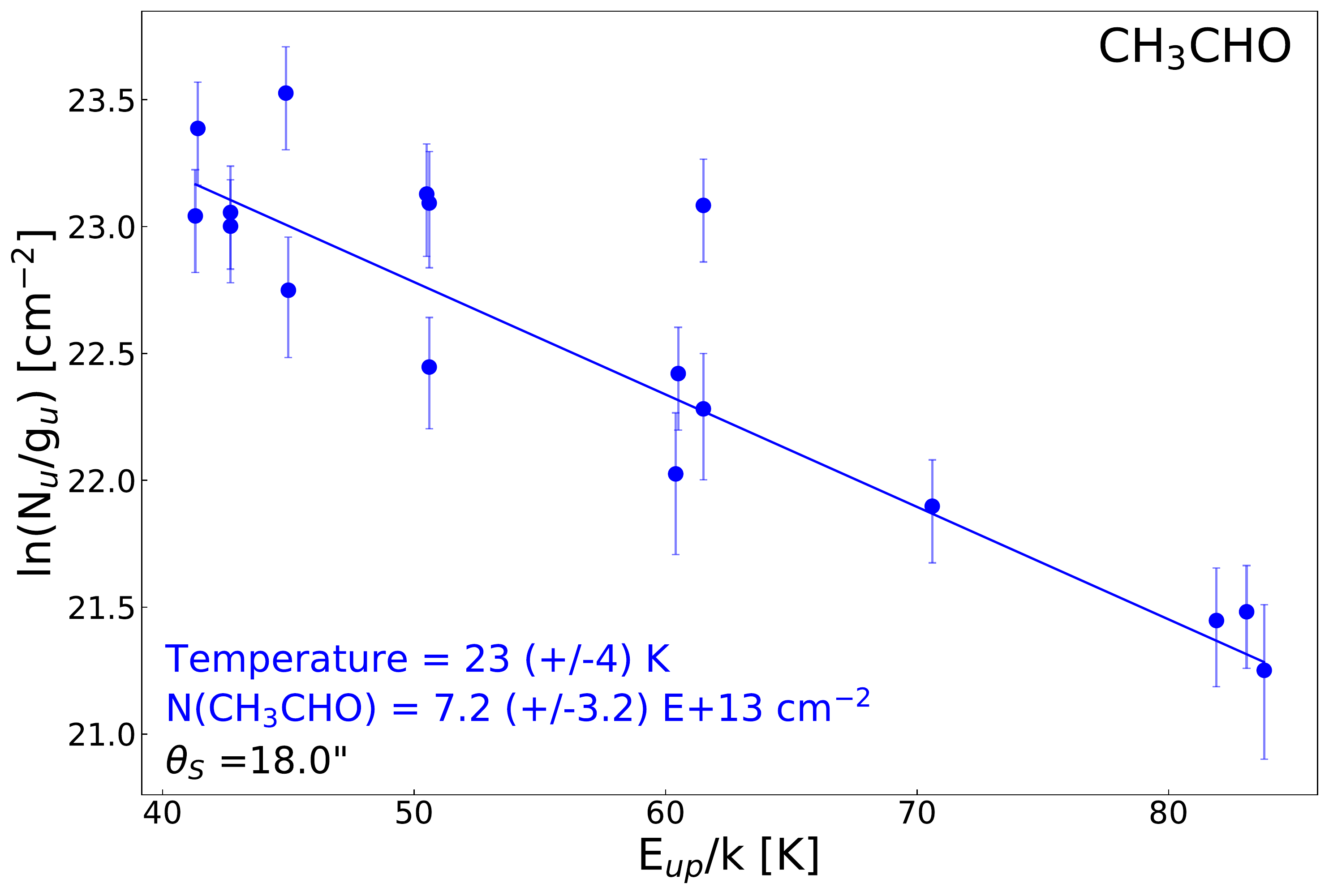}
    \includegraphics[scale=0.25]{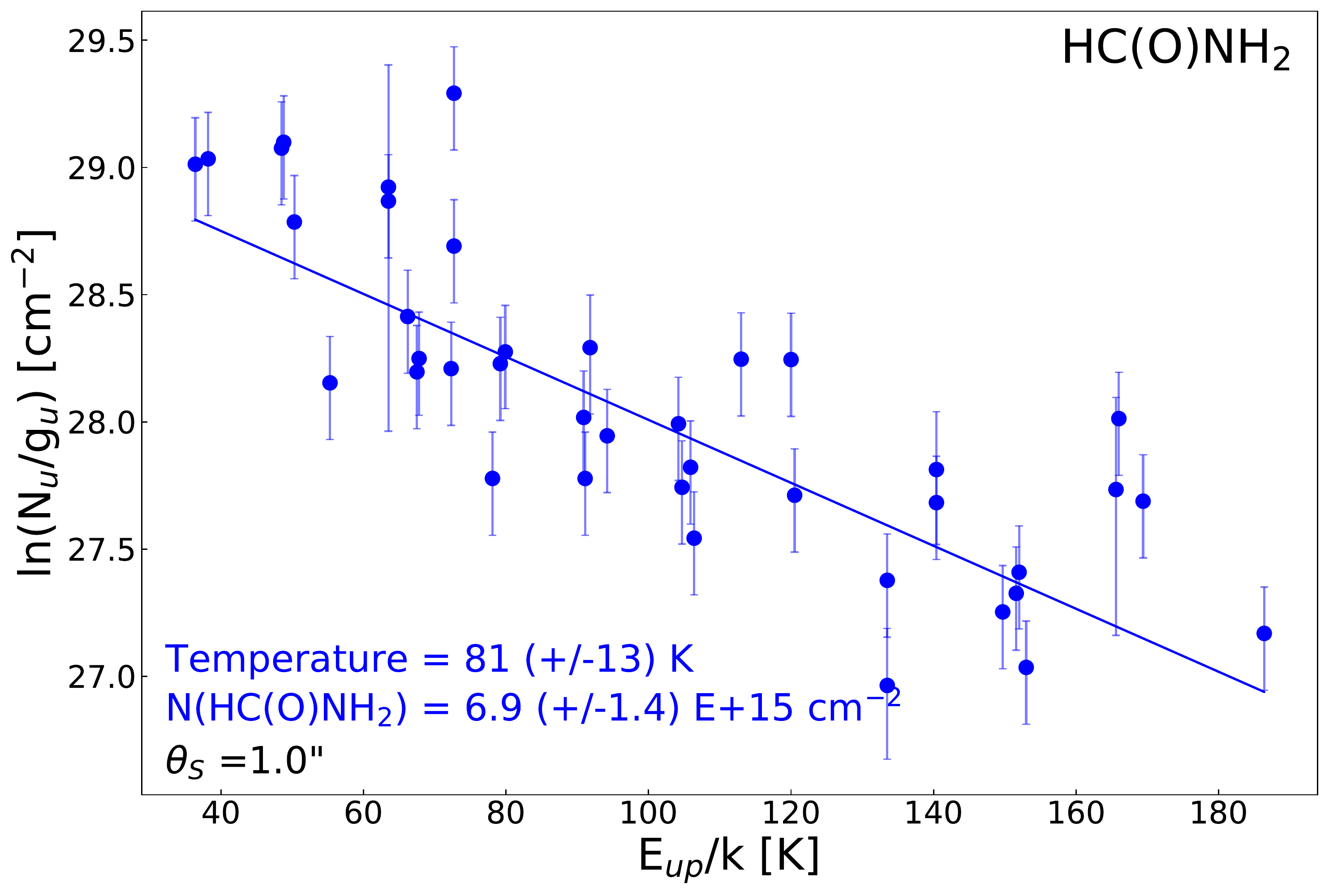}
    \caption{Rotational diagrams of the COMs CH$_3$OH, CH$_3$OCH$_3$, CH$_3$CHO, HC(O)NH$_2$, and CH$_3$OCHO. The three rotational diagrams of methanol and dimethyl ether represent the rotational diagram of the three components (cold and warm components of the envelope and shocks) traced by these molecules.}
    \label{diag_COMs1}
\end{figure*}
\begin{figure*}[h!]
    \centering
    \includegraphics[scale=0.25]{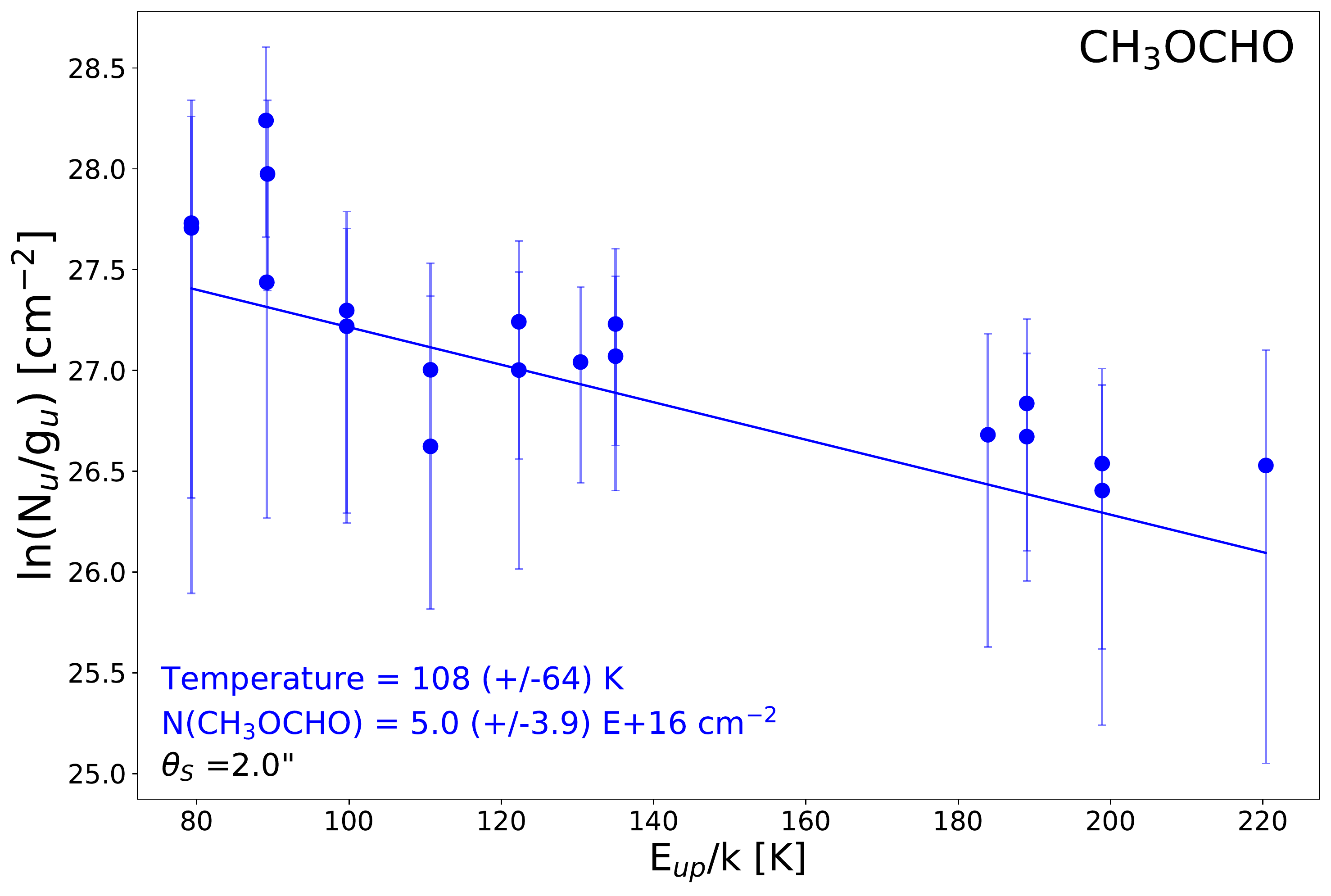}
    \includegraphics[scale=0.25]{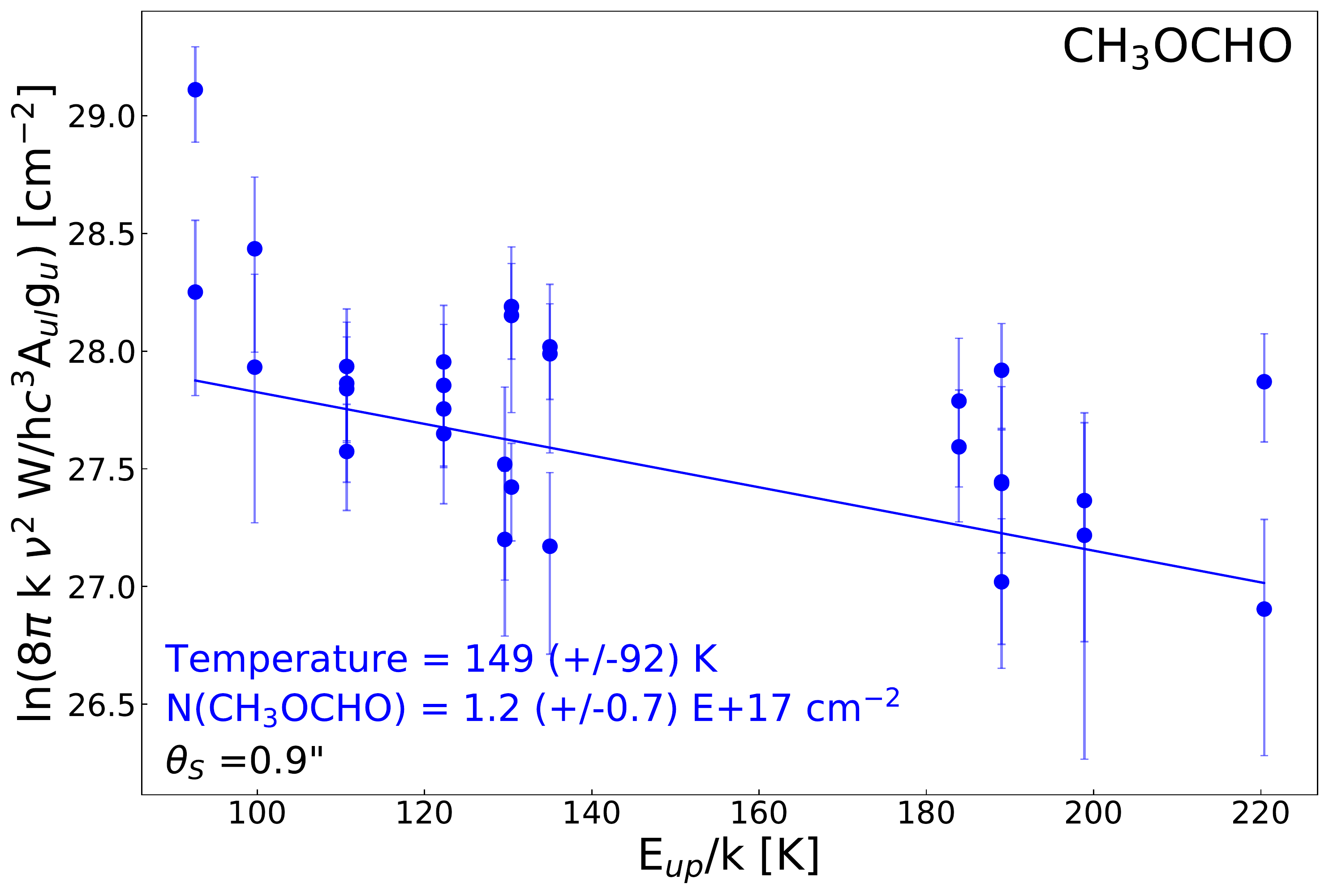}
    \includegraphics[scale=0.25]{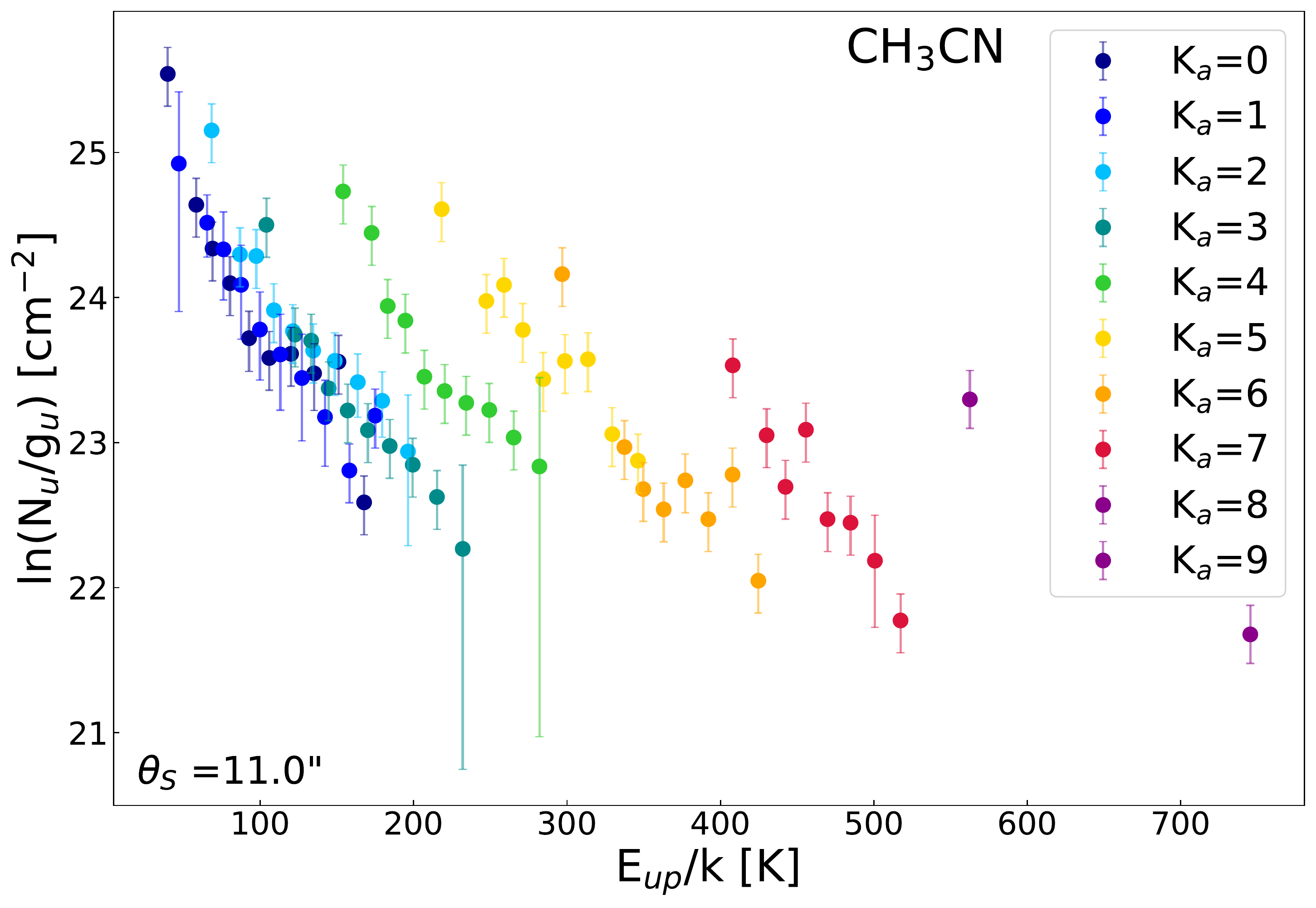}
    \includegraphics[scale=0.25]{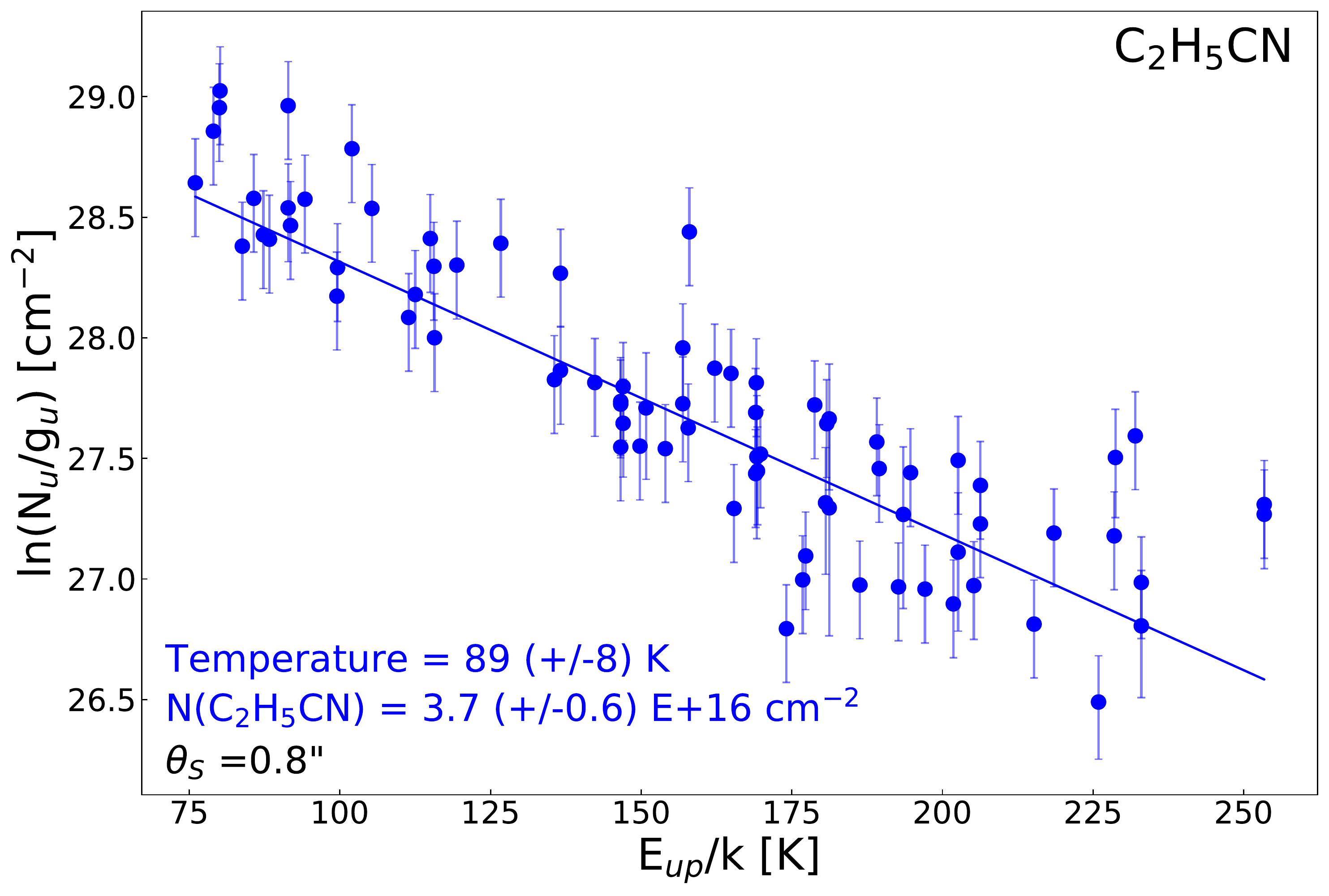}
    \includegraphics[scale=0.25]{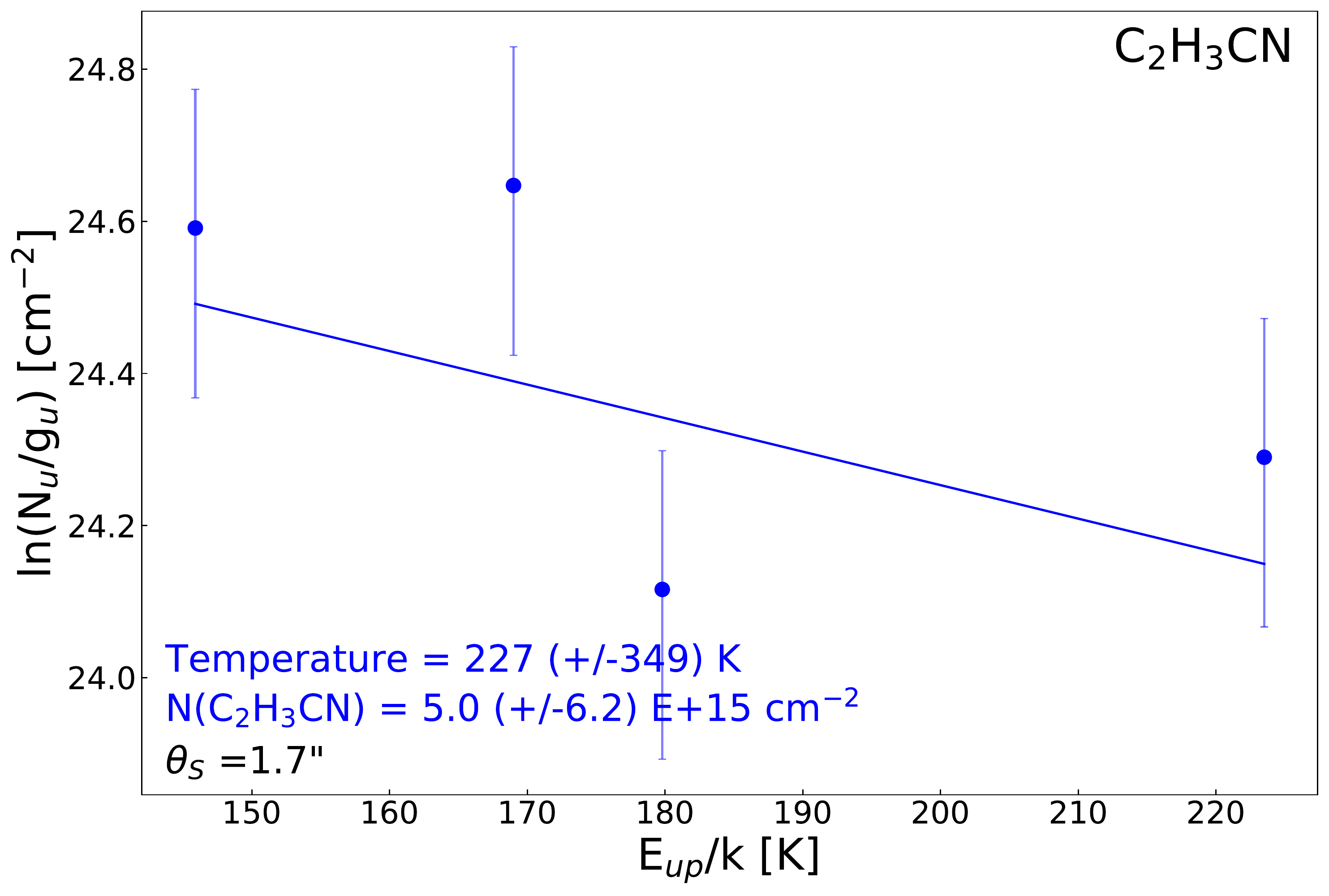}
    \includegraphics[scale=0.25]{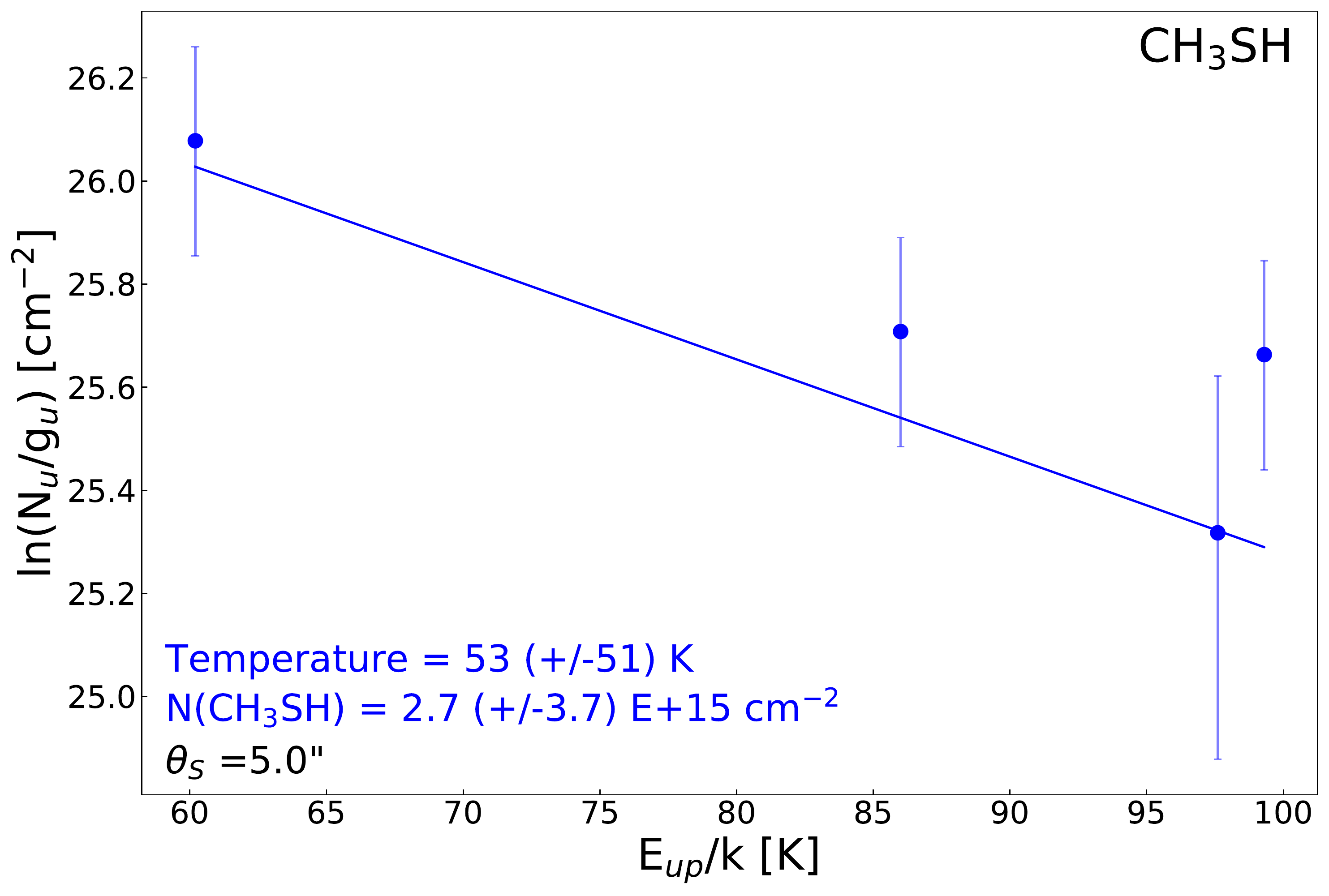}
    \caption{Rotational diagrams of the COMs CH$_3$CHO, CH$_3$CN, C$_2$H$_5$CN, C$_2$H$_3$CN, and CH$_3$SH.}
    \label{diag_COMs2}
\end{figure*}

\section{Spectral survey of \mysou}
\label{app:SpecSurvey}

\begin{sidewaysfigure*}
\includegraphics[width=0.9\linewidth]{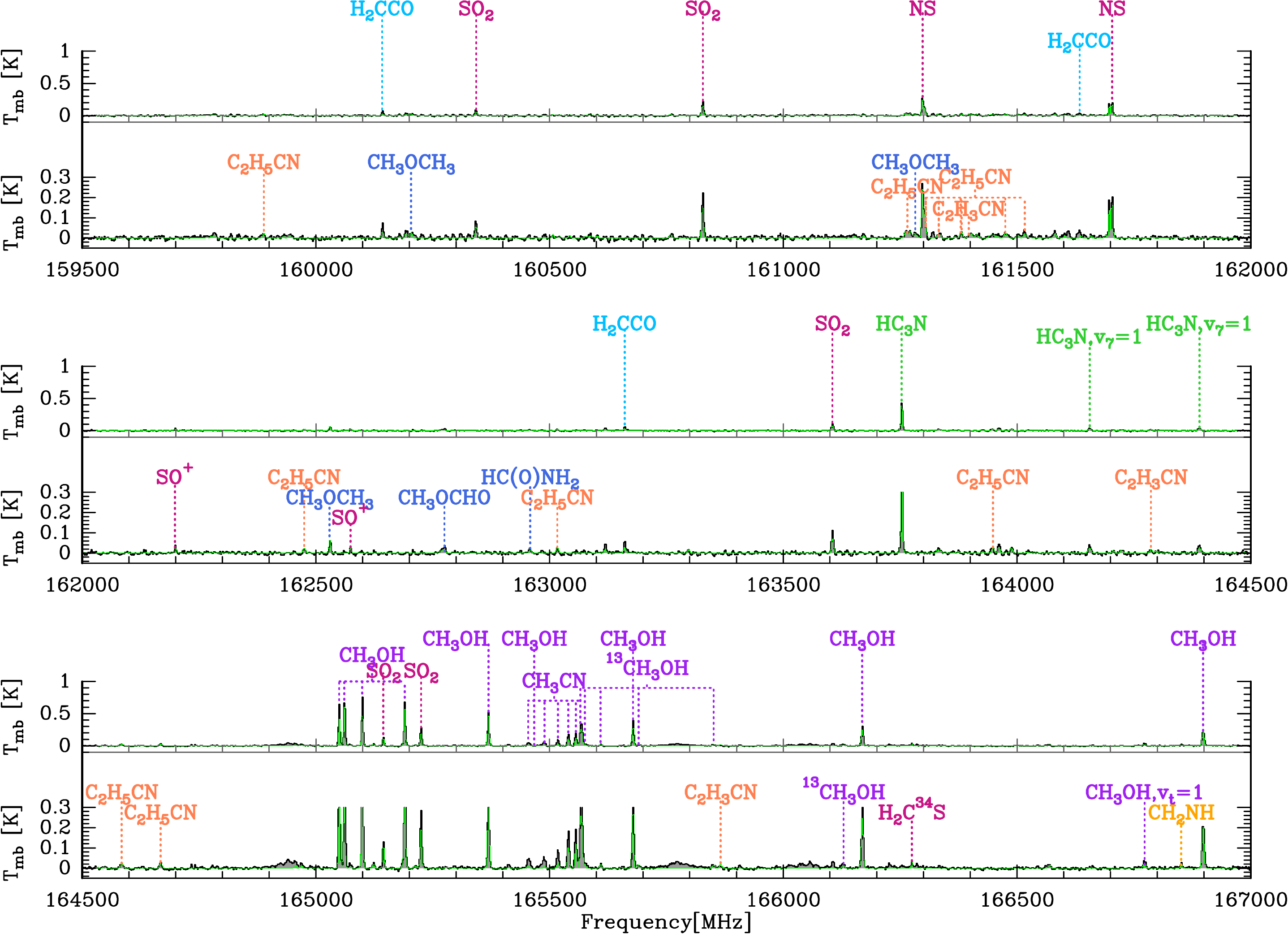}
\caption{Spectrum towards \mysou. The grey filled histogram shows the spectra and the coloured labels the detected transitions of all the light molecules and some transitions of the COMs. The green line represents the LTE model obtained with Weeds. The upper panel for each frequency range shows the labels for the small molecules, and the lower panel corresponds to a zoom in temperature for  better visibility of the COMs that are labeled.}
\label{SpecSurvey_mol}
\end{sidewaysfigure*}

\begin{sidewaysfigure*}
\ContinuedFloat
\includegraphics[width=0.9\linewidth]{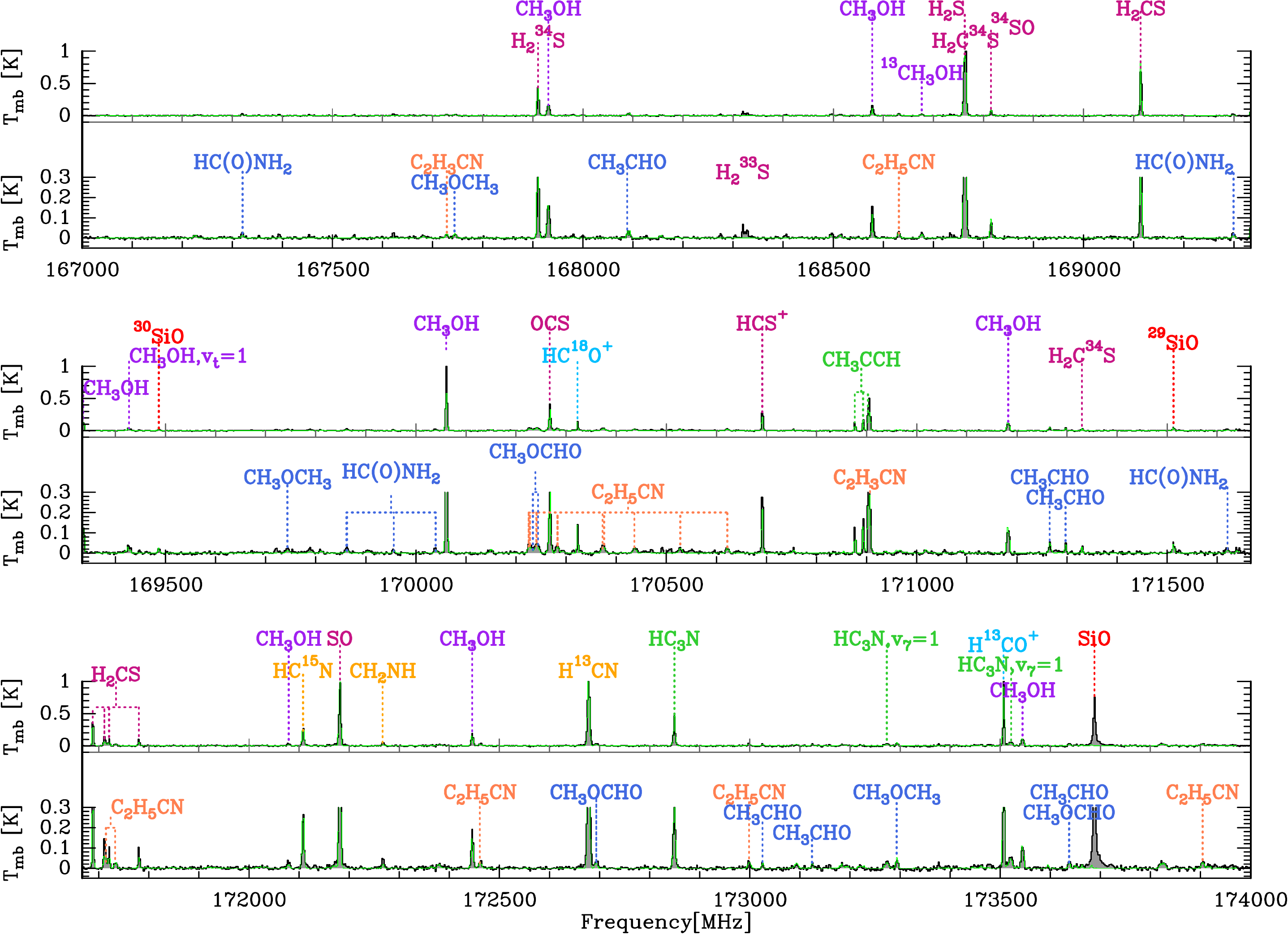}
\caption{Continued.}
\label{SpecSurvey_mol}
\end{sidewaysfigure*}

\begin{sidewaysfigure*}
\ContinuedFloat
\includegraphics[width=0.9\linewidth]{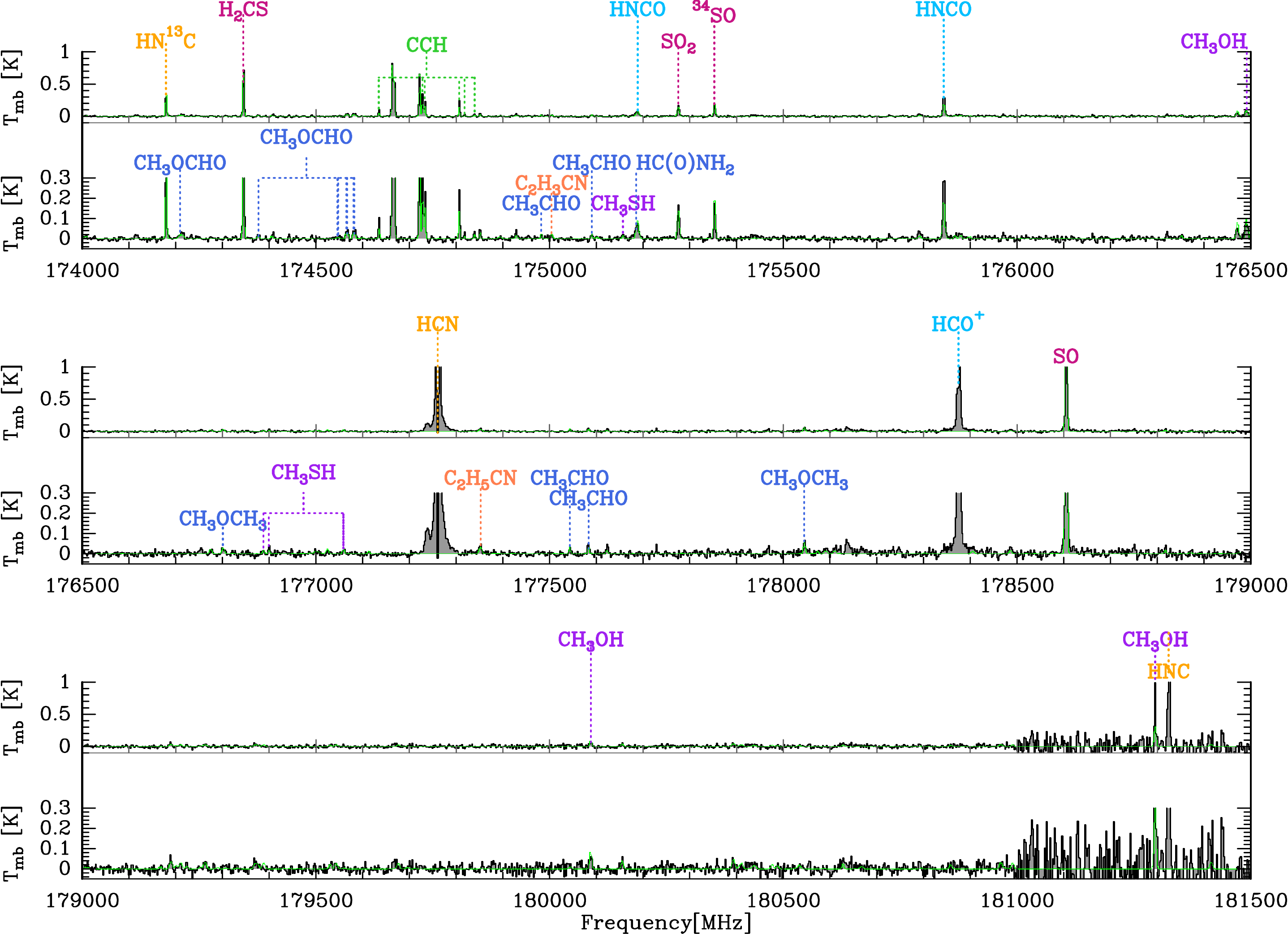}
\caption{Continued.}
\label{SpecSurvey_mol}
\end{sidewaysfigure*}

\begin{sidewaysfigure*}
\ContinuedFloat
\includegraphics[width=0.9\linewidth]{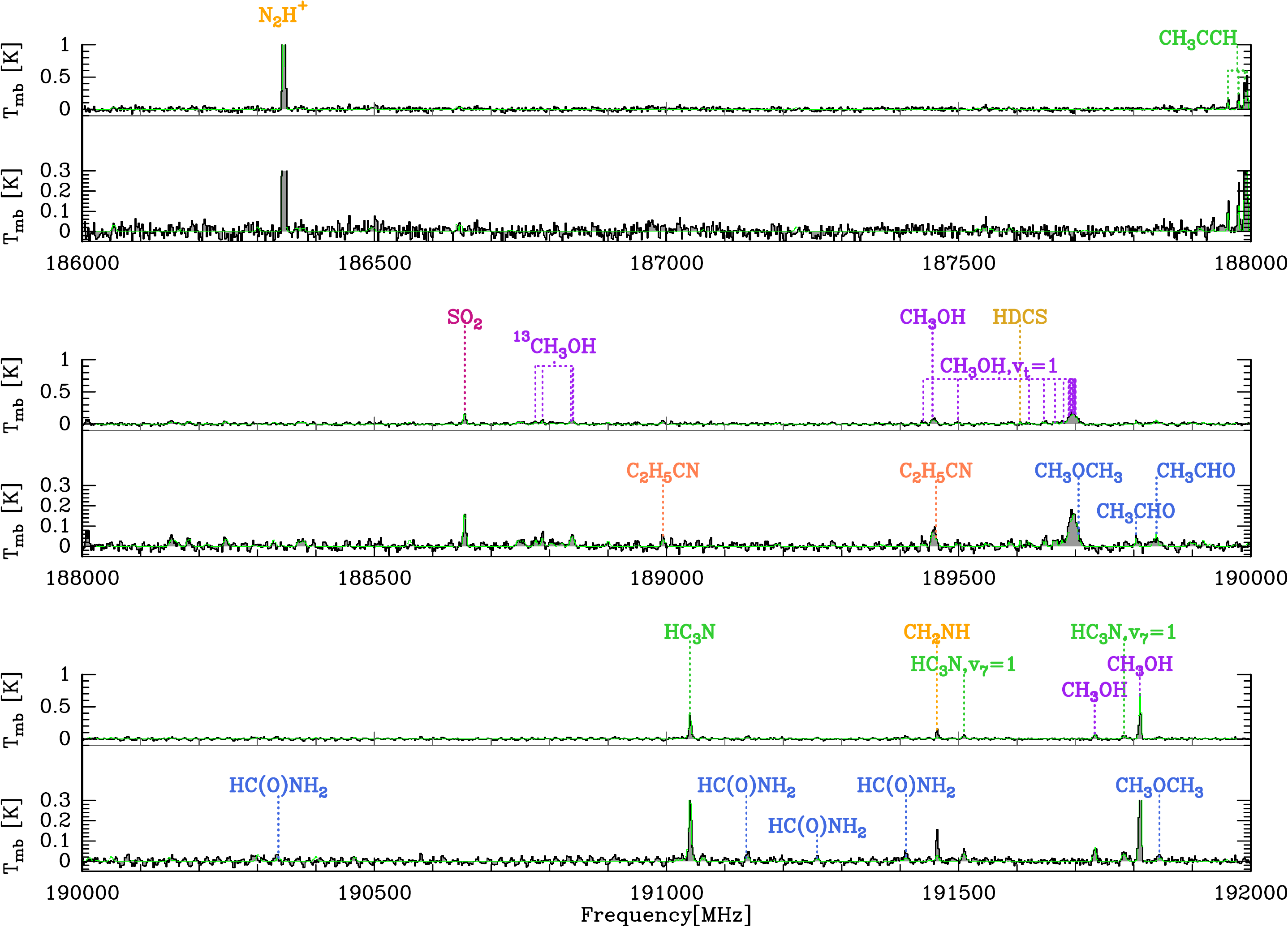}
\caption{Continued.}
\label{SpecSurvey_mol}
\end{sidewaysfigure*}

\begin{sidewaysfigure*}
\ContinuedFloat
\includegraphics[width=0.9\linewidth]{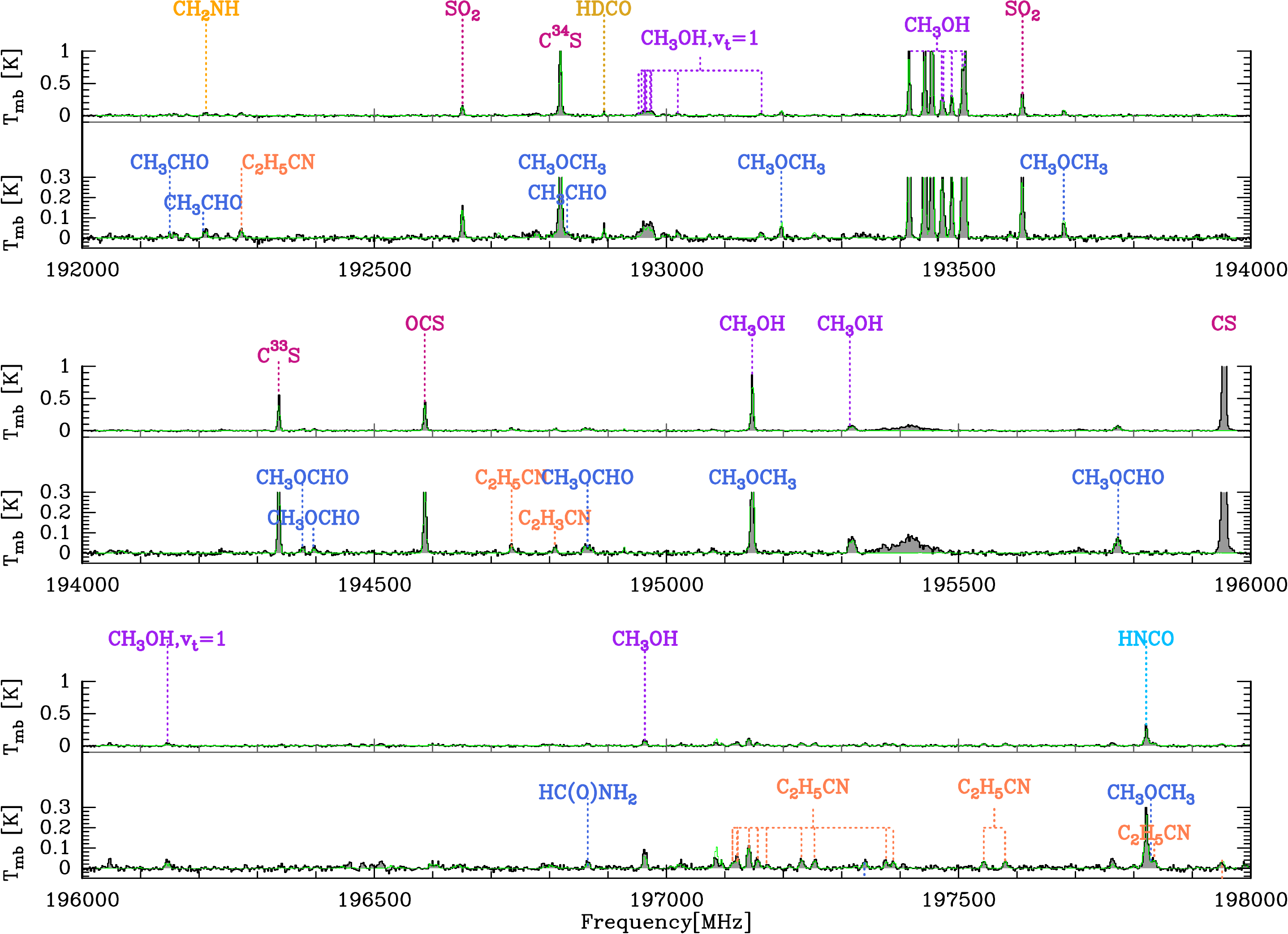}
\caption{Continued.}
\label{SpecSurvey_mol}
\end{sidewaysfigure*}

\begin{sidewaysfigure*}
\ContinuedFloat
\includegraphics[width=0.9\linewidth]{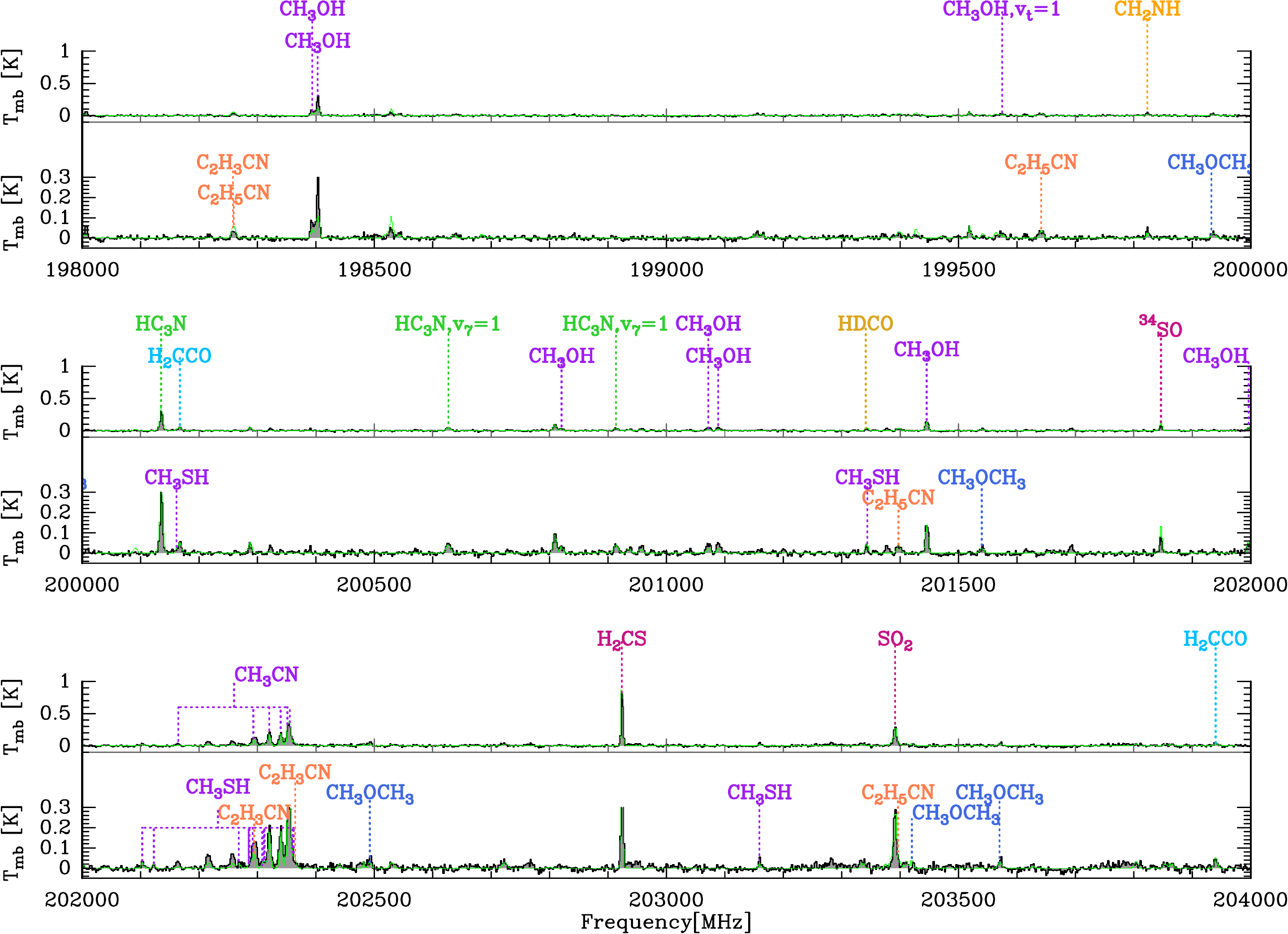}
\caption{Continued.}
\label{SpecSurvey_mol}
\end{sidewaysfigure*}

\begin{sidewaysfigure*}
\ContinuedFloat
\includegraphics[width=0.9\linewidth]{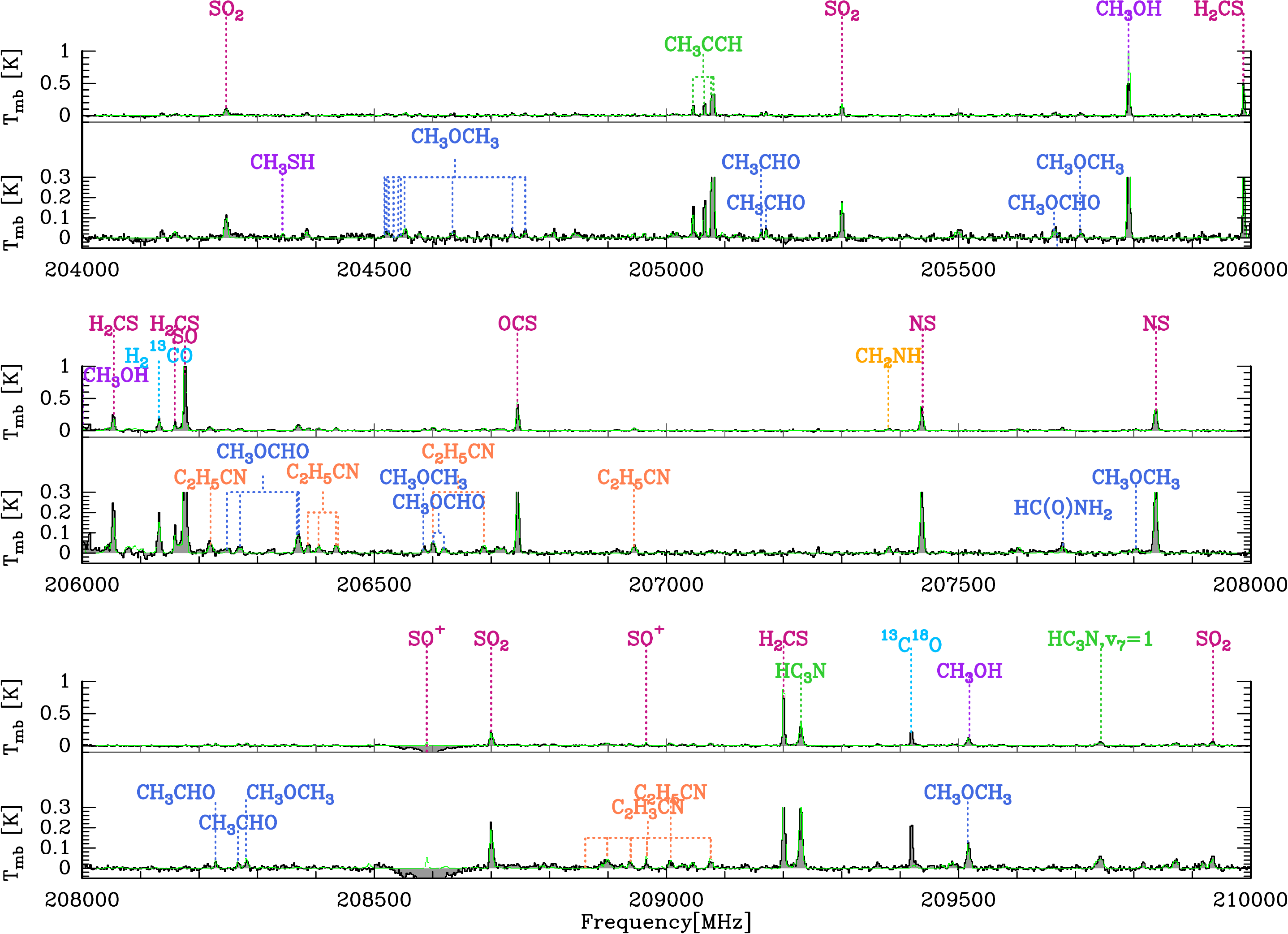}
\caption{Continued.}
\label{SpecSurvey_mol}
\end{sidewaysfigure*}

\begin{sidewaysfigure*}
\ContinuedFloat
\includegraphics[width=0.9\linewidth]{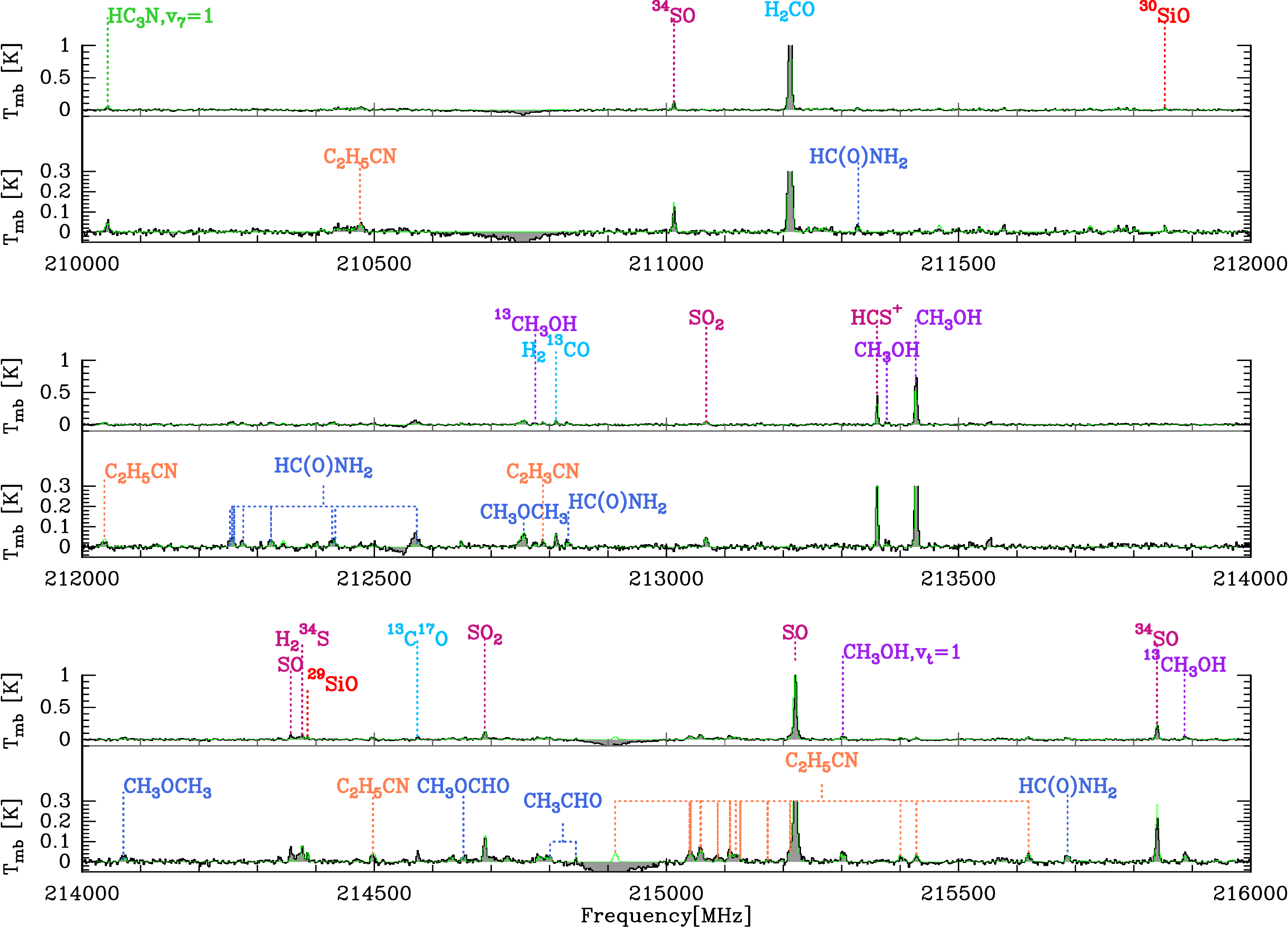}
\caption{Continued.}
\label{SpecSurvey_mol}
\end{sidewaysfigure*}

\begin{sidewaysfigure*}
\ContinuedFloat
\includegraphics[width=0.9\linewidth]{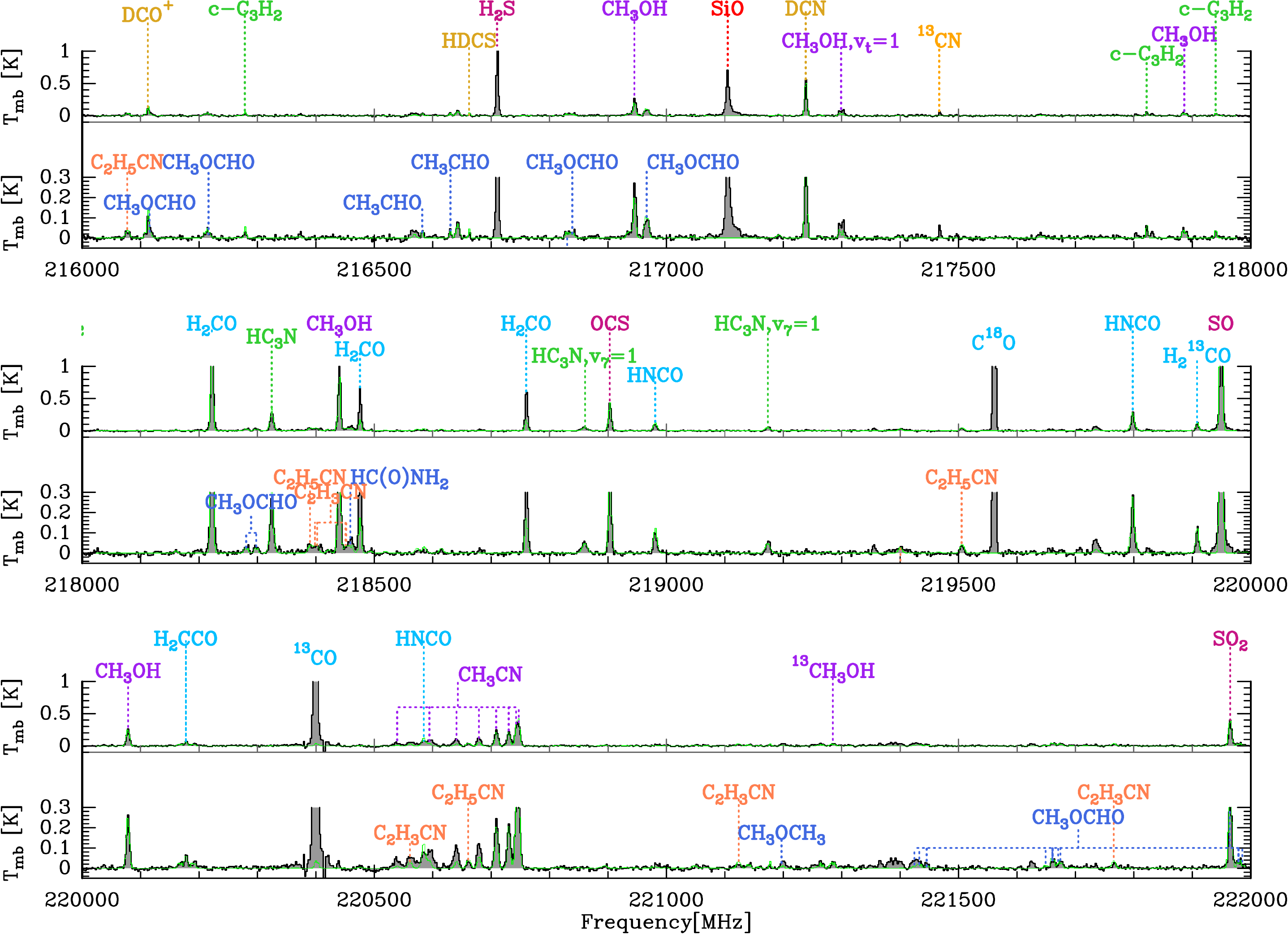}
\caption{Continued.}
\label{SpecSurvey_mol}
\end{sidewaysfigure*}

\begin{sidewaysfigure*}
\ContinuedFloat
\includegraphics[width=0.9\linewidth]{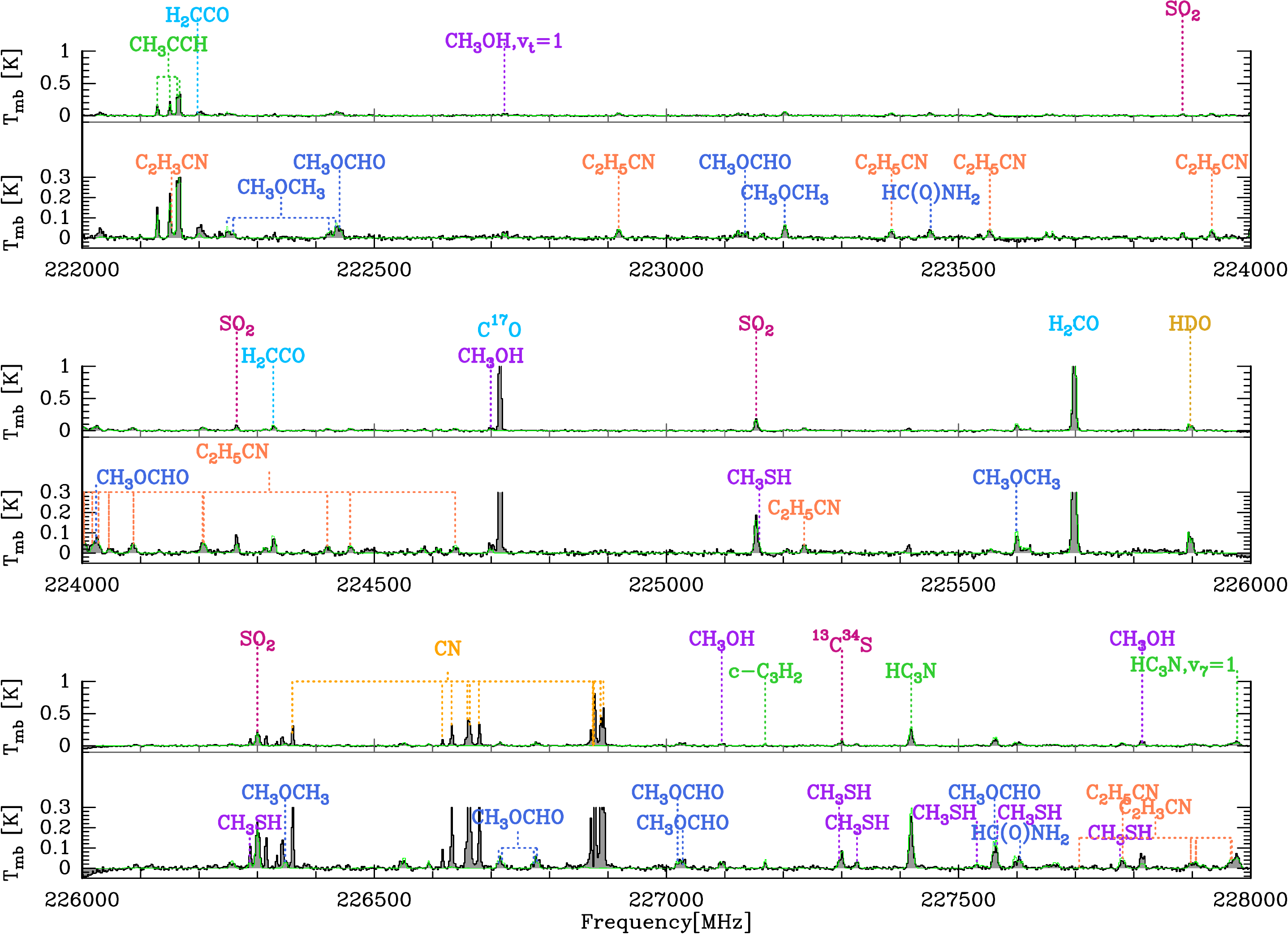}
\caption{Continued.}
\label{SpecSurvey_mol}
\end{sidewaysfigure*}

\begin{sidewaysfigure*}
\ContinuedFloat
\includegraphics[width=0.9\linewidth]{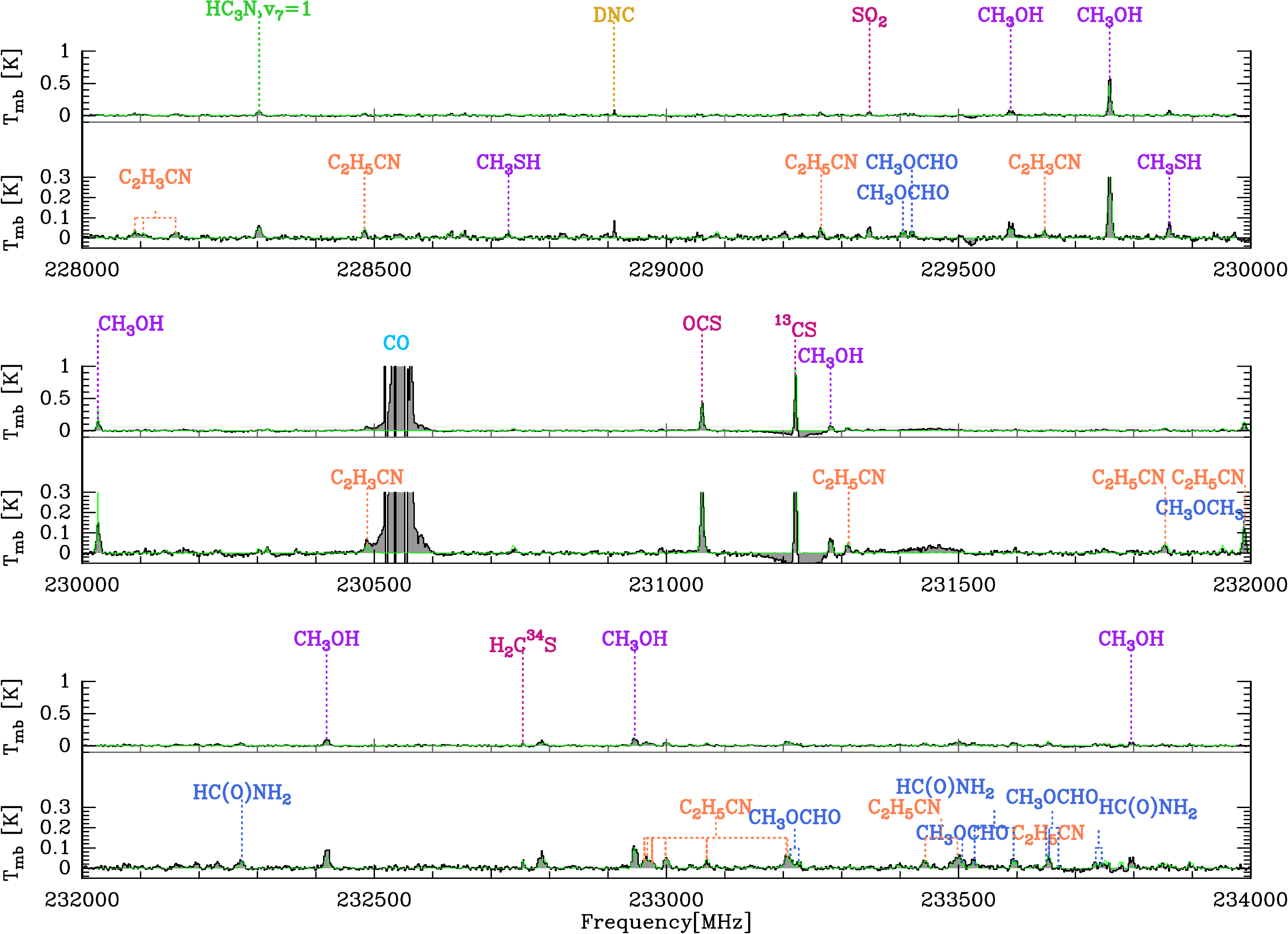}
\caption{Continued.}
\label{SpecSurvey_mol}
\end{sidewaysfigure*}

\begin{sidewaysfigure*}
\ContinuedFloat
\includegraphics[width=0.9\linewidth]{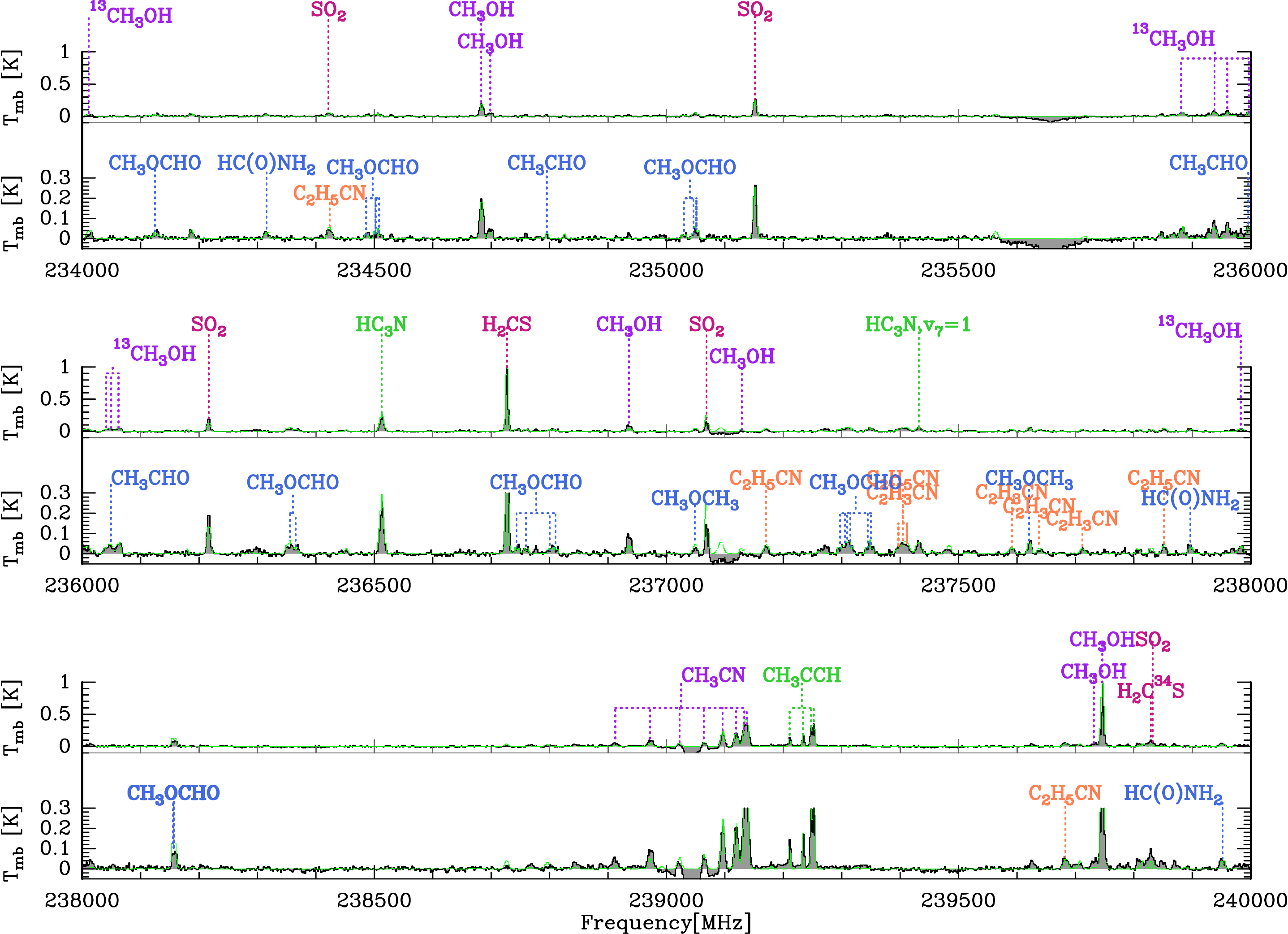}
\caption{Continued.}
\label{SpecSurvey_mol}
\end{sidewaysfigure*}

\begin{sidewaysfigure*}
\ContinuedFloat
\includegraphics[width=0.9\linewidth]{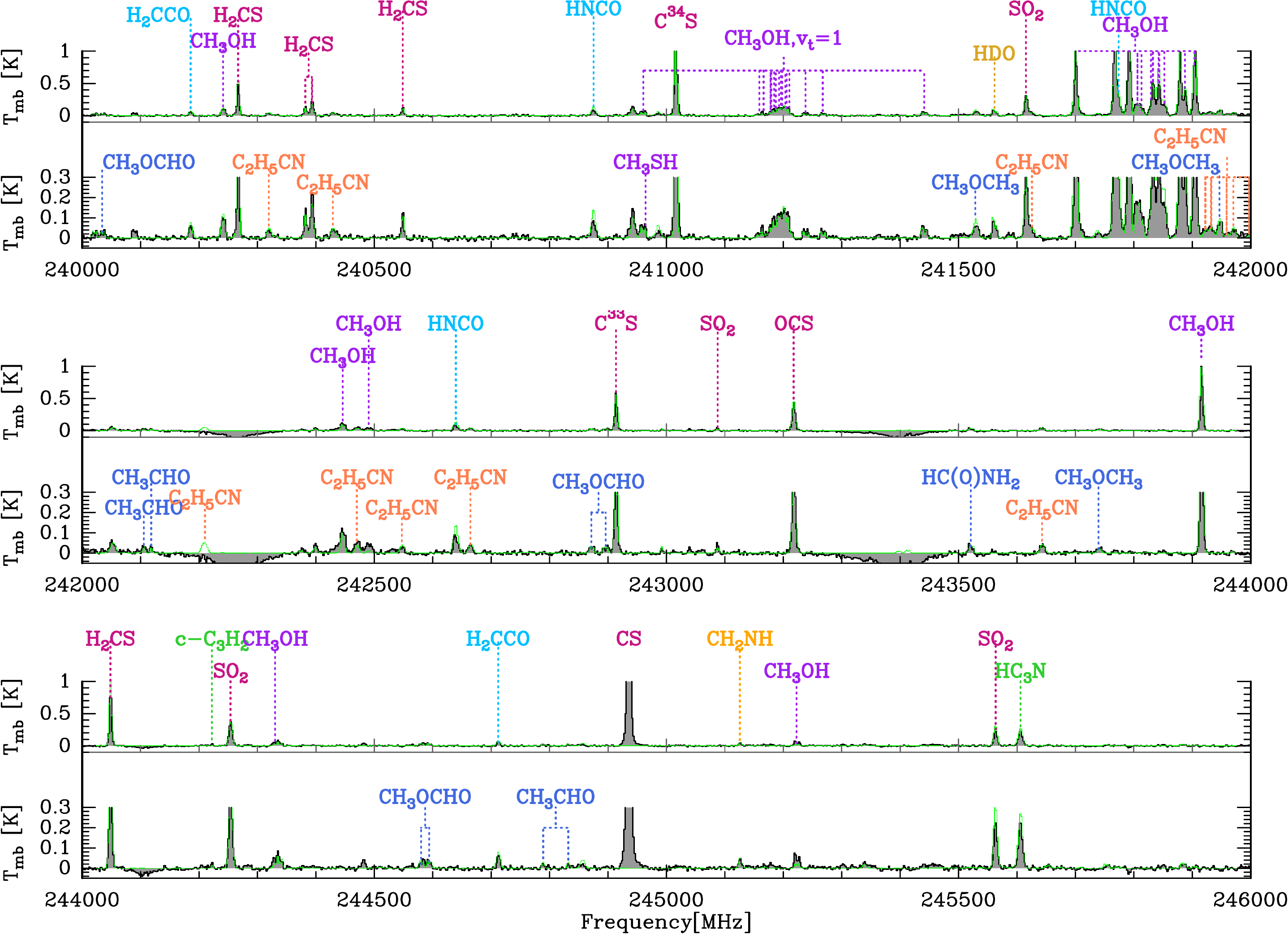}
\caption{Continued.}
\label{SpecSurvey_mol}
\end{sidewaysfigure*}

\begin{sidewaysfigure*}
\ContinuedFloat
\includegraphics[width=0.9\linewidth]{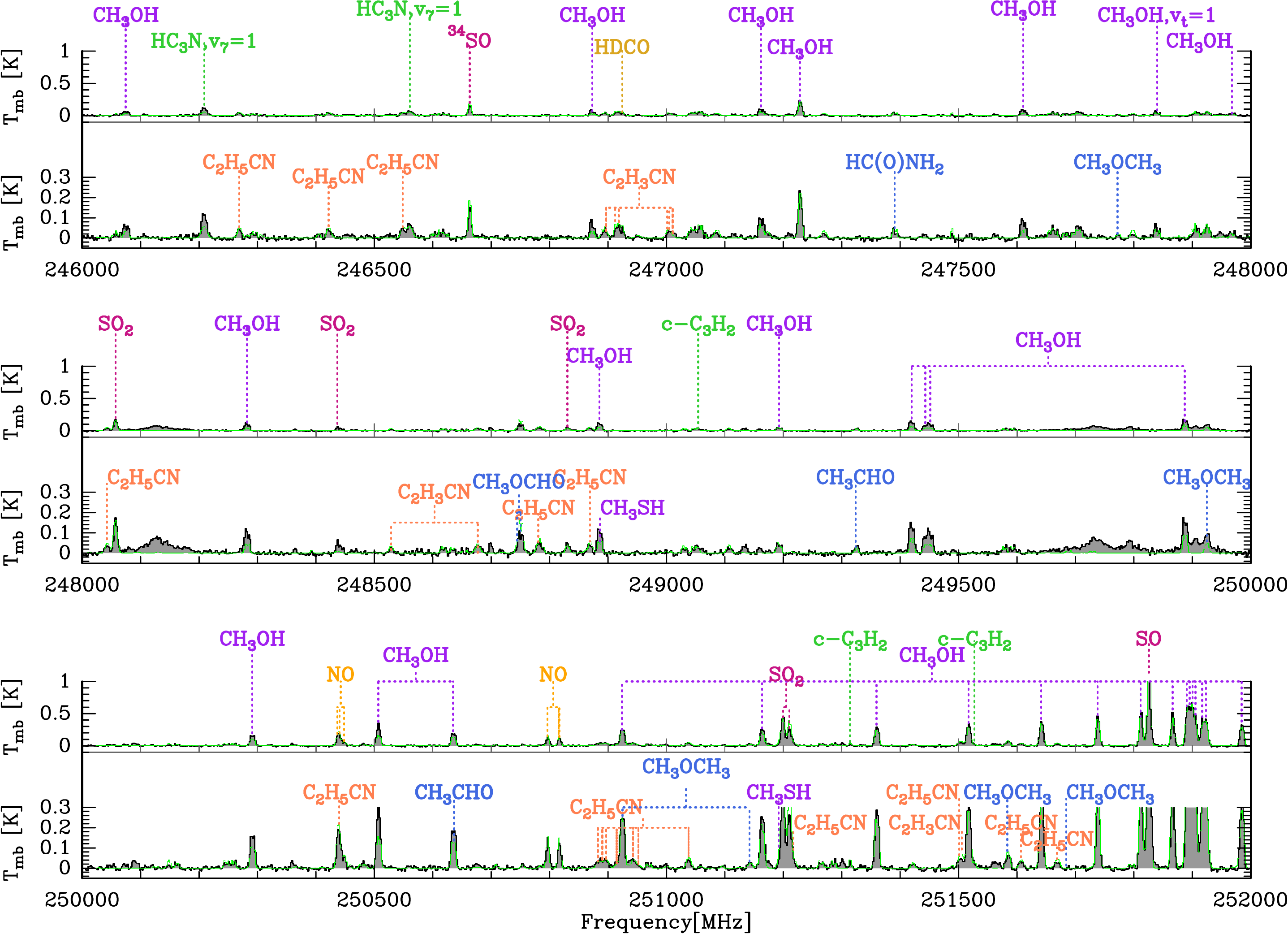}
\caption{Continued.}
\label{SpecSurvey_mol}
\end{sidewaysfigure*}

\begin{sidewaysfigure*}
\ContinuedFloat
\includegraphics[width=0.9\linewidth]{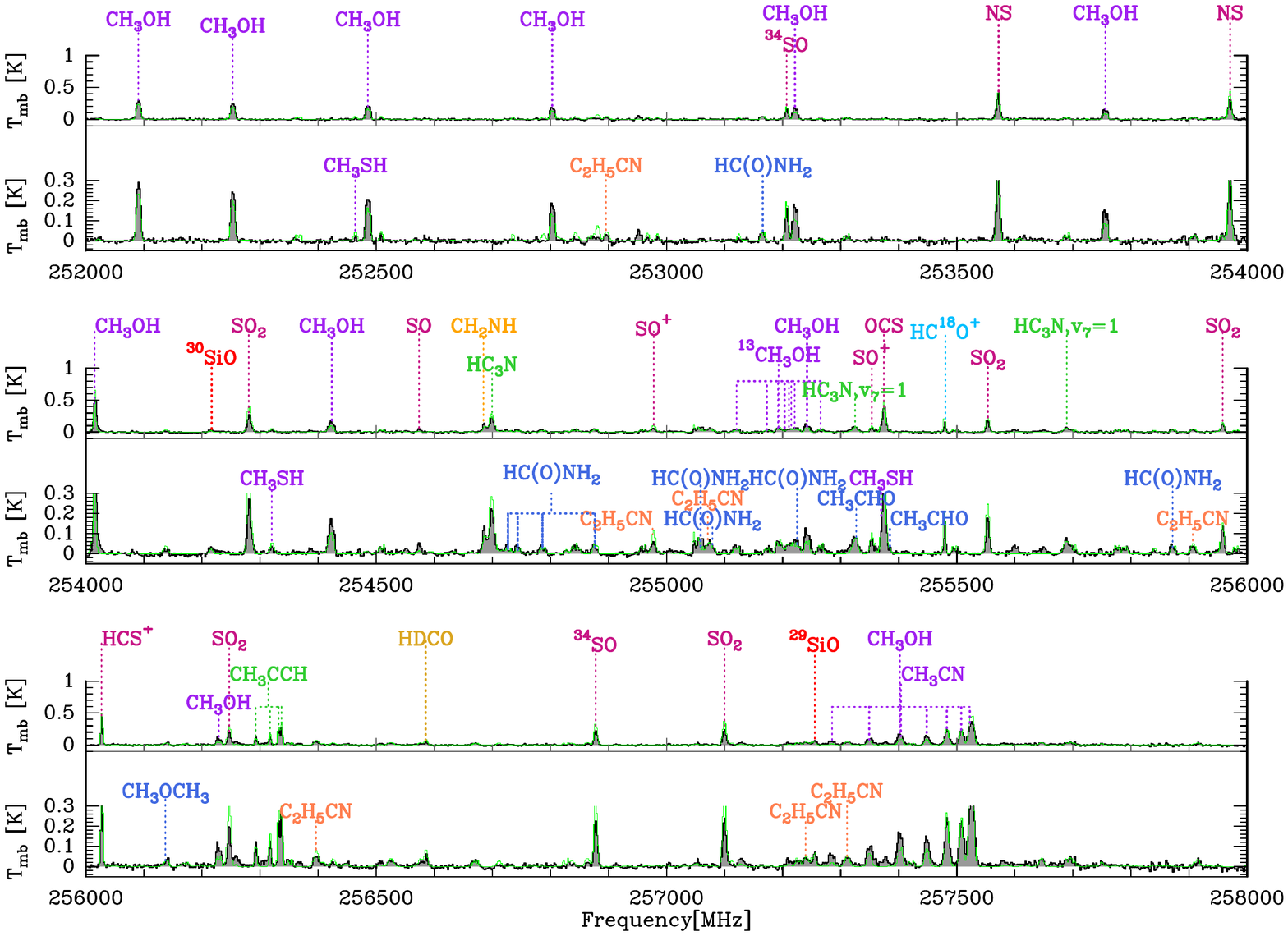}
\caption{Continued.}
\label{SpecSurvey_mol}
\end{sidewaysfigure*}

\begin{sidewaysfigure*}
\ContinuedFloat
\includegraphics[width=0.9\linewidth]{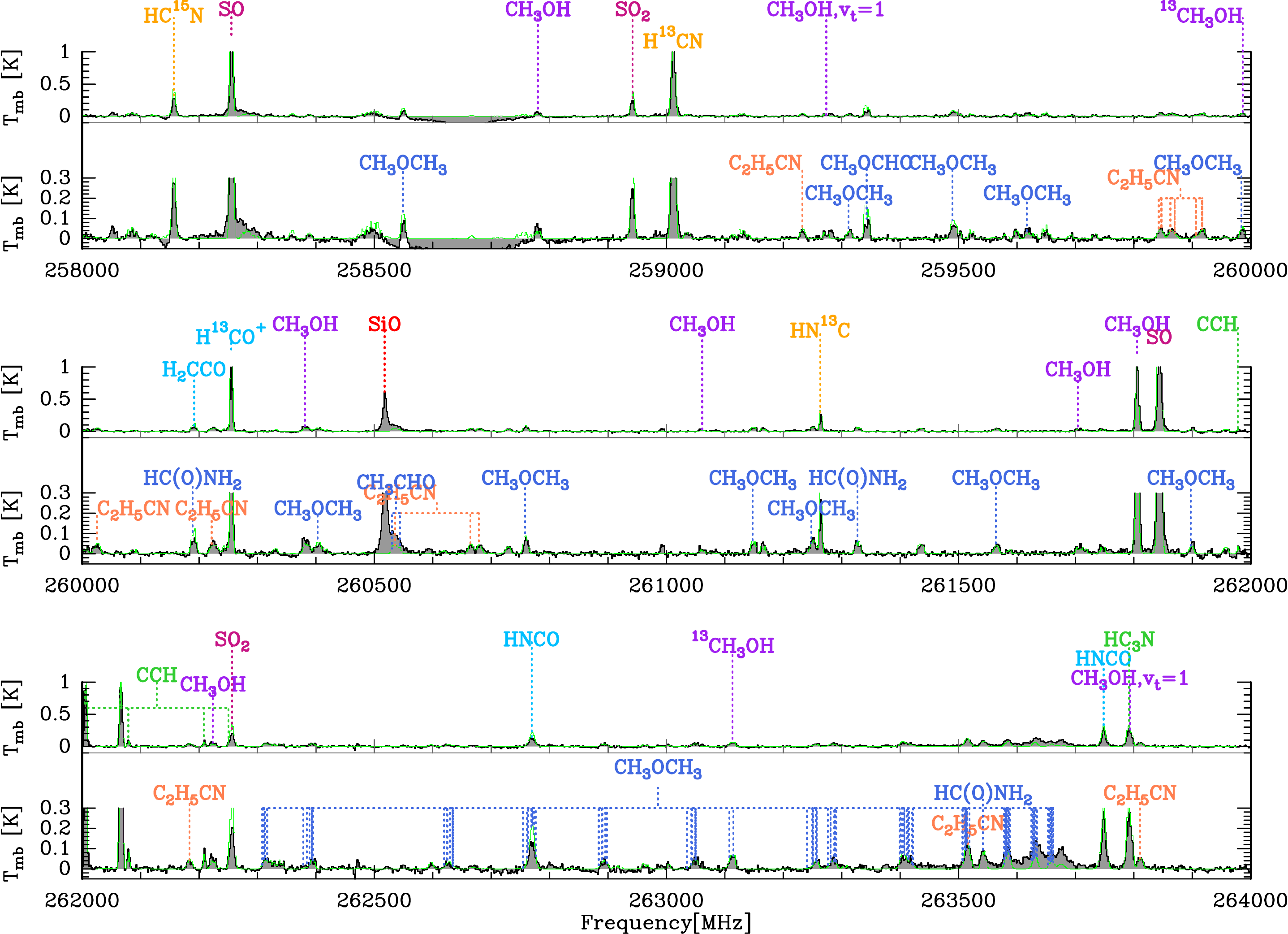}
\caption{Continued.}
\label{SpecSurvey_mol}
\end{sidewaysfigure*}

\begin{sidewaysfigure*}
\ContinuedFloat
\includegraphics[width=0.9\linewidth]{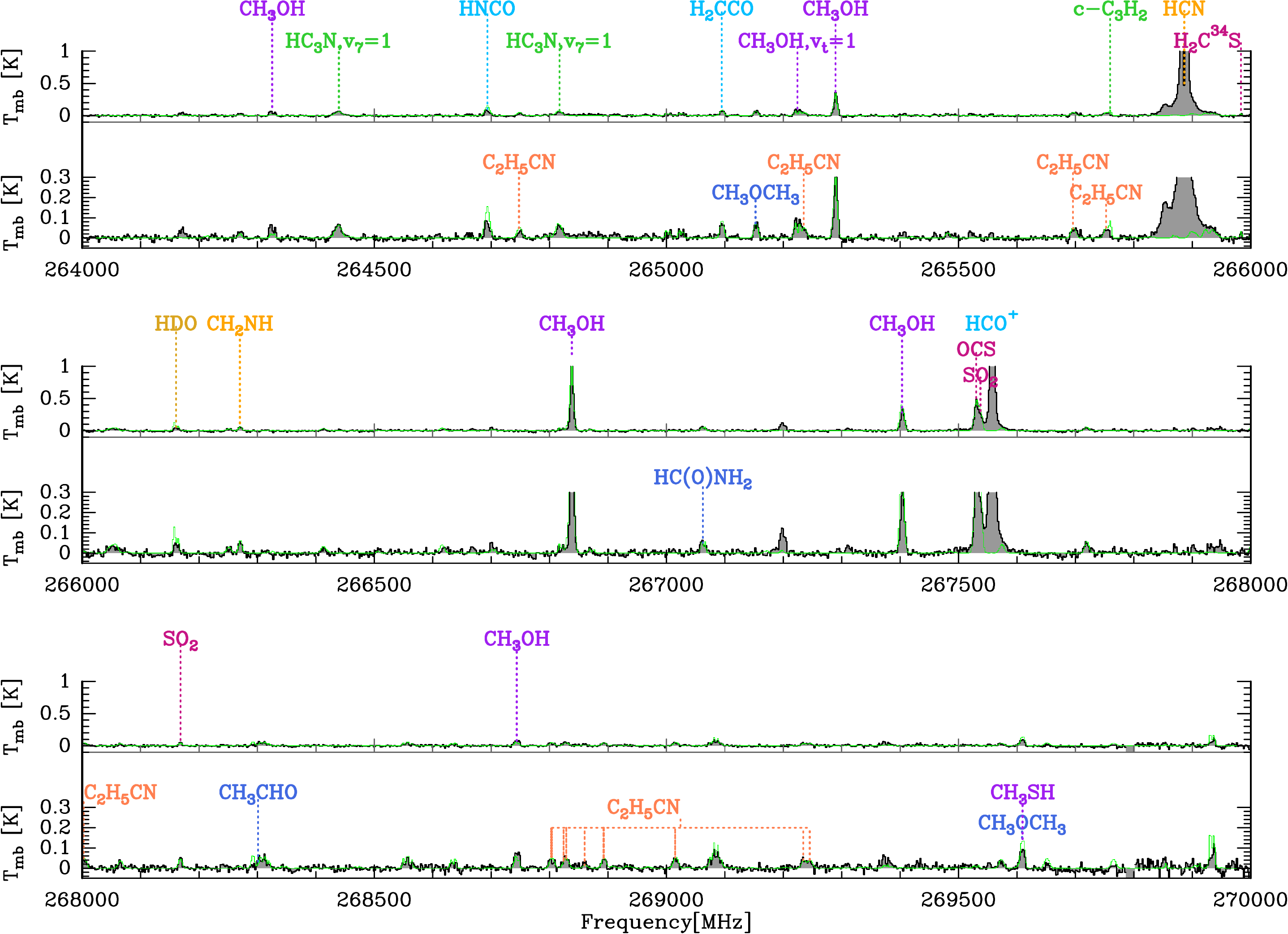}
\caption{Continued.}
\label{SpecSurvey_mol}
\end{sidewaysfigure*}

\begin{sidewaysfigure*}
\ContinuedFloat
\includegraphics[width=0.9\linewidth]{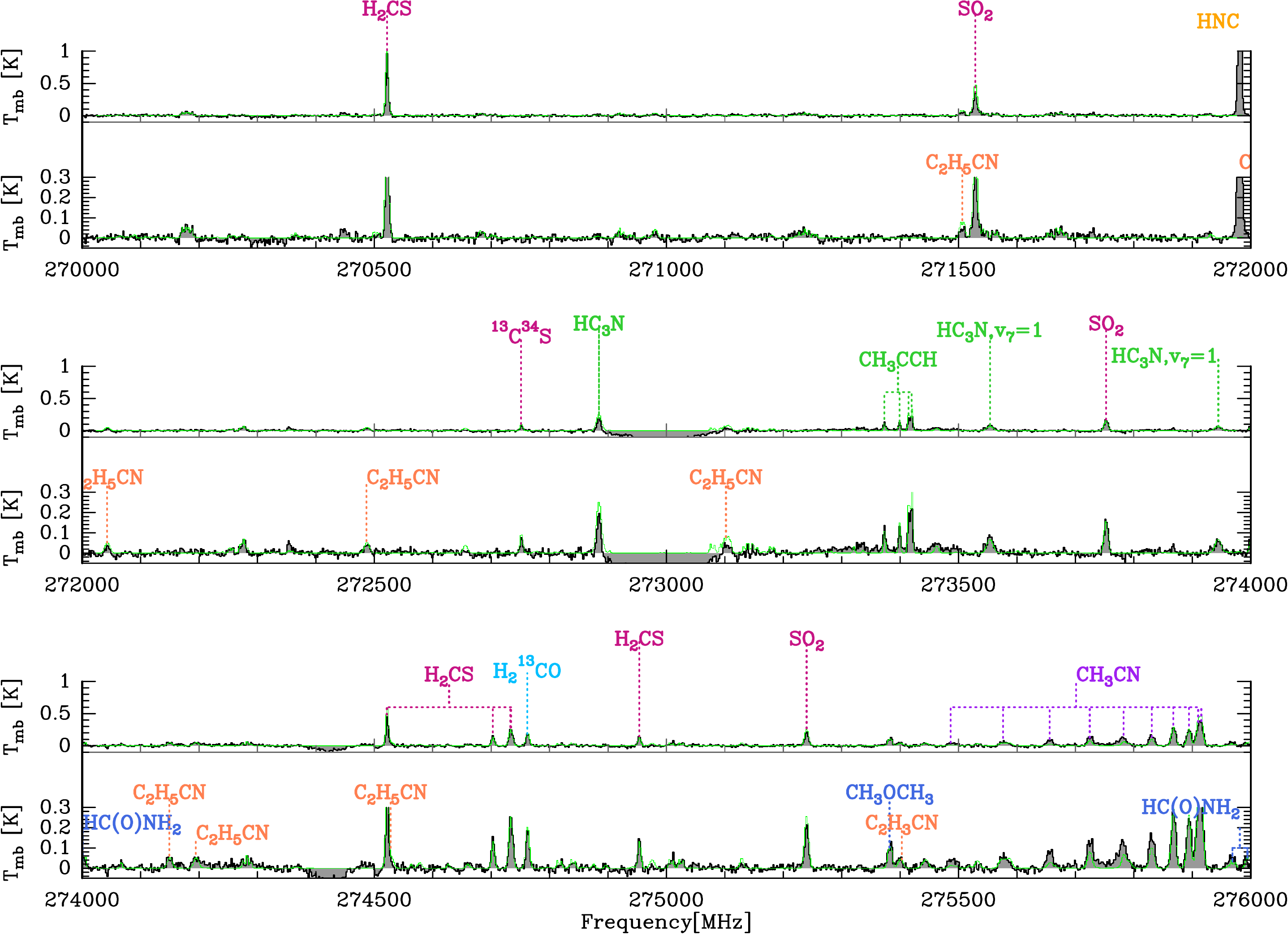}
\caption{Continued.}
\label{SpecSurvey_mol}
\end{sidewaysfigure*}

\begin{sidewaysfigure*}
\ContinuedFloat
\includegraphics[width=0.9\linewidth]{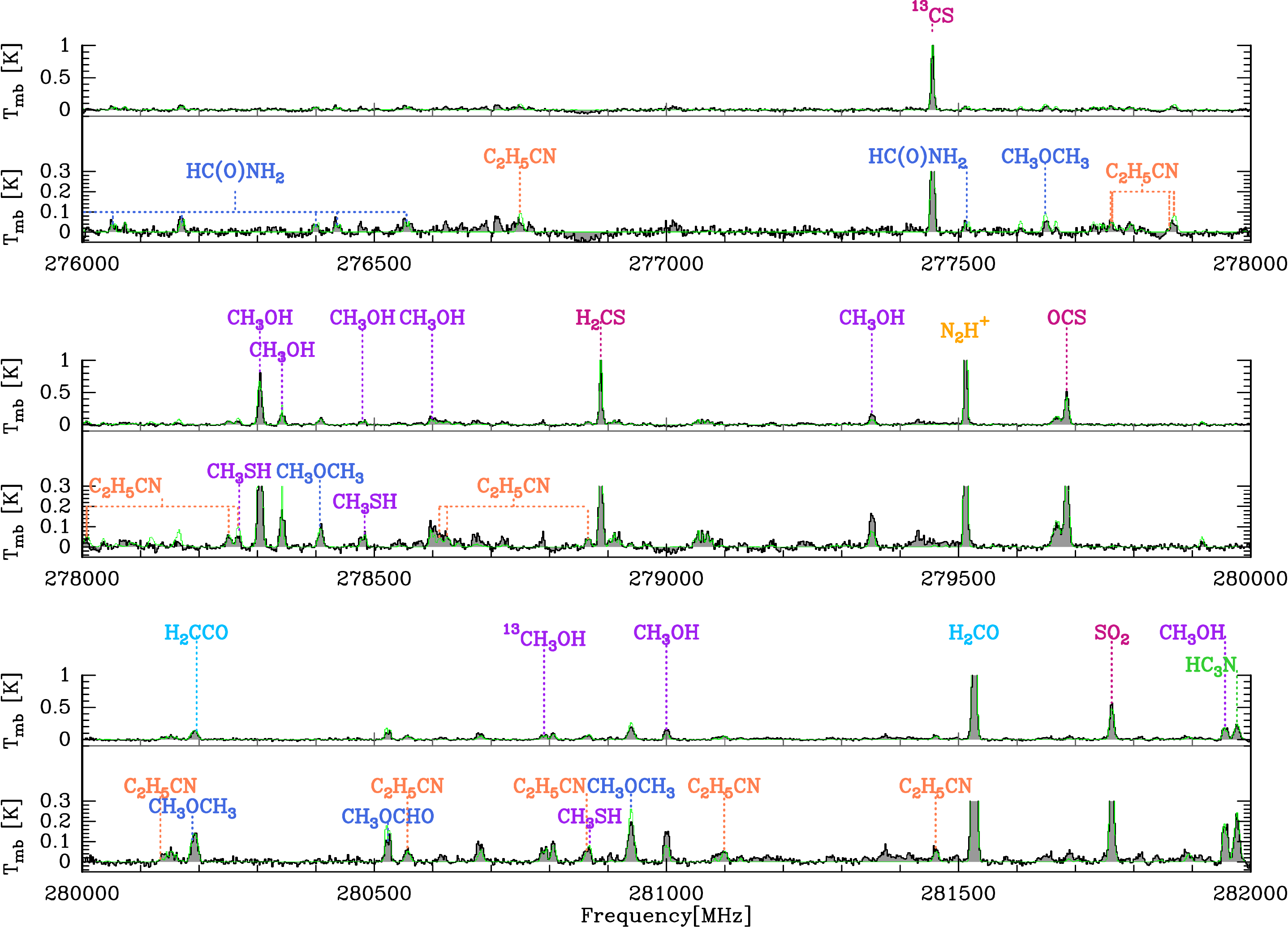}
\caption{Continued.}
\label{SpecSurvey_mol}
\end{sidewaysfigure*}

\begin{sidewaysfigure*}
\ContinuedFloat
\includegraphics[width=0.9\linewidth]{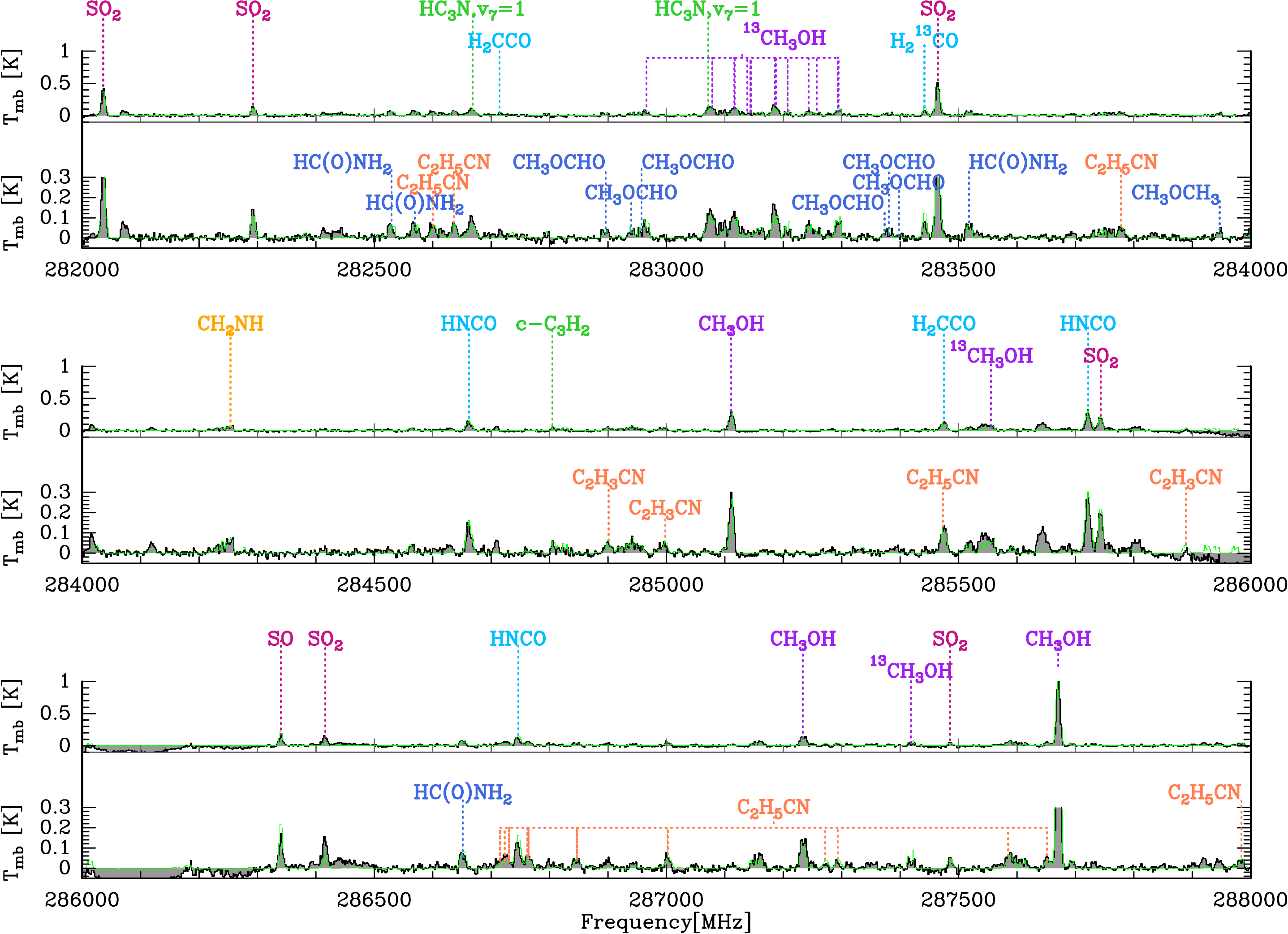}
\caption{Continued.}
\label{SpecSurvey_mol}
\end{sidewaysfigure*}

\begin{sidewaysfigure*}
\ContinuedFloat
\includegraphics[width=0.9\linewidth]{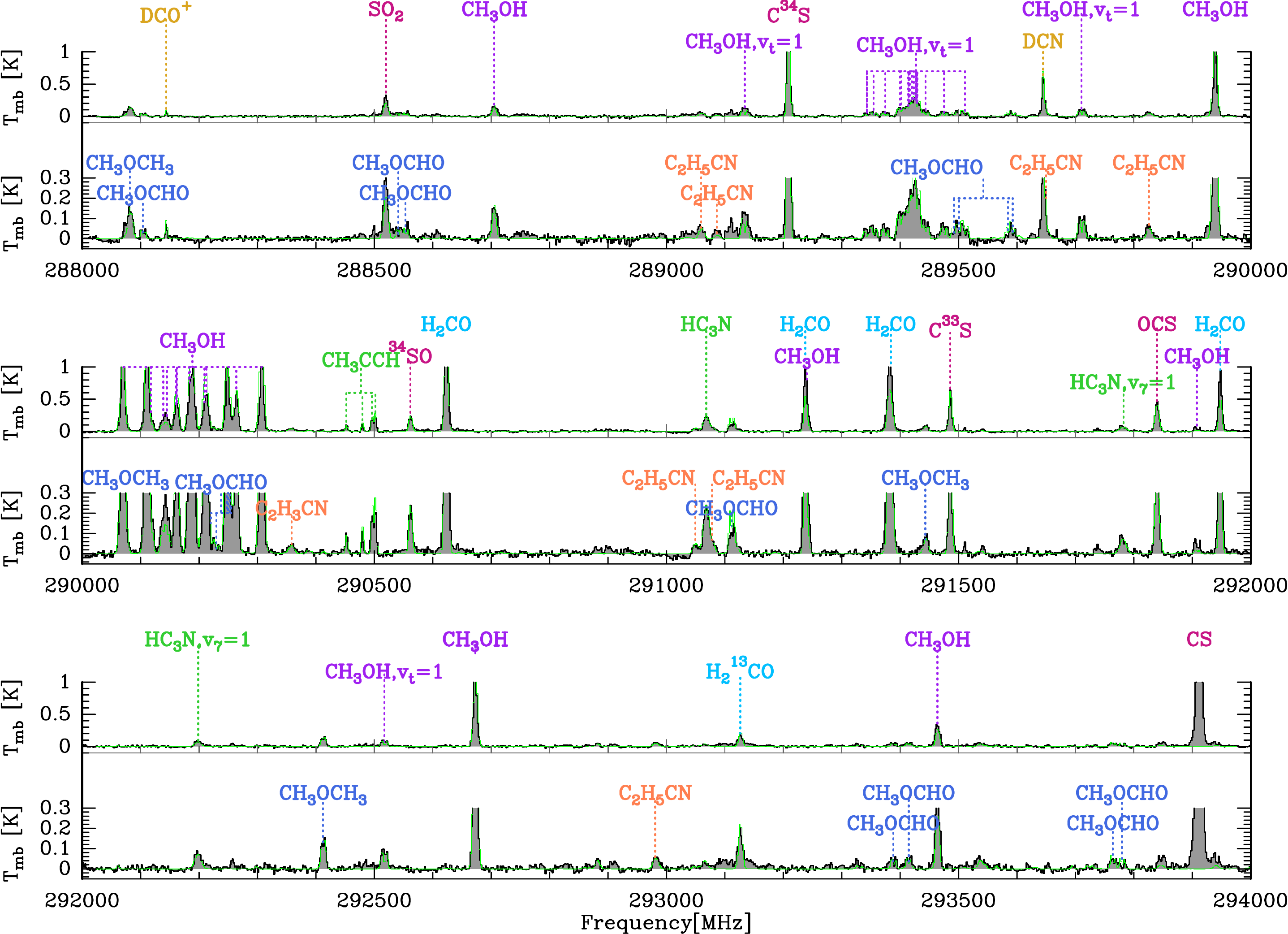}
\caption{Continued.}
\label{SpecSurvey_mol}
\end{sidewaysfigure*}

\begin{sidewaysfigure*}
\ContinuedFloat
\includegraphics[width=0.9\linewidth]{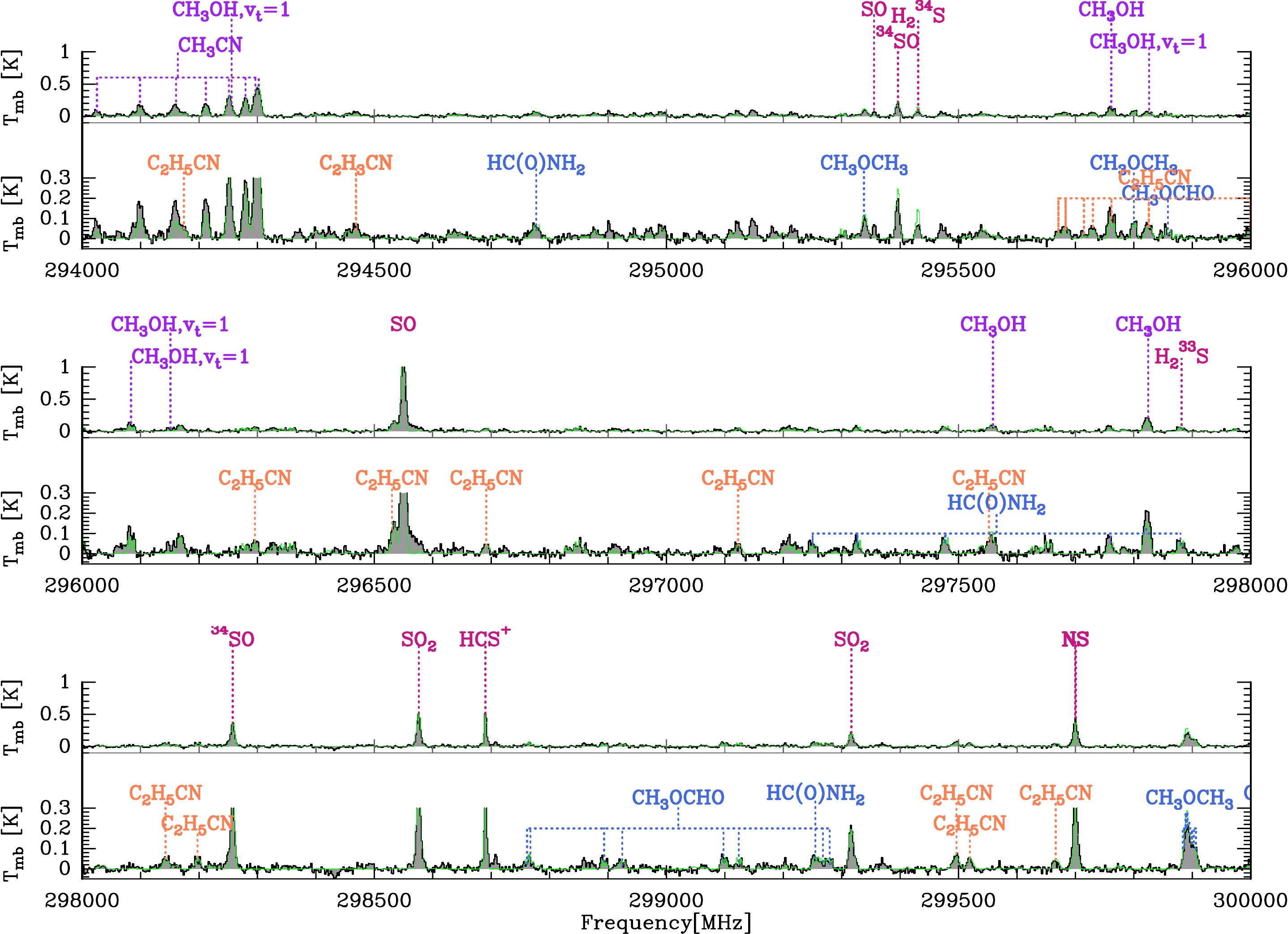}
\caption{Continued.}
\label{SpecSurvey_mol}
\end{sidewaysfigure*}

\begin{sidewaysfigure*}
\ContinuedFloat
\includegraphics[width=0.9\linewidth]{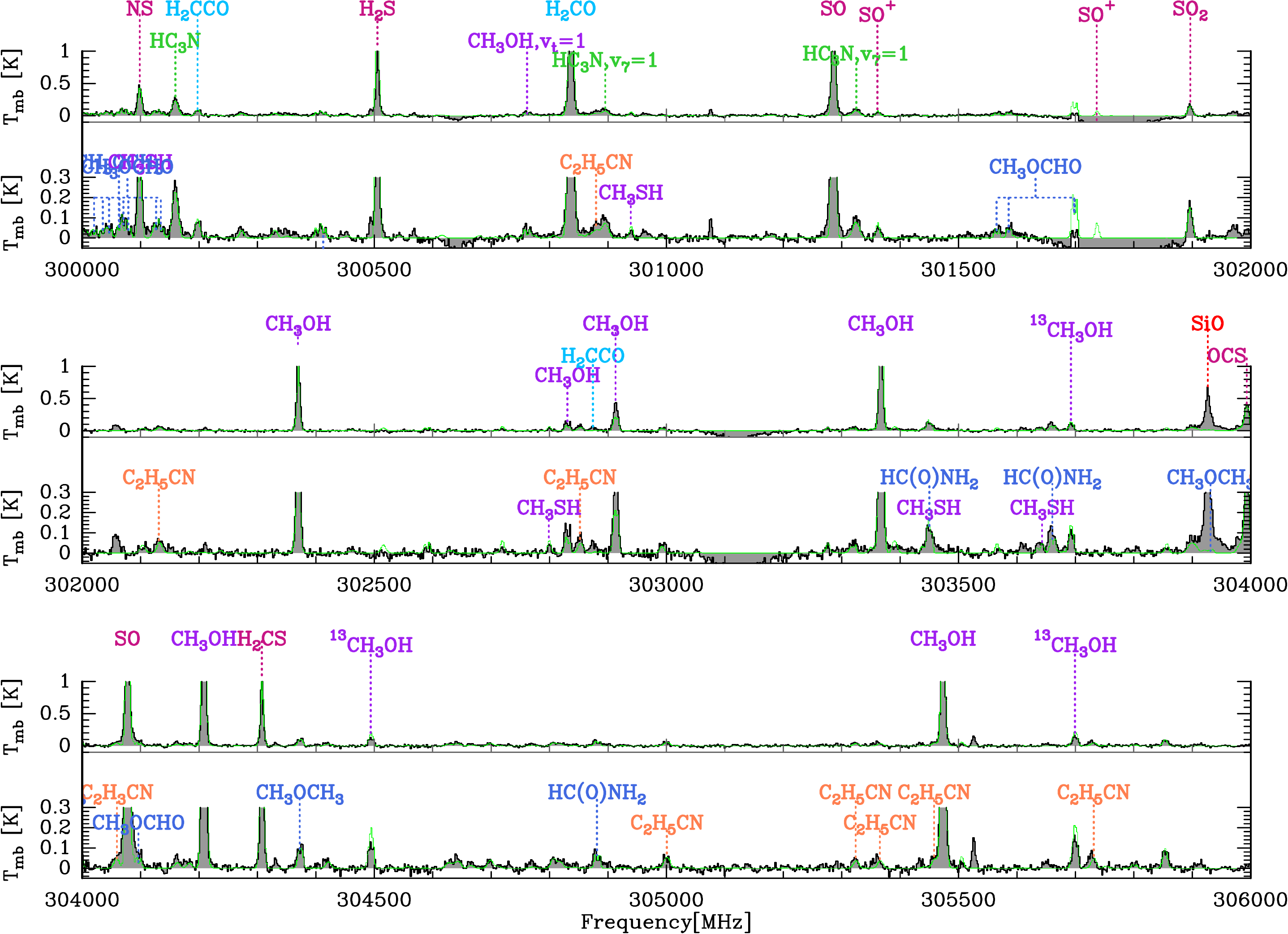}
\caption{Continued.}
\label{SpecSurvey_mol}
\end{sidewaysfigure*}

\begin{sidewaysfigure*}
\ContinuedFloat
\includegraphics[width=0.9\linewidth]{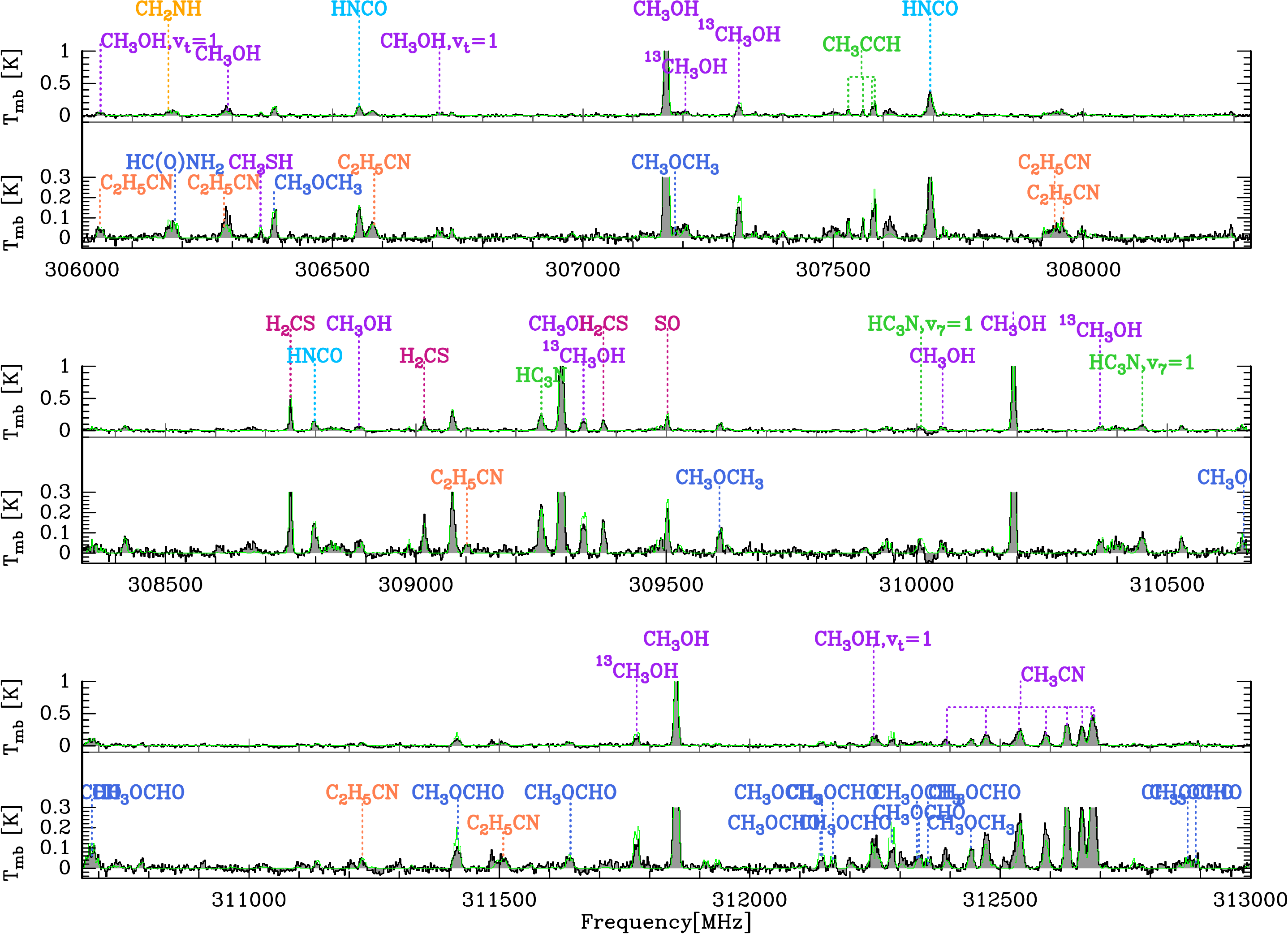}
\caption{Continued.}
\label{SpecSurvey_mol}
\end{sidewaysfigure*}

\begin{sidewaysfigure*}
\ContinuedFloat
\includegraphics[width=0.9\linewidth]{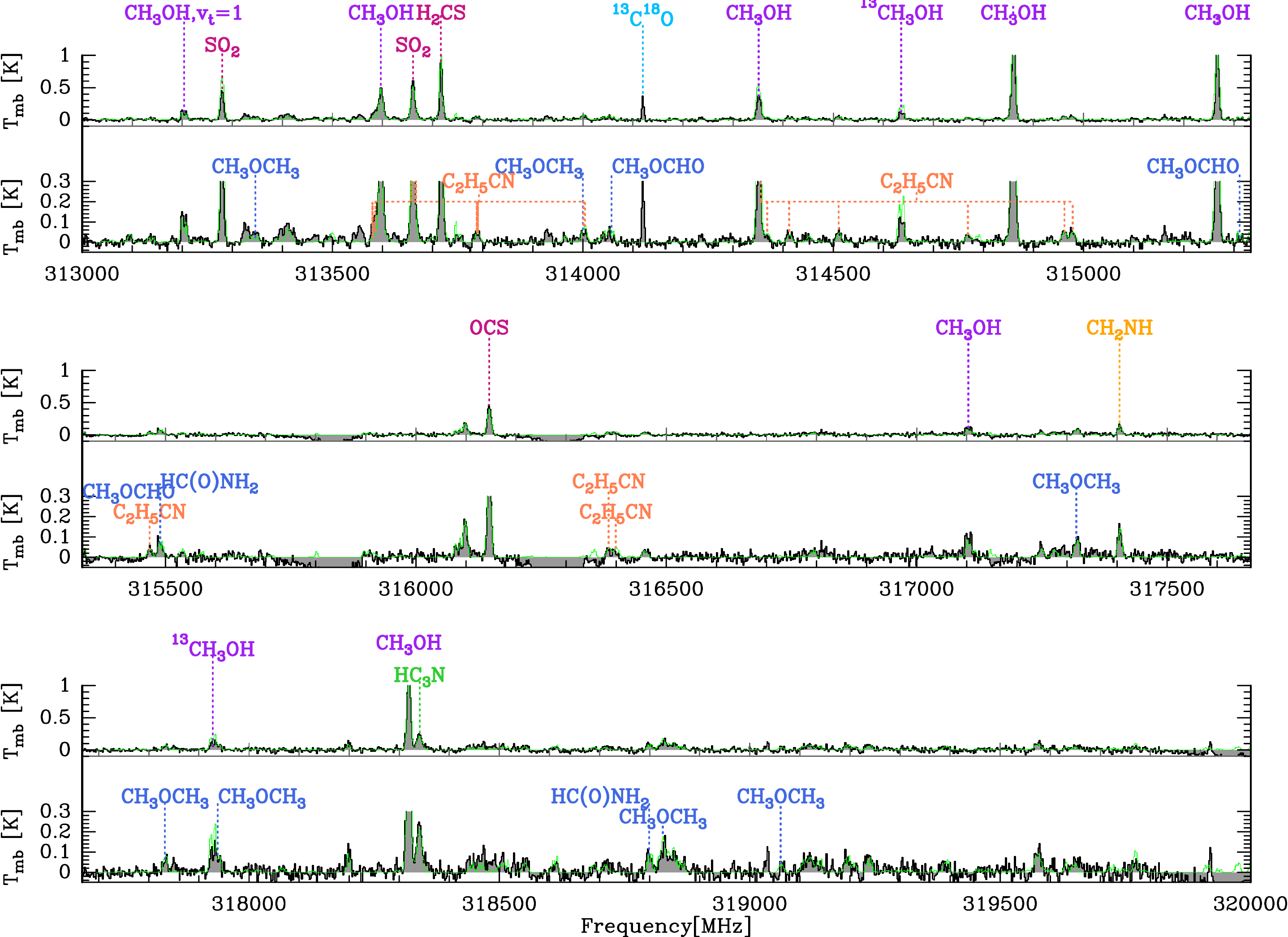}
\caption{Continued.}
\label{SpecSurvey_mol}
\end{sidewaysfigure*}

\begin{sidewaysfigure*}
\ContinuedFloat
\includegraphics[width=0.9\linewidth]{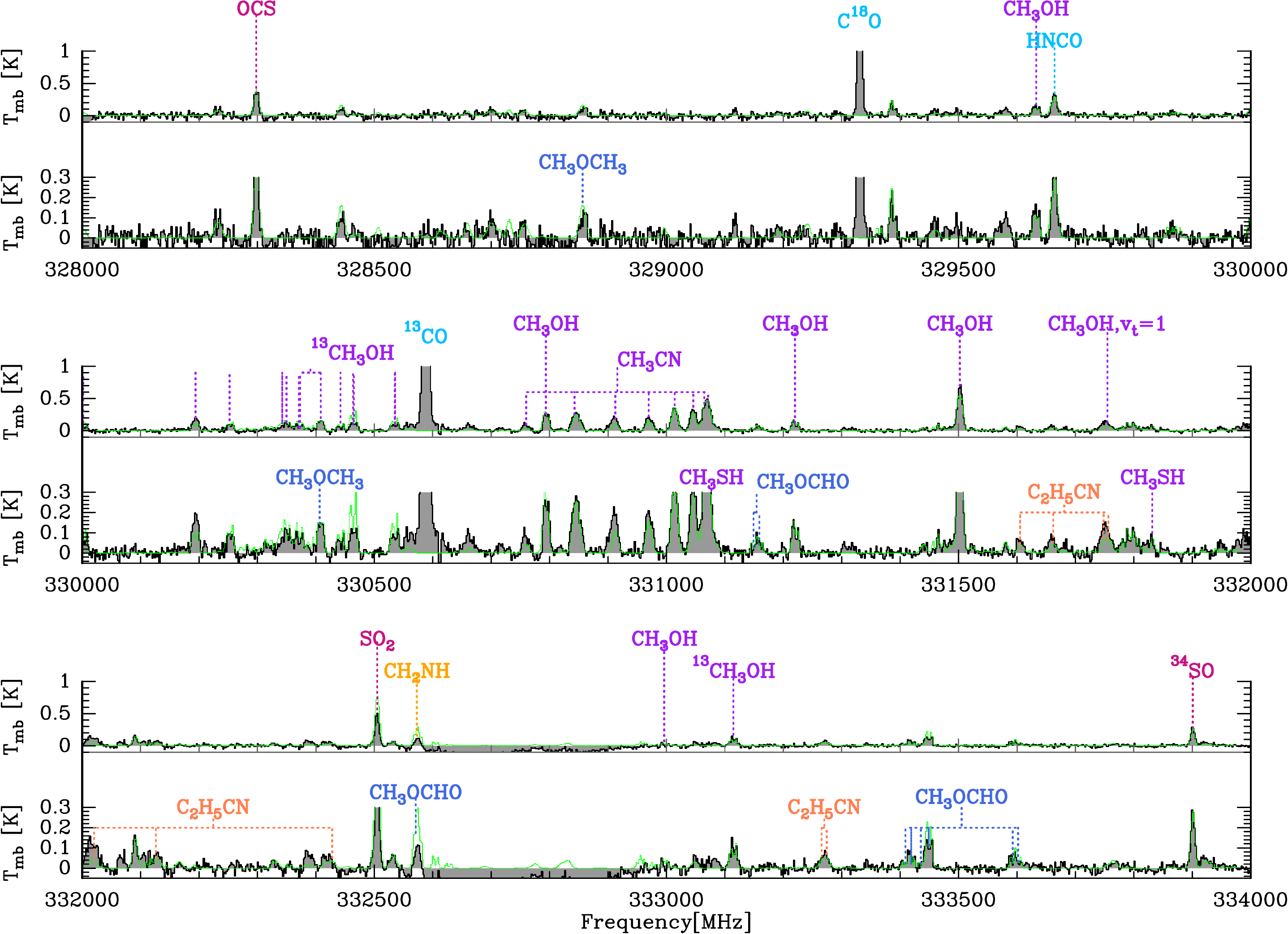}
\caption{Continued.}
\label{SpecSurvey_mol}
\end{sidewaysfigure*}

\begin{sidewaysfigure*}
\includegraphics[width=0.9\linewidth]{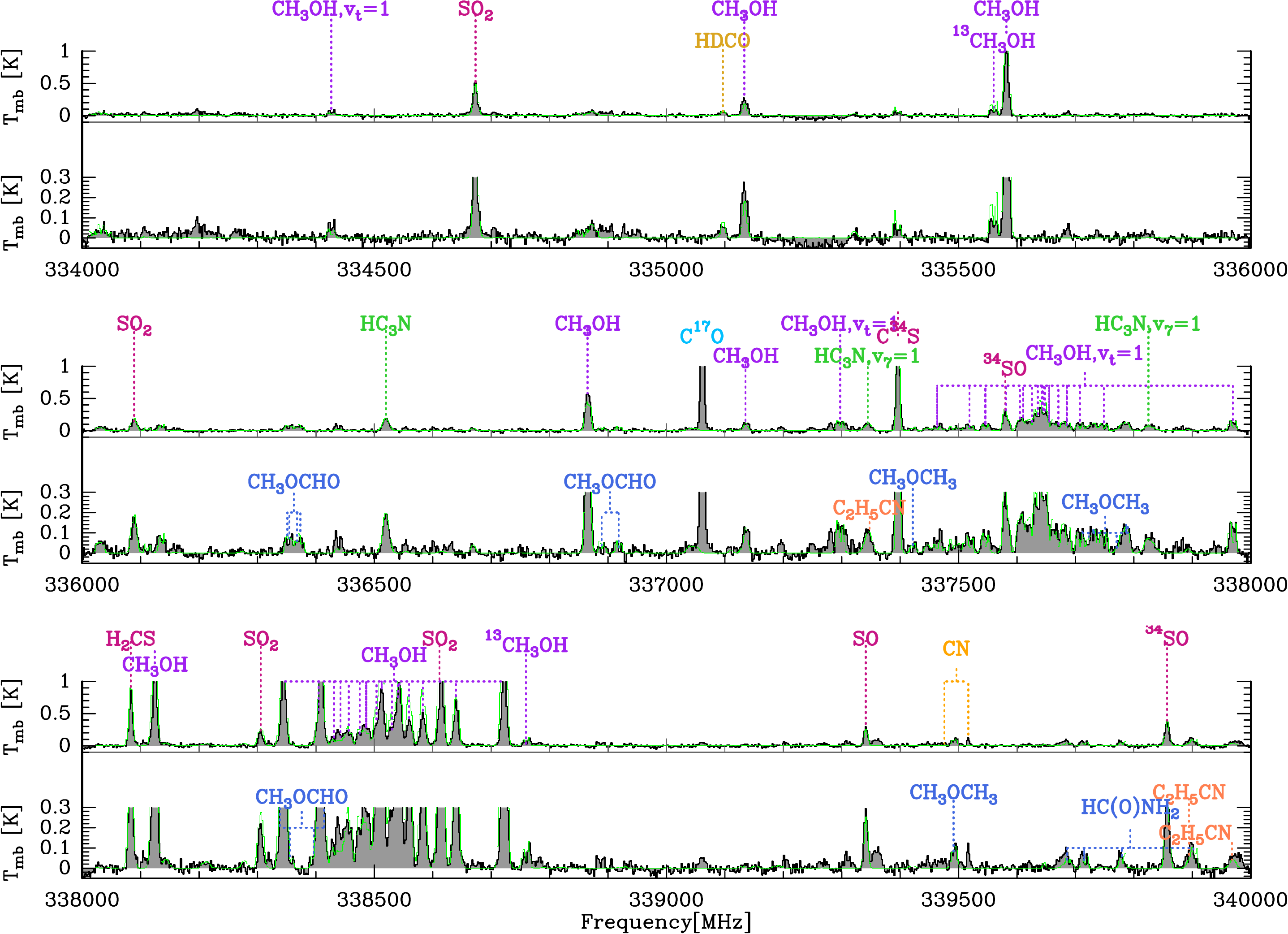}
\caption{Continued.}
\label{SpecSurvey_mol}
\end{sidewaysfigure*}

\begin{sidewaysfigure*}
\ContinuedFloat
\includegraphics[width=0.9\linewidth]{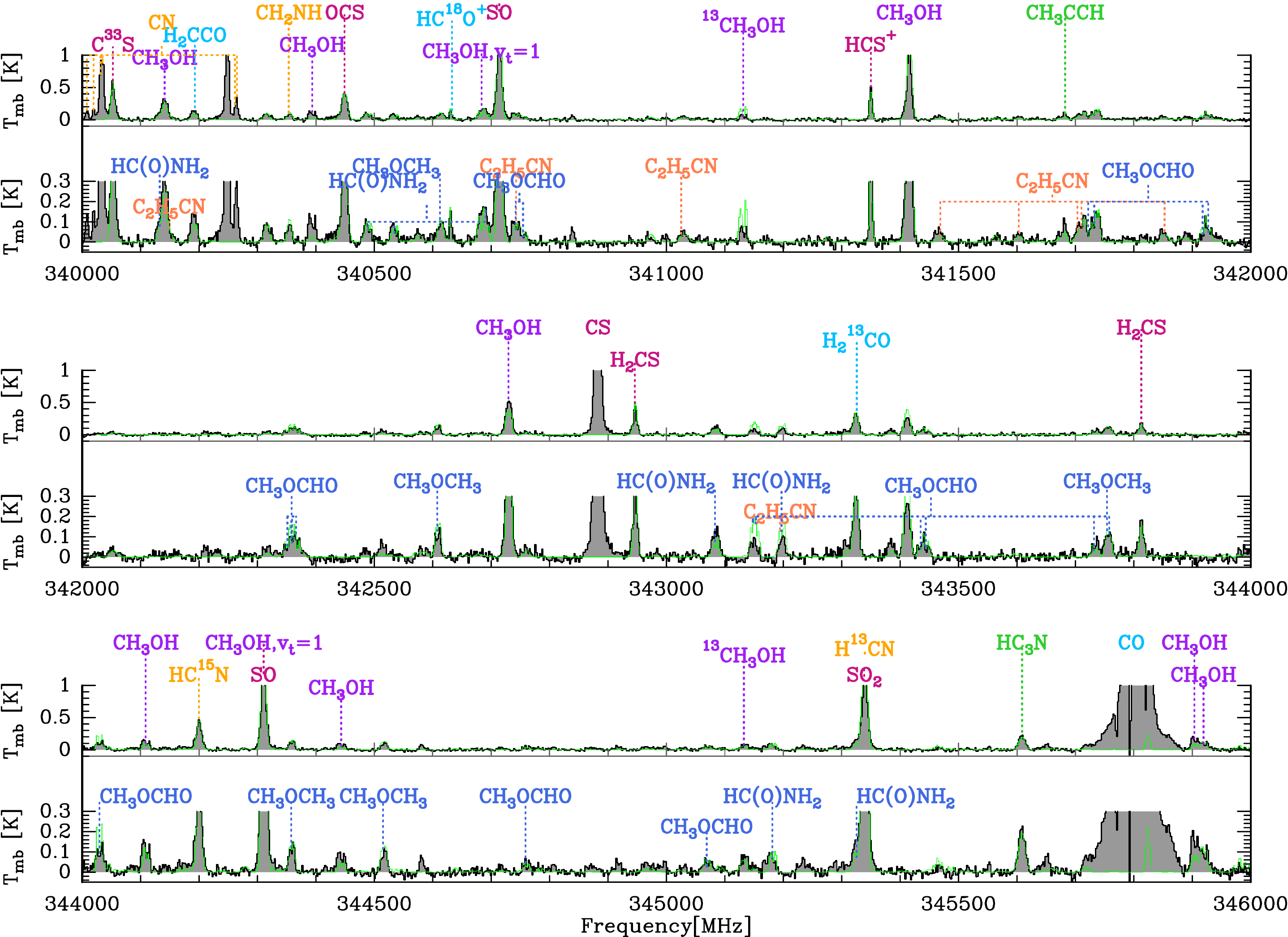}
\caption{Continued.}
\label{SpecSurvey_mol}
\end{sidewaysfigure*}

\begin{sidewaysfigure*}
\ContinuedFloat
\includegraphics[width=0.9\linewidth]{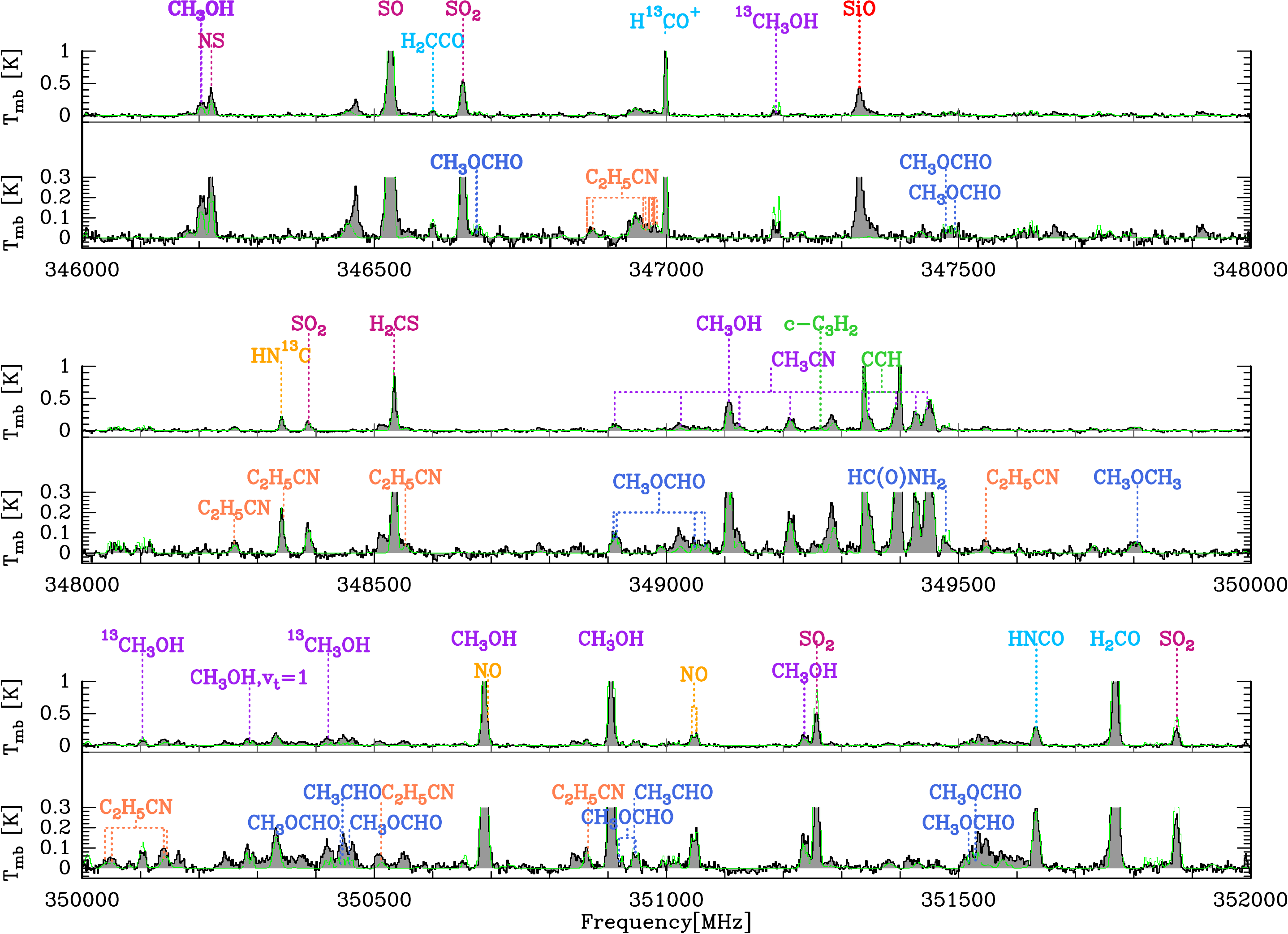}
\caption{Continued.}
\label{SpecSurvey_mol}
\end{sidewaysfigure*}

\begin{sidewaysfigure*}
\ContinuedFloat
\includegraphics[width=0.9\linewidth]{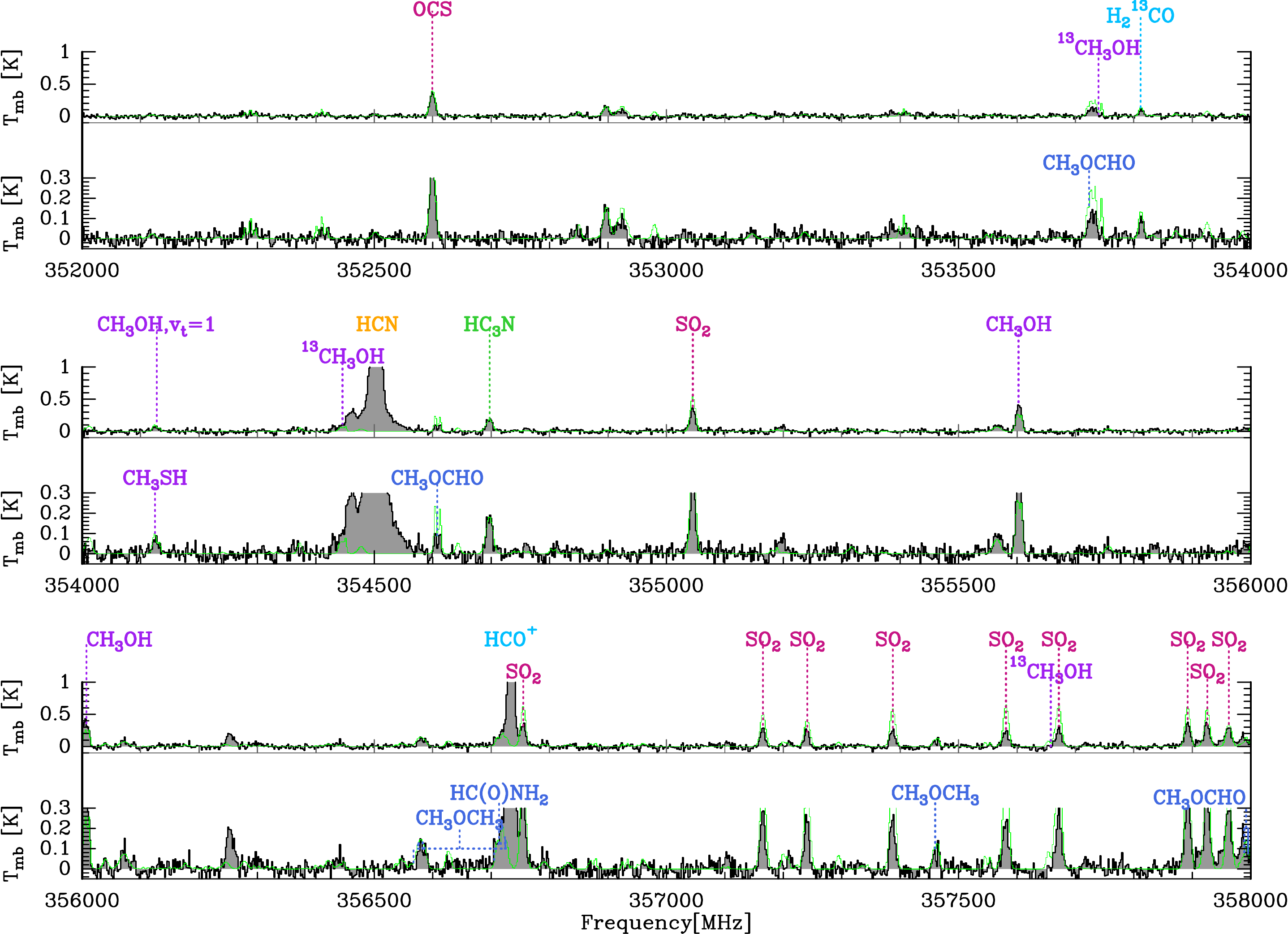}
\caption{Continued.}
\label{SpecSurvey_mol}
\end{sidewaysfigure*}

\begin{sidewaysfigure*}
\ContinuedFloat
\includegraphics[width=0.9\linewidth]{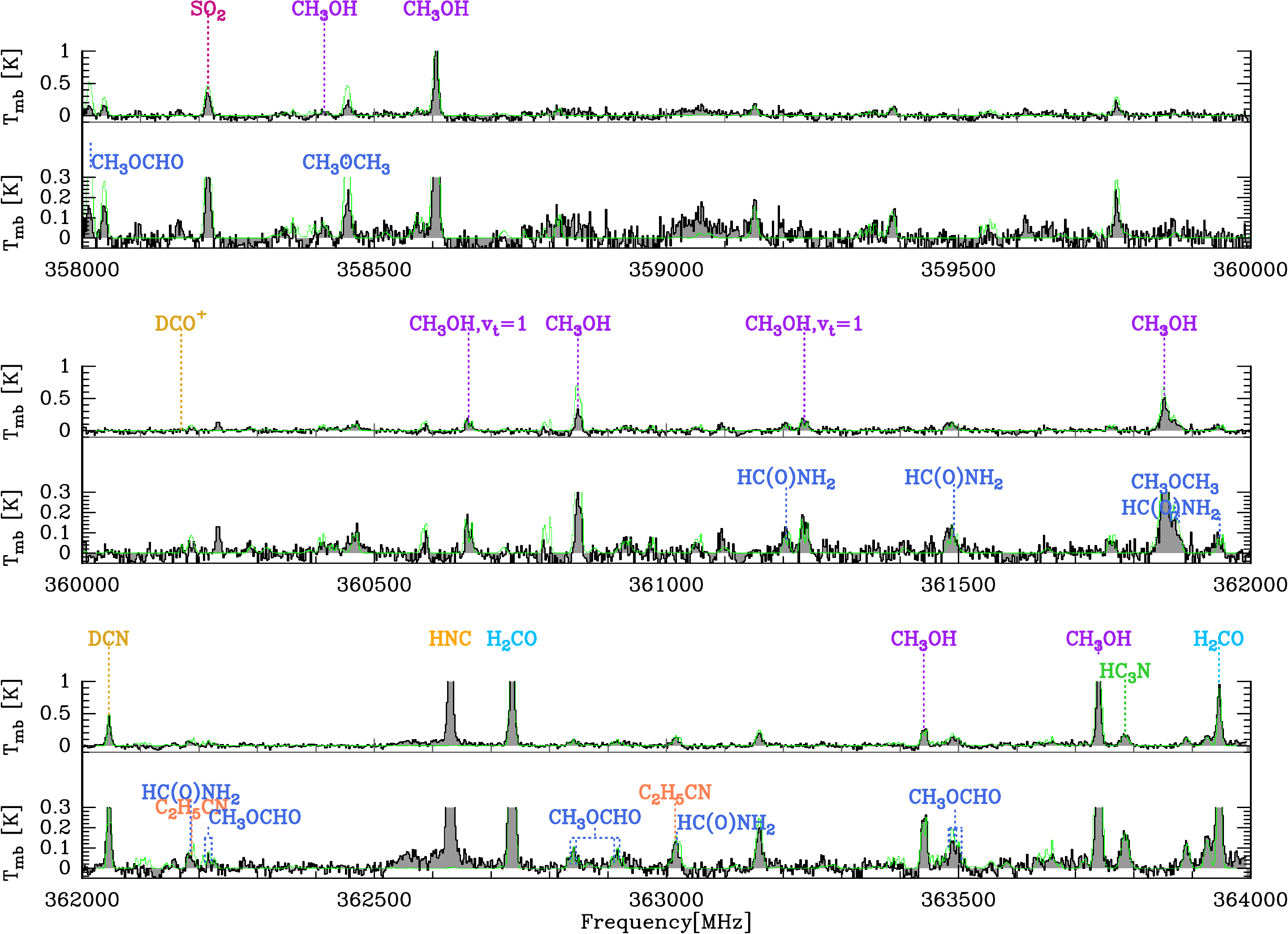}
\caption{Continued.}
\label{SpecSurvey_mol}
\end{sidewaysfigure*}

\begin{sidewaysfigure*}
\ContinuedFloat
\includegraphics[width=0.9\linewidth]{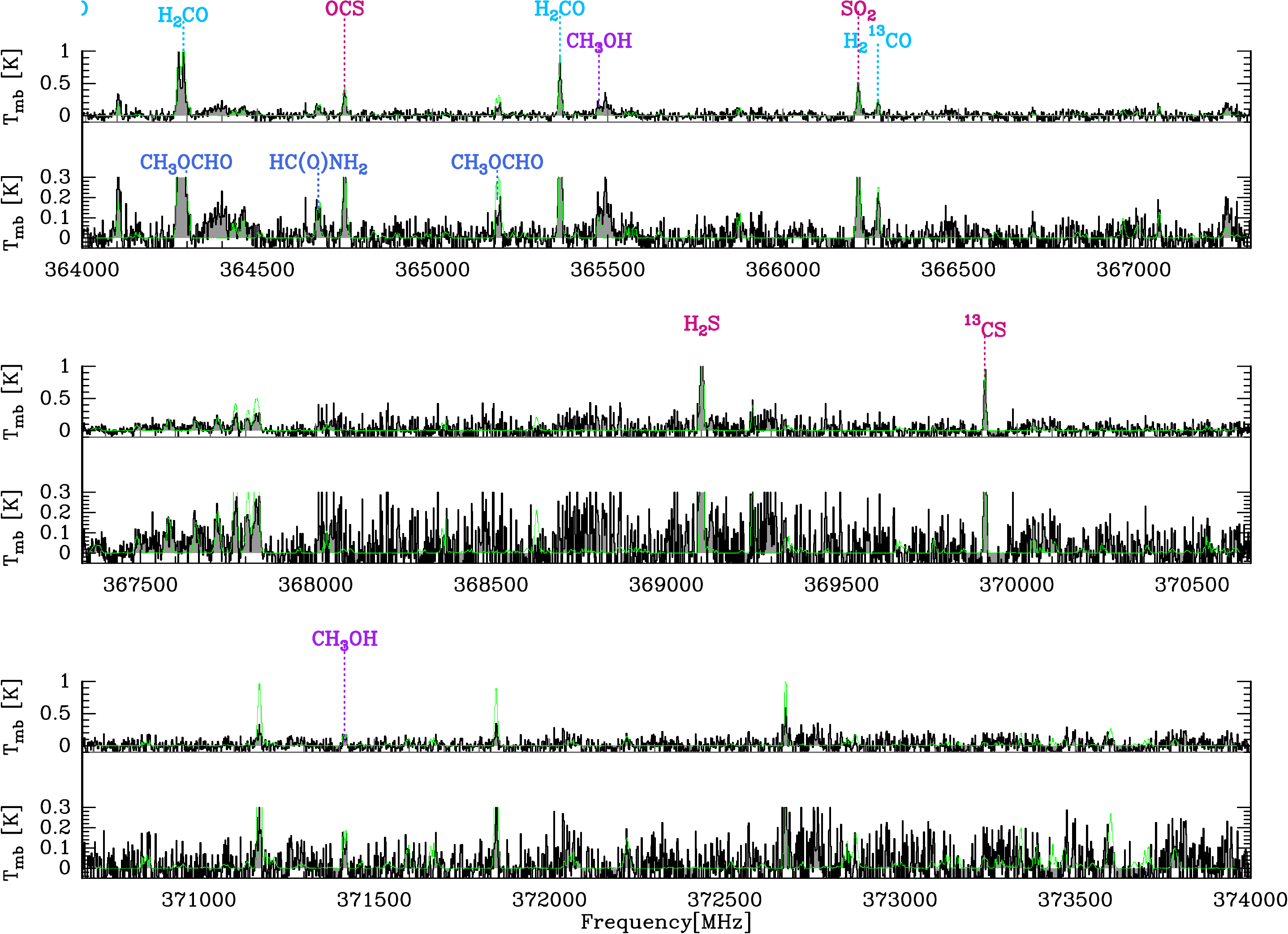}
\caption{Continued.}
\label{SpecSurvey_mol}
\end{sidewaysfigure*}

\end{appendix}

\clearpage
\onecolumn

\end{document}